\documentclass[a4paper,11pt]{book}

\usepackage{lipsum}

\pdfoutput=1 
\usepackage[margin=1.0in]{geometry}
\usepackage[utf8]{inputenc}
\usepackage[titletoc]{appendix}
\usepackage{indentfirst}
\usepackage{amsfonts}
\usepackage{amsmath}
\usepackage{amssymb}
\usepackage{amsthm}
\usepackage{dsfont}
\usepackage{graphicx}
\usepackage{enumerate}
\usepackage{hyperref}
\usepackage{amscd}
\usepackage{nccmath}
\usepackage{cancel}
\usepackage{fancyhdr}
\usepackage[english]{babel}
\usepackage{tikz}
\usepackage{tabls}
\usepackage{import}
\usepackage{csquotes}
\usepackage{mathtools}
\usepackage{subfig}
\usepackage{tensor}
\usepackage{float}
\usepackage{IEEEtrantools}

\usepackage[numbers,sort&compress]{natbib}
\usepackage{chngcntr}
\counterwithin{equation}{chapter}

\usepackage{emptypage}

\allowdisplaybreaks

\usepackage{color}
    \definecolor{darkgreen}{rgb}{0,0.7,0}
    \definecolor{darkblue}{rgb}{0,0,0.6}
    \definecolor{purple}{rgb}{0.4,.2,0.7}

\newcommand{\white}[1]{\textcolor{white}{#1}}

\newcommand{\ssc}{\scriptscriptstyle}

\hypersetup{
colorlinks = true,
linkcolor = darkblue,
urlcolor = darkblue,
citecolor = blue,
}

\newcommand{\df}{\mathrm{d}}
\DeclareMathOperator{\tr}{Tr}

\DeclareMathOperator{\csch}{csch}
\DeclarePairedDelimiter{\ceil}{\lceil}{\rceil}
\def\ie{{\it i.e.} }
\def\eg{{\it e.g.} }

\newcommand{\bF}[1]{\bar{\mathcal{F}}_{#1}}
\newcommand{\lpeff}{\ell_{ \ssc \rm P}^{\text{eff}}}
\newcommand{\lnr}{^{\mathrm{L}}}
\newcommand{\RR}{\mathcal{R}_2}
\newcommand{\RRR}{\mathcal{R}_3}
\newcommand{\Ls}{L_\star}
\newcommand{\lp}{\ell_{\mt P}}
\newcommand{\mt}[1]{\textrm{\tiny #1}}

\makeatletter
\newcommand{\dal}{\mathop{\mathpalette\dal@\relax}}
\newcommand{\dal@}[2]{%
  \begingroup
  \sbox\z@{$\m@th#1\square$}%
  \dimen0=\fontdimen8
    \ifx#1\displaystyle\textfont\else
    \ifx#1\textstyle\textfont\else
    \ifx#1\scriptstyle\scriptfont\else
    \scriptscriptfont\fi\fi\fi3
  \makebox[\wd\z@]{%
    \hbox to \ht\z@{%
      \vrule width \dimen0
      \kern-\dimen0
      \vbox to \ht\z@{
        \hrule height \dimen0 width \ht\z@
        \vss
        \hrule height 2\dimen0
      }%
      \kern-2.5\dimen0
      \vrule width 2.5\dimen0
    }%
  }%
  \endgroup
}
\makeatother

\newcommand{\req}[1]{eq.~(\ref{#1})}

\begin{document}

\pagenumbering{gobble}


%
%







\begin{titlepage}
\vspace*{0cm}
\begin{center}
\Huge
\textbf{Brane-Worlds and \\ Higher-Derivative Gravities}
\vspace*{1.5cm}

\Large
Memòria presentada per optar al grau de doctor per la \\ 
\LARGE
Universitat de Barcelona\\
\vspace*{0.5cm}
\Large
Programa de doctorat en Física\\

\vspace*{1cm}
Autor\\
\LARGE
\textsc{Quim Llorens Giralt}\\
\vspace*{4cm}

\Large
Director\\
\textsc{Roberto Alejandro} \\ \textsc{Emparan García de Salazar}\\
\vspace*{0.5cm}
Codirector\\
\textsc{Pablo Bueno Gómez}\\
\vspace*{0.5cm}
Tutor\\
\textsc{Joan Soto Riera}\\

\vspace*{2.5cm}
\includegraphics[width=0.6\textwidth]{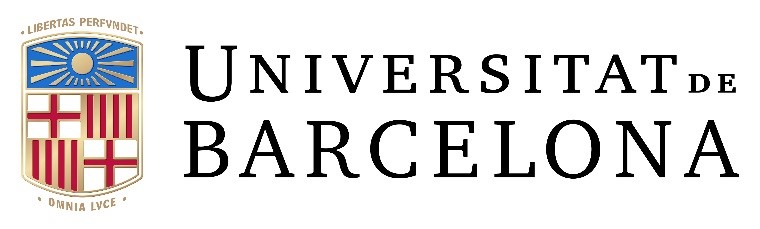}

\end{center}
\end{titlepage}


\clearpage
\thispagestyle{empty}
\hfill
\clearpage


\begin{quote} 
\flushright 
\white{\textit{``What about second breakfast?''}, \\ \medskip 
\textsc{Pippin Took}, \ \\ \textit{The Lord of the Rings}}
\end{quote}

\linespread{1.15}

\vspace{3.5cm} 

\begin{quote} 
\flushright 
\textit{``There is a time for any fledgling artist \\ where one's taste exceeds one's abilities. \\ The only way to get through this period \\ is to make things anyway.''} \\ \medskip
\textsc{Gabrielle Zevin}, \ \\ \textit{Tomorrow, and Tomorrow, and Tomorrow}
\end{quote}
\vspace*{\fill}


\clearpage
\thispagestyle{empty}
\hfill
\clearpage


\pagestyle{fancy}
\fancyhf{}
\renewcommand{\headrulewidth}{0pt}
\cfoot{\thepage}



\newpage
\pagenumbering{roman}
\setcounter{page}{1}
\vspace*{2cm}
\huge
\textbf{Acknowledgements}
\normalsize
\vspace*{2cm}

A PhD is often compared to a marathon --—a long challenge that demands endurance and determination, but most of all, plenty of support.
As I cross the finish line, I wish to thank all those who have helped me on this journey, by providing assistance, guiding me, and encouraging me through the most demanding parts of this race.
Moltes gràcies a tothom.

Primero de todo, me gustaría agradecer a mi director, Roberto Emparan, su contagiosa pasión por la física, su enorme paciencia, y por siempre creer en mí.
He disfrutado y aprendido mucho de nuestras conversaciones honestas y abiertas, tanto las de física ---desde los pequeños detalles técnicos hasta las grandes ideas del Universo--- como especialmente todas las otras. 

Quiero dar las gracias también a mi colaborador y codirector, Pablo Bueno, por desatascar mi tesis y empujarme a participar en sus proyectos.
No es casualidad que mis artículos de la tesis sean con Pablo.
Gracias por los consejos, y por ser un ejemplo de constancia y eficiencia.

M'agradaria agrair a la resta de professors del grup, en Tomeu Fiol, en David Mateos, en Jaume Garriga, i en Jorge Casalderrey, el fet d'haver creat un equip de recerca tan estimulant.
He après moltíssim dels nostres journal clubs, seminaris, i àpats junts.
These acknowledgements extend to all postdocs who have been in the group during these last four years: Tomás Andrade, for introducing me to brane-worlds, Antonia Frassino and her constant smile, Robie Hennigar and his witty sense of humour, Pablo Cano and his mathematical dexterity, and Yago Bea, Alexandre Serantes, Juan Pedraza, Jorge Rocha, Benson Way, and Ryotaku Suzuki.

My fellow graduate students have been my training partners throughout this journey. 
I am grateful to Raimon, Javi, David, Alan, and Jairo, for giving me advice on how to survive a thesis.
To Marija, for being the tough but truly caring older sister I never had.
A Mikel, por sus consejos, su cariño, y sus bromas de mal gusto.
To the new generation, Mingda, Pau, Pedro, and Jéssica, who have brought fresh energy to the group, and to our many visitors, Itzi, Masaya, Mauro, Martín, Théo, and Musfar, who I hope to see again soon.
Gràcies, Lucía, 
per ser un exemple de disciplina i fortalesa, i 
per preguntar-nos dubtes de català, de mecànica o del Mathematica.
Una abraçada gegant a la família del Balcón Milenario, tot i que últimament no aparegui tant per la planta 7, i a la resta de doctorands de la Facultat. Gràcies Mar, Dani, Carlos, Marcel, Andreu, Carla, Mireia, i molts més. 
Gràcies Toni per ser tan crack, i gràcies Óscar per ser un referent tant en la carrera acadèmica com en carreres esportives.

L'Ana i en Jordi es mereixen un paràgraf a part.
Tenir-los al costat ha fet que aquesta experiència fos molt més agradable i divertida.
Les bromes, els memes, les pizzes al Majestic, els àpats a l'atri, els palitos, les excursions a la màquina, els problemes a pissarra, i un llarg etcètera.
Us trobaré molt a faltar a Würzburg, encara que trobi amb qui jugar a la botifarra.
Jordi, ets un geni, i estic convençut que sabràs trobar el teu camí.
Ana, tens totes les qualitats necessàries per menjar-te el món, segueix endavant sense por.

I want to express my deepest gratitude to Mukund Rangamani and Veronika Hubeny for a wonderful experience at the QMAP in Davis, California. Thanks also to Shruti, Christian, Julio, and Taro, Umut, Ilija, y mi argentino favorito ---después de Messi---, Lucas. 
Spending three months in such an enriching scientific environment was a turning point in my PhD.

Many thanks to my other co-authors during my PhD, Javier Moreno, Guido van der Velde, and Sergio Aguilar-Gutiérrez, and to the collaborators with whom I have discussed projects and ideas, Andy Svesko, Jani Kastikainen, Sanjit Shashi, David Pereñíguez, and many more.

Thanks to Dominik Neuenfeld, for his trust in me, and for three amazing weeks ---and two future years--- in Würzburg. I am really looking forward to this new adventure.

També m'agradaria mencionar alguns dels meravellosos professors o mentors que he tingut al llarg de tota la meva educació, i que m'han inspirat i empès a seguir fent ciència.
La Sílvia Jansà i en Jordi Miralles, que em vau portar a les Cangur i em vau convèncer d'apuntar-me a Joves i Ciència, on vaig tenir la sort d'aprendre d'en Kike Herrero i la Mercè Romero; la Marta Domènech, en Jordi Dantart, la Mercè Talló, i l'Anna Ballaró, per inscriure'm a les Olimpíades i fer que gaudís tant del Batxillerat; en Lluís Mañosa, l'Àngels Ramos, l'Ignasi Mundet, en Joan Carles Naranjo, i en Jordi Ortín, per guiar-me durant el grau; to Julian Merten, for a fantastic summer in Oxford and my first scientific paper; i a en Pau Figueras, per recomanar-me de tornar a casa a fer la tesi amb en Roberto.
I moltíssima més gent, la llista completa de la qual seria massa llarga com per incloure aquí.
He estat realment afortunat.

M'agradaria agrair a tot el personal d'administració de l'ICCUB i la Facultat haver-me fet la vida més fàcil. En especial a l'Anna Argudo, per aguantar tots els meus e-mails.

La llista d'amics amb qui he tingut la fortuna de compartir aquests anys és, per sort, ben llarga.
Gràcies a en Sergi, en Jasper, la Paula, la Laura, i en Joan, per fer del pis de Manso una llar. Als Matefísics, JiCs, Fellows, i Quinters, amb qui he compartit ciència i alegria. Als de Trail i del grup de Joves del CEC, per allunyar-me de les pantalles.
Thanks also to Elyse, Meiji, and the In`N'Out group ---Cass, Giulia, Leo, Fabian, Brenda, Flo, Lorenzo, and many more---, for making me feel at home, even though I was so far away.
To Brian, Elba, and especially Heather, who I really wish lived closer.
I Anna, Clara, Marc, Miki, Eu, Carla, Jan, Joan, Meri, gràcies per ser-hi sempre. Quina sort haver-nos fet grans junts.

Finalment, vull agrair a la família el seu amor i suport incondicional. A la família de Bordils, per fer-me sentir com un més des del primer minut. A la de Blanes, on m'hi sento més a casa que enlloc més. A l'Anna i la Cris, per apreciar-me tant, i especialment a la Rosa, la Mariona, i la Maria Cinta, per cuidar-me a banda i banda de l'Atlàntic. A les àvies Maria i Mercè, per ser les meves fans número u.
A la Palmira, qui m'entén com ningú, i òbviament als meus pares, a qui vull dedicar aquesta tesi, sense els quals res d'això seria possible. Us estimo molt, tinc una sort immensa.

I a la Irina, qui fa que la meva vida sigui molt més emocionant. Créixer junts m'ha fet la persona que sóc avui. Quines ganes de gaudir de moltes més aventures plegats. T'estimo.

\newpage

This work was supported by the Spanish Ministry of Universities through FPU grant No. FPU19/04859.
We also acknowledge financial support from MICINN grants PID2019-105614GB-C22 and PID2022-136224NB-C22, AGAUR grants 2017-SGR 754 and 2021-SGR 872, and State Research Agency of MICINN through the ``Unit of Excellence María de Maeztu 2020-2023'' award to the Institute of Cosmos Science CEX2019-000918-M.
The research stay at the QMAP in Davis was supported by the ``Ayuda de Movilidad para Estancias Breves FPU 2023'' EST23/00620, while the research visit at the JMU in Würzburg was supported by the COST Action 22113 ``Fundamental challenges in theoretical physics'', Short-Term Scientific Mission E-COST-GRANT-CA22113-8f5f0659.



\cleardoublepage


\newpage
\vspace*{2cm}
\huge
\textbf{Resum}
\normalsize
\vspace*{2cm}


Aquesta tesi presenta dos temes rellevants dins el camp de la física teòrica d'altes energies, les teories de gravetat amb derivades superiors i els models món-brana, des d'una perspectiva hologràfica.

La Part \ref{part:HDGs} de la tesi tracta sobre teories de gravetat amb derivades superiors.
En primer lloc, estudiem teories de curvatura superior en espaitemps de tres dimensions. Presentem les seves equacions del moviment i n'analitzem l'espectre de pertorbacions lineals en espaitemps màximament simètrics. També identifiquem totes les teories tridimensionals que satisfan un teorema-c hologràfic, així com totes les gravetats quasitopològiques generalitzades en tres dimensions.
A continuació, estudiem un cas més general, el de les gravetats amb derivades superiors arbitràries en espaitemps de qualsevol dimensió.
Caracteritzem l'estructura de les seves equacions lineals en espaitemps màximament simètrics i n'analitzem l'espectre lineal en espaitemps de Minkowski.

A la Part \ref{part:BWs} de la tesi presentem els models món-brana, en què una brana de codimensió-1 talla un espaitemps Anti-de Sitter a prop de la seva frontera.
Revisem resultats coneguts anteriorment sobre la localització la gravetat a la brana, i els ampliem, generalitzant-los a dimensió arbitrària, per tots tres tipus de branes màximament simètriques. 
Interpretem els resultats obtinguts des de dues perspectives diferents, tant des de l'estudi de pertorbacions gravitatòries a l'interior d'Anti-de Sitter, com des de la seva reinterpretació hologràfica: a la brana, hi tenim una teoria efectiva de gravetat amb derivades superiors acoblada a una teoria quàntica de camps efectiva.
A continuació, afegim un terme DGP a l'acció de brana, i definim quins valors pot prendre la seva constant d'acoblament, més enllà dels quals la teoria és patològica.

La Part \ref{part:HDGsBWs} de la tesi combina els dos temes descrits anteriorment per obtenir nous resultats sobre les propietats de la teoria de gravetat induïda a la brana.
En primer lloc, ensenyem com calcular els termes amb derivades superiors de la teoria induïda. Posteriorment, procedim a estudiar-la per si sola, desacoblant-la de la teoria quàntica a la brana.
Demostrem que, a cada ordre en derivades, la teoria compleix un teorema-c hologràfic.
Finalment, n'estudiem l'espectre de pertorbacions lineals en espaitemps de Minkowski, tenint en compte tota la sèrie completa de termes amb derivades superiors.
A part del gravitó sense massa usual, hi descobrim una torre infinita de partícules fantasma massives de spin-2.

Els nostres resultats permeten entendre millor l'holografia en mons-brana i les seves aplicacions com a model de gravetat semiclàssica dins del marc de la correspondència AdS/CFT.


\cleardoublepage


\newpage
\vspace*{2cm}
\huge
\textbf{Abstract}
\normalsize
\vspace*{2cm}

This thesis explores and combines, through the lens of the holographic correspondence, two relevant topics in the field of gravitational high-energy theory: higher-derivative theories of gravity and brane-world models.

Part \ref{part:HDGs} of the thesis deals with higher-derivative gravities. 
First, we focus on three-dimensional higher-curvature gravities. We present their equations of motion and study their spectrum when linearized around maximally symmetric spacetimes. We also identify all three-dimensional higher-curvature gravities satisfying a holographic c-theorem, and all three-dimensional Generalized Quasitopological Gravities. 
Then, we move on to the more general case of studying arbitrary higher-derivative gravities in spacetimes with any number of dimensions. We uncover the structure of their linearized equations in maximally symmetric spacetimes and describe their spectrum of gravitational perturbations in Minkowski space.

In Part \ref{part:BWs}, we introduce Karch-Randall brane-world models, in which a codimension-one brane sits near the boundary of an AdS bulk.
We review and extend previously known results on the localization of gravity on the brane, both from the study of bulk metric perturbations and from their reinterpretation through brane-world holography ---an induced higher-derivative theory of gravity coupled to a cut-off CFT on the brane.
We then add a DGP term on the brane action and, through studying the localization of gravity on the brane, we establish bounds for its coupling constant, beyond which the theory presents pathologies.

Part \ref{part:HDGsBWs} of the thesis draws from both previous topics and combines them to derive new results describing the properties of the induced gravity theory on the brane.
First, we show how to calculate the higher-derivative terms of the induced theory, and then, we proceed to study it on its own, decoupling it from the cut-off CFT on the brane.
We prove that, at each and every curvature order, the theory satisfies a holographic c-theorem. 
Finally, we study its linearization around Minkowski space, taking into account the full series of higher-derivative terms.
Besides the presence of the usual massless graviton, we uncover an infinite tower of massive spin-2 ghosts, which signals at an instability of the theory when it is decoupled from the holographic CFT on the brane.

Our findings shed light on brane-world holography and its applications as a model for semiclassical gravity within AdS/CFT.

\clearpage
\thispagestyle{empty}
\hfill
\clearpage



\newpage
\vspace*{2cm}
\huge
\textbf{Preface}
\normalsize
\vspace*{2cm}

This is a manuscript-style thesis, mostly based on previously published papers, some parts of which have been included here almost verbatim.
The author names in the publications are ordered alphabetically.
The thesis also contains a substantial amount of unpublished work.

In Part \ref{part:HDGs}, Chapter \ref{chp:3DHCGs} is an adapted version of P. Bueno, P. A. Cano, Q. Llorens, J. Moreno, G. van der Velde, ``Aspects of three-dimensional higher-curvature gravities'', Class. Quant. Grav. 39.12 (2022), p. 125002 \cite{Bueno:2022lhf}; while the results in Chapter \ref{chp:HDGs} were published in S. E. Aguilar-Gutiérrez, P. Bueno, P. A. Cano, R. A. Hennigar, Q. Llorens, ``Aspects of higher-curvature gravities with covariant derivatives'', Phys. Rev. D 108.12 (2023), p. 124075 \cite{Aguilar-Gutierrez:2023kfn}.
Some results in which the author did not contribute directly have been removed from both chapters.

Part \ref{part:BWs} of this thesis is unpublished work.
Chapter \ref{chp:ReviewBWs} presents an original review on brane-world models, along with an extension and improvement of known results.
Chapter \ref{chp:BWsWithDGP} is unpublished, original work done by the author alone, although it has considerably benefited from conversations with R. Emparan ---who proposed this research idea---  and D. Neuenfeld.

In Part \ref{part:HDGsBWs}, Chapters \ref{chp:Alg} and \ref{chp:cTheorem} are composed mostly of results from P. Bueno, R. Emparan, Q. Llorens, ``Higher-curvature gravities from braneworlds and the holographic c-theorem'', Phys. Rev. D 106.4 (2022), p. 044012. \cite{Bueno:2022log}, while Chapter \ref{chp:Spectrum} is again mostly based on the last section of \cite{Aguilar-Gutierrez:2023kfn}.


\clearpage
\thispagestyle{empty}
\hfill
\clearpage


\pagenumbering{arabic}
\setcounter{page}{1}

\tableofcontents


\cleardoublepage
\renewcommand{\headrulewidth}{1pt}


\addcontentsline{toc}{part}{Introduction}
\lhead{}
\rhead{Introduction}

\chapter*{Introduction}
\label{chp:Introduction}

General Relativity (GR) is one of the most successful theories in the history of physics, both theoretically and experimentally \cite{Barack:2018yly, Will:2005yc, Will:2014kxa, LIGOScientific:2016lio, LIGOScientific:2018dkp}.
Yet, we know that it cannot describe gravity at its most fundamental level, since it does not agree with the principles of quantum mechanics.
In GR, ``spacetime tells matter how to move, matter tells spacetime how to curve'' \cite{wheeler2010geons}. However, in quantum mechanics, particles do not have a well-defined position in spacetime.
Instead, they are described by a superposition of the many possible positions that the particle could take.
What is the imprint of this superposition onto the geometry of spacetime? 
Should we also consider spacetime as being in a superposition of possible geometries?\footnote{Yes.} And if so, how?

After all, GR is a classical field theory, so we could try to canonically quantize it, just as one does with electromagnetism. 
The quantum field theory of gravitons ---small fluctuations of spacetime--- on a fixed background is consistent and can be used to make predictions for low-energy physics \cite{Donoghue:1995cz, Burgess:2003jk}, but it breaks down at high energies, since the theory can be shown to be perturbatively non-renormalizable \cite{Feynman:1995}. 


One could then consider GR to be a low-energy effective field theory (EFT), with the Einstein-Hilbert term being the first of an infinite series of operators involving a growing number of derivatives of the metric \cite{Endlich:2017tqa}. 
Through dimensional analysis, their couplings would be controlled, up to $\mathcal{O}(1)$ factors, by the cut-off of the theory, perhaps at an energy scale lower than the Planck scale.
From this point of view, it seems relevant to characterise the features of classical gravity in regimes in which GR is expected to receive higher-derivative corrections \cite{PabloPablo4,Arciniega:2018tnn}.

This is the field of higher-derivative gravity, which we will study in Part \ref{part:HDGs} of this thesis. 
We will consider diffeomorphism-invariant theories whose action is built from the metric and its derivatives.
A first step might be considering theories whose Lagrangian is built from arbitrary contractions of the metric and its Riemann tensor, 
\begin{equation}\tag*{(I.1)}
    I_{\text{HCG}} = \frac{1}{16\pi G_N} \int \df^D x \sqrt{-g}\, \mathcal{L}(g_{ab},R_{abcd})\,.
\end{equation}
We will call these theories \textit{higher-curvature} theories of gravity (HCGs), and we will study them in detail, for $D=3$, in Chapter \ref{chp:3DHCGs}. 
Since the Riemann tensor is second order in derivatives of the metric, the equations of motion of these theories are, in general, of quartic order. 

However, in the spirit of generality, we can ---and from an EFT perspective, we should--- also consider Lagrangians built from general contractions of the metric, its Riemann tensor, and its Levi-Civita covariant derivative.
\begin{equation}\tag*{(I.2)}
    I_{\text{HDG}} = \frac{1}{16\pi G_N} \int \df^D x \sqrt{-g}\, \mathcal{L}( g_{ab},R_{abcd},\nabla_a)\, .
\end{equation}
We will call these ---more general--- theories \textit{higher-derivative} theories of gravity (HDGs).\footnote{
This distinction is not widespread within the community, but we will make it in this thesis, since we will be considering both cases.
}
Although their equations of motion can now be of arbitrary order in derivatives, we will still be able to make some progress, as we will show in Chapter \ref{chp:HDGs} of this thesis.
A longer introduction into the field of general higher-derivative gravities is given at the beginning of this same chapter.

In both chapters of Part \ref{part:HDGs}, we will first derive general results, valid for \textit{any} (three-dimensional) HCG or ($D$-dimensional) HDG theory, respectively. 
Then, we will study two particular cases of higher-derivative gravities: theories satisfying a simple holographic c-theorem \cite{Sinha:2010ai, Paulos:2010ke}, and Generalized Quasi-topological gravities (GQTs) \cite{Hennigar:2017ego,PabloPablo3,Bueno:2022res,Moreno:2023rfl}.
Theories satisfying a holographic c-theorem, also known as Cosmological Gravities \cite{Moreno:2023arp}, are theories which admit FLRW-like solutions whose equation for the scale factor is second order in derivatives. 
GQTs are higher-derivative theories of gravity which admit static and spherically symmetric solutions characterized by only one function, that is, they fulfil $g_{tt}g_{rr} = -1$.
In this thesis, we will characterise all three-dimensional theories which fulfil either requirement, and also those which (trivially) satisfy both requirements.\footnote{It has recently been shown that there exist HCG theories which non-trivially fulfil both the Cosmological and the GQT requirements for $D \geq 4$ and at any order in curvature \cite{Moreno:2023arp}.} 
We will prove that all $D$-dimensional GQTs propagate only the massless spin-2 graviton around maximally symmetric spacetimes, and we will also show the first known examples of four-dimensional GQTs with explicit covariant derivatives of the Riemann tensor in the Lagrangian.

In studying these particular sets of higher-derivative theories, we will put the EFT perspective on the side and study them mostly for their mathematical significance, rather than their relevance in our quest for quantum gravity.
However, it has been proven that any higher-curvature gravitational effective action is equivalent, via metric redefinitions, to some GQT \cite{Bueno:2019ltp, Moreno:2023rfl}.
It is not clear whether that is also true for all higher-derivative gravitational effective actions, but evidence suggests that this may also be the case.


Although the study of higher-derivative theories of gravity allows us to explore what may happen at energies somewhat higher than the energies up to which we should trust GR, the EFT description must still break down at ---or before--- the Planck scale.
At this scale, we expect the appearance of new degrees of freedom which UV-complete the theory.
For example, Fermi's theory of the weak interaction, which was also perturbatively non-renormalizable, only had predictive power up to $\lesssim 100 \ \text{GeV} $, where new physics appeared ---in that case, the $W^{\pm}$ and $Z$ gauge bosons \cite{Peskin:1995ev}, whose mass is precisely $80.3692 \pm 0.0133 \ \text{GeV}$ and $91.1880 \pm 0.0020  \ \text{GeV}$ \cite{ParticleDataGroup:2024cfk}, respectively.


One possibility could be that the graviton might be composite, and that we should resolve it into two spin-one gauge bosons. This possibility, however, was ruled out long ago by the Weinberg-Witten theorem \cite{Weinberg:1980kq} ---assuming that the graviton lives in the same spacetime as its parent gauge bosons \cite{Horowitz:2006ct}.


Another route to explore might be considering that gravity is not fundamental and need not be quantized, but that it emerges as a mean field approximation of a quantum field theory on an arbitrary background.
This is the idea behind induced gravity, put forward by Sakharov almost sixty years ago \cite{Sakharov:1967pk, Frolov:2001xr}, in which classical GR emerges as the 1-loop effective action of a quantum field theory on an arbitrary geometry.


There are many other approaches to quantum gravity, but perhaps the most famous and fruitful one has been the possibility of resolving the graviton and its interactions ---as well as all other particles and fundamental interactions--- into extended objects: strings.
String theory was first proposed as a theory for the strong interaction, but it was soon discovered that its spectrum contained a massless spin-2 mode, a graviton \cite{Green:1987mn}.
The field has massively evolved since then, and it is now understood that there are many other extended objects in the theory, with different dimensionality ---such as D-branes, on which fundamental strings can end \cite{Polchinski:1996na}.
It has also been shown that there are five consistent (supersymmetric) string theories containing both bosons and fermions, superstring theories \cite{Green:1987mn, Green:1987sp}, all regarded to be different limits of a yet unknown theory called M-theory \cite{Witten:1995ex, Witten:1995zh}.
Their low-energy effective actions indeed show higher-derivative terms weighted by powers of the inverse string tension \cite{Gross:1986mw,Green:1997tv,Frolov:2001xr}.

One of the phenomenological problems of these superstring theories is that, in order for them not to have quantum anomalies, they must live on ten-dimensional (or eleven-dimensional) spacetimes \cite{Green:1987sp}.
Therefore, six of the spatial dimensions in superstring theory need to be hidden away in order to explain a four-dimensional world like ours.
One way of doing so is by making the extra dimensions compact and small.
But we could also consider the case in which the Standard Model particles are confined to a four-dimensional subspace ---a brane--- within the ten-dimensional theory.
However, gravity must necessarily feel all dimensions, since it is a geometric theory. 
Therefore, we then need a mechanism to confine gravity as well. 
Besides compactification, it is possible to localize gravity on the brane either thanks to the higher-dimensional geometry, or by adding an explicit Einstein-Hilbert term on the brane.

The first proposal goes back to Randall and Sundrum (RS) \cite{Randall:1999vf}.
They considered a four-dimensional purely-tensional flat brane sitting on a five-dimensional Anti-de Sitter (AdS) bulk, close to its asymptotic boundary, and realized that, thanks to the bulk warping factor, a bulk graviton zero-mode localizes on the brane.
This is the topic of brane-world models, which we will consider in Part \ref{part:BWs} of this thesis.
The second proposal was put forth by Dvali, Gabadadze, and Porrati (DGP) \cite{Dvali:2000hr}, and consists in placing a four-dimensional flat brane, with an Einstein-Hilbert term in its action, in a five-dimensional Minkowski bulk.
The RS model reproduces four-dimensional gravity on the brane at long distances but not at high energies, while the DGP model does the opposite \cite{Garriga:1999yh, Tanaka:2003zb}.
In Chapter \ref{chp:ReviewBWs}, we will examine RS brane-worlds and their generalization to (A)dS branes, known as Karch-Randall (KR) brane-worlds \cite{Karch:2000ct}.
We will call them KRS brane-worlds, or simply, brane-worlds.
In the subsequent Chapter \ref{chp:BWsWithDGP}, instead of reconsidering the original DGP set-up, we will investigate the case of DGP branes sitting on an AdS bulk, as an extension of the KR framework.
In both cases, we will study the localization of gravity on the brane, and we will use these results to put bounds on the allowed DGP coupling on KR branes.
A more complete introduction to both topics can be found at the beginning of their respective chapter.


This mismatch between the observed number of dimensions of our Universe and the required spacetime dimensions of superstring theories is not the only challenge that these theories are facing. Even though they help clarify some of their qualities, even string theory struggles in fully describing GR's most simple and fascinating solutions: black holes (BHs).

The classical laws of BH mechanics \cite{Hawking:1971vc, Bardeen:1973gs}, when compared to the laws of thermodynamics, suggest that BHs have an entropy proportional to their area and a temperature proportional to their surface gravity \cite{Bekenstein:1972tm, Bekenstein:1973ur, Bekenstein:1974ax}.
If BHs have a temperature, they must radiate through black-body radiation, which seems to contradict their classical definition as regions of spacetime from which nothing ---not even light--- can escape.
Fifty years ago, however, through semiclassically studying quantum fields around the event horizon of a BH, Hawking showed that this is indeed the case: BH horizons do radiate, with a perfect black-body spectrum at a temperature $T_H$ \cite{Hawking:1974sw, Hawking:1975vcx}.
Moreover, Hawking's calculation was able to fix the proportionality constants and show that BHs have an entropy and a temperature precisely given by
\begin{align}\tag*{(I.3)}\label{SBH}
    S_{BH} & = \frac{A}{4G_N}\,, \\ T_{H} & = \frac{\kappa}{2\pi}\,, \tag*{(I.4)}\label{TH}
\end{align}
where $A$ and $\kappa$ are the area and surface gravity of the event horizon, respectively, $G_N$ is Newton's constant, and we are using units in which $c = \hbar = k_B = 1$.
A couple of years later, Gibbons and Hawking rederived these formulas using a semiclassical saddle-point approximation on the gravitational path integral, in which one considers the path integral of Euclidean metrics fulfilling the required boundary conditions of the problem \cite{Gibbons:1976ue}.
Moreover, they showed that these formulas also apply to other kinds of horizons in GR, such as cosmological horizons \cite{Gibbons:1977mu}.
Indeed, even acceleration horizons present such a temperature \cite{Unruh:1976db}.
Therefore, equation \ref{TH} is an equation about QFT in curved spacetime.
But, as we will now see, equation \ref{SBH} is our first formula of quantum gravity, and it has been shaping most of the research in the field ever since, as it raises several fundamental questions.

One issue that arises from interpreting both equations is that, since BHs radiate, they lose energy, and so they must eventually evaporate and disappear \cite{Hawking:1974sw}.
The laws of quantum mechanics tell us that no information is lost in unitary time evolution, but the black-body radiation of an evaporating BH carries no information but the temperature of the horizon \cite{Hawking:1976ra}.
What happens then to the information of whatever objects that may have fallen into a BH?
This paradox is known as the Black Hole Information Problem.
Nowadays, there is strong evidence that information must eventually leak out of the evaporating BH \cite{Almheiri:2020cfm}, as we will explain later. It seems that quantum entanglement and some form of non-locality play an important role in the physics behind this process \cite{Giddings:2022jda}, but the precise details of the mechanism which resolves the paradox are still not completely known.

Another conceptual problem that stems from these formulas is the fact that, in statistical mechanics, entropy describes the number of microstates available to a given thermodynamic macrostate.
In classical GR, however, black holes are fully described by just a handful of parameters ---mass, charge, and angular momentum---, the famous No-Hair Theorem \cite{Israel:1967wq, Israel:1967za}. 
From which microscopical degrees of freedom does the BH entropy come from? What are the possible microstates of a BH?
In string theory, it is possible to account for the microstates of some extremal, supersymmetric BHs \cite{Strominger:1996sh}, but not of the four-dimensional Kerr BH, for example, which is believed to describe all BHs in our Universe.\footnote{Recently, it has been shown that the entropy of many kinds of BHs can be accounted for by an infinite family of microstates semiclassically described by dust shells in the BH interior. One then finds the correct dimensionality of the Hilbert space through a Euclidean path integral calculation, which yields the desired result if one takes into account subleading wormhole saddles \cite{Balasubramanian:2022gmo,Balasubramanian:2022lnw,Climent:2024nuj}. These microstates, however, are non-generic, so there is still much to be understood from this issue.}
And, most importantly, in most thermodynamic set-ups, the entropy of a system scales with its volume, and not its area. Do the microscopic degrees of freedom describing a BH live only on its event horizon, and not its interior? And if so, why?

Now, consider a spherical region in spacetime with area $A$, within which there are quantum fields with some energy, and with some entropy greater than $A/4G$.
If we were to collapse all this energy into a BH, its horizon area would be smaller than the original area of the region $A$, and so we would would decrease the total entropy of the Universe. 
Thus, in order not to violate the second law of thermodynamics, the maximum entropy that quantum fields can have in some region of space is the entropy of a BH of that same size \cite{Bekenstein:1980jp,Casini:2008cr}.

Since this bound 
is proportional to the area of the spacetime region, one may conjecture that, in fact, any fundamental quantum gravity description of a region of spacetime must be encoded in some degrees of freedom at the boundary of the region.
This is known as the holographic principle, pioneered by 't Hooft and Susskind \cite{tHooft:1993dmi, Susskind:1994vu, Bousso:2002ju}.

The holographic principle became concrete in 1997, when Maldacena posited the AdS/CFT correspondence \cite{Maldacena}, which asserts that quantum gravity ---string theory--- on a $(d+1)$-dimensional AdS spacetime is equivalent to a $d$-dimensional conformal field theory (CFT) living on a spacetime with no gravity, at the conformal boundary of the AdS bulk.
The correspondence provided the first non-perturbative definition of string theory ---previously, there were only definitions of string theory that were perturbative in the string coupling.

The gauge/gravity duality has grown beyond its initial form within string theory into one of the most active and fruitful fields within theoretical physics.
Nowadays, it is understood that the duality is more general, and that many non-Abelian quantum field theories, not necessarily fully conformal, describe gravitational effective field theories in asymptotically AdS spacetimes \cite{Gubser, Witten, Horowitz:2006ct}.
Conversely, one may also think about using, through AdS/CFT, holographic higher-derivative gravities as toy models of conformal field theories. Indeed, universal properties valid for completely general CFTs have been discovered through such explorations \cite{Kats:2007mq,Brigante:2007nu,Buchel:2008ae,Myers:2010xs,Camanho:2010ru,Mezei:2014zla,Bueno1,Bueno:2018yzo,Bueno:2022jbl}.

The duality has many far-reaching consequences.
It was suggested by Ryu and Takayanagi \cite{Ryu:2006bv, Ryu:2006ef}, and later proven by Lewkowycz and Maldacena \cite{Lewkowycz:2013nqa}, that the entanglement entropy of boundary subregions of the CFT can be computed by the area of an extremal bulk surface homologous to the boundary subregion through a formula that has the same form as the entropy of BHs. 
This result suggests that spacetime itself may emerge from quantum entanglement \cite{VanRaamsdonk:2010pw}.

Another consequence of the AdS/CFT duality is that, a priori, it solves the BH Information Paradox: since the boundary theory is a perfectly unitary non-gravitating QFT, the BHs in AdS dual to their thermal states must also evolve unitarily.
However, it was not until 2019 that precise calculations showed how to reproduce a unitary Page curve for evaporating large AdS black holes \cite{Almheiri:2020cfm}. The key was realizing that the correct way of computing entropies in semi-classical gravity is by using quantum extremal surfaces \cite{Penington:2019npb, Almheiri:2019hni, Almheiri:2019yqk, Rozali:2019day, Chen:2020hmv, Chen:2020uac}, in what has become known as the island formula, which consists in taking into account additional subleading saddles ---spacetime Euclidean wormholes--- to the Euclidean gravitational path integral.

Many of the clues that led to these recent breakthroughs were inspired \cite{Almheiri:2019hni,Geng:2020qvw} by what is known as brane-world holography.
It turns out that one can also apply the holographic duality to KRS brane-worlds: the AdS$_{d+1}$ bulk ending on an end-of-the-world (EOW) brane is dual to an effective $d$-dimensional gravitational theory coupled to a cut-off CFT$_d$ on the brane \cite{deHaro:2000vlm, Neuenfeld:2021wbl}.
Since the brane is at a finite distance from the asymptotic boundary, we can think of brane-world models as imposing a UV cut-off to the dual CFT. Doing so, we obtain a holographic realization of Sakharov's aforementioned induced gravity proposal \cite{Sakharov:1967pk}, as the gravitational theory on the brane is induced by integrating out the CFT degrees of freedom above the cut-off.

Brane-world holography has also been used to describe (quantum) three-dimensional black holes interacting with strongly coupled CFTs \cite{Emparan:1999fd, Emparan:1999wa, Emparan:2000rs,Emparan:2020znc, Emparan:2022ijy, Emparan:2023dxm, Feng:2024uia, Climent:2024nuj}.

As we will explain in detail in Chapters \ref{chp:ReviewBWs} and \ref{chp:Alg}, the induced gravity theory on the brane is not simply Einstein gravity but a higher-derivative theory of gravity.
In Part \ref{part:HDGsBWs} of this thesis, we will use the mathematical tools for higher-derivative theories developed in Part \ref{part:HDGs}, and insight on brane-worlds from Part \ref{part:BWs}, to thoroughly study the higher-derivative theory of gravity on the brane.

We hope that our results from this thesis will provide insights into the inner workings of brane-world holography, allowing us to expand and clarify its uses as a model for semi-classical gravity within AdS/CFT.


\paragraph{Summary of this thesis.}
Let us end this introduction with a summary of this thesis.
It consists of three main parts. The first two are disjoint from each other, and can be read in any order, while the third combines results from both previous parts and should be read at the end.

Part \ref{part:HDGs} deals with higher-derivative theories of gravity.
In Chapter \ref{chp:3DHCGs}, we present new results involving general higher-curvature gravities in three dimensions. The most general Lagrangian of that kind can be written as a function of $R,\mathcal{S}_2,\mathcal{S}_3$, where $R$ is the Ricci scalar, $\mathcal{S}_2\equiv \tilde R_{a}^b \tilde R_b^a$, $\mathcal{S}_3\equiv \tilde R_a^b \tilde R_b^c \tilde R_c^a$, and $\tilde R_{ab}$ is the traceless part of the Ricci tensor. First, we provide a general formula for the exact number of independent order-$n$ densities, $\#(n)$. This satisfies the identity $\#(n-6)=\#(n)-n$. Then, we show that, linearized around a general Einstein solution, a generic order-$n\geq 2$ density can be written as a linear combination of $R^n$, which by itself would not propagate the generic massive graviton, plus a density which by itself would not propagate the generic scalar mode, $ R^n-12n(n-1)R^{n-2}\mathcal{S}_2$, plus $\#(n)-2$ densities which contribute trivially to the linearized equations. Then, we provide a recursive formula as well as a general closed expression for order-$n$ densities which non-trivially satisfy a holographic c-theorem, clarify their relation with Born-Infeld gravities and prove that the scalar mode is always absent from their spectrum. We show that, at each order $n \geq 6$, there exist $\#(n-6)$ densities which satisfy the holographic c-theorem in a trivial way and that all of them are proportional to a single sextic density $\Omega_{(6)}\equiv 6 \mathcal{S}_3^2-\mathcal{S}_2^3$. Next, we show that there are also $\#(n-6)$ order-$n$ Generalized Quasi-topological (GQT) gravities in three dimensions, all of which are ``trivial'', since they do not contribute to the metric function equation. Remarkably, the set of such densities coincides exactly with the one of theories trivially satisfying the holographic c-theorem. We comment on the meaning of $\Omega_{(6)}$ and its relation to the Segre classification of three-dimensional metrics.

In Chapter \ref{chp:HDGs}, we study, in arbitrary dimensions, higher-derivative theories of gravity built from contractions of the metric, its Riemann tensor, and its Levi-Civita covariant derivative, $\mathcal{L}(g^{ab},R_{abcd},\nabla_a)$. We show the structure of the linearized equations of these theories on maximally symmetric backgrounds, and we characterise their linearized spectrum on Minkowski space. Then, we study GQTs involving covariant derivatives of the Riemann tensor. We argue that they always have second-order linearized equations on maximally symmetric backgrounds, and that they display an Einsteinian spectrum.
Focusing on four spacetime dimensions, we present the first examples of densities of this type, involving eight and ten derivatives of the metric.

In Part \ref{part:BWs}, we introduce Karch-Randall brane-world models, in which a $d$-dimensional brane sits near the boundary of an AdS$_{d+1}$ bulk.
In Chapter \ref{chp:ReviewBWs}, we start by reviewing the localization of gravity on the brane. We linearly perturb the bulk and show how a bulk zero-mode localizes on the brane. We extend the known results by presenting them in a new formulation that allows one to deal with the three different maximally symmetric brane geometries at once, and by generalizing them to an arbitrary number of dimensions $d \geq 3$. For the case of AdS branes, we improve on the formula of the graviton mass as a function of the brane position, and present new formulas describing the mass of the higher overtones.
Finally, we reinterpret these results through brane-world holography.
We integrate the bulk to obtain an effective description of the brane dynamics in terms of brane variables as a higher-derivative theory of induced gravity coupled to a cut-off CFT on the brane.

In Chapter \ref{chp:BWsWithDGP}, we then add a DGP term ---an explicit Einstein-Hilbert term--- on the brane action. 
We extend the results from the previous chapter to study how the localization of gravity on the brane changes with the presence of this extra term.
This allows us to establish bounds for the DGP coupling constant, beyond which the theory presents pathologies: either the position of the brane ceases to be well-defined, or its spectrum presents a tachyonic mode.
We again perform brane-world holography to reinterpret these results from the brane perspective.
We finish by commenting on the possibility of adding further ---higher-derivative--- terms to the brane action.

Part \ref{part:HDGsBWs} of the thesis combines results from both previous parts of the thesis to describe the properties of the higher-derivative theory of induced gravity on the brane.
In Chapter \ref{chp:Alg}, we show how to compute the higher-derivative gravitational densities that are induced from holographic renormalization in AdS$_{d+1}$. In the previous brane-world construction, these densities define the $d$-dimensional higher-derivative gravitational theory on the brane. 
Inevitably, there is some redundancy between Chapters \ref{chp:ReviewBWs} and \ref{chp:Alg} concerning the definition of the induced gravity theory on the brane, but we have chosen not to remove it so that each chapter can be read independently.

In Chapter \ref{chp:cTheorem}, we show that the CFT$_{d-1}$ dual to the $d$-dimensional induced gravity theory satisfies a holographic c-theorem in general dimensions, since at every order in derivatives, the densities in the action satisfy c-theorems on their own. 
We find that, in these densities, the terms that affect the monotonicity of the holographic c-function are algebraic in the curvature, and do not involve covariant derivatives of the Riemann tensor. 
We examine various other features of the holographically induced higher-curvature densities, such as the presence of reduced-order traced equations, and their connection to Born-Infeld-type gravitational Lagrangians.

Finally, in Chapter \ref{chp:Spectrum}, we study the linearized spectrum on flat space of these induced brane-world gravities. We show that the effective quadratic action for the full tower of higher-derivative terms in the induced gravity action can be written explicitly in a closed form in terms of Bessel functions. We use this result to compute the propagator of metric perturbations around Minkowski and its pole structure in various dimensions, always finding infinite towers of ghost modes, as well as tachyons and more exotic modes in some cases.




\newpage
\subsection*{Notation}

We use units such that $c = \hbar = k_B = 1$, our metric sign convention is $(- + \cdots +)$. 

If we are doing brane-world holography, we use $d$ to denote the number of spacetime dimensions of the brane, while the bulk is a ($d+1$)-dimensional asymptotically AdS spacetime.
Otherwise, we use $D$ as our number of spacetime dimensions.

We generally use $g$ to denote our spacetime metric.
However, if we are doing (brane-world) holography, we use $G$ as our bulk metric and $g$ as our induced metric on the brane.

We use abstract index notation.
Capital Latin indices $M,N,...$ denote bulk tensor equations, while early-alphabet indices $a,b,c,...$ denote brane tensor equations, or otherwise general tensor equations in chapters in which we are not doing brane-world holography.
These equations are valid in any basis.
We use Greek indices $\mu,\nu,...$ to denote bulk tensor components with respect to some bulk basis, while middle-alphabet indices $i,j,k,...$ denote brane or general tensor components with respect to some basis.

\bigskip

\begin{tabular}{ l l }
 BH & Black Hole.   \\ 
 GR & General Relativity \\
 (A)dS & (Anti-)de Sitter spacetime.   \\
 $G_N$ & Newton's constant. \\
 $g$ & Determinant of the metric $g_{ab}$. \\
 EOW & End-of-the-world (brane). \\
 RS & Randall-Sundrum (brane-world). \\
 KR & Karch-Randall (brane-world). \\
 DGP & Dvali-Gabadadze-Porrati (term). \\
 RT & Ryu-Takayanagi (holographic entanglement entropy).\\
 GQT & Generalized Quasi-topological (gravity).
\end{tabular}

\cleardoublepage
\part{Higher-Derivative Gravities}\label{part:HDGs}
\lhead{Chapter 1}
\rhead{Three-Dimensional Higher-Curvature Gravities}

\chapter{Three-Dimensional Higher-Curvature Gravities}
\label{chp:3DHCGs}


\section{Introduction}\label{sec:Intro3D}

Gravity becomes simpler when we go down to three dimensions.
Firstly, the Weyl tensor vanishes identically, implying that all curvatures are Ricci curvatures. This means that all solutions of three-dimensional Einstein gravity are locally equivalent to maximally symmetric backgrounds and that no gravitational waves propagate. In spite of this, global differences between spacetimes do appear and prevent the theory from being ``trivial'', even at the classical level. In particular, in the presence of a cosmological constant, the theory admits black hole solutions \cite{Banados:1992wn,Banados:1992gq} which, despite important differences with their higher-dimensional counterparts, do share many of their properties ---including the existence of event and Cauchy horizons, thermodynamic properties, holographic interpretation, etc. 

The local equivalence of all classical solutions allows for a characterization of the phase space of the theory \cite{Witten:1988hc}.  In addition ---up to non-negligible details--- three-dimensional Einstein gravity is classically equivalent to a Chern-Simons gauge theory \cite{Achucarro:1986uwr}. From a holographic point of view \cite{Maldacena,Witten,Gubser}, these qualitative changes with respect to higher dimensions are manifest in the distinct nature of conformal field theories in two dimensions. In fact, while the observation that the symmetry algebra of AdS$_3$ spaces is generated  by two copies of the conformal algebra in two dimensions \cite{Brown:1986nw} is often considered to be  a precursor of AdS/CFT, the nature of the putative holographic theory ---or ensemble of theories--- dual to pure Einstein gravity is still subject of debate \cite{Maloney:2007ud,Keller:2014xba,Benjamin:2019stq,Alday:2019vdr,Cotler:2020ugk,Maxfield:2020ale}.

The above simplifications also affect higher-curvature modifications of Einstein gravity. In particular, all theories can be constructed exclusively from contractions of the Ricci tensor, which reduces the number of independent densities drastically. Similarly, the usual arguments for considering higher-curvature corrections ---which involve their appearance in the form of infinite towers of terms coming from stringy corrections--- do not make much sense in three-dimensions. This is because all non-Riemann curvatures can be removed via field redefinitions, and hence one is left again with Einstein gravity ---plus cosmological constant and a possible gravitational Chern-Simons term \cite{Gupta:2007th}. However, there is a different reason to consider higher-curvature gravities with non-perturbative couplings in three dimensions. This is the fact that, as opposed to Einstein gravity, they can give rise to non-trivial local dynamics. This appears in the form of a massive graviton and/or a scalar mode ---see \eg \cite{Gullu:2010sd}. 

By far, the best known higher-curvature modification of Einstein gravity in three dimensions is the so-called ``New Massive Gravity'' (NMG) \cite{Bergshoeff:2009hq}.
At the linearized level, the theory describes a massive graviton  with the same dynamics of a Fierz-Pauli theory. In addition, the theory is distinguished by possessing second-order traced equations \cite{Oliva:2010zd}, by admitting a holographic c-theorem \cite{Sinha:2010ai}, and by admitting a Chern-Simons description \cite{Hohm:2012vh}. Unfortunately, demanding unitarity of the bulk theory spoils the unitarity of the boundary theory and viceversa \cite{Bergshoeff:2009aq}, a problem which has been argued to be unavoidable for general higher-curvature theories sharing the spectrum of NMG \cite{Gullu:2010vw}.
 
Moving from quadratic to higher orders, one can use some of the above criteria to select special theories. One possibility is to demand that the corresponding theories admit a holographic c-theorem \cite{Sinha:2010ai,Paulos:2010ke}. Alternatively, one can look for additional theories which admit a Chern-Simons description \cite{Afshar:2014ffa,Bergshoeff:2014bia,Bergshoeff:2021tbz}. A different route involves considering special $D\rightarrow 3$ limits of higher-dimensional theories with special properties \cite{Alkac:2020zhg}. Often, the densities resulting from these different approaches coincide with each other.   Alternative routes include \cite{Banados:2009it,Paulos:2012xe,Bergshoeff:2013xma,Bergshoeff:2014pca,Alkac:2017vgg,Alkac:2018eck,Ozkan:2018cxj,Afshar:2019npk}. 
 
While higher-curvature modifications of three-dimensional Einstein gravity have been studied extensively by now, most of the results are only valid for the lowest curvature orders or for particular theories ---see \eg \cite{Paulos:2010ke,Gurses:2011fv,Gurses:2015zia} for exceptions. In this chapter, we present a collection of new results for general-order higher-curvature theories. Without further ado, let us summarize them.


\begin{itemize}
  \item In Section \ref{counting}, we obtain a formula for the exact number of independent order-$n$ densities, $\#(n)$. This is given by
  \begin{equation}
    \#(n) = \ceil[\Big]{\frac{n}{2} \left(\frac{n}{6}+1 \right)+\epsilon}\, ,
  \end{equation}
  where $\ceil[]{x}$ is the usual ceiling function and $\epsilon$ is any positive number such that $\epsilon\ll 1$. The function $\#(n)$ satisfies the interesting recursive relation $\#(n-6)=\#(n)-n$, which says that the number of order-$n$ densities minus $n$ equals the number of densities of six orders less.
  \item In Section \ref{sec:eomEs}, we present the equations of motion for a general higher-curvature gravity and the algebraic equations these reduce to when evaluated for Einstein metrics. We also make a few comments about single-vacuum theories.
  \item In Section \ref{sec:lineq}, we obtain the linearized equations of a general higher-curvature gravity around an Einstein spacetime as a function of the effective Planck length and the masses of the new spin-2 and spin-0 modes generically propagated. Formulas for such physical parameters are obtained for a general theory. Using these results, we show that the most general order-$n$ density can be written as 
  \begin{equation}
    \mathcal{L}_{(n)}=\alpha_n R^n+ \beta_n [R^n-12 n (n-1) R^{n-2}\mathcal{S}_2]+\mathcal{G}_{(n)}^{\rm trivial}\, ,
  \end{equation}
  where: the first term is a density which by itself does not propagate the massive spin-2 mode but which does propagate the scalar one (for $n\geq 2$), the second term is a density which by itself does not propagate the scalar mode but which does propagate the spin-2 mode, and the third term, $\mathcal{G}_{(n)}^{\rm trivial}$ ---which involves $\#(n)-2$ densities--- does not contribute at all to the linearized equations.
  \item In Section \ref{sec:ctheorem3D}, we study higher-curvature theories which satisfy a holographic c-theorem. First, we provide a recursive formula for densities which satisfy it in a non-trivial fashion ---\ie they contribute non-trivially to the c-function. This is given by
  \begin{equation}
    \mathcal{C}_{(n)}=\frac{4(n-1)(n-2)}{3n(n-3)}\left(\mathcal{C}_{(n-1)}\mathcal{C}_{(1)}-\mathcal{C}_{(n-2)}\mathcal{C}_{(2)}\right)\, ,
  \end{equation}
  which allows one to obtain an order-$n$ density of that kind from the two immediately lower order ones. Then, we solve the recurrence explicitly and provide a general explicit formula for  $\mathcal{C}_{(n)}$. We argue that there are $\#(n)-(n-1)$ densities of order $n$ which satisfy the holographic c-theorem. Of those, $\#(n)-n$ are trivial in the sense of making no contribution to the c-function. These start appearing for $n \geq 6$. We show that all such ``trivial'' densities are proportional to the sextic density
  \begin{equation}
    \Omega_{(6)}\equiv 6\mathcal{S}_3^2-\mathcal{S}_2^3\, ,
  \end{equation}
  so that the most general order-$n$ density satisfying the holographic c-theorem can be written as
  \begin{equation}
    \mathcal{L}_{(n)}^{\rm c-theorem}=\alpha_n \mathcal{C}_{(n)}+\Omega_{(6)}\cdot \mathcal{L}^{\rm general}_{(n-6)}\, ,
  \end{equation}
  where $ \mathcal{L}^{\rm general}_{(n-6)}$ is the most general linear combination of order-$(n-6)$ densities. In addition, we show that if a theory satisfies the holographic c-theorem, then it does not include the scalar mode in its spectrum. Finally, we study the relation of $\mathcal{C}_{(n)}$ with the general term obtained from expanding the Born-Infeld gravity Lagrangian of \cite{Gullu:2010pc}.
  \item In Section \ref{sec:GQTss}, we explore the possible existence of Generalized Quasi-topological gravities in three dimensions. We show that there exist $\#(n)-n$ theories of that kind, and that all of them are ``trivial'' ---in the sense of making no contribution to the equation of the black hole metric function--- and again proportional to the same sextic density $ \Omega_{(6)}$ that appeared in the previous section.
  \item In Section \ref{sec:Ommm6}, we make some comments on the relation between $\Omega_{(6)}$ and the Segre classification of three-dimensional spacetimes. We explain why the prominent role played by this density in the identification of ``trivial'' densities of the types studied in the previous two sections could have been expected ---at least to some extent.
\end{itemize}


\paragraph{Notation and conventions:} Throughout this chapter we consider higher-curvature theories constructed from contractions of the Ricci tensor and the metric. When referring to generic Lagrangian densities, we use the notation $\mathcal{L}\equiv \mathcal{L}(g_{ab},R_{ab})$, and we express the gravitational constant in terms of the Planck length $\ell_{\rm \ssc P}\equiv 8\pi G_N$.  We choose to work always with a negative cosmological constant, which we denote in terms of the action length scale $L$, so that  $-2\Lambda\equiv 2/L^2$. The Anti-de Sitter$_3$ (AdS$_3$) radius is denoted by $L_{\star}$, and sometimes we use the notation $\chi_{0}\equiv L^2/L_{\star}^2$, so that $\chi_{0}=1$ for Einstein gravity.
We will often consider Lagrangians which involve an Einstein gravity plus cosmological constant part, plus a general function of the three basic densities which span the most general higher-curvature invariants in three-dimensions. Those three invariants can be alternatively chosen to be $\{R,\mathcal{R}_2\equiv R_{a}^b R_{b}^a, \mathcal{R}_3\equiv R_a^bR_b^cR_c^a\}$ or $\{R,\mathcal{S}_2\equiv \tilde R_{a}^b \tilde R_{b}^a, \mathcal{S}_3\equiv \tilde R_a^b \tilde R_b^c \tilde R_c^a\}$, where $\tilde R_{ab}$ is the traceless part of the Ricci tensor. Consequently, we will often consider general functions of either set of densities, which we will denote respectively by $\mathcal{F}\equiv \mathcal{F}(R,\mathcal{R}_2,\mathcal{R}_3)$ and $\mathcal{G}\equiv \mathcal{G}(R,\mathcal{S}_2,\mathcal{S}_3)$. The different invariants are often classified attending to their curvature order $n$, corresponding to the number of Ricci tensors involved in their definition. Generic order-$n$ densities are denoted $\mathcal{L}_{(n)}$  and the most general linear combination of order-$n$ densities is denoted $\mathcal{L}_{(n)}^{\rm general}$. We use the notation $\mathcal{G}_X\equiv \partial \mathcal{G}/\partial X$, $\mathcal{G}_{X,X}\equiv \partial^2 \mathcal{G}/\partial X^2$ and so on to denote partial derivatives. Expressions with a bar denote evaluation of the invariants on an Einstein background metric,  $\bar X \equiv X(\bar R, \bar{\mathcal{R}}_2,\bar{\mathcal{R}}_3)$. In the case of terms  which require taking derivatives with respect to some of the arguments, it is understood that the derivatives are taken first, and the resulting expression is then evaluated on the background. Hence, for instance, $\bar{\mathcal{F}}_R \equiv \left. [\partial \mathcal{F}/\partial R]\right|_{R= \bar R,\mathcal{R}_2= \bar{\mathcal{R}}_2,\mathcal{R}_3=\bar{\mathcal{R}}_3}$.


\section{Counting higher-curvature densities}\label{counting}

In this section, we compute the exact number of independent densities of order $n$ constructed from arbitrary contractions of the Riemann tensor and the metric.  The vanishing of the Weyl tensor in three dimensions reduces the analysis to theories constructed from contractions of the Ricci tensor and the metric,  $\mathcal{L}\left(g_{ab},R_{ab}\right)$. Additionally, the existence of Schouten identities  implies that the most general higher-curvature action can be written as \cite{Paulos:2010ke,Gurses:2011fv}
\begin{equation}\label{action3}
S_{(\mathcal{R})}=\frac{1}{2\ell_{\rm \ssc P}} \int \df ^3x \sqrt{|g|} \mathcal{L}_{(\mathcal{R})}\, , \quad \mathcal{L}_{(\mathcal{R})}\equiv \frac{2}{L^2}+R+\mathcal{F}\left(R,\mathcal{R}_2,\mathcal{R}_3 \right) \, ,
\end{equation}
where we chose a negative cosmological constant and we defined 
\begin{equation}
\mathcal{R}_2\equiv R_a^bR_b^a\, , \quad \mathcal{R}_3\equiv R_a^bR_b^c R_c^a\, .
\end{equation}
Often we will assume $\mathcal{F}\left(R,\mathcal{R}_2,\mathcal{R}_3 \right) $ to be either an analytic function of its arguments, or a series of the form
\begin{equation}\label{Fseries}
\mathcal{F}\left(R,\mathcal{R}_2,\mathcal{R}_3 \right) =\sum_{i,j,k}L^{2(i+2j+3k-1)}\,  \alpha_{ijk}  R^i \mathcal{R}_2^j \mathcal{R}_3^k\, ,
\end{equation}
for some dimensionless coefficients $\alpha_{ijk}$.

As we just mentioned, the ``Schouten identities'' drastically reduce the number of independent densities of a given order, leaving \req{action3} as the most general case. Those identities take the form \cite{Paulos:2010ke}
\begin{equation}\label{schou}
\delta_{b_1 \dots b_n}^{a_1\dots a_n} R_{a_1}^{b_1}R_{a_2}^{b_2}\cdots R_{a_n}^{b_n}=0\, ,\quad \text{for} \quad n>3\, ,
\end{equation}
where $\delta_{b_1 \dots b_n}^{a_1\dots a_n}$ is the (totally antisymmetric) generalized Kronecker delta. These identities rely on the fact that totally antisymmetric tensors with ranks higher than $3$ vanish identically in $D=3$. From \req{schou}, it follows that the cyclic contraction of $n>3$ Riccis can be written in terms of lower-order densities, and hence the generality of \req{action3}. One finds, for instance
\begin{align}
R_a^bR_b^cR_c^dR_d^a &=  \frac{1}{6} R^4+ \frac{4}{3} R \mathcal{R}_3 + \frac{1}{2} \mathcal{R}_2^2- \mathcal{R}_2 R^2 \, , \\  R_a^bR_b^cR_c^dR_d^e R_e^a&= \frac{1}{6}R^5+\frac{5}{6}\left(\mathcal{R}_3\mathcal{R}_2+\mathcal{R}_3 R^2-\mathcal{R}_2 R^3\right)\, , \\
R_a^bR_b^cR_c^dR_d^e R_e^fR_f^a&= \frac{1}{12}R^6+\mathcal{R}_3\mathcal{R}_2 R+\frac{1}{3}\mathcal{R}_3 R^3-\frac{1}{4}\mathcal{R}_2 R^4 -\frac{3}{4}\mathcal{R}_2^2 R^2+\frac{1}{4}\mathcal{R}_2^3+ \frac{1}{3} \mathcal{R}_3^2\, . 
\end{align}

It is often convenient to use a basis of invariants involving the traceless part of the Ricci tensor,
\begin{equation}\label{tracelessR}
\tilde R_{ab}\equiv R_{ab}-\frac{1}{3}g_{ab} R\, .
\end{equation}
Then, we can define 
\begin{equation}
\mathcal{S}_2\equiv \tilde R_a^b \tilde R_b^a=\mathcal{R}_2-\frac{1}{3}R^2\, , \quad  \mathcal{S}_3\equiv \tilde R_a^b \tilde R_b^c \tilde R_c^a=\mathcal{R}_3-R \mathcal{R}_2+\frac{2}{9}R^3\, ,
\end{equation}
and alternatively write the most general theory replacing $\mathcal{F}\left(R,\mathcal{R}_2,\mathcal{R}_3 \right) $ by $\mathcal{G}\left(R,\mathcal{S}_2,\mathcal{S}_3 \right)$ in \req{action3} \cite{Gurses:2011fv}, namely
\begin{equation}\label{action4}
S_{(\mathcal{S})}=\frac{1}{2\ell_{\rm \ssc P}} \int \df ^3x \sqrt{|g|} \mathcal{L}_{(\mathcal{S})}\, , \quad \mathcal{L}_{(\mathcal{S})}\equiv \frac{2}{L^2}+R+\mathcal{G}\left(R,\mathcal{S}_2,\mathcal{S}_3 \right) \, .
\end{equation}
We will write the polynomial version of $\mathcal{G}$ as
\begin{equation}\label{Gseries}
\mathcal{G}(R,\mathcal{S}_2,\mathcal{S}_3)=\sum_{i,j,k}L^{2(i+2j+3k-1)}\,  \beta_{ijk}  R^i \mathcal{S}_2^j \mathcal{S}_3^k\, .
\end{equation}
While \req{action3}  feels like a  more natural choice from a higher-dimensional perspective, it turns out that many formulas simplify considerably when expressed in terms of $\tilde R_{ab}$ instead. We will try to present most of our results in both bases.

Let us consider the case in which the theory is a power series of the building blocks $R,\mathcal{R}_2,\mathcal{R}_3$ as in \req{Fseries} (or, alternatively, $R,\mathcal{S}_2,\mathcal{S}_3$ as in \req{Gseries}). The order $n$ of a certain combination of scalar invariants is related to the powers of the individual components through $n=i+2j+3k$. One finds the following possible invariants at the first orders,
\begin{align}
R\, , \quad \text{for}\quad &n=1\, ,\\
R^2\, , \quad \mathcal{R}_2\, , \quad \text{for}\quad &n=2\, ,\\
R^3\, , \quad R \mathcal{R}_2\, , \quad \mathcal{R}_3\, ,\quad \text{for}\quad &n=3\, ,\\
R^4\, , \quad R^2 \mathcal{R}_2\, , \quad R \mathcal{R}_3\, , \quad \mathcal{R}_2^2 \, , \quad \text{for}\quad &n=4\, ,\\
R^5\, , \quad R^3 \mathcal{R}_2\, , \quad R^2 \mathcal{R}_3\, , \quad R \mathcal{R}_2^2 \, , \quad \mathcal{R}_2 \mathcal{R}_3\, , \quad \text{for}\quad &n=5\, ,\\
R^6\, , \quad R^4 \mathcal{R}_2\, , \quad R^3 \mathcal{R}_3\, , \quad R^2 \mathcal{R}_2^2 \, , \quad R \mathcal{R}_2 \mathcal{R}_3\, , \quad \mathcal{R}_2^3\, ,\quad \mathcal{R}_3^2\, , \quad \text{for}\quad &n=6\, ,
\end{align}
and so on. Then, the function $\#(n)$ counting the number of invariants of order $n$ takes the values $\#(1)=1$, $\#(2)=2$, $\#(3)=3$, $\#(4)=4$, $\#(5)=5$, $\#(6)=7$. 

In order to find the explicit form of $\#(n)$ as a function of $n$, we can proceed as follows. If we understand the number of elements constructed from powers of $R$ alone up to order $n$ as the coefficients of a power series, we can define the generating function $f^{(R)}(x)$ as 
\begin{equation}
f^{(R)}(x)\equiv \frac{1}{1-x}\sim 1+x+x^2+x^3+\dots\, ,
\end{equation}
\ie such that the right-hand-side, which is the Maclaurin series of the left-hand-side, has coefficient $1$ for all powers. This is because at every order $n$ there is a single density we can construct with $R$ alone, namely, $R^n$. Now, if we want to do the same for $\mathcal{R}_2$, we need to take into account that the corresponding coefficients should be $1$ when $n$ is even, and $0$ otherwise. We define then
\begin{equation}
f^{(\mathcal{R}_2)}(x)\equiv  \frac{1}{1-x^2}\sim 1+x^2+x^4+x^6+\dots\, 
\end{equation}
Following the same reasoning for $\mathcal{R}_3$, we define
\begin{equation}\label{eq:GenFR}
f^{(\mathcal{R}_3)}(x)\equiv  \frac{1}{1-x^3}\sim 1+x^3+x^6+x^9+\dots\, 
\end{equation}  
Now, we can obtain $\#(n)$ as the coefficient of the Maclaurin series corresponding to the generating function which results from the product of the three generating functions previously defined, namely
 \begin{equation}
 f^{(R)}(x) f^{(\mathcal{R}_2)}(x) f^{(\mathcal{R}_3)}(x)=\frac{1}{(1-x)(1-x^2)(1-x^3)}\sim \sum_n \#(n) x^n \, .
 \end{equation}
The result can be written explicitly as
\begin{equation}\label{num1}
\#(n)=\frac{1}{72}\left[47+(-1)^n9+6n(6+n)+16\cos\left(\frac{2n\pi}{3}\right)\right]\, .
\end{equation}
This gives the exact number of independent  three-dimensional order $n$ densities. It is easy to verify that this yields the same values obtained above for the first $n$'s. Note that $\#(n)$ is not an analytic function, but it is still easy to see that, for $n\gg 1$, it goes as
\begin{equation}\label{apro}
 \#(n)\sim \frac{n}{2} \left(\frac{n}{6}+1 \right)\,.
\end{equation}
The fact that $ \#(n)$ scales with $\sim n^2$ for large $n$ had been previously observed in \cite{Paulos:2010ke}.
  
It can be shown that $\#(n)$ can be alternatively written exactly (for integer $n$, which is the relevant case) as
\begin{equation}\label{num2}
  \#(n) = \ceil[\Big]{\frac{n}{2} \left(\frac{n}{6}+1 \right)+\epsilon}\, ,
\end{equation}
where $\ceil[]{x}\equiv {\rm min} \left\{k \in \mathbb{Z}\, |\, k\geq x \right\}$ is the usual ceiling function and $\epsilon$ is any positive number such that $\epsilon \ll 1$. For instance, at order $n=1729$, one has $\#(1729)=249985$ independent densities, as one can easily verify both from \req{num1} or \req{num2}. 
  
The function $\#(n)$ satisfies several relations which connect its values at different orders. A particularly suggestive one is the recursive relation
\begin{align}\label{properttt}
   \#(n-6) &=  \#(n)-n\, ,
\end{align}
which connects the number of densities of a given order with the number of densities  of six orders less. This follows straightforwardly from the general expression of $\#(n)$ in \req{num2}. We will use this relation in Sections \ref{sec:ctheorem3D} and \ref{sec:GQTss} to prove a couple of results concerning the general form of densities which trivially satisfy a holographic c-theorem and of densities which belong to the Generalized Quasi-topological class.


\section{Equations of motion and Einstein solutions}\label{sec:eomEs}
The equations of motion of a general higher-curvature theory constructed from arbitrary contractions of the Ricci scalar and the metric can be written as \cite{Padmanabhan:2011ex}
\begin{equation}\label{eomsss}
\mathcal{E}_{ab}\equiv P_{a}^{c}R_{b c} -\frac{1}{2}g_{a b}\mathcal{L}-\nabla_{(a}\nabla_c P_{b)}^c+\frac{1}{2}\dal P_{a b}+\frac{1}{2}g_{a b} \nabla_c\nabla_d P^{c d}=0\,,
\end{equation}
where
\begin{equation}
    P^{a b}\equiv \left.\frac{\partial \mathcal{L}}{\partial R_{a b}}\right\vert_{g^{c d}}\,.
\end{equation}
In our three-dimensional case, when written as in \req{action3}, the explicit form of these equations reads
\begin{equation}\label{eq:EOM2}
\begin{aligned}
\mathcal{E}_{ab}^{(\mathcal{R})}\equiv &+R_{a b}(1+\mathcal{F}_R)-\frac{1}{2}g_{a b}\left(R+\frac{2}{L^2}+\mathcal{F}\right)+\left(g_{a b}\dal - \nabla_{a}\nabla_{b}\right) \mathcal{F}_R\\ &+ 2 \mathcal{F}_{\mathcal{R}_2}  R_{a}^{c}R_{c b}+ 3 \mathcal{F}_{\mathcal{R}_3}R_{a}^{c}R_{c d}R_{b}^{d}+g_{a b}\nabla_{c}\nabla_{d}\left(\mathcal{F}_{\mathcal{R}_2}R^{c d}+\frac{3}{2}\mathcal{F}_{\mathcal{R}_3}R^{c f}R_{f}^{d}\right)\\
&+\dal \left( \mathcal{F}_{\mathcal{R}_2}R_{a b}+\frac{3}{2}\mathcal{F}_{\mathcal{R}_3}R_{a}^{c}R_{c b}\right)-2\nabla_{c}\nabla_{(a}\left(R_{b)}^{c}\mathcal{F}_{\mathcal{R}_2}+\frac{3}{2}R_{b)}^{d}R_{d}^{c}\mathcal{F}_{\mathcal{R}_3}\right)=0 \, ,
\end{aligned}
\end{equation}
In the $R,\mathcal{S}_2,\mathcal{S}_3$ basis, the equations of motion read instead \cite{Gurses:2011fv}
\begin{equation}\label{eq:EOM3}
\begin{aligned}
\mathcal{E}_{ab}^{(\mathcal{S})}\equiv &+\left(\tilde{R}_{a b}+\frac{1}{3}g_{a b}R\right)-\frac{1}{2}g_{a b}\left(R+\frac{2}{L^2}+\mathcal{G}\right)+2 \mathcal{G}_{\mathcal{S}_2}  \tilde{R}_{a}^{c}\tilde{R}_{c b}+ 3 \mathcal{G}_{\mathcal{S}_3}\tilde{R}_{a}^{c}\tilde{R}_{c d}\tilde{R}_{b}^{d}\\ &+\left(g_{a b}\dal - \nabla_{a}\nabla_{b}+\tilde{R}_{a b}+\frac{1}{3}g_{a b}R\right) \left(\mathcal{G}_R-\mathcal{G}_{\mathcal{S}_3}\mathcal{S}_2\right)+g_{a b}\nabla_{c}\nabla_{d}\left(\mathcal{G}_{\mathcal{S}_2}S^{c d}+\frac{3}{2}\mathcal{G}_{\mathcal{S}_3}S^{c f}\tilde{R}_{f}^{d}\right)\\
&+\left(\dal+\frac{2}{3}R\right) \left( \mathcal{G}_{\mathcal{S}_2}\tilde{R}_{a b}+\frac{3}{2}\mathcal{G}_{\mathcal{S}_3}\tilde{R}_{a}^{c}\tilde{R}_{c b}\right)-2\nabla_{c}\nabla_{(a}\left(\tilde{R}_{b)}^{c}\mathcal{G}_{\mathcal{S}_2}+\frac{3}{2}\tilde{R}_{b)}^{d}\tilde{R}_{d}^{c}\mathcal{G}_{\mathcal{S}_3}\right)=0 \, .
\end{aligned}
\end{equation}

Solutions of Einstein gravity plus cosmological constant can be easily embedded in the general higher-curvature theory 
\req{action3} or \req{action4}. These include, for instance, pure AdS$_3$ and the BTZ black hole. Indeed, consider Einstein metrics of the form 
\begin{equation}\label{constantr}
\bar R_{ab}=-\frac{2}{L_{\star}^2} \bar g_{ab}\, .
\end{equation}
In that case, one has
\begin{equation}\label{RR2R3S2S3}
\bar R=-\frac{6}{L_{\star}^2}\, , \quad \bar{\mathcal{R}}_2=\frac{12}{L_{\star}^ 4}\, , \quad   \bar{\mathcal{R}}_3=-\frac{24}{L_{\star}^ 6}\, , \quad \bar{\mathcal{S}}_2=0\, , \quad \bar{\mathcal{S}}_3=0\, .
\end{equation}
Hence, \req{constantr} satisfies the equations of motion \req{eq:EOM2} provided
\begin{equation} \label{sks}
\frac{6}{ L^2}-\frac{6}{L_{\star}^2}\left[1-2\bar{\mathcal{F}}_R+\frac{8}{L_{\star}^2 } \bar{\mathcal{F}}_{\mathcal{R}_2}-\frac{2 4}{L_{\star}^4 } \bar{\mathcal{F}}_{\mathcal{R}_3} \right] +3\bar{\mathcal{F}}=0
\, ,
\end{equation}
is satisfied. In the alternative formulation in terms of traceless Ricci tensors, the analogous equation is considerably simpler and reads \cite{Gurses:2011fv}\begin{equation}\label{vacuS}
\frac{6}{ L^2}-\frac{6}{L_{\star}^2}\left[1- 2 \bar{\mathcal{G}}_R \right] +3\bar{\mathcal{G}}=0\,.
\end{equation}
For Einstein gravity, this simply reduces to $L^2=L_{\star}^2$, which just says that the AdS$_3$ radius coincides with the cosmological constant scale. In general, \req{sks} and \req{vacuS} are equations for the quotient $\chi_{0}\equiv L^2/L_{\star}^2$. 
If the series form (\ref{Fseries}) is assumed, then \req{sks} takes the form 
\begin{equation}
1-\chi_{0}+\sum_{n} a_n  \chi_{0}^n=0\, ,\quad \text{where}\quad a_n\equiv (-1)^n 6^{n-1} (3-2n) \sum_{j,k} \frac{\alpha_{n-2j-3k,j,k} }{3^{j+2k}} \, .
\end{equation}
Similarly, \req{vacuS} takes the form
\begin{equation}
1-\chi_{0}+\sum_{n} b_n  \chi_{0}^n=0\, ,\quad \text{where}\quad b_n\equiv (-1)^n 6^{n-1} (3-2n) \beta_{n00} \,,
\end{equation}
where observe that terms involving $\mathcal{S}_2$ and $\mathcal{S}_3$ make no contribution to the equation.

On general grounds, the above polynomial equations will possibly have several positive solutions for $\chi_{0}$, so the corresponding theories will possess several AdS$_3$ vacua.  Finding higher-curvature theories with a single vacuum in three and higher dimensions has been subject of study of numerous papers ---see \eg \cite{Crisostomo:2000bb,Gullu:2015cha,Karasu:2016ifk} and references therein. In the present case, a complete analysis of the conditions which lead to a single vacuum can be easily performed in a case-by-case basis, but not so much for a completely general theory, so we will not pursue it here.  Let us nonetheless make a couple of comments. First, observe that all extensions of Einstein gravity with terms involving either $\mathcal{S}_2$ and/or $\mathcal{S}_3$ will have a single vacuum, since for those, the Einstein gravity solution $\chi_{0}=1$ will be the only one.  A different possibility for single-vacuum theories would correspond to an order-$n$ degeneration of the solutions of the above polynomial equations, \ie to the cases in which these become
\begin{equation}
\left(1-\frac{\chi_0}{n}\right)^n=0\, .
\end{equation} 
Observe that this  involves $n-1$ conditions for a theory containing densities of order $n$ and lower and these will necessarily mix couplings of different orders. In particular, for a theory written in the $\{R,\mathcal{S}_2,\mathcal{S}_3 \}$ basis involving densities of order up to $n$, these read
\begin{equation}
\beta_{i00}=\binom{n}{i}\frac{1}{n^i 6^{i-1}(3-2i)}\, , \quad i=2,\dots, n\,.
\end{equation}
Hence, a Lagrangian of the form
\begin{align}
\mathcal{L}_{(n)}^{\rm single\, \, vac.}&=\frac{2}{L^2}+R+\sum_{i=2}^n \binom{n}{i}\frac{L^{2(i-1)}}{n^i 6^{i-1}(3-2i)}R^i + \mathcal{S}_2 h_2(R,\mathcal{S}_2,\mathcal{S}_3)+ \mathcal{S}_3 h_3(R,\mathcal{S}_2,\mathcal{S}_3)\, ,
\end{align}
where $h_{2,3}$ are any analytic functions of their arguments will have a single AdS$_3$ vacuum.


\section{Linearized equations}\label{sec:lineq}
The linearized equations of motion around maximally symmetric backgrounds of higher-curvature gravities involving general contractions of the Riemann tensor and the metric  were obtained in \cite{PabloPablo,Bueno:2016ypa} ---see also \cite{Tekin1,Tekin2}. The resulting expression was expressed in terms of four parameters, $a$, $b$, $c$ and $e$, and a simple method for computing such coefficients for a given theory was also provided, along with the connection between them and the relevant physical parameters ---namely, the effective Newton constant and the masses of the additional modes. In this section, we apply this method to a general higher-curvature theory in three dimensions and classify theories according to the content of their linearized spectrum.

Let $g_{ab}=\bar g_{ab}+h_{ab}$ where the background metric is an Einstein spacetime satisfying \req{constantr} and 
$h_{ab}\ll 1$, $\forall a,b=0,1,2$. Then, restricted to a general three-dimensional higher-curvature gravity of the form \req{action3}, the equations of motion of the theory read, at leading order in the perturbation \cite{Bueno:2016ypa}  
\begin{equation}\label{line}
\frac{1}{4\ell_{\ssc \rm P}}\mathcal{E}\lnr_{ab}\equiv \left[e+c\left(\bar{\dal}+\frac{2}{L_{\star}^2}\right)\right]G\lnr_{ab}+(2b+c)\left(\bar{g}_{ab}\bar{\dal}-\bar{\nabla}_a\bar{\nabla}_b\right)R\lnr-\frac{1}{L_{\star}^2}\left(4b+c\right)\bar{g}_{ab}R\lnr=\frac{1}{4}T_{ab}\lnr\, ,
\end{equation}
where we included a putative matter stress-tensor for clarity purposes and where the linearized Einstein and Ricci tensors, and Ricci scalar read
 \begin{IEEEeqnarray}{ll}
G\lnr_{ab}&=R\lnr_{ab}-\frac{1}{2}\bar{g}_{ab}R\lnr+\frac{2}{L_{\star}^2} h_{ab}\, ,\\
R\lnr_{ab}&=\bar{\nabla}_{\left(a\right|}\bar{\nabla}_{c}h\indices{^c_{\left|b\right)}}-\frac{1}{2}\bar{\dal}h_{ab}-\frac{1}{2}\bar{\nabla}_a\bar{\nabla}_b h-\frac{3}{L_{\star}^2} h_{ab}+\frac{1}{L_{\star}^2} h \bar{g}_{ab}\, ,\\
R\lnr &=\bar{\nabla}^a\bar{\nabla}^b h_{ab}-\bar{\dal} h+\frac{2}{L_{\star}^2} h\, .
\end{IEEEeqnarray}
In higher dimensions, there is an additional parameter ---denoted ``$a$'' in \cite{Bueno:2016ypa}--- appearing in the linearized equations. However, this turns out to be non-zero only for densities which involve Riemann curvatures, and so we have $a=0$ for all three-dimensional theories. For a generic higher-curvature theory in that case, \req{line} describes three propagating degrees of freedom, corresponding to a massive ghost-like spin-2 mode plus a spin-0 mode.  The parameters $e$, $c$ and $b$ above can be related to the effective Planck length $\ell_{\ssc \rm P}^{\rm eff}$ and the masses (squared) of such modes, which we denote $m_g^2$ and $m_s^2$, as
\begin{equation}\label{phypara}
\ell_{\ssc \rm P}^{\rm eff}=\frac{1}{4e}\, , \quad m_g^2=-\frac{e}{c}\, , \quad m_s^2=\frac{e+\frac{8}{L_{\star}^2}(3b+c)}{3c+8b}\, .
\end{equation}
In subsection \ref{subsec:classifi} we explain how to compute these parameters for a general higher-curvature theory and do this explicitly in our three-dimensional context.

In terms of the physical quantities, the linearized equations read
\begin{equation}
\begin{aligned}\label{linEQ}
\frac{\ell_{\ssc \rm P}^{\rm eff}}{\ell_{\ssc \rm P}} m_g^2\cdot \mathcal{E}\lnr_{ab}\equiv &+ \left(m_g^2-\frac{2}{L_{\star}^2}-\bar{\dal}\right)G\lnr_{ab}- \frac{1}{L_{\star}^2}\left(\frac{m_g^2+m_s^2-\frac{2}{L_{\star}^2}}{2(m_s^2-\frac{3}{L_{\star}^2})} \right) \bar g_{ab} R\lnr \\  &+ \left(\frac{m_g^2-m_s^2+\frac{4}{L_{\star}^2}}{4(m_s^2-\frac{3}{L_{\star}^2})} \right) (\bar g_{ab} \bar \dal- \bar\nabla_a\bar\nabla_b)R\lnr=\ell_{\ssc \rm P}^{\rm eff}m_g^2\cdot T\lnr_{ab}\, .
\end{aligned}
\end{equation}


\subsection{Physical modes}\label{subsec:physmod}
From what we have said so far, it is not obvious that \req{linEQ} describes the aforementioned modes of masses $m_s$, $m_g$. In order to see this, it is convenient to decompose the metric perturbation as
\begin{equation}
h_{ab}=\hat{h}_{ab}+\frac{\bar\nabla_{\langle a}\bar\nabla_{b\rangle} h}{\left(m_s^2-\frac{3}{L_{\star}^2}\right)}+\frac{1}{3}h \bar{g}_{ab}\, ,
\end{equation}
where $\langle ab\rangle$ denotes the traceless part, and $\hat{h}_{ab}$ is transverse and traceless, satisfying 
\begin{equation}
\bar{g}^{ab}\hat{h}_{ab}=0\, ,\quad \bar\nabla^{a}\hat h_{ab}=0\, ,
\end{equation}
where the second condition is imposed using gauge freedom. Let us note that this decomposition fails in the special case $m_s^2=\frac{3}{L_{\star}^2}$. In that situation, it is not possible to decouple the trace and traceless parts of $h_{ab}$. However, from \req{phypara}, it follows that in this case $m_g^2=-1/L_\star^2$, so the spin-2 mode is a tachyon. Hence, we will assume that $m_s^2\neq\frac{3}{L_{\star}^2}$ to avoid this problematic situation. 

Then, the trace and the traceless part of the linearized equations become, respectively, \cite{Bueno:2016ypa}
\begin{align}\label{hmode}
\frac{2}{L_{\star}^2} \frac{\left(1+\frac{1}{m_g^2L_{\star}^2}\right)}{\left(m_s^2-\frac{3}{L_{\star}^2}\right)} (\bar \dal - m_s^2)h &= \ell_{\ssc \rm P}^{\rm eff}T^{\rm L}\, , \\ \label{hmode2}
\frac{1}{2m_g^2} \left(\bar \dal +\frac{2}{L_{\star}^2} \right)\left(\bar \dal +\frac{2}{L_{\star}^2}-m_g^2\right) \hat{h}_{ab}&= \ell_{\ssc \rm P}^{\rm eff} T^{\rm{L,eff}}_{\langle ab\rangle}\, ,
\end{align}
where $T^L\equiv \bar g^{ab}T_{ab}^L$ and 
\begin{equation}
T^{\rm{L,eff}}_{\langle ab\rangle}\equiv T^{\rm{L}}_{\langle ab\rangle}-\frac{L_{\star}^2}{2}\frac{\left(\bar \dal +\frac{1}{L_{\star}^2}-m_g^2 \right)}{\left(m_g^2+\frac{1}{L_{\star}^2} \right)}\bar\nabla_{\langle a}\bar\nabla_{b\rangle} T^{\rm L}\, .
\end{equation}
Eq. (\ref{hmode}) describes a spin-0 mode corresponding to the trace of the perturbation. On the other hand, \req{hmode2} can be further rewritten by defining
\begin{equation}
\hat h_{ab}\equiv \hat h_{ab}^{(m)} +\hat h_{ab}^{(M)}\, ,\quad \text{where} \quad \hat h_{ab}^{(m)}\equiv -\frac{1}{m_g^2}\left[ \bar \dal +\frac{2}{L_{\star}^2}-m_g^2\right] \hat h_{ab}\, , \quad \hat h_{ab}^{(M)}\equiv \frac{1}{m_g^2}\left[ \bar \dal +\frac{2}{L_{\star}^2}\right] \hat h_{ab}\, , 
\end{equation}
as
\begin{align}
-\left(\bar \dal +\frac{2}{L_{\star}^2}\right)\hat h_{ab}^{(m)} &= \ell_{\ssc \rm P}^{\rm eff} T^{\rm{L,eff}}_{\langle ab\rangle} \, , \\ \label{spin2m}
\left(\bar \dal +\frac{2}{L_{\star}^2}-m_g^2\right)\hat h_{ab}^{(M)} &=  \ell_{\ssc \rm P}^{\rm eff} T^{\rm{L,eff}}_{\langle ab\rangle}\, .
\end{align}
These describe two traceless spin-2 modes which couple to matter with opposite signs. However, as opposed to higher dimensions, only the massive one is propagating in $D=3$. The would-be massless spin-2 mode is pure gauge (whenever $T^{L}_{ab}=0$) in this number of dimensions  ---see \eg \cite{Deser:1983tn,Nakasone:2009bn,Myung:2011bn,Moynihan:2020ejh}.\footnote{Massless and massive gravitons in $D$ dimensions propagate  $\frac{D(D-3)}{2}$ and $\frac{(D+1)(D-2)}{2}$ degrees of freedom, respectively, which means $0$ and $2$ degrees of freedom respectively for $D=3$.}  Hence, the relevant equations are \req{hmode} and \req{spin2m}, which describe a maximum of three degrees of freedom ---one from the scalar mode and two from the spin-2 one--- propagated around Einstein solutions by higher-curvature gravities in the most general case. 

When $\ell_{\ssc \rm P}^{\rm eff}>0$, the massive graviton is a ghost and the scalar mode has positive energy, but since there is no massless graviton, one could also consider $\ell_{\ssc \rm P}^{\rm eff}<0$, so that the massive graviton has positive energy and the scalar is a ghost. As we will see below, there are theories that only propagate either the scalar mode or the massive spin-2 mode, and these can be made unitary by taking $\ell_{\ssc \rm P}^{\rm eff}>0$ or $\ell_{\ssc \rm P}^{\rm eff}<0$, respectively. An example of the latter is NMG, as introduced in \cite{Bergshoeff:2009hq}, in which the Ricci scalar appears with the ``wrong'' sign in the action, hence implying $\ell_{\ssc \rm P}^{\rm eff}<0$.


\subsection{Identification of physical parameters}
Given a higher-curvature theory, one can linearize its equations and deduce the values of the parameters $b,c,e$ (and consequently $\ell_{\rm \ssc P}^{\rm eff},m_g^2,m_s^2$) by comparing them with the above general expressions. A much faster way of performing this  identification was proposed in \cite{Bueno:2016ypa}, which we adapt here to our three-dimensional setup. One starts by replacing all Ricci tensors in the Lagrangian by
\begin{equation}
R^{\rm aux}_{ab}=-\frac{2}{L_{\star}^2}g_{ab}+\alpha (x -1)k_{ab} \, ,
\end{equation}
where $x$ is an arbitrary integer constant and the symmetric tensor $k_{ab}$ is defined such that $k_a^a\equiv x$ and $k_a^b k_b^c=k_a^c$. Then, the parameters can be unambiguously extracted from the general formulas \cite{Bueno:2016ypa}
\begin{equation}
\left. \frac{\partial \mathcal L(R^{\rm aux}_{ab})}{\partial \alpha}\right|_{\alpha=0}=2e\, x(x-1)\, , \quad \left. \frac{\partial^2 \mathcal L(R^{\rm aux}_{ab})}{\partial \alpha^2}\right|_{\alpha=0}=4x(x-1)^2 (c+ b x)\, .
\end{equation}
It is straightforward to do this for our general three-dimensional actions. When the theory is expressed in terms of the traceless Ricci tensor as in \req{action4}, the resulting parameters take a particularly simple form
\begin{equation}\label{GGL}
e=\frac{1}{4\ell_{\ssc \rm P}} [1+\bar{\mathcal{G}}_R]\, , \quad b=\frac{1}{4\ell_{\ssc \rm P}} \left[\frac{1}{2}\bar{\mathcal{G}}_{R,R}-\frac{1}{3} \bar{\mathcal{G}}_{\mathcal{S}_2} \right]\, , \quad c=\frac{1}{4\ell_{\ssc \rm P}} \bar{\mathcal{G}}_{\mathcal{S}_2}\, ,
\end{equation}
where recall that we are using the notation $\mathcal{G}_X\equiv \partial \mathcal{G}/\partial X$, $\mathcal{G}_{X,X}\equiv \partial^2 \mathcal{G}/\partial X^2$ and the bar means that we are evaluating the resulting expressions on the background geometry, which is implemented through \req{RR2R3S2S3}. 
If we assume that the Lagrangian allows for a polynomial expansion, it is useful to decompose $\mathcal{G}$ in the following way,
\begin{equation}\label{GS2}
\mathcal{G}(R,\mathcal{S}_2,\mathcal{S}_3)=f(R)+\mathcal{S}_2 g(R)+\mathcal{G}_{\rm triv}\,,
\end{equation}
where
\begin{equation}
\mathcal{G}_{\rm triv}\equiv \mathcal{S}_2^2 h(R,\mathcal{S}_2)+\mathcal{S}_3 l(R,\mathcal{S}_2,\mathcal{S}_3)
\end{equation}
includes all terms which do not contribute to the linearized equations around any constant curvature solution. That is the case of any density involving any power of $\mathcal{S}_2$ greater or equal than two and any power of $\mathcal{S}_3$ (different from zero).
With the Lagrangian expressed in this way, the background equation, \req{vacuS}, reduces to
\begin{equation}
\frac{6}{L^2}-\frac{6}{L_{\star}^2}[1-2 \bar f_R]+3\bar f=0\, ,
\end{equation}
and using eq. \eqref{phypara} we find the physical quantities of the linearized spectrum, 
\begin{align}\label{ppog}
\ell_{\rm \ssc P}^{\rm eff}=\frac{\ell_{\rm \ssc P}}{[1+\bar f_R]}\, ,\quad
m_g^2=-\frac{[1+\bar f_R]}{ \bar g}\, ,\quad
m_s^2=\frac{[1+\bar f_R]+\frac{12}{L_{\star}^2} \bar f_{R,R}}{4\bar f_{R,R}+\frac{1}{3}\bar g}\, .
\end{align}
When expressed explicitly in terms of the gravitational couplings in an expansion of the form (\ref{Gseries}) these read 
\begin{equation}
\begin{aligned}\label{ppog2}
\ell_{\rm \ssc P}^{\rm eff}=&\frac{\ell_{\rm \ssc P}}{\left[1+\sum_i\beta_{i00} i(-6 \chi_{0})^{i-1} \right]}\, ,\quad
m_g^2=-\frac{\left[ 1+\sum_i\beta_{i00} i(-6\chi_{0})^{i-1} \right]}{\Ls^2\sum_i\beta_{i10}(-6)^i}\,,\\
m_s^2=&\frac{\left[1-\sum_i\beta_{i00} i (2 i-3)(-6\chi_{0})^{n-1} \right]}{4\Ls^2\sum_i(-6 \chi_{0})^{i-2}  [(i-1) i \beta_{i00}+3\beta_{i10}]}\, .
\end{aligned}
\end{equation}

With the above expressions at hand, it is straightforward to classify the different theories according to the presence or absence of the massive graviton and scalar modes in their spectrum. Before doing so, let us present in passing the expressions analogous to \req{ppog} when the analysis is performed for a theory expressed in the
$\{R,\mathcal{R}_2,\mathcal{R}_3\}$ basis instead. In that case, the equations become more involved and a decomposition of the form (\ref{GS2}) is not available. We have  
\begin{equation}
\begin{aligned}\label{555}
\lpeff=&\frac{\lp\Ls^4}{\Ls^4(1+\bF{R})-4\Ls^2\bF{\RR}+12\bF{\RRR}}\, , \quad
m_g^2=\frac{\Ls^4(1+\bF{R})-4\Ls^2\bF{\RR}+12\bF{\RRR}}{6\Ls^2\bF{\RRR}-\Ls^4\bF{\RR}}\, , \\
m_s^2=&\frac{3}{\Ls^2}+\left[\Ls^8\left(1+\bF{R}\right)-5\Ls^6\bF{\RR}-18\Ls^4\bF{\RRR}\right]/\left[3\Ls^8\bF{\RR}-18\Ls^6\bF{\RRR}+4\Ls^8\bF{R,R}\right.\\
&\left.-32\Ls^6\bF{R,\RR}+96\Ls^4\bF{R,\RRR}-384\Ls^2\bF{\RR,\RRR}+4\Ls^8\bF{R,R}+64\Ls^4\bF{\RR,\RR}+576\bF{\RRR,\RRR}\right]\, .
\end{aligned}
\end{equation}
The polynomial form is straightforward to obtain from these expressions (and as ugly as one may anticipate).


\subsection{Classification of theories}\label{subsec:classifi}
The decomposition (\ref{GS2}) and \req{ppog} make it very simple to classify all theories depending on the mode content of their linearized spectrum. The three sets of theories we consider are: theories which are equivalent to Einstein gravity at the linearized level, theories which do not propagate the massive graviton, and theories which do propagate the scalar mode.

\subsubsection{Einstein-like theories}
A first group of densities are those for which $m_g^2,m_s^2\rightarrow \infty$, namely, densities in whose spectrum both the massive graviton and the scalar mode are absent. These are theories which, at the level of the linearized equations, are identical to Einstein gravity ---up to, at most, a change in the effective Planck length.  As we mentioned earlier, a large set of densities do not contribute whatsoever to the linearized equations. These are given by 
\begin{equation}\label{gtriv}
\mathcal{G}|_{\text{trivial linearized equations}}= \mathcal{S}_2^2 h(R,\mathcal{S}_2)+\mathcal{S}_3 l(R,\mathcal{S}_2,\mathcal{S}_3)\, .
\end{equation}
It is not difficult to see that there are $\#(n)-2$ densities of this kind at order $n$. Namely, all order-$n$ densities but those of the forms $R^n$ and $\mathcal{S}_2 R^{n-2}$ contribute trivially to the linearized equations. While there are no ``trivial'' densities for $n=1,2$, they start to proliferate for $n\geq 3$, becoming the vast majority at higher orders. As it turns out, these ``trivial'' densities are the only Einstein-like theories which exist beyond Einstein gravity itself. The reason is that removing both the massive graviton and the scalar from the spectrum amounts at imposing $c=b=0$, which implies $\bar g=\bar f_{R,R}=0$. These are on-shell conditions, but if we want to avoid relations between densities of different orders, we must force them to hold for any value of $\bar R$. Hence, the conditions become $g(R)=f_{R,R}(R)\equiv 0$, whose only non-trivial solution besides \req{gtriv} is Einstein gravity plus a cosmological constant.  Hence, most higher-curvature densities have in fact trivial linearized equations.

It is a remarkable ---and exclusively three-dimensional--- fact that Einstein gravity is unique in this sense. Observe that starting in four dimensions and for higher $D$ there are generally several Einstein-like densities with non-trivial linearized equations at each curvature order. Examples are Lovelock \cite{Lovelock1,Lovelock2} and some $f(\text{Lovelock})$ densities \cite{Love}, Einsteinian cubic gravity \cite{PabloPablo}, Quasi-topological \cite{Quasi2,Quasi,Dehghani:2011vu,Cisterna:2017umf} and Generalized quasi-topological gravities \cite{Hennigar:2017ego,PabloPablo3,PabloPablo4,Bueno:2019ycr}, among others \cite{Li:2017ncu,Karasu:2016ifk,Li:2017txk}.

\subsubsection{Theories without massive graviton}
Theories for which  $m^2_g\rightarrow\infty$ do not propagate the massive graviton. In terms of our parameters $e$, $b$ and $c$, this condition is given by $c=0$.
From \req{ppog} it is clear that this set of theories are those with $\bar g\equiv g(\bar R)=0$. Again, in order to impose this condition at each curvature order we must demand $g(R)\equiv 0$. Hence, the most general (polynomial) density which makes a non-trivial contribution to the linearized equations and which does not propagate the massive graviton in three-dimensions is $f(R)$ gravity
\begin{equation}
\mathcal{G}|_{\text{no massive graviton}}=f(R)\, ,
\end{equation}
Obviously, at  order $n$ there is $1$ such density, corresponding to $R^n$. Of course, one can obtain more complicated densities satisfying the $m^2_g\rightarrow\infty$ condition by combining some of the trivial Einstein-like densities with the $f(R)$ ones. Hence, there are actually $\#(n)-1$ independent densities which do not propagate the massive graviton at order $n$.

For comparison, observe that in $D\geq 4$ there is a large set of higher-curvature theories which do not have the massive graviton in their spectrum. This is the case, in particular, of all $f(\text{Lovelock})$ theories \cite{Love} ---the set also includes all the Einstein-like theories mentioned in the last paragraph of the previous subsubsection. 

\subsubsection{Theories without scalar mode}
The condition for the scalar mode to be absent from the spectrum, $m_s^2\rightarrow \infty$,  reads instead $3c+8b=0$, which is satisfied by theories for which $12\bar f_{R,R}+\bar g=0$. From this we learn that the most general class theories of this kind contributing non-trivially to the linearized equations reads
\begin{equation}
\mathcal{G}|_{\text{no scalar mode}}=f(R)-12f_{R,R}(R)\mathcal{S}_2\, ,\qquad (f_{R,R}(R)\neq 0)\,.
\end{equation}
Again, there is a single order-$n$ density of this kind, corresponding to
\begin{align}
\mathcal{G}^{(n)}|_{\text{no scalar mode}} &=R^n-12 n (n-1) R^{n-2}\mathcal{S}_2 \, , \\  &= [1+4n(n-1)]R^n-12 n (n-1) R^{n-2}\mathcal{R}_2\, .
\end{align}
For $n=2$, the above density is nothing but the New Massive Gravity one \cite{Bergshoeff:2009hq}.  Once again, we can combine the above order-$n$ densities with the $\#(n)-2$ ``trivial'' densities to obtain additional densities which do not propagate the scalar mode. There are then $\#(n)-1$ densities which do not propagate the scalar mode at each order.

Another property of NMG is that it fulfils a holographic c-theorem. In the following section, we will see that, in $D = 3$, all higher-curvature theories which fulfil a simple holographic c-theorem do not propagate the scalar mode.

In higher dimensions, a prototypical example of a theory which satisfies this condition (as well as trivially satisfying a hologarphic c-theorem) is conformal gravity \cite{Hassan:2013pca,Bueno:2016ypa}, which can be thought of as a natural $D$-dimensional extension of NMG. 

In sum, in $D=3$, at any order  $n\geq 2$ we can always decompose the most general linear combination of higher-curvature densities as a sum of a term which by itself would not propagate the massive graviton, plus a term which by itself would not propagate the scalar mode, plus  $\#(n)-2$ densities which do not contribute to the masses of any of them. 


\section{Theories satisfying a holographic c-theorem}\label{sec:ctheorem3D}
Interesting extensions of Einstein and New Massive Gravities to higher orders can be obtained by demanding that the corresponding densities satisfy a simple holographic c-theorem \cite{Sinha:2010ai,Paulos:2010ke}. This set of theories is defined by the property that they yield second-order equations when evaluated on the ansatz
\begin{equation}\label{eq:RGmetric}
\df s^2=\df \rho^2+a(\rho)^2[-\df t^2+\df x^2]\,.
\end{equation}
Supplementing the action with an appropriate stress-tensor, the metric can be made to interpolate between  two asymptotic AdS$_3$ regions \cite{Freedman:1999gp,Girardello:1998pd} which, from the CFT point of view, would represent IR and UV fixed points. Intermediate values of the holographic coordinate are then interpreted as representing the RG flow between both CFTs.

The idea behind the holographic c-theorem\footnote{The c-theorem for general $2d$ CFTs has been proven in \cite{Zamolodchikov:1986gt,Casini:2017vbe}.} involves constructing a function $c(\rho)$ which decreases monotonously along the RG flow, as we move from the UV to the IR. In the present holographic context, the fixed points can be chosen to be  $\rho_{\rm UV}= +\infty$  and   $\rho_{\rm IR}=-\infty$,
so a function satisfying
\begin{equation}
    c'(\rho) \geq 0  \quad \forall\, \rho\, ,
\end{equation}
does the job.
Now, the usual holographic c-theorem construction involves considering a function $c(\rho)$ such  that  $c'(\rho)$ is  proportional to the combination of stress-tensor components $T_t^t-T_{\rho}^{\rho}$. Then, imposing that the stress-tensor satisfies  the null energy   condition, such combination has a sign, namely,
\begin{equation}
    T_t^t-T_{\rho}^{\rho}  \overset{\rm  \ssc NEC}{\leq} 0\, .
\end{equation}
Therefore, any $c(\rho)$ such that  $c'(\rho)\propto -(T_t^t-T_{\rho}^{\rho})$, up to an overall positive-definite constant, satisfies the requirement.

For theories of the type considered above, it is straightforward to construct an appropriate c-function such  that \cite{Freedman:1999gp,Myers:2010xs,Myers:2010tj} 
\begin{equation}
    c'(\rho)=-\frac{a^2}{8G a'^{2}}\,[T_t^t-T_{\rho}^{\rho}]\, .
\end{equation}
This can be obtained from the Wald-like \cite{Wald:1993nt} formula \cite{Sinha:2010ai,Myers:2010tj}
\begin{equation}
    c(\rho)\equiv \frac{\pi a}{2 a'}\frac{\partial \mathcal{L}}{\partial R^{t\rho}\,_{t\rho}}\, ,
\end{equation}
where the Lagrangian derivative components are evaluated on \req{eq:RGmetric}. By construction, $c(\rho)$ coincides with the Virasoro central charges of the fixed-point theories.

As argued in \cite{Paulos:2010ke}, demanding second-order equations for the ansatz \req{eq:RGmetric} for a set of order-$n$ densities amounts at imposing $n-1$ conditions. The idea is to consider the on-shell evaluation of the corresponding Lagrangian densities and impose that neither terms involving derivatives of $a(\rho)$ higher than two nor terms involving powers of $a''(\rho)$ higher than one appear in the resulting expression. This enforces the corresponding equations of motion to be second-order and that a simple c-function can be defined from the above formulas.

As we have shown, there are $\#(n)$ independent densities at order $n$, which means that there are $\#(n)-(n-1)$ independent order-$n$ densities which satisfy a simple holographic c-theorem. Hence, for $n=1,\dots,5$, there is a single such density at each order, but degeneracies start to appear at order six. As observed in \cite{Paulos:2010ke}, it is always possible to write the corresponding linear combination of order-$n$ densities satisfying a simple holographic c-theorem as a single density which has a non-trivial on-shell action when evaluated on \req{eq:RGmetric}, plus densities which simply vanish when evaluated on such ansatz. Hence, we learn that there are $\#(n)-n$ independent  order-$n$ densities which are trivial on the \req{eq:RGmetric} ansatz. Remarkably, as we show below, all such densities of arbitrary orders turn out to be proportional to a single sextic density, which identically vanishes on the metric \eqref{eq:RGmetric}. As for the densities which contribute non-trivially to the holographic c-function we find a new recursive formula which allows for the construction of the corresponding order-$n$ density from the order-$(n-1)$, the order-$(n-2)$, the Einstein gravity and the NMG densities. The recurrence can be solved explicitly, and so we are able to provide an explicit formula for a general order density which non-trivially satisfies the holographic c-theorem. Finally, we explore the relation between such general order density and the one resulting from the expansion of previously proposed Born-Infeld gravities which also satisfy the holographic c-theorem. Naturally, the relation always involves densities trivially satisfying the holographic c-theorem. 

\subsection{Recursive formula}
As we have mentioned, at each order there is a single possible functional dependence on $a(\rho)$ of the on-shell action of theories satisfying the holographic c-theorem. Then, up to terms which do not contribute when evaluated on \req{eq:RGmetric}, there is a unique  such density at each curvature order. The on-shell expressions for $R,\mathcal{S}_2,\mathcal{S}_3$ read
\begin{align}
   \left. R\right|_a 
    = -\frac{2(a'^2 +2 a a'') }{ a^2} \, , \quad
  \left.  \mathcal{S}_2\right|_a 
    = \frac{2(a'^2 - a a'')^2}{3 a^4} \, , \quad
  \left.  \mathcal{S}_3\right|_a 
    = \frac{2(a'^2 - a a'')^3}{9 a^6}\, .
\end{align}
As observed in \cite{Paulos:2010ke}, the on-shell Lagrangian of densities satisfying the holographic c-theorem in a non-trivial fashion follows the simple pattern 
\begin{equation}\label{eq:Lnonshell}
\left. \mathcal{C}_{(n)}\right|_{a}=\left(\frac{a'}{a}\right)^{2(n-1)}\left[\frac{a''}{a}+\frac{3-2n}{2n}\left(\frac{a'}{a}\right)^{2}\right]\, .
\end{equation}
With this choice of normalization, the first three densities read
\begin{align}
\mathcal{C}_{(1)}&=-\frac{1}{4}R\, ,\\
\mathcal{C}_{(2)}&=+\frac{3 R^2}{16}-\frac{\mathcal{R}_2}{2}\\ &=+\frac{R^2}{48}-\frac{\mathcal{S}_2}{2}\, ,\\
\mathcal{C}_{(3)}&=-\frac{17 R^3}{48}+\frac{3 R \mathcal{R}_2}{2}-\frac{4 \mathcal{R}_3}{3}\\ &=-\frac{R^3}{432}+\frac{R \mathcal{S}_2}{6}-\frac{4 \mathcal{S}_3}{3}\, .
\end{align}
Now, an easy way to prove that instances of non-trivial densities actually exist at arbitrarily high orders is by finding a recursive relation.
Since, essentially, these densities are defined by the form of their on-shell Lagrangian on the RG-flow metric (\ref{eq:RGmetric}), we can try to derive such recursive relations by using \req{eq:Lnonshell}. We find the particularly simple relation,
\begin{equation}\label{recuu}
\mathcal{C}_{(n)}=\frac{4(n-1)(n-2)}{3n(n-3)}\left(\mathcal{C}_{(n-1)}\mathcal{C}_{(1)}-\mathcal{C}_{(n-2)}\mathcal{C}_{(2)}\right)\, .
\end{equation} 
This expression allows us to generate holographic c-theorem satisfying densities of arbitrary orders once we know $\mathcal{C}_{(1)}$, $\mathcal{C}_{(2)}$ and $\mathcal{C}_{(3)}$, which are given above. Since $\mathcal{C}_{(4)}$ and $\mathcal{C}_{(5)}$ are unique, this formula should give precisely those densities. This is indeed the case, and we find
\begin{align}
    \mathcal{C}_{(4)} & =+ \frac{41R^4}{384} - \frac{3R^2\mathcal{R}_2}{8} + \frac{2R\mathcal{R}_3}{3} - \frac{\mathcal{R}_2^2}{2}
    \\ &=+ \frac{R^4}{3456} - \frac{R^2\mathcal{S}_2}{24} + \frac{2R\mathcal{S}_3}{3} - \frac{\mathcal{S}_2^2}{2},
\end{align}
and
\begin{align}
    \mathcal{C}_{(5)} & =+ \frac{61R^5}{960} - \frac{7R^3\mathcal{R}_2}{12} + \frac{2R^2\mathcal{R}_3}{15} + \frac{7R\mathcal{R}_2^2}{5} - \frac{16\mathcal{R}_2\mathcal{R}_3}{15}
    \\ & = -\frac{R^5}{25920} + \frac{R^3\mathcal{S}_2}{108} - \frac{2R^2\mathcal{S}_3}{9} + \frac{R\mathcal{S}_2^2}{3} - \frac{16\mathcal{S}_2\mathcal{S}_3}{15},
\end{align}
which agree with the results previously reported in \cite{Sinha:2010ai,Paulos:2010ke}. On the other hand, for $n\ge 6$, the recursion produces a single representative non-trivial density. For example, for $n=6$ ---which is the order at which degeneracies start to appear due to the existence of densities trivially satisfying the holographic c-theorem--- we find from the recursive formula
\begin{align}
    \mathcal{C}_{(6)} & = -\frac{1103 R^6}{20736} + \frac{115R^4\mathcal{R}_2}{288} - \frac{19R^3\mathcal{R}_3}{81} - \frac{71R^2\mathcal{R}_2^2}{108} + \frac{8R\mathcal{R}_2\mathcal{R}_3}{9} - \frac{10 \mathcal{R}_2^3}{27}
   \\ & = +\frac{R^6}{186624} - \frac{5R^4\mathcal{S}_2}{2592} + \frac{5R^3\mathcal{S}_3}{81} - \frac{5R^2\mathcal{S}_2^2}{36} + \frac{8R\mathcal{S}_2\mathcal{S}_3}{9} - \frac{10 \mathcal{S}_2^3}{27}\, .
\end{align}


\subsection{General formula for order-n densities}
Interestingly, it is possible to solve the two-term recurrence relation (\ref{recuu}) analytically and obtain an explicit expression for the order-$n$ density non-trivially satisfying the holographic c-theorem. The result, which takes a simpler form in terms of the $\{R,\mathcal{S}_{2},\mathcal{S}_3\}$ set, reads,
\begin{equation}\label{eq:cdensityn}
\begin{aligned}
\mathcal{C}_{(n)}=\frac{3 (-1)^{n}}{4\cdot 6^n n}\Bigg\{&\left(R+\sqrt{24\mathcal{S}_2}\right)^{n-1}\left(R-(n-1)\sqrt{24\mathcal{S}_2}\right)\left(1-\sqrt{6}\frac{\mathcal{S}_3}{\mathcal{S}_2^{3/2}}\right)\\
+&\left(R-\sqrt{24\mathcal{S}_2}\right)^{n-1}\left(R+(n-1)\sqrt{24\mathcal{S}_2}\right)\left(1+\sqrt{6}\frac{\mathcal{S}_3}{\mathcal{S}_2^{3/2}}\right)
\Bigg\}\, .
\end{aligned}
\end{equation}
Even though this expression may look odd because it depends in a non-polynomial way on the densities, it does reduce to a polynomial expression when we evaluate it for any integer $n\ge 1$. One can check this by expanding the $\left(R\pm \sqrt{24\mathcal{S}_2}\right)^{n-1}$ terms using the binomial coefficients.  In particular, note that this formula is even under the exchange $\mathcal{S}_2^{1/2}\rightarrow- \mathcal{S}_2^{1/2}$, and therefore $\mathcal{S}_2^{1/2}$ always appears with even powers (\textit{i.e.}, there are no square roots). Explicitly, the result of this expansion reads
\begin{equation}\label{CnFormula}
\begin{aligned}
\mathcal{C}_{(n)}=\frac{3 (-1)^{n}}{2\cdot 6^n n}\Bigg\{&\sum_{k=0}^{\lfloor \frac{n}{2} \rfloor}(24 \mathcal{S}_2)^k R^{n-2k}\left[\binom{n-1}{2k}-(n-1)\binom{n-1}{2k-1}\right]\\
&-288\mathcal{S}_3\sum_{k=0}^{\lfloor\frac{n-3}{2}\rfloor}(24 \mathcal{S}_2)^k R^{n-3-2k}\left[\binom{n-1}{2k+3}-(n-1)\binom{n-1}{2k+2}\right]\Bigg\}\, ,
\end{aligned}
\end{equation}
which is valid whenever $n\in \mathbb{N}$. 

Interestingly, the density \eqref{eq:cdensityn} can also be applied for non-integer $n$, since it always yields the result \eqref{eq:Lnonshell} when evaluated on the metric \eqref{eq:RGmetric}, and therefore it yields second-order equations for the RG-flow metric.  Hence, these Lagrangians provide a generalization of the holographic c-theorem-satisfying densities for arbitrary real values of $n$.


\subsection{All densities with a trivial c-function emanate from a single sextic density }
For the first five curvature orders, there exists a single density which satisfies the holographic c-theorem condition. Now, for $n=6$, there exists an additional density,
\begin{align}\label{Omeg6}
\Omega_{(6)} &\equiv 6 \mathcal{S}_3 ^2-\mathcal{S}_2 ^3\\ &=\frac{1}{3} \left[R^6-9R^4\mathcal{R}_2+8R^3\mathcal{R}_3+21R^2\mathcal{R}_2^2-36R\mathcal{R}_2\mathcal{R}_3-3\mathcal{R}_2^3+18\mathcal{R}_3^2\right]\, ,
\end{align}
with the property of being identically vanishing when evaluated on the c-theorem metric (\ref{eq:RGmetric}) and which therefore does not
contribute to the equations of motion for that ansatz. 

An immediate consequence is that any product of $\Omega_{(6)} $ with any other density also satisfies the holographic c-theorem trivially.
 Therefore, for $n\ge 6$ we have, at least, the following set of densities which satisfy the holographic c-theorem
\begin{equation}\label{cThLs}
\mathcal{L}_{(n)}^{\rm c-theorem}=\alpha_n \mathcal{C}_{(n)}+\Omega_{(6)}\cdot \mathcal{L}^{\rm general}_{(n-6)}\, ,
\end{equation}
where $\mathcal{L}^{\rm general}_{(n-6)}$ is the general Lagrangian of order $n-6$ in the curvature. Remarkably, these are all the densities of this type that exist. 

This can be proven as follows. First, observe that there exist $\#(n-6)$ densities of order $n-6$. Hence, there exists the same number of order-$n$ densities in the set $\Omega_{(6)} \cdot \mathcal{L}^{\rm general}_{(n-6)}$. Now, as observed earlier, there exist $\#(n)-(n-1) $ independent order-$n$ densities which satisfy the holographic c-theorem, one of which does so in a non-trivial fashion. The latter can be chosen to be $\mathcal{C}_{(n)}$ and we are left with $\#(n)-n$ independent densities which trivially satisfy the holographic c-theorem. Now, invoking the result in \req{properttt}, we observe that this number exactly matches the number of densities in the set $\Omega_{(6)} \cdot \mathcal{L}^{\rm general}_{(n-6)}$.

 In sum, $\mathcal{L}_{(n)}^{\rm c-theorem}$ as defined above is the most general higher-curvature order-$n$ density satisfying the holographic c-theorem and all densities satisfying it in a trivial fashion emanate from the sextic density $\Omega_{(6)}$. This is a rather intriguing result which suggests that there may be something more fundamentally special about this density. As a matter of fact, this will not be the last time we encounter it.


\subsection{Absence of scalar mode in the spectrum}
An immediate consequence of \req{cThLs} is that none of the densities trivially satisfying the holographic c-theorem contributes to the linearized equations around an Einstein metric. This is because all densities involved take the form $\Omega_{(6)} \cdot \mathcal{L}^{\rm general}_{(n-6)}$  and therefore belong to the set $\mathcal{G}|_{\text{trivial linearized equations}}$ as defined in \req{gtriv}, since $\Omega_{(6)}$ can be written in terms of only $\mathcal{S}_2$ and $\mathcal{S}_3$ as seen in \req{Omeg6}.

On the other hand, we can use our previous results to prove that densities which satisfy the holographic c-theorem in a non-trivial fashion do not incorporate the scalar mode in their spectrum. This latter property seems to have been observed in certain particular cases \cite{Afshar:2014ffa} but we have found no general proof in the literature.

We saw in section \ref{sec:lineq} that the condition for the absence of the scalar mode in the linearized spectrum, $m_s^2 \to \infty$, was satisfied by theories of the form
\begin{equation}
\mathcal{G}(R,\mathcal{S}_2,\mathcal{S}_3) = f(R) + \mathcal{S}_2 g(R) + \mathcal{G}_{\rm triv}\,,
\end{equation}
for which
\begin{equation}\label{12frr}
    12 \bar{f}_{R,R} + \bar{g} = 0\,.
\end{equation}

For theories where $\mathcal{G}(R,\mathcal{S}_2,\mathcal{S}_3)$ is a polynomial, as the ones we are considering, this cancellation must occur order by order. At any given order $n$, the only possible forms of $f$ and $g$ are $f_{(n)}(R) = \lambda_{(n)} R^n$ and $g_{(n)}(R) = \mu_{(n)}R^{n-2}$ for some constants $ \lambda_{(n)}$ and $ \mu_{(n)}$.
Therefore, at any order $n$, condition \eqref{12frr} reads 
\begin{equation}\label{cTnoS}
    12n(n-1)\lambda_{(n)} + \mu_{(n)} = 0\,,
\end{equation}
where $\lambda_{(n)} = \beta_{n00}$ is the coefficient in front of the $R^n$ term and $\mu_{(n)} = \beta_{(n-2)10}$ is the coefficient in front of the $\mathcal{S}_2R^{n-2}$ term in the series expansion of $\mathcal{G}(R,\mathcal{S}_2,\mathcal{S}_3)$ given in \req{Gseries}.

Now, it is easy to see that our $\mathcal{C}_{(n)}$ densities fulfil this condition.
Expanding eq. \eqref{CnFormula} and keeping only the terms with $k = 0,1$ in the first sum, we see
\begin{equation}
    \mathcal{C}_{(n)} = \frac{3(-1)^n}{2\cdot6^nn} \Big\{ R^n - 12n(n-1) \mathcal{S}_2 R^{n-2} + \cdots \Big\}\,,
\end{equation}
and so 
\begin{equation}
    \lambda_{(n)} = \frac{3(-1)^n}{2\cdot6^nn}\,, \quad \quad \mu_{(n)} = - 12n(n-1)\frac{3(-1)^n}{2\cdot6^nn}\,,
\end{equation}
which clearly fulfil condition \eqref{cTnoS}. This proves that all theories satisfying the holographic c-theorem have a linearized spectrum which does not include the scalar mode.


\subsection{Born-Infeld gravity }

It was proposed in \cite{Gullu:2010pc} that New Massive Gravity could also be extended through a Born-Infeld gravity theory with Lagrangian density
\begin{equation}\label{BINMG}    \mathcal{L}_{\text{BI-NMG}} = \sqrt{\det \left( \delta_a^b + \frac{\sigma}{m^2} G_a^b \right) }  - \left( 1 - \frac{\Lambda}{2m^2} \right)\,,
\end{equation}
where $G_{ab} = R_{ab} - \frac{1}{2}g_{ab}R$ is the Einstein tensor and $\sigma = \pm 1$. This theory reproduces NMG when expanded to quadratic order in the curvature. Then, after \cite{Sinha:2010ai} proved that both NMG and the cubic order term of eq. \eqref{BINMG} admitted a holographic c-function, it was soon proven in \cite{Gullu:2010st} that the full theory also satisfied a simple holographic c-theorem of the same kind as the one described in the previous subsections. The cancellations on the on-shell evaluation of these theories required by the c-theorem construction occur order by order, and so the theory defined by eq. \eqref{BINMG} generates an infinite number of higher-curvature densities which non-trivially fulfil a holographic c-theorem at any truncated order \cite{Alkac:2018whk}.

Now, in view of our results, it would be interesting to know whether the terms generated by the expansion of \eqref{BINMG} order by order, which we shall call $\mathcal{B}_{(n)}$, are the same ones as the non-trivial densities $\mathcal{C}_{(n)}$ generated by the recursive formula \eqref{recuu}.
Following what we have just learned in the previous section, that should indeed be the case for $n = 1,\ldots,5$. For $n\geq6$ we expect both sets of densities to coincide up to ``trivial'' densities, and we find that to be the case.

Let us expand the density \eqref{BINMG}. We set $\sigma = 1$ and $m^2 = 1$ for simplicity, as they can be easily restored by dimensional analysis. In 3 dimensions the determinant of any matrix can be computed as 
\begin{equation}
    \det (A) = \frac{1}{6} \left[ \left( \tr (A) \right)^3 - 3 \tr (A) \tr (A^2) + 2 \tr(A^3) \right]\,.
\end{equation}
In our case, we have $A = \mathds{1} + g^{-1}G$, which gives
\begin{equation}
    \det (\mathds{1} + g^{-1}G ) = 1 + \frac{-1}{2}R + \frac{1}{4} \mathcal{T}_2 + \frac{1}{24}\mathcal{T}_3\,,
\end{equation}
where we have defined
\begin{align}
    \mathcal{T}_2 & \equiv R^2 - 2 \mathcal{R}_2 = \frac{1}{3} R^2 - 2 \mathcal{S}_2\,,
    \\
    \mathcal{T}_3 & \equiv  R^3 - 6 R \mathcal{R}_2 + 8 \mathcal{R}_3 = \frac{-1}{9}R^3 + 2 R \mathcal{S}_2 + 8 \mathcal{S}_3\,.
\end{align}
We can now simply Taylor expand the square root, $\sqrt{1+x} = \sum_{m=0}^{\infty} \binom{1/2}{m} x^m$, with $x = \det (\mathds{1} + g^{-1}G )-1$ and then collect the relevant terms at each order $n$ to build $\mathcal{B}_{(n)}$. The result is the following,
\begin{equation}\label{Bn}
    \mathcal{B}_{(n)} = \hat{\sum} \binom{1/2}{i+j+k} \frac{(i+j+k)!}{i!j!k!}\left( \frac{-1}{2}R \right)^i\left( \frac{1}{4}\mathcal{T}_2 \right)^j \left( \frac{1}{24} \mathcal{T}_3 \right)^k\,,
\end{equation}
where the sum is performed over the indices $i,j,k$ that fulfil the integer partition $n = i+2j+3k$.

The lowest order densities given by the formula above are
\begin{equation}
    \mathcal{B}_{(1)}  = \mathcal{C}_{(1)}\,,\quad 
    \mathcal{B}_{(2)}   = \frac{1}{2}\mathcal{C}_{(2)}\,, \quad 
    \mathcal{B}_{(3)}  =  -\frac{1}{8}\mathcal{C}_{(3)}\,, \quad 
    \mathcal{B}_{(4)}  = \frac{1}{16}\mathcal{C}_{(4)}\,, \quad 
    \mathcal{B}_{(5)}  = -\frac{5}{128}\mathcal{C}_{(5)}\,,
\end{equation}
which are indeed proportional to the densities $\mathcal{C}_{(n)}$ found previously through the recursion relation \eqref{recuu}, as expected.
At the next orders, however, eq. \eqref{Bn} gives a different non-trivial density than the one given by the recursion relation \eqref{recuu}. Following eq. \eqref{cThLs}, we see that the relationship between the densities $\mathcal{B}_{(n)}$ and $\mathcal{C}_{(n)}$ at orders $n \geq 6$ is given by
\begin{equation}
    \mathcal{B}_{(n)} = (-1)^n \frac{(2n-5)!!}{(2(n-1))!!} \mathcal{C}_{(n)} + \Omega_{(6)} \cdot \mathcal{L}_{(n-6)}\,,
\end{equation}
for some particular densities $\mathcal{L}_{(n-6)}$.
For example,
\begin{align}
    \mathcal{B}_{(6)} & = \frac{7}{256} \mathcal{C}_{(6)} - \frac{1}{432} \Omega_{(6)}\,, \\
     \mathcal{B}_{(7)} & = -\frac{21}{1024} \mathcal{C}_{(7)} - \frac{R}{576} \Omega_{(6)}\,, \\
     \mathcal{B}_{(8)} & = \frac{33}{2048} \mathcal{C}_{(8)} - \frac{11R^2 + 24\mathcal{S}_2}{13824} \Omega_{(6)}\,.
\end{align}
Hence, both $\mathcal{C}_{(n)} $ and $\mathcal{B}_{(n)} $ provide sets of order-$n$ densities which non-trivially satisfy the holographic c-theorem. While the $\mathcal{C}_{(n)} $ are distinguished by the property of satisfying the simple recurrence relation (\ref{recuu}), the $\mathcal{B}_{(n)} $ have the property of corresponding to the general term in the expansion of the Born-Infeld theory (\ref{BINMG}). Both sets are equal up to terms which identically vanish in the holographic c-theorem ansatz which, as we have seen, are all proportional to the density $\Omega_{(6)}$.

Another Born-Infeld theory has been proposed as a non-minimal extension of NMG \cite{Alkac:2018whk}, with Lagrangian density
\begin{equation}\label{nMBI}
    \mathcal{L}_{\text{nM-BI}} = \sqrt{\det \left( \delta_a^b - \frac{2}{m^2} P_a^b + \frac{1}{m^4} P_a^c P_c^b \right) }  - \left( 1 - \frac{\Lambda}{2m^2} \right)\,,
\end{equation}
where $P_a^b = R_a^b - \frac{1}{4}\delta_a^b R$ is the Schouten tensor. The full theory also allows for a holographic c-function. However, when expanded order by order using a similar method as the one described above, we see that it does not produce an infinite number of higher-curvature densities which non-trivially fulfil a holographic c-theorem. At order $n=2$ and $n=3$ we obtain terms proportional to $\mathcal{C}_{(2)}$ and $\mathcal{C}_{(3)}$, as expected, but the terms with $n \geq 4$ all trivialize due to the Schouten identities described in section 2. 
Therefore, the density \eqref{nMBI} is equivalent to the much simpler density 
\begin{equation}\label{nMBIv2}
    \mathcal{L} = R - 2\Lambda + \frac{2}{m^2} \left( \mathcal{R}_2 - \frac{3}{8}R^2 \right) + \frac{1}{m^4} \left( \frac{17}{48} R^3 - \frac{3}{2}R \mathcal{R}_2 + \frac{4}{3} \mathcal{R}_3 \right)\,.
\end{equation}


\section{Generalized Quasitopological gravities}\label{sec:GQTss}
A different classification criterion which has attracted a lot of attention in higher dimensions entails considering higher-curvature theories which admit generalizations of the $D$-dimensional Schwarzschild black hole characterized by a single function, \ie satisfying $g_{tt}g_{rr}=-1$. Theories of this kind have been coined ``Generalized quasitopological gravities'' (GQTs) \cite{Hennigar:2017ego,PabloPablo3,Bueno:2019ycr}, and include Quasitopological \cite{Quasi2,Quasi,Dehghani:2011vu,Cisterna:2017umf,Myers:2010jv} and Lovelock gravities \cite{Lovelock1,Lovelock2} as particular cases. 

Given a $D$-dimensional higher-curvature gravity with Lagrangian density $\mathcal{L}(R_{abcd},g^{ef})$, let  $L_{f}$ be the effective Lagrangian 
obtained from the evaluation of $\sqrt{|g|}\mathcal{L}$ in the ansatz
\begin{equation}\label{fmetric}
\df s^2=f(r)\df t^2+\frac{\df r^2}{f(r)}+\df r^2 \df \Omega_{(D-2)}^2 \, .
\end{equation}
Then, we say that $\mathcal{L}$ is a GQT gravity if
\begin{equation}\label{eq:GQTcond}
\frac{\partial L_f}{\partial f}-\frac{\df}{\df r}\frac{ \partial L_f}{\partial f'}+\frac{\df^2}{\df r^2}\frac{\partial L_f}{\partial f''}=0\, ,
\end{equation} 
namely, if the Euler-Lagrange equation of $f(r)$ identically vanishes \cite{PabloPablo3}. This is equivalent to requiring that $L_f$ is a total derivative, \ie 
\begin{equation} \label{totder}
L_f= \frac{\df F_0}{\df r} \, , \quad \text{ for some function} \quad F_0\equiv F_0[r,f(r),f'(r)]\, .
\end{equation}
Theories satisfying these requirements satisfy a number of interesting properties, such as admitting non-hairy generalizations of the Schwarzschild AdS$_D$ solution characterized by a single function or possessing a linearized  spectrum around maximally symmetric backgrounds identical to the Einstein gravity one. For a more detailed summary of the properties satisfied by GQTs see \eg \cite{Bueno:2019ltp}. In the same reference it has been proven that any gravitational effective action in $D\geq 4$ can be mapped via field redefinitions to a GQT.

Here we are interested in exploring the possible existence of GQTs in three dimensions. In order to do that, we need to determine the set of densities for which eq. (\ref{eq:GQTcond}) holds, if any. As a first step, we need to evaluate our fundamental building-block densities on such ansatz. Defining the quantities
\begin{equation}
A\equiv \frac{f''}{2},\quad B\equiv -\frac{f'}{2r},\quad \psi\equiv \frac{1-f}{r^2},
\end{equation}
which are the only functional dependences on $f(r)$ appearing in the Riemann tensor, we find
 \begin{align}
 R\big|_f= -2A+4B\, , \quad  \mathcal{S}_2\big|_f =\frac{2}{3}(A+B)^2 \, , \quad \mathcal{S}_3\big|_f =\frac{2}{9}(A+B)^3 \, , 
 \end{align}
 and 
\begin{equation}
\mathcal{R}_2\big|_f=2(A^2-2A B+3B^2)\, , \quad \mathcal{R}_3\big|_f =-2(A^3-3A^2 B+3A B^2+5B^3) \, .
\end{equation}
We immediately observe the absence of $\psi$ in these expressions, which means that three-dimensional on-shell Lagrangians do not depend on the function $f(r)$ explicitly, but only on its first and second derivatives.
Hence, in this case the GQT condition \eqref{eq:GQTcond} becomes simpler, namely,
\begin{equation}\label{integ}
\frac{ \partial L_f}{\partial f'}=\frac{\df}{\df r}\frac{\partial L_f}{\partial f''}+c\, ,
\end{equation}
where $c$ is an integration constant.

Evaluating on-shell a general order-$n$ density in the $\{R,\mathcal{S}_2,\mathcal{S}_3\}$ basis, we find
\begin{align}
L_{(n),f} &=r L^{2(n-1)} \sum_{j,k} \beta_{n-2j-3k,j,k} \frac{(-1)^{n-2j-3k}}{6^j 36^k} \left[ f''+\frac{2f'}{r}\right]^{n-2j-3k} \left[f''-\frac{f'}{r}\right]^{2j+3k} \\
& =r L^{2(n-1)} \sum_{j,k,l,m} c_{jklm} {f''}^{(l+m)} \left[\frac{f'}{r}\right]^{n-m-l}\,,
\end{align}
where $L_{(n),f}\equiv \sqrt{g}\mathcal{L}_{(n)}|_{f}$ and where we used the binomial expansion twice in the second line and defined the constants
\begin{equation}
c_{jklm}\equiv \frac{ \beta_{n-2j-3k,j,k}}{6^{j+2k}}  (-1)^{n-m} 2^{n-2j-3k-l}\binom{n-2j-3k}{l}\binom{2j+3k}{m} \, .
\end{equation}
The combination $l+m$ takes integer values from $0$ to $n$, and hence $L_{(n),f}$ can be written as a linear combination of terms with different powers of  $f''$ taking such values. Now, in order for  $L_{(n),f}$ to be a total derivative as required by \req{totder}, we need to impose that all terms involving powers of $f''$ higher than one vanish. This implies imposing $n-1$ conditions on the coefficients $\beta_{n-2j-3k,j,k}$. Once this is done, we are left with
\begin{align}
L_{(n),f} = g_1[r,f'(r)] + g_2[r,f'(r)] f''(r)\, , 
\end{align}
where
\begin{equation}
g_1\equiv r L^{2(n-1)} \sum_{j,k}c_{jk00} \left[\frac{f'}{r}\right]^{n}\, , \quad g_2\equiv r L^{2(n-1)} \sum_{j,k}(c_{jk10} +c_{jk01})  \left[\frac{f'}{r}\right]^{n-1}\, .
\end{equation}
However, the fact that $L_{(n),f}$ is linear in $f''(r)$ does not guarantee that $L_{(n),f}$ is a total derivative. In order for this to be the case, we need to impose the additional condition  given by \req{integ} which, in terms of $g_1$ and $g_2$ becomes 
\begin{equation}
\frac{\partial g_1}{\partial f'}=\frac{\partial g_2}{\partial r}\, .
\end{equation}
Explicitly, this condition becomes
\begin{equation}\label{condime}
n \sum_{j,k}c_{jk00} = (n-2) \sum_{j,k} (c_{jk10}+c_{jk01})\, ,
\end{equation}
which in terms of the original $\beta_{ijk}$ coefficients reads,
\begin{equation}
\sum_{j,k} \frac{\beta_{n-2j-3k,j,k}}{6^{j+2k}} 2^{n-2j-3k-1}[2-n+6j+9k]=0\, .
\end{equation}
Adding this to the $n-1$ conditions imposed earlier, we find a total of $n$ conditions to be imposed to $L_{(n),f} $ in order for it to be a GQT density. Hence, we have $\#(n)-n=\#(n-6)$ GQT densities at order $n$. 

\subsection{All GQT densities emanate from the same sextic density }
 Interestingly, the number of order-$n$ GQT densities exactly coincides with the number of densities trivially satisfying the holographic c-theorem. More remarkably, the two sets of densities are, in fact, identical: the special sextic density $\Omega_{(6)}$ defined in \req{Omeg6} as the source of all densities trivially satisfying the holographic c-theorem turns out to be also the source of all GQT densities. Indeed, it is not difficult to see that 
 \begin{equation}
\left.  \Omega_{(6)}\right|_f=0\, ,
 \end{equation}
 which means that all densities involving $\Omega_{(6)}$ identically vanish and are therefore ``trivial'' GQT densities ---in the sense that they make no contribution to the equation of $f(r)$. Since there are $\#(n-6)$ of such densities, we learn that in fact all GQT densities in three dimensions are ``trivial'' and proportional to $\Omega_{(6)}$, 
  \begin{equation}
\mathcal{L}_{(n)}^{\rm GQT}=\Omega_{(6)}\cdot \mathcal{L}^{\rm general}_{(n-6)}\, ,
\end{equation}
where $\mathcal{L}^{\rm general}_{(n-6)}$ is the most general order-$(n-6)$ density.

In sum, we learn that, in three dimensions, there exist no non-trivial GQTs. This situation is very different from higher-dimensions: in $D=4$ there exists one independent non-trivial GQT density for every $n\geq 3$ whereas for $D\geq 5$ there actually exist $n-1$ independent inequivalent GQT densities for every $n$ ---namely, there exist $n-1$ densities of order $n$ each of which makes a functionally different contribution to the equation of $f(r)$ \cite{Bueno:2022res}. As a matter of fact, the triviality of the three-dimensional case unveiled here is not so surprising given that all higher-curvature theories admit the BTZ solution (since it is locally AdS$_3$) ---as opposed to non-trivial GQTs in higher dimensions, which admit modifications of Schwarzschild as solutions, but not Schwarzschild itself. %


\section{A ``mysteriously'' simple sextic density}\label{sec:Ommm6}
We have seen that all GQT densities as well as all densities trivially satisfying the holographic c-theorem  emanate from a single sextic density, $\Omega_{(6)}$, defined in \req{Omeg6}. The reason for such occurrence can be understood as follows. As we saw earlier, when evaluated on-shell on \req{eq:RGmetric} and \req{fmetric} respectively, the densities $R,\mathcal{S}_2,\mathcal{S}_3$ read\footnote{As a matter of fact, these two ans\"atze have been previously considered simultaneously before in the four-dimensional case \cite{Arciniega:2018fxj,Arciniega:2018tnn,Cisterna:2018tgx} in a cosmological context. The reason is that the condition for demanding a simple holographic c-theorem can be alternatively understood as the condition that the equations of motion for the scale factor in a standard Friedmann-Lema\^itre-Robertson-Walker ansatz  are second order \cite{Sinha:2010pm}.   }
\begin{align}
  \left. R\right|_a      &=   -\frac{2(a'^2 +2 a a'') }{ a^2} \, , \quad
  \left.  \mathcal{S}_2\right|_a 
    = \frac{2(a'^2 - a a'')^2}{3 a^4} \, , \quad
  \left.  \mathcal{S}_3\right|_a 
    = \frac{2(a'^2 - a a'')^3}{9 a^6}\, .\\
\left. R\right|_f &=-\frac{1}{r}\left(r f''+2f'\right) \, , \quad  \left. \mathcal{S}_2\right|_f=\frac{1}{6r^2}\left(r f''- f'\right)^2 \, , \quad  \left.  \mathcal{S}_3\right|_f=\frac{1}{36r^3}\left(r f''- f'\right)^3 \, .
\end{align}
Observe that $\mathcal{S}_2$ and $\mathcal{S}_3$ have in both cases the same functional dependence on $f(r)$ and $a(\rho)$, respectively, up to a power, whereas $R$ has a different dependence from the other two densities in both cases. Now, for $n\leq 5$, there is no way to construct linear combinations of the various order-$n$ densities such that the resulting expression identically vanishes. This is not the case for $n=6$. In that case, $\mathcal{S}_2^3$ and $\mathcal{S}_3^2$ have exactly the same functional dependence on $f(r)$ and $a(\rho)$ respectively, and a particular linear combination of them can be found such that it identically vanishes. This combination is precisely $\Omega_{(6)}$ in both cases,
\begin{equation}
\Omega_{(6)} = 6 \mathcal{S}_3 ^2-\mathcal{S}_2 ^3\, , \quad \left. \Omega_{(6)} \right|_{a}=\left. \Omega_{(6)} \right|_{f}=0\, .
\end{equation}
It is obvious that any density multiplied by $\Omega_{(6)}$ will similarly vanish for these two ans\"atze. One could wonder what happens for other values of $n$ such as $n=12,18,\dots$, for which there is again a match in the functional dependence of the seed densities to the corresponding powers. It is however easy to see that in those cases the combinations which vanish are precisely the ones given by powers of $\Omega_{(6)}$.

It is natural to wonder whether $\Omega_{(6)}$ may actually vanish identically for general metrics. This is however not the case. For instance, for a general static black hole ansatz 
\begin{equation}
\df s^2=-N^2(r)f(r)\df t^2 +\frac{ \df r^2}{f(r)}+r^2 \df \phi^2\, ,
\end{equation}
one finds that  $\Omega_{(6)}$ is a complicated function of $f(r)$ and $N(r)$. 

As it turns out, the particular linear combination $6\mathcal{S}_3 ^2-\mathcal{S}_2^3$ appearing in $\Omega_{(6)}$ is, in fact, connected to the Segre classification of three-dimensional spacetimes \cite{Hall,Sousa:2007ax}. This classification consists in characterizing the different types of metrics according to the eigenvalues of the traceless Ricci tensor $\tilde R_{ab}$. There exist three large sets of metrics which are precisely characterized by the relative values of  $6\mathcal{S}_3 ^2$ and $\mathcal{S}_2^3$, namely \cite{Chow:2009km,Gurses:2011fv,Chow:2009vt}: 
\begin{align}
\rm{Group\,\, 1:} \quad &6\mathcal{S}_3 ^2=\mathcal{S}_2^3=0\, , \quad &[\text{Type-O, Type-N, Type-III}]\,, \\ \label{group2}
\rm{Group\,\, 2:} \quad &6\mathcal{S}_3 ^2=\mathcal{S}_2^3 \neq 0\, , \quad &[\text{Type-D}_s \text{, Type-D}_t \text{, Type-II}]\,, \\
\rm{Group\,\,  3:} \quad &6\mathcal{S}_3 ^2 \neq \mathcal{S}_2^3\, , \quad &[\text{Type-I}_{\mathbb{R}}, \text{Type-I}_{\mathbb{C}}]\,.
\end{align}
The first group, which in the ---perhaps more familiar--- Petrov notation includes Type-O, Type-N and Type-III spacetimes corresponds to spacetimes such that both $\mathcal{S}_3$ and $\mathcal{S}_2$ vanish. The second group, which includes Type-D and Type-II spacetimes is the one corresponding to metrics which have non-vanishing $6\mathcal{S}_3^2$ and $\mathcal{S}_2^3$ but such that they are equal to each other, \ie such that $\Omega_{(6)}=0$. Finally, spacetimes of Type-I  have a non-vanishing $\Omega_{(6)}$. 

Metrics of the Group 1 have traceless Ricci tensors which can be written as
\begin{align}
&\tilde R_{ab}=0\,, \quad &[\text{Type-O}]\, , \\ 
&\tilde R_{ab}= s \lambda_a \lambda_b\, , \quad &[\text{Type-N}]\, ,\\
&\tilde R_{ab}= 2 s \xi_{(a}\lambda_{b)}\, , \quad & [\text{Type-III}]\, ,
 \end{align}
where 
\begin{equation}
\quad g^{ab}\lambda_a\lambda_b=0\, , \quad  g^{ab}\xi_a\xi_b=1\, , \quad  g^{ab}\lambda_a\xi_b=0\, , \quad s=\pm 1 \,.
\end{equation}
On the other hand, for metrics of the Group 2 we have 
\begin{align}\label{typeD}
&\tilde R_{ab}=p(x^a) \left[g_{ab}-\frac{3}{\sigma} \xi_a \xi_b\right]\,, \quad &[\text{Type-D}_{s,t}]\, , \\ 
&\tilde R_{ab}=p(x^a) \left[g_{ab}-\frac{3}{\sigma} \xi_a \xi_b\right] + s \lambda_a\lambda_b\,, \quad &[\text{Type-II}]\, , 
 \end{align}
where $p(x^a)$ are scalar functions and
\begin{equation}
\quad g^{ab}\lambda_a\lambda_b=0\, , \quad  g^{ab}\xi_a\xi_b=\sigma=\pm 1\, , \quad  g^{ab}\lambda_a\xi_b=0\, , \quad s=\pm 1 \,.
\end{equation}
Finally, metrics of the Group 3 satisfy
\begin{align}
&\tilde R_{ab}=p(x^a) \left[g_{ab}-3 \xi_a \xi_b\right] -q(x^a) [\lambda_a\lambda_b+\nu_a\nu_b]\,, \quad &[\text{Type-I}_{\mathbb{R}}]\, , \\ 
&\tilde R_{ab}=p(x^a) \left[g_{ab}-3 \xi_a \xi_b\right] -q(x^a) [\lambda_a\lambda_b-\nu_a\nu_b]\,, \quad &[\text{Type-I}_{\mathbb{C}}]\, ,
\end{align}
where $p(x^a)$ and $q(x^a)$ are scalar functions (such that $q\neq \pm 3 p$ for Type-I$_{\mathbb{R}}$) and where
\begin{equation}
\quad g^{ab}\lambda_a\lambda_b=0\, , \quad  g^{ab}\xi_a\xi_b=1\, , \quad  g^{ab}\nu_a\nu_b=0\, ,\quad  g^{ab}\lambda_a\xi_b=g^{ab}\lambda_a\nu_b=0\, , \quad g^{ab}\lambda_a\nu_b=-1\, .
\end{equation}

For the single-function black hole metric and the holographic c-theorem metric, one finds that the traceless Ricci tensor satisfies \req{typeD} with
\begin{align}
p(r)=\frac{f'(r)-r f''(r)}{6r}\, , \quad \xi_a =r \delta_{a\phi} \, ,\quad \text{and} \quad
p(\rho)=\frac{a''(\rho) a(\rho)-a'(\rho)^2}{3a(\rho)^2}\, , \quad \xi_a = \delta_{ar} \, ,
\end{align}
respectively. Hence, both spacetimes are of Type-D$_s$ and from \req{group2} it follows that $\Omega_{(6)}=0$ in both cases.

From this we learn that the appearance of $\Omega_{(6)}$ as a distinguished density was to be expected for the classes of metrics considered here, and that a similar phenomenon is likely to occur for all metrics of the Types D and II.\footnote{For various papers classifying and obtaining explicit solutions of the Groups 1 and 2 for three-dimensional higher-curvature gravities, see \cite{Chow:2009km,Gurses:2011fv,Chow:2009vt,Ahmedov:2010uk,Ahmedov:2011yd,Ahmedov:2012di,Alkac:2016xlr,Alkac:2017rxr}.} Still, the fact that all densities satisfying the holographic c-theorem in a trivial fashion and that all GQTs are proportional to this density was far from obvious in advance.

We believe it would be interesting to further study the properties of $\Omega_{(6)}$, understood as a higher-curvature density. 
Its equations of motion can be easily computed using expression \eqref{eq:EOM3}, and read
\begin{align}
\frac{\mathcal{E}_{ab}^{\Omega_{(6)}}}{3}=&-\frac{1}{6}g_{ab}\left(6\mathcal{S}_3^2-\mathcal{S}_2^3\right)-2\mathcal{S}_2^2\tilde{R}_a^c\tilde{R}_{bc}+12\mathcal{S}_3\tilde{R}_a^c\tilde{R}_{cd}S^d_b\notag\\
&-4\left(g_{ab}\Box-\nabla_a\nabla_b+\tilde{R}_{ab}+\frac{1}{3}g_{ab}R\right)\mathcal{S}_2\mathcal{S}_3-g_{ab}\nabla_c\nabla_d\left(\mathcal{S}_2^2\tilde{R}^{cd}-6\mathcal{S}_3\tilde{R}^{cf}\tilde{R}_f^d\right)\notag\\
&-\left(\Box+\frac{2}{3}R\right)\left(\mathcal{S}_2^2\tilde{R}_{ab}-6\mathcal{S}_3\tilde{R}_a^c\tilde{R}_{bc}\right) +2\nabla_c\nabla_{(a}\left(\tilde{R}_{b)}^c\mathcal{S}_2^2-6\tilde{R}_{b)}^d\tilde{R}_d^c\mathcal{S}_3\right)\, .
\end{align}
These are identically satisfied by all metrics belonging to the Groups 1 and 2, but not for Type-I metrics. It would be then within such set that non-trivial solutions would arise. 


\section{Final comments}
In this chapter, we have presented several new results involving general higher-curvature gravities in three dimensions. 
A summary of our findings can be found at the end of Section \ref{sec:Intro3D}. Let us close with a couple of possible directions which would be in the spirit of the results presented here.

In \cite{Alkac:2020zhg} it was shown that removing the terms involving Weyl tensors of the $D$-dimensional Lovelock densities and taking the $D\rightarrow 3$ limit in the remainder Ricci parts, one is left with three-dimensional densities which precisely match the ones satisfying a holographic c-theorem \cite{Sinha:2010ai,Paulos:2010ke}. The procedure is applied to the $n=2,3,4$ densities, for which no ``trivial'' densities exist. It would be interesting to explore which particular combination of non-trivial densities is selected by this procedure for higher-order densities and what their relation is with the densities $\mathcal{C}_{(n)}$ and $\mathcal{B}_{(n)}$ identified here.

In this paper we have shown that no non-trivial GQT gravities exist in three dimensions. The situation changes when matter fields are included in the game. Following higher-dimensional inspiration \cite{Cano:2020qhy}, in \cite{Bueno:2021krl} they found a family of non-trivial theories linear in the Ricci tensor coupled to a scalar field which becomes a total derivative when evaluated in \req{fmetric} with a magnetic ansatz for the scalar. These ``electromagnetic quasi-topological gravities'' possess solutions which are continuous extensions of the BTZ black hole ---some of which describe regular black holes without any fine-tuning of parameters. 
 In this context, we expect that more theories of the ``electromagnetic generalized quasi-topological''  class should exist when terms involving more Ricci curvatures are considered.
 
Another venue involves the problem of finding theories with reduced-order traced equations. As we mentioned in the introduction, this was explored in \cite{Oliva:2010zd}, where it was shown that NMG is the only quadratic and/or cubic density which has traced equations of second order. Within the same group of densities, it was shown that $C_{abc}C^{abc}$, where $C_{abc}$ is the Cotton tensor, is the only one which has third-order traced equations. The condition is essentially related to the vanishing of the terms involving two explicit covariant derivatives in the field equations (\ref{eomsss}). For theories involving no explicit covariant derivatives in the action  ---as in \req{action3} --- the trace of the equations (\ref{eq:EOM2}) reads
\begin{equation}
-\frac{1}{2}R-\frac{3}{L^2}-\frac{3}{2}\mathcal{F} + R \mathcal{F}_R +2\mathcal{R}_2 \mathcal{F}_{\mathcal{R}_2}+3 \mathcal{R}_3 \mathcal{F}_{\mathcal{R}_3} +\nabla_a\nabla_b Y^{a b}=0,
\end{equation}
where 
\begin{equation}
Y^{a b}\equiv R^{ab}\mathcal{F}_{\mathcal{R}_2}+\frac{3}{2}R^{ac}R_c ^b \mathcal{F}_{\mathcal{R}_3}+g^{ab}\left(2 \mathcal{F}_R+ R \mathcal{F}_{\mathcal{R}_2}+\frac{3}{2}\mathcal{R}_2 \mathcal{F}_{\mathcal{R}_3}\right).
\end{equation}
Hence, theories with traced equations of second order would be those for which the rank-two symmetric tensor $Y^{ab}$ is conserved. As it turns out, for $n=2$ the only theory of this kind is NMG, precisely because $Y^{ab}\propto G^{ab}$, with $G^{ab}$ the Einstein tensor. The question of whether there is any higher order gravity other than NMG with second order traced equations of motion is that of whether it is possible to build a conserved tensor $Y^{ab}$ at $n\geq 3$. A natural candidate would be the tensor appearing in the equations of motion of the most general theory of order $n-1$, which is automatically conserved. However, the analysis of  \cite{Oliva:2010zd} shows that the only possibility for $n=3$ would be the $C_{abc}C^{abc}$ density, which involves explicit covariant derivatives in the action and is therefore excluded from the analysis. Since $Y^{ab}$ itself does not involve explicit covariant derivatives, the same question in $n=4$  would necessarily require the existence of densities of order $n= 3$ whose equations of motion are free from explicit covariant derivatives. The analysis of   \cite{Oliva:2010zd} disproves this possibility and therefore the only chance would be that some other divergence-free rank-two symmetric tensor cubic in curvatures ---not corresponding to the equations of motion of any covariant density--- exists. We believe that such a tensor does not exist  ---which would mean that no other theories with reduced-order traced equations exist among $\mathcal{L}(g^{ab},R_{ab})$ theories--- but we have not found a proof of this fact.
\cleardoublepage
\lhead{Chapter 2}
\rhead{D-Dimensional Higher-Derivative Gravities}

\chapter{D-Dimensional Higher-Derivative Gravities}
\label{chp:HDGs}


\section{Introduction}

Despite the spectacular list of experimental successes of general relativity, there are good reasons to explore alternatives to Einstein's theory. Firstly, it is expected that the Einstein-Hilbert action is the first in an infinite series of terms involving an increasing number of derivatives of the metric \cite{Endlich:2017tqa}. 
This can be seen explicitly within the string theory framework, where the new terms appear weighted by powers of the inverse string tension  \cite{Gross:1986mw,Green:1997tv,Frolov:2001xr}. Additionally, holographic higher-derivative gravities can be used, through AdS/CFT \cite{Maldacena,Witten}, as toy models of conformal field theories (CFTs) which, being inequivalent from their Einsteinian counterparts, can sometimes be used to unveil new universal properties valid for completely general CFTs \cite{Kats:2007mq,Brigante:2007nu,Buchel:2008ae,Myers:2010xs,Camanho:2010ru,Mezei:2014zla,Bueno1,Bueno:2018yzo,Bueno:2022jbl}.

From a different perspective, it is important to characterise the possible existence (or lack thereof) of universal features of classical gravity in regimes in which the Einsteinian description is expected to receive higher-derivative corrections \cite{PabloPablo4,Arciniega:2018tnn}. In order to do this, it is often convenient to consider particular classes of higher-derivative gravities displaying certain special properties. The list includes: quadratic \cite{Alvarez-Gaume:2015rwa,Lu:2015cqa}, Lovelock \cite{Lovelock1,Lovelock2,Boulware:1985wk,Padmanabhan:2013xyr}, Quasi-topological \cite{Quasi2,Quasi,Dehghani:2011vu,Ahmed:2017jod,Cisterna:2017umf} and Generalized Quasi-topological gravities (GQTs) \cite{Hennigar:2017ego,PabloPablo3,Bueno:2022res,Moreno:2023rfl}, among others \cite{Lu,Karasu:2016ifk,Li:2017ncu,Love}. All of these belong to the subset of theories built from contractions of the Riemann tensor and the metric, which we call higher-curvature gravities. In particular, GQTs ---which are characterised by admitting ``single function'' static and spherically symmetric solutions (see Section~\ref{gqT}) as well as possessing second-order equations on maximally symmetric backgrounds ---  have been shown to provide a basis for general gravitational effective actions built from general contractions of the Riemann tensor and the metric: any $\mathcal{L}(g^{ab},R_{abcd})$ theory can be mapped order by order, via a field redefinition, to certain GQTs \cite{Bueno:2019ltp}.  

Although seemingly less likely, it is also possible that deviations from Einstein gravity are eventually measured in unexpected situations (\eg beyond the effective field theory regime) and it is important to have alternative predictions which can be tested \cite{Cayuso:2023xbc}.  Along this direction, there have been numerous attempts at constructing alternatives to general relativity which are compatible with all current observations and internally consistent. This includes again quadratic theories \cite{Stelle:1976gc,Salvio:2018crh,Salvio:2019ewf}, $f(R)$ models \cite{Sotiriou}, as well as non-local gravities which, by including an infinite number of derivatives in the action, can be made free of ghosts \cite{Tomboulis:1997gg,Biswas:2005qr,Biswas:2010zk,Biswas:2011ar,Modesto:2011kw,Biswas:2013cha,Modesto:2014lga,Frolov:2015bta}.  
Non-local gravities are particular instances of the general set of theories which will be the subject of study in the present chapter, namely, diffeomorphism-invariant theories constructed from general contractions of the Riemann tensor and its covariant derivatives,
\begin{equation}\label{GeneralAction}
    I=\frac{1}{16\pi G_N} \int \df^D x \sqrt{-g}\, \mathcal{L}( g_{ab},R_{abcd},\nabla_a)\, .
\end{equation}
We call these ---more general--- theories higher-derivative gravities.
As a matter of fact, terms involving covariant derivatives of the Riemann tensor generically appear in gravitational effective actions \cite{Ruhdorfer:2019qmk,Li:2023wdz}. A scenario in which this is apparent corresponds to the so-called brane-world gravities \cite{Randall:1999ee,Randall:1999vf,Karch:2000ct}, which we will study in detail in Part III of this thesis. These are effective gravitational theories defined on the world volume of codimension-one branes inserted on higher-dimensional Anti-de Sitter spacetimes.
Originally introduced with phenomenological motivations, they have received a lot of attention recently in the holographic context ---see \eg  \cite{Chen:2020uac,Chen:2020hmv,Emparan:2020znc}. 

In this chapter, we present the first examples of GQT gravities with covariant derivatives. Analogously to their ``polynomial'' counterparts, we show that they have second-order linearized equations on maximally symmetric backgrounds\footnote{This provides a counterexample to the conjecture of \cite{Edelstein:2022lco} regarding the absence of theories with covariant derivatives of the curvature possessing an Einsteinian spectrum.} and that they admit black hole solutions characterized by a single function, $g_{tt} g_{rr}=-1$. Focusing on four dimensions, we find that the lowest-order instances of  GQT densities involve eight derivatives of the metric. However, we observe that all such theories admit the Schwarzschild metric as a solution, and therefore do not give rise to new solutions when considered as corrections to general relativity. The first GQT density with covariant derivatives which does correct the Schwarzschild solution occurs at tenth order in derivatives of the metric ---see \req{L1049} below for its explicit form. 

The analysis of the linearized spectrum of GQTs is performed after obtaining some general results on the linearization of general higher-derivative theories with covariant derivatives around maximally symmetric spacetimes. We present general formulas which allow for the computation of the linearized equations around flat space of a given general higher-derivative theory from its effective quadratic action. Using these results, we show that 
GQTs belong to the family of theories which do not include scalar modes in their linearized spectrum.
Indeed, in Subsection \ref{subsec:linearGQT}, we prove a stronger result: GQTs have an Einsteinian spectrum around maximally symmetric backgrounds.
Later, in Chapter \ref{chp:Spectrum}, we will see that brane-world gravities do not propagate scalar modes either, but they do propagate massive spin-2 modes.

The structure of the chapter is the following. Section \ref{lini} contains some comments on the structure of the linearized equations of  general higher-derivative gravities with covariant derivatives on general maximally symmetric backgrounds, a characterization of the structure of poles of the metric propagator on Minkowksi spacetime.
In Section \ref{gqT} we construct the first GQTs with covariant derivatives in four spacetime dimensions. 
We conclude in Section \ref{conclu} with some comments on future directions.
Appendix \ref{chp:App-basis} contains the complete list of the higher-derivative invariants at each order in derivatives up to eight, as well as a non-exhaustive set at order ten,  which we have used in Section \ref{gqT} of this chapter.


\section{Linearized higher-derivative gravities with covariant derivatives}\label{lini}
In this section we analyse the structure of the linearized equations for a general theory of the form (\ref{GeneralAction}) in general dimensions.
We derive their general form on a maximally symmetric background and then, focusing on the Minkowski case, we identify the precise relation between the effective quadratic action and the linearized equations, classifying the different theories according to the modes propagated. In particular, we identify a set of generalisations of a particular type of quadratic density involved in the definition of the so-called ``critical gravities'' ---which have the peculiarity of propagating no scalar modes. This set of theories will include, as particular instances, both the new GQTs theories, presented in Section \ref{gqT}, and the brane-world theories, as we will show in Chapter \ref{chp:Spectrum}.
 
Before starting, let us point out that many of the results presented in this section have appeared in different forms in previous literature. For example, the linearization on maximally symmetric backgrounds of general $\mathcal{L}(g^{ab},R_{abcd},\nabla_a)$ theories has been studied in-depth for the four-dimensional case in \cite{Biswas:2013kla,Conroy:2014eja,Biswas:2016etb,Edholm:2018wjh,Biswas:2016egy,Dengiz:2020xbu, Kolar:2023mkw}.

We are interested in gravity theories of the form (\ref{GeneralAction}).
Sometimes it is convenient to split the Lagrangian as follows
\begin{equation}
 \mathcal{L}( g^{ab},R_{abcd},\nabla_a)=\frac{(D-1)(D-2)}{\ell^2}+R+ \mathcal{L}_{\rm R}( g^{ab},R_{abcd})+ \mathcal{L}_{\nabla}( g^{ab},R_{abcd},\nabla_a)\, ,
\end{equation}
where we included an explicit Einstein-Hilbert plus (negative) cosmological constant piece, $\mathcal{L}_{\rm R}$ includes terms which do not involve covariant derivatives, and $\mathcal{L}_{\nabla}$ includes terms which contain at least one covariant derivative of the Riemann tensor.
The equations of motion for this theory can be written as \cite{Iyer:1994ys}
\begin{equation}\label{NLEEs}
\mathcal{E}_{ab}\equiv T_{a}\,^{cde}R_{b cde}-\frac{1}{2}g_{ab}\mathcal{L}-2 \nabla^c\nabla^d T_{a c d b}=0\,,
\end{equation}
where
\begin{equation}
T^{abcd}\equiv \left[ \frac{\partial \mathcal{L}}{\partial R_{abcd}} - \nabla_{a_1} \frac{\partial \mathcal{L}_{\nabla}}{\partial \nabla_{a_1} R_{abcd}}+\dots+ (-1)^m \nabla_{(a_1}\dots \nabla_{a_m)}\frac{\partial \mathcal{L}_{\nabla}}{\partial  \nabla_{(a_1}\dots \nabla_{a_m)}R_{abcd}} \right]\, .
\end{equation}
In the case of maximally symmetric backgrounds with metric $\bar g_{ab}$, the Riemann tensor is
\begin{equation}
\bar R_{abcd}= -\frac{2}{\ell_\star^2} \bar g_{a[c}\bar g_{d]b}\, ,
\end{equation}
where $\ell_\star^2$ has dimensions of length$^2$ and it is a positive number in the case of an AdS$_D$ background, a negative number in the case of dS$_D$, and infinite for Minkowski.  In order for $\bar g_{ab}$ to be a solution of $\mathcal{L}( g^{ab},R_{abcd},\nabla_a)$, the equations of motion impose the algebraic equation \cite{Bueno:2016ypa}
\begin{equation}
1-\chi+\frac{\ell^2}{(D-1)(D-2)}\left[\mathcal{L}_{\rm R}(\chi)-\frac{2\chi}{D}\mathcal{L}_{\rm R}'(\chi)\right]=0\, ,
\end{equation}
 where $\chi \equiv \ell^2/\ell_{\star}^2$,  $\mathcal{L}_{\rm R}(\chi)$ stands for the on-shell evaluation of the corresponding Lagrangian on the maximally symmetric background, and $\mathcal{L}_{\rm R}'(\chi)\equiv \df \mathcal{L}_{\rm R}(\chi)/\df \chi$ is also evaluated on-shell. Observe that the piece of the Lagrangian involving covariant derivatives of the Riemann tensor makes no contribution to this equation, which follows from $\bar \nabla_a \bar g_{bc}=0$. Naturally, for Einstein gravity, the above equation simply imposes the condition $\chi=1$. For a Lagrangian built from polynomials of the Riemann tensor  involving densities up to order $n$ in the curvature, the above equation is an order-$n$ algebraic equation for $\chi$, which will in general have many possible solutions, depending on the values of the corresponding higher-derivative couplings.


\subsection{Linearized equations}

Let us now consider the linearized equations of a general theory of the form given by \req{GeneralAction} around a maximally symmetric background. We expand the metric as
\begin{equation}
g_{ab}=\bar g_{ab}+  h_{ab}\, ,
\end{equation}
where $h_{ab}$ is a small perturbation. Every relevant object built from the metric can then be expanded to the desired order in the perturbation as $T=T^{(0)}+T^{(1)}+T^{(2)}+\mathcal{O}(h^3)$. 

Given a particular theory, we have two routes to derive its linearized equations. On the one hand, we can  take the full non-linear equations and expand each of the terms to linear order in the perturbation. Alternatively, we can expand the action to second order in the perturbation and derive the linearized equations from the first variation. 
As we have seen, the full non-linear equations \eqref{NLEEs} of a theory like  (\ref{GeneralAction}) have a rather complicated form. However, it is not difficult to argue that the most general form of the linearized equations is much simpler. To see this, we start by characterizing all possible terms that may arise in the linearized equations. Doing this amounts to classifying all symmetric tensors of 2 indices built from $R_{abcd}^{(1)}$, $\bar g_{ab}$ and $\bar \nabla_a$ which are linear in the metric perturbation. 

Let us start with a few observations. First, observe that the linearized Riemann tensor $R_{abcd}^{(1)}$ is linear in $h_{ab}$, and therefore all possible terms will have a single Riemann tensor, possibly acted upon with covariant derivatives and with various indices contracted.
Another observation is that all terms must necessarily contain an even number of covariant derivatives, since $\nabla_a$ is the only available object with an odd number of indices.  In addition, note that all Riemann tensors will actually appear in the form of Ricci tensors. This is because: a) any term involving exclusively metrics and Riemann tensors reduces to Ricci tensors or vanishes, since at most two of the indices can remain uncontracted; b) any term involving covariant derivatives and Riemann tensors reduces to covariant derivatives and Ricci tensors. Indeed,  when only two indices are left uncontracted, a tensor of the form
\begin{equation}
\nabla_a\nabla_b R^{(1)}_{cdef}\, ,
\end{equation}
reduces to one of the following four possibilities: $\nabla^c\nabla^d R^{(1)}_{cadb},\, \Box R^{(1)}_{ab},\, \nabla_a \nabla_b R^{(1)},\, 0 $. In addition, using the second Bianchi identity it follows that the first possibility can only give rise to a linear combination of the second and the third, plus higher-order terms in $h_{ab}$. 
We therefore conclude that the most general possible term will come from contracting all but two indices in an expression of the form
\begin{equation}
\Box^l \nabla_{c_1}\nabla^{c_2} \dots \nabla_{c_{2m-1}}\nabla^{c_{2m}} R^{(1)}_{ab}\,,
\end{equation}
where $c_i \neq c_j \, \forall i\neq j$. Contracting $2m$ of the indices, we immediately see that the only three possibilities are in fact
\begin{equation}
\bar g_{ab} \Box^l R^{(1)}\, , \quad \nabla_a\nabla_b \Box^l R^{(1)}\, , \quad \Box^lR^{(1)}_{ab}\, .
\end{equation}
We then conclude that the linearized equations of a general $\mathcal{L}(g^{ab},R_{abcd},\nabla_a)$ theory around maximally symmetric backgrounds will always take the form
\begin{equation}\label{eq:Eab general msb}
\mathcal{E}_{ab}\equiv \sum_{l=0}\ell^{2l} \left[ \alpha_l  \bar \Box^l G^{(1)}_{ab}+ \beta_l  \bar \Box^l R^{(1)} \bar g_{ab} +\gamma_{l+1} \ell^2 \Box^l [\bar g_{ab}\bar \Box- \bar\nabla_a\bar\nabla_b]  R^{(1)} \right] = 0\, ,
\end{equation}
for certain dimensionless constants $\alpha_l,\beta_l,\gamma_l$ which will be related to the gravitational couplings, and where we rearranged some of the terms for later convenience. Implicitly, we have assumed that the theory involves a polynomial dependence on the covariant derivatives. Relaxing this requirement, would yield the more general form
\begin{equation}\label{eq:Eab general msb2}
\mathcal{E}_{ab}\equiv \left[ f_1(\ell^2 \bar \Box) G^{(1)}_{ab}+  f_2(\ell^2 \bar \Box) R^{(1)} \bar g_{ab} +   f_3(\ell^2 \bar \Box)  [\bar g_{ab}\bar \Box- \bar\nabla_a\bar\nabla_b]R^{(1)} \right] = 0\, ,
\end{equation}
for certain functions $f_1, f_2, f_3$. The form of the equations can be further constrained by noting that the tensor $\mathcal{E}_{ab}$ must be divergence-free, that is, $\bar\nabla^{a}\mathcal{E}_{ab}=0$. By commuting $\bar{\nabla^{a}}$ and $\bar\Box$, one can show that the divergence reads
\begin{equation}
\begin{aligned}
\bar\nabla^{a}\mathcal{E}_{ab}=\Bigg\{&f_2\left(\ell^2\tilde \Box\right)+\frac{D-1}{D}\left[f_{3}\left(\ell^2\tilde \Box\right)\tilde\Box-f_{3}\left(\ell^2\hat \Box\right)\left(\bar\Box-\frac{1}{\ell_{\star}^2}\right)\right]\\
&+\frac{2-D}{2D}\left[f_{1}\left(\ell^2\tilde \Box\right)-f_{1}\left(\ell^2\hat\Box\right)\right]\Bigg\}\bar\nabla_{b}R^{(1)}\, ,
\end{aligned}
\end{equation}
where
\begin{equation}
\hat\Box=\bar\Box-\frac{D+1}{\ell_{\star}^2}\, ,\quad \tilde\Box=\bar\Box+\frac{D-1}{\ell_{\star}^2}\, .
\end{equation}
Therefore, the function $f_2$ is not free, but it depends on $f_1$ and $f_3$ by
\begin{equation}\label{f2relation}
\begin{aligned}
f_2\left(\ell^2\bar\Box\right)&=\frac{D-2}{2D}\Big[f_1\left(\ell^2\bar\Box\right)-f_1\left(\ell^2\bar\Box+2D\chi\right)\Big]
\\
&+\frac{D-1}{D\ell^2}\Big[f_3\left(\ell^2\bar\Box+2D \chi\right)\left(\ell^2\bar\Box-\chi D\right)-f_3\left(\ell^2\bar\Box\right)
\ell^2\bar\Box\Big]\, ,
\end{aligned}
\end{equation}
where we recall that $\chi=\ell^2/\ell_{\star}^2$. Observe that in the case of flat space, $f_2$ vanishes. 

In the case of theories which do not involve covariant derivatives, it is known that the most general form of the linearized equations is captured by a general quadratic action in the Riemann tensor. Something similar happens for a general $\mathcal{L}(g^{ab},R_{abcd},\nabla_a)$ theory. Indeed, in that case, the most general quadratic action reads
\begin{equation}
    \label{effq}
    \mathcal{L}_{\rm eff} = \lambda\left[\frac{(D-1)(D-2)}{\ell^2}+R+ \ell^2 R F_1(\ell^2 \bar \Box) R + \ell^2 R_{ab}   F_2(\ell^2 \bar \Box) R^{ab}+\ell^2 R_{abcd} F_3(\ell^2\bar \Box)R^{abcd}\right]\,, 
\end{equation}
for certain functions $F_1,F_2,F_3$. It is then possible to relate these functions to the $f_{1}$ , $f_{2}$ , $f_{3}$ of the linearized equations \eqref{eq:Eab general msb2}. Such relations are quite cumbersome in the case of (A)dS backgrounds, as we illustrate in Subsection~\ref{adsbck} for the simple case of $F_i \propto \Box$. However, they simplify greatly in the case of Minkowski backgrounds, which we analyse now.


\subsection{Minkowski background}\label{minko}

First, let us note that, if we are considering linearization around flat space, then the term $R^{abcd} \Box^n R_{abcd}$ in \req{effq} is not independent of the other two quadratic terms.
Indeed, integrating by parts and using the Bianchi identities, one can show that the following relation is true up to a total derivative \cite{Edelstein:2022lco},
\begin{align}
    R_{abcd} \Box R^{abcd} \cong  4 R^{bd} \Box R_{bd} - R \Box R + \mathcal{O}(R_{abcd}^3)\,.
\end{align}
A similar calculation follows for any number $n$ of box operators, since the remaining $\Box^{n-1}$ operators simply introduce extra $\mathcal{O}(R_{abcd}^3)$ terms upon commutation.
Therefore, for all $n \geq 0$,
\begin{equation}\label{BoxnGB}
    R \Box^n R - 4 R^{ab} \Box^n R_{ab} + R^{abcd} \Box^n R_{abcd} = \text{Total Derivative} + \mathcal{O}(R_{abcd}^3) \,.
\end{equation}
Since the $\mathcal{O}(R_{abcd}^3)$ terms do not contribute to the linearized equations on Minkowski space, we could redefine out $F_3$ in $\mathcal{L}_{\text{eff}}$ without loss of generality.
However, we will not do so, since given any general Lagrangian $\mathcal{L}( g^{ab},R_{abcd},\nabla_a)$, finding its effective quadratic action $\mathcal{L}_{\rm eff}$ for linearized perturbations around flat space will just consist in dropping all terms cubic or higher in curvature.

The linearized equations around Minkowski space for the quadratic Lagrangian \eqref{effq} read
\begin{eqnarray} \notag
&&\frac{\lambda}{2} \left\{ \left[1+\left[4F_3(\ell^2 \bar \Box)+F_2(\ell^2 \bar \Box)\right] \ell^2\bar\Box \right] G_{ab}^{(1)} \right. \\ && \left.-\ell^2\left[ 2F_1(\ell^2 \bar \Box)+F_2(\ell^2 \bar \Box)+2F_3(\ell^2 \bar \Box)\right][ \bar\nabla_a\bar\nabla_b-\bar g_{ab}\bar \Box]R^{(1)} \right\} =0\, . \label{lineareqs}
\end{eqnarray}
where
\begin{align}
G_{ab}^{(1)}&=-\frac{1}{2}\bar\Box h_{ab}+\bar\nabla_{(a|}\bar\nabla^{c}h_{c|b)}-\frac{1}{2}\bar\nabla_{a}\bar\nabla_{b}h-\frac{1}{2}\bar g_{ab} R^{(1)}\, ,\\
R^{(1)}&=\bar\nabla^{a}\bar\nabla^{b}h_{ab}-\bar \Box h\, ,
\end{align}
are the linearized Einstein tensor and Ricci scalar, respectively.
These linearized equations can be obtained immediately using the result found in \cite{Bueno:2016ypa} for theories which do not involve covariant derivatives of the Riemann tensor. The idea is to use the same relations between the quadratic action couplings and the constant parameters ($a,b,c,e$) appearing in such equations, but now promoting the constants to functions of $\ell^2\bar \Box$. 

The trace of the equations reads
\begin{equation}
    -\frac{\lambda}{4} \left[(D-2)-\ell^2 \bar \Box  \left[4F_3(\ell^2 \bar \Box)+D F_2(\ell^2\bar \Box)+4(D-1)F_1(\ell^2 \bar\Box) \right] \right] R^{(1)}=0\,,
\end{equation}
and their traceless part is given by
\begin{eqnarray}\notag
&&
\frac{\lambda}{2} \left\{  \left[1+\left[4F_3(\ell^2 \bar \Box)+F_2(\ell^2 \bar \Box)\right]\ell^2\bar\Box \right] R^{(1)}_{\langle a b\rangle }\right. \\ && \left.- \ell^2\left[ 2F_1(\ell^2 \bar \Box)+F_2(\ell^2 \bar \Box)+2F_3(\ell^2 \bar \Box)\right] \bar\nabla_{\langle a}\bar\nabla_{b\rangle}R^{(1)} \right\}=0\, .
\end{eqnarray}
Observe now that for theories satisfying the condition
\begin{equation}\label{keyko}
4F_3(\ell^2 \bar \Box)+D F_2(\ell^2\bar \Box)+4(D-1)F_1(\ell^2 \bar\Box)=0\, ,
\end{equation}
the trace equation becomes second order and simply reads
\begin{equation}\label{oop}
-\frac{\lambda}{4}(D-2) R^{(1)}=0\, ,
\end{equation}
which is nothing but the Einstein gravity result. In the case in which $F_i=c_i$ are constants, this condition (\ref{keyko}) selects a linear combination of quadratic terms which appear in the so-called ``critical gravities'' in general dimensions---see \eg \cite{Lu:2011zk,Maldacena:2011mk,Bergshoeff:2009hq,Oliva:2011xu,Hassan:2013pca,Kan:2013moa,Anastasiou:2016jix}. In particular, the quadratic action reduces in that case to
\begin{equation}\label{confi}
\mathcal{L}_{\rm eff}=\lambda \left[ \frac{(D-1)(D-2)}{\ell^2}+R + \ell^2 c_3 \mathcal{X}_4 + \ell^2 (c_1-c_3) \left(R^2-\frac{4(D-1)}{D}R_{ab}R^{ab} \right)  \right]\, ,
\end{equation}
where $\mathcal{X}_4\equiv R^2-4R_{ab}R^{ab}+R_{abcd}R^{abcd}$ is the Gauss-Bonnet density, and the second term can be written as a linear combination of  $\mathcal{X}_4$ and the Weyl tensor squared. For this theory, the linearized spectrum on a general maximally symmetric background is known to involve the usual massless graviton and the massive one, but not the scalar mode. This is also the case for theories satisfying \req{keyko} with non-constant functions. As we will see later, both Generalized Quasi-topological and brane-world gravities belong to that class.

In order to study the physical modes propagated by the metric perturbation, let us now fix the harmonic gauge, which amounts to setting
\begin{equation}
\bar\nabla^a h_{ab}= \frac{1}{2}\bar\nabla_b h\, .
\end{equation} 
Then, the linearized Einstein tensor and Ricci scalar become
\begin{equation}
G_{ab}^{(1)}=-\frac{1}{2}\bar\Box h_{ab}+\frac{1}{4}\bar g_{ab} \bar\Box h\, , \qquad R^{(1)}=-\frac{1}{2}\bar\Box h\, .
\end{equation}
For theories satisfying \req{keyko}, the trace equation (\ref{oop}) imposes $\bar\Box h=0$. Using the residual gauge freedom $h_{ab}\rightarrow h_{ab}+\nabla_{(a}\xi_{b)}$, with $\Box\xi_a=0$, we can set $h=0$. Therefore, the trace of the perturbation has no dynamics and indeed there are no scalar modes. On the other hand, the traceless part of the equations becomes
\begin{eqnarray}\label{tracelesss}
-\frac{\lambda}{4} \left[1+\left[4F_3(\ell^2 \bar \Box)+F_2(\ell^2 \bar \Box)\right]\ell^2\bar\Box \right] \bar \Box h_{\langle a b \rangle}=0\, .
\end{eqnarray}
By performing the Fourier transform in this expression, which amounts to $\Box\rightarrow -k^2$, we can read off the propagator
\begin{equation}
P(k)=\frac{4}{\lambda k^2\left[1-   \ell^2 k^2 \left(4F_3(-\ell^2 k^2)+F_2(-\ell^2 k^2)\right)\right]}\, .
\end{equation}
Poles of the propagator inform about the degrees of freedom of the theory. For each pole, $k^2=-m^2$ indicates the mass. Thus, imaginary poles correspond to massive modes, while real poles are tachyonic modes. On the other hand, the residue of each pole tells us about the energy carried out by the corresponding mode. A positive residue ---like the massless graviton one, $k^2=0$--- corresponds to positive energy, and viceversa for a negative residue. For constant functions, $F_i=c_i$, we have the poles
\begin{equation}
m^2=0\, , \quad m_g^2=-\frac{1}{(4c_3+c_2)\ell^2}\, ,
\end{equation}
corresponding to the anticipated massless and massive graviton, respectively, and in agreement with the result of \cite{Bueno:2016ypa,Sisman:2011gz}. The next to simplest case corresponds to
$F_i(\ell^2 \bar \Box)=c_i+b_i \ell^2\bar \Box $. For that, one finds 
\begin{equation}
m^2=0\, , \quad m_{\pm}^2=-\frac{(c_2+4c_3)\pm \sqrt{ (c_2+4c_3)^2 - 4 (b_2+4b_3)}}{2(b_2+4b_3)\ell^2} \, ,
\end{equation}
which correspond, in addition to the usual massless graviton, to two new massive gravitons.

An additional simplification occurs for theories such that, besides \req{keyko}, also satisfy the condition $F_3(\ell^2 \bar\Box)=-F_2(\ell^2 \bar\Box)/4$. Those two conditions can then be rewritten as 
\begin{equation}\label{gbb}
F_1(\ell^2 \bar\Box)=F_3(\ell^2 \bar\Box)=-F_2(\ell^2 \bar\Box)/4 \, ,
\end{equation}
and, in that case, the linearized equations reduce to
\begin{equation}
\frac{\lambda}{2} G_{ab}^{(1)}=0\, ,
\end{equation}
namely, to the usual linearized Einstein equation.
Hence, for theories whose effective action satisfies the pair of conditions (\ref{gbb}), the linearized equations on Minkowski space are identical to the Einstein gravity ones ---or, in other words, the higher-derivative densities do not contribution at all to the linearized equations. 
This fact was expected from \req{BoxnGB} and our discussion at the beginning of this subsection, but it is good to see that it holds even when the functions $F_i$ cannot be written as a series expansion.
Gauss-Bonnet gravity is a particular instance of this kind of gravities, which corresponds to setting all functions equal to constants, but the set of higher-derivative theories with this property contains infinitely many densities with an arbitrarily large number of covariant derivatives. We will see later that Generalized Quasi-topological gravities fall within this category (not so brane-world gravities). 


\subsection{AdS background}\label{adsbck}
Here we present the explicit linearized equations of motion around an AdS background for the simplest examples of the theories of the form $\mathcal{L}(g^{ab},R_{abcd},\nabla_a)$.
We will see that there is no straightforward relation between the functions $F_i$ in the effective quadratic Lagrangian \eqref{effq}, and the functions $f_i$ in the linearized equations \eqref{eq:Eab general msb}.

First, we consider the effective quadratic theory arising from purely polynomial theories,
\begin{equation} 
    \mathcal{L}_{(0)} = \lambda \left(R - 2 \Lambda_0\right) + \alpha_{(0)} R^2 +\beta_{(0)}  R_{ab} R^{ab}+\gamma_{(0)} R_{abcd} R^{abcd} \, .
\end{equation}
For this theory, the linearized equations were computed in~\cite{Bueno:2016ypa} and read
\begin{align}
    \mathcal{E}_{ab}^{(1)}=&  \left[\frac{\lambda}{2} - \frac{2}{\ell_\star^2}\left( D(D-1) \alpha_{(0)} + (D-2) \beta_{(0)} + (D-3)(D-4)\gamma_{(0)} \right)  + \beta_{(0)} \bar{\Box} \right] G_{ab}^{(1)} 
    \nonumber
    \\
    &+ \left[2\alpha_{(0)} + \beta_{(0)} \right] \left[\bar{g}_{ab} \bar{\Box} - \bar{\nabla}_a \bar{\nabla}_{b} \right] R^{(1)} - \frac{1}{\ell_\star^2} \left[2(D-1)\alpha_{(0)} + \beta_{(0)} \right] \bar{g}_{ab} R^{(1)} \,.
\end{align}
Now, let us consider the effective action involving one d'Alembertian acting on curvature,
\begin{equation}  
    \mathcal{L}_{(1)} =\alpha_{(1)} R\Box  R+\beta_{(1)}  R_{ab}\Box R^{ab}+\gamma_{(1)} R_{abcd} \Box R^{abcd} \, .
\end{equation}
For this theory, we see that the linearized equations of motion take the form
\begin{align}
    \mathcal{E}_{ab}^{(1)}=& \left[ -\frac{8}{\ell_\star^4}\gamma_{(1)}  (D-3) D  + \frac{2}{\ell_\star^2} (\beta_{(1)} + 2 \gamma_{(1)} (5 - D) ) \bar{\Box} + (\beta_{(1)} +4 \gamma_{(1)} ) \bar{\Box}^2 \right] G_{ab}^{(1)}
    \nonumber
    \\
    &   
    + \left(2 \alpha_{(1)} + \beta_{(1)} + 2 \gamma_{(1)} \right) \left[\bar{g}_{ab} \bar{\Box} - \bar{\nabla}_a \bar{\nabla}_b \right] \bar{\Box} R^{(1)} + \frac{2 (D-2) (4 \gamma_{(1)} + \beta_{(1)})}{\ell_\star^2} \bar{\nabla}_a \bar{\nabla}_b R^{(1)}
    \nonumber 
    \\
    &- \frac{(D-1)(2 \alpha_{(1)} + \beta_{(1)} + 2 \gamma_{(1)})}{\ell_\star^2}  \bar{g}_{ab} \bar{\Box} R^{(1)} - \frac{4(D-2)(D-3)}{\ell_\star^2} \gamma_{(1)}  \bar{g}_{ab} R^{(1)} \, .
\end{align}
Note that the linearized field equations of this six-derivative action involves terms with two, four, and six derivatives.

Therefore, we see that there is a mixing of orders that prevents us from writing an easy relation between the coefficients of the effective quadratic action and the coefficients of the linearized equations. 
It may be that the linearized equations can be simplified in alternative gauges, e.g.,~\cite{Kolar:2023mkw}, but we have not pursued this any further. 


\section{Generalized Quasi-topological gravities}\label{gqT}

In this section, we present the first examples of (four-dimensional) Generalized Quasi-topological (GQT) densities involving covariant derivatives of the Riemann tensor. In $D=4$, in the absence of covariant derivatives, it has been shown that there exists a unique non-trivial GQT density at each curvature order. Here we show that the landscape of GQT theories is modified considerably by allowing covariant derivatives of the Riemann tensor to appear in the action. In particular, while we find no new densities at four- and six-derivative(s of the metric) orders, we obtain four new inequivalent GQTs at eight-derivative order. Of these, only one possesses an integrated equation for $f(r)$ which is of second order in derivatives, two of them have third-order equations, and the remaining one has an integrated fourth-order equation for the metric function. In all cases, we find that the Schwarzschild solution is also a solution of these theories. As a consequence, coupling Einstein gravity to these theories does not give rise to new spherically symmetric black hole solutions. Extending the analysis to ten-derivative order, we find new examples which do not  admit Schwarzschild as a solution. For those, the coupling to Einstein gravity does produce new non-trivial modifications of the Schwarzschild black hole. 

\subsection{Definition}

Let us start by recalling the basic definition and properties of GQTs. 
Consider a general static and spherically symmetric (SSS) spacetime parametrized by two functions,
$N(r)$ and $f(r)$,
\begin{equation}
\label{SSS}
\df s^2_{N, f}=-  N(r)^2 f(r) \df t^2+\frac{\df  r^2}{f(r)}  + r^2\df \Omega^2_{(D-2)}\, ,
\end{equation}
where $\df \Omega^2_{(D-2)}$ is the  $(D-2)$-dimensional sphere metric. The following comments extend, with minor modifications, to the cases in which the horizon is  hyperbolic or planar instead. The expressions below will incorporate those cases through a parameter  denoted $k$ which will take the values $+1,0,-1$, respectively for the spherical, planar and hyperbolic cases.

For a given higher-derivative invariant of order $2m$ in derivatives of the metric and involving $p$ covariant derivatives of the Riemann tensor, $\mathcal{R}_{(2m,p)}$, let  $S_{N,f}$ and $L_{N,f} $ be, respectively, the  effective on-shell action and Lagrangian resulting from the evaluation of  $\sqrt{|g|}\mathcal{R}_{(2m,p)}$ in the ansatz (\ref{SSS}), namely,
\begin{equation}\label{ansS}
L_{N,f}\equiv \left. N(r)   r^{D-2}   \mathcal{R}_{(2m,p)}\right|_{N,f}\, ,   \quad  S_{N,f}\equiv   \Omega_{(D-2)}\int \df t  \int  \df r L_{N,f} \, ,
\end{equation}
where we performed the trivial integral over the angular directions,   $\Omega_{(D-2)}\equiv   2\pi^{\frac{D-1}{2}}/\Gamma[\frac{D-1}{2}]$. We denote by $L_f\equiv L_{1,f}$ and $S_f\equiv S_{1,f}$ the expressions resulting from setting $N=1$  in $L_{N,f}$. 
Now, solving the full nonlinear equations of motion for a metric of the form (\ref{SSS}) can be shown to be equivalent to solving  the Euler-Lagrange equations of $S_{N,f}$ associated to $N(r)$ and $f(r)$ \cite{Palais:1979rca,Fels:2001rv,Deser:2003up,PabloPablo4}, namely,
\begin{equation}
    \left.\mathcal{E}^{ab}\right|_{N,f}\equiv \left. \frac{1}{\sqrt{|g|}}  \frac{\delta S}{\delta g^{ab}}  \right|_{N,f}=0 \quad   \Leftrightarrow   \quad \frac{\delta S_{N,f}}{\delta N}= \frac{\delta S_{N,f}}{\delta f}=0\,.
\end{equation}
We say that  $\mathcal{R}_{(2m,p)}$ is a GQT density if the Euler-Lagrange equation of $f(r)$ associated to  $S_f$ is identically   vanishing, namely,  if
\begin{equation} \label{GQTGcond}
\frac{\delta S_{f}}{\delta f}=0\, ,   \quad \forall \, \, f(r)\, .
\end{equation}
This condition is equivalent to  asking $L_f$ to be a total derivative,  
\begin{equation}\label{condd2}
L_f =T_0'\, ,
\end{equation}
for a certain function  $T_0(r,f(r),f'(r),\dots , f^{(p+1)})$.

Thus, the variation with respect to $f(r)$ of the on-shell action $S_f$ determines whether a given density is of the  GQT class. When that is the case, the full non-linear equations of $\mathcal{R}_{(2m,p)}$ reduce to a single equation for $f(r)$ which can be integrated once. This integrated equation can be obtained from the variation of $L_{N,f}$ with respect to $N(r)$ as \begin{equation}\label{eqf}
\left.\frac{\delta S_{N,f}}{\delta N}\right|_{N=1}=0\, \quad \Leftrightarrow   \quad   \text{equation of}\quad f(r)\, .
\end{equation}
Let us see this in more detail. As explained in \cite{PabloPablo3}, whenever \req{condd2} holds, the effective Lagrangian  $L_{N,f}$ takes the form
\begin{equation}\label{fofwo}
L_{N,f}=N T_0' +  N' T_1 +  N'' T_2+ \dots + N^{(p+2)}T_{p+2} +\mathcal{O}(N'^2/N)\, ,
\end{equation}
where $T_{1},T_2,\dots,T_{p+2}$  are functions of $f(r)$ and its derivatives (up to $f^{(p+2)}$), and $\mathcal{O}(N'^2/N)$ is a sum of  contributions  which are all at least quadratic in  derivatives of   $N(r)$. Integrating by parts, one finds
\begin{equation}
S_{N,f} = \Omega_{(D-2)}  \int \df t \int \df r \left[N\left(T_0 + \sum_{j=1}^{p+2} (-1)^{j} T_{j}^ {(j-1)}\right)' +\mathcal{O}(N'^2/N) \right]\, .
\end{equation}
Therefore, we can write every term involving one power of $N(r)$  or  its derivatives as a certain  product of $N(r)$ and a total derivative which depends on $f(r)$ alone. Then, eq.~(\ref{eqf}) equates this total derivative to zero. Integrating it once, we are left with \cite{PabloPablo3}
\begin{equation} \label{eqqqf}
\mathcal{F}_{\mathcal{R}_{(2m,p)}}  \equiv  T_0 + \sum_{j=1}^{p+2} (-1)^{j} T_{j}^ {(j-1)}=\frac{M}{\Omega_{(D-2)}}\, ,
\end{equation}
where the integration constant was written in terms of the ADM mass of the solution \cite{Arnowitt:1960es,Arnowitt:1960zzc,Arnowitt:1961zz,Deser:2002jk}.

In sum, given some linear combination of GQT densities, the equation satisfied by $f(r)$ can be obtained from  $L_{N,f}$ as defined in \req{ansS} by identifying the functions $T_{\{ j \}}$ from \req{fofwo}. The order of the integrated equation $\mathcal{F}_{\mathcal{R}_{(2m,p)}} $ is at least two orders less than the one of the equations determining $f(r)$ and $N(r)$ in the most general case, namely, 
\begin{equation}
\mathcal{F}_{\mathcal{R}_{(2m,p)}}=\mathcal{F}_{\mathcal{R}_{(2m,p)}}(r,f,f',\dots,f^{(2p+2)})\,.
\end{equation}
In particular, when $p=0$, corresponding to the case without covariant derivatives of the Riemann tensor, the integrated equation is at most second-order in derivatives of $f(r)$. In that case, one can see that the integrated equations are either of order $0$ in derivatives ---these are called simply ``Quasi-topological'' theories \cite{Quasi2,Quasi,Dehghani:2011vu,Ahmed:2017jod,Cisterna:2017umf}, which includes Lovelock theories \cite{Lovelock1,Lovelock2} as particular cases--- or, alternatively, of order $2$.  As we will see in a moment, the actual order of the integrated equations that we will find in our new GQT densities with covariant derivatives will be considerably lower than the $2p+2$ upper bound.
 
We will say that two GQT densities $\{ \mathcal{R}^{I}_{(2m,p)} ,\mathcal{R}^{II}_{(2m,p)} \}$ are ``inequivalent'' (as far as  SSS  solutions are  concerned) whenever the quotient of their respective integrated equations is not constant, namely,
\begin{equation}
\mathcal{R}^{I}_{(2m,p)} \quad \text{inequivalent from} \quad \mathcal{R}^{II}_{(2m,p)} \quad \Leftrightarrow \quad  \frac{\mathcal{F}_{\mathcal{R}^I_{(2m,p)}}(r,f,f',\dots,f^{(2p+2)}) }{\mathcal{F}_{\mathcal{R}^{II}_{(2m,p)}} (r,f,f',\dots,f^{(2p+2)})} \neq \text{constant} \, .
\end{equation}
Otherwise we will call them ``equivalent''. Two equivalent densities differ by densities which make no contribution whatsoever to the integrated equation of $f(r)$. Those densities are ``trivial'' as far as SSS solutions are concerned. 

In the $p=0$ case, it has been argued that: there exist no (non-trivial) GQTs in $D=3$ \cite{Bueno:2022lhf}; there exists a single inequivalent GQT density at each curvature order $m$ in $D=4$ whose integrated equation is a differential equation of order $2$ \cite{Moreno:2023rfl}; there exists a single inequivalent Quasi-topological density at each curvature order $m$ in $D\ge 5$ whose integrated equation is algebraic  \cite{Bueno:2022res}; there exist $(m-2)$ inequivalent GQT densities at each  curvature order in $D\ge 5$ whose integrated equation is a differential equation of order $2$  \cite{Bueno:2022res,Moreno:2023rfl}.


\subsection{Linear spectrum}\label{subsec:linearGQT}
A remarkable property of all GQTs built from polynomial curvature invariants is that their linear spectrum on maximally symmetric backgrounds is devoid of ghosts. In fact, the linearized equations of motion are proportional to those of Einstein gravity on the same background. For polynomial GQTs, the second-order nature of the linearized equations was first verified explicitly in case-by-case examples ---see \eg \cite{Quasi2,Quasi, PabloPablo, Hennigar:2017ego, Ahmed:2017jod}. It was subsequently proven that the single-metric-function condition that defines GQTs also implies the linearization is second-order in general~\cite{PabloPablo3} ---c.f. page 102 of \cite{CanoMolina-Ninirola:2019uzm} for the most up-to-date version of this proof. 
Here, we show that this result in fact holds for all GQTs, including those that contain covariant derivatives of the curvature (and hence have equations of motion of order greater than four). 

The idea behind the proof consists in considering a metric perturbation within the single-function static spherically symmetric ansatz. Thus, we start by considering the metric \eqref{SSS} with $N(r)=1$. For convenience, let us rewrite this metric as 
\begin{equation}
ds^2=-f(r)\df u^2-2 \df r \df u+r^2\df\Omega_{(D-2)}^2\, ,
\end{equation}
where $u=t+r_{*}$, with $r_*$ being the tortoise coordinate, defined by $\df r_{*}=\df r/f(r)$. One can show that, in this coordinate system, the GQT condition \eqref{GQTGcond} is equivalent to the vanishing of the $rr$ component of the equations of motion, that is, 
\begin{equation}\label{GQTGcond2}
\mathcal{E}_{rr}=0\, ,\quad \forall f(r)\, .
\end{equation}
We can then take $f(r)$ to be 
\begin{equation}\label{radialpert}
f(r)=1+\frac{r^2}{\ell_{\star}^2}+h(r)\, ,\quad h(r)\ll 1\, ,
\end{equation}
corresponding to a maximally symmetric vacuum plus a small perturbation $h_{ab}$ given by
\begin{equation}\label{radialperthab}
h_{uu}=h(r)\, .
\end{equation} 
Then, the idea is to impose the condition \eqref{GQTGcond2} at the level of the linearized equations by using this perturbation. We know that, in general, the linearized equations are given by \eqref{eq:Eab general msb} for certain coefficients $\alpha_l$, $\beta_l$ and $\gamma_l$. Let us for instance assume that our theory has sixth-order equations of motion --- so that only the coefficients with $l\le 2$ are nonzero --- and let us set $D=4$. We get, after a direct evaluation of \eqref{eq:Eab general msb} on \eqref{radialperthab},
\begin{equation}\label{Errlinear}
\begin{aligned}
\mathcal{E}_{rr}^{(1)}&=\alpha_1 \ell^2\left(-\frac{4 h}{r^4}+\frac{2 h''}{r^2}\right)+\gamma _1\ell^2 \left(-\frac{12 h}{r^4}+\frac{6 h''}{r^2}-\frac{4 h^{(3)}}{r}-h^{(4)}\right)\\
&+\alpha _2 \ell^4 \left(h \left(-\frac{32}{r^6}+\frac{56}{L^2 r^4}\right)+\frac{32 h'}{r^5}+\left(-\frac{16}{r^4}-\frac{28}{L^2 r^2}\right) h''+\frac{16 h^{(3)}}{L^2
   r}+\left(\frac{4}{L^2}+\frac{4}{r^2}\right) h^{(4)}\right)\\
   &+\gamma _2 \ell^4\left(h \left(-\frac{80}{r^6}+\frac{120}{L^2 r^4}\right)+\frac{80
   h'}{r^5}+\left(-\frac{40}{r^4}-\frac{60}{L^2 r^2}\right) h''+\frac{40 h^{(3)}}{L^2 r}+\left(-\frac{20}{L^2}+\frac{10}{r^2}\right) h^{(4)}\right.\\
&\left.+\left(-\frac{6}{r}-\frac{12
   r}{L^2}\right) h^{(5)}+\left(-1-\frac{r^2}{L^2}\right) h^{(6)}\right)\, .
   \end{aligned}
\end{equation}
Then, the GQT condition \eqref{GQTGcond2} implies that this expression above must vanish for any choice of $h(r)$. Clearly, this only happens if $\alpha_1=\alpha_2=\gamma_1=\gamma_2=0$, since all terms are linearly independent. The same conclusion follows in general dimensions and if the theory has higher-order equations of motion. In the latter case, \req{Errlinear} will include $\alpha_l$- and $\gamma_l$-terms with higher $l$, but these are all linearly independent because they contain different numbers of derivatives of $h$ and/or different radial dependence. 

In conclusion, \eqref{GQTGcond2} implies the vanishing of all the $\alpha_{l}$ and $\gamma_{l}$ except for $\alpha_0$, corresponding to the coefficient of the linearized Einstein tensor. Finally, the relation \eqref{f2relation} implies the vanishing of the $\beta_{l}$ coefficients. Therefore, the linearized equations must be proportional to the linearized Einstein tensor. 

In the next subsection, we will obtain some explicit new GQTs in four dimensions. Following our results from Subsection \ref{minko}, one can easily obtain their effective quadratic Lagrangians by dropping all but curvature-squared terms. It is then easy to see that they all fulfil conditions \eqref{gbb}, thus confirming that they have an Einsteinian spectrum, at least around flat space.


\subsection{Classification of four-dimensional theories}
\label{sec:4dTheories}

In this subsection, we will classify all possible GQT Lagrangians in $D=4$, based on the number of derivatives of the metric appearing in the action. In the case of four and six derivatives, the result is in line with previous considerations~\cite{Quasi2, Quasi, Hennigar:2017ego, PabloPablo3}: nothing new beyond those theories constructed from the polynomial invariants is found. However, the cases of eight and ten derivatives reveals new features not seen before. 

Let us briefly summarize the methodology. At a given derivative order, we construct the most general Lagrangian density by performing a linear combination of all higher-derivative invariants that appear at that order:
\begin{equation}
    \mathcal{L}_{(2m)} = \sum_{i} c_{(i)}^{(2m,p)} \mathcal{R}^{(i)}_{(2m,p)}\,.
\end{equation}
Here, $2m$ refers to the number of derivatives of the metric appearing in the term, while the $c_{(i)}^{(2m,p)}$'s are constants. The densities $ \mathcal{R}^{(i)}_{(2m,p)} $ involve contractions of the Riemann tensor and its covariant derivatives. In appendix \ref{chp:App-basis} we present a generating set of these invariants for up to eight derivatives of the metric. The action is then evaluated on a single-function SSS metric ansatz and we impose \req{GQTGcond}, namely, that the Euler-Lagrange equation for $f(r)$ vanishes. This leads to constraints on the $c_{(i)}^{(2m,p)}$'s such that the resulting theory is of the GQT  type. 

Let us make a few further comments regarding the densities involving derivatives of the curvature. In general it is possible to reduce the number of invariants that make non-trivial contributions to the equations of motion by integrating by parts and utilizing the Bianchi identities. However, we have not pursued this option here. The reasons are simply because, at high-order in derivatives, there are so many terms that it would be impractical to do so. Furthermore, as will be obvious below, it is not necessary to do this  to understand the effects of these terms. Therefore, in constructing our actions at the four, six, and eight-derivative levels, we include all possible terms at a given order (as listed in the appendix). On the other hand, in the case of ten-derivative theories our analysis will not be exhaustive.

\subsubsection{Two-derivative actions}

\noindent For completeness, we include here the two-derivative sector, which is simply Einstein gravity, 
\begin{equation}
\mathcal{L}_{(2,0)}^{(1)}= R \, .
\end{equation}
The integrated equation for the metric function is given by
\begin{equation} \label{EEEin}
\mathcal{F}^{(1)}_{(2,0)} = -2 r (f-k) \, .
\end{equation}

\subsubsection{Four-derivative actions}

\noindent There are no non-trivial four-derivative GQT actions in four-dimensions.

\subsubsection{Six-derivative actions}

\noindent There is a single non-trivial six-derivative GQT action in four-dimensions. The action for this theory may be taken to be that of Einsteinian Cubic Gravity  \cite{PabloPablo}
\begin{equation}
\mathcal{L}_{(6,0)}^{(1)}=+12 \tensor{R}{_{a}^{c}_{b}^{d}}\tensor{R}{_{c}^{e}_{d}^{f}}\tensor{R}{_{e}^{a}_{f}^{b}}+R_{ab}^{cd}R_{cd}^{ef}R_{ef}^{ab}-12R_{abcd}R^{ac}R^{bd}+8R_{a}^{b}R_{b}^{c}R_{c}^{a}\, ,
\end{equation}
whose integrated equation for the metric function $f(r)$ reads \cite{Hennigar:2016gkm,PabloPablo2}
\begin{align}
\mathcal{F}_{(6,0)}^{(1)} &= - \frac{4}{r^2} \bigg[ r f f'' \left(rf' + 2(k-f) \right) - \frac{f'}{3} \left(r^2 f'^2 + 3 r k f' + 6 f(k-f) \right)\bigg] \, .
\end{align}

\subsubsection{Eight-derivative actions}

\noindent There are five non-trivial eight-derivative GQTG actions in four-dimensions. The first of these possibilities may be taken to be that given by the standard polynomial invariants ---see \eg \cite{PabloPablo4}. However, the additional four theories require terms involving covariant derivatives of the Riemann tensor. Of these, a single combination can be formed such that the integrated equations are second-order, while the remaining three involve higher-derivatives of the metric function. As examples of actions that give rise to each of the new sets of GQTGs, the following choices may be made:
\begin{align}
\mathcal{L}_{(8,0)}^{(1)} =&+ \tensor{R}{^{pqrs}} \tensor{R}{_p^t_r^u} \tensor{R}{_t^v_q^w} \tensor{R}{_{uvsw}}  - \frac{13}{5} \tensor{R}{^{pqrs}} \tensor{R}{_{pq}^{tu}} \tensor{R}{_r^v_t^w} \tensor{R}{_{svuw}}  - \frac{1}{8} \tensor{R}{^{pqrs}} \tensor{R}{_{pq}^{tu}} \tensor{R}{_{tu}^{vw}} \tensor{R}{_{rsvw}}  
\nonumber\\
&+ \frac{1}{5} R \tensor{R}{^{pqrs}} \tensor{R}{_q^t_s^u}\tensor{R}{_{tpur}}  \, ,
\\
\mathcal{L}_{(8,2)}^{(2)} =&+ \tensor{R}{^{pqrs}} \tensor{R}{^t_p^u_r ^{;v}} \tensor{R}{_{tqus;v}} - \tensor{R}{^{pq;r}} \tensor{R}{^{st}_p^u} \tensor{R}{_{stqr;u}} + 2 \tensor{R}{^{pqrs}} \tensor{R}{_p ^{tuv}} \tensor{R}{_{qtru;sv}} + \tensor{R}{^{pq}} \tensor{R}{^{rs;t}} \tensor{R}{_{rtsp;q}} 
\nonumber\\
&- 2 \tensor{R}{^{pq;r}} \tensor{R}{_p ^{s;t}} \tensor{R}{_{qsrt}} + \tensor{R}{^{pq;rs}} \tensor{R}{^t_p^u_r}\tensor{R}{_{tqus}} 
-\tensor{R}{^{pq}} \tensor{R}{^{rs}_{;p}} \tensor{R}{_{rs;q}} - \frac{1}{2} \tensor{R}{^{;p}} \tensor{R}{^{qr;s}} \tensor{R}{_{pqrs}}\nonumber\\
& + \tensor{R}{^{pq}} \tensor{R}{^{rs}_{;q}^t} \tensor{R}{_{prst}} 
- \frac{1}{2} \Box  \tensor{R}{^{pq}} \tensor{R}{^{rst}_p} \tensor{R}{_{rstq}}\, ,
\\
\mathcal{L}_{(8,4)}^{(3)} =&+ 21 R^{pq;rst} R_{prqs;t} - 12 \Box R^{pq} \Box R_{pq} - 12 R^{pq} R^{rs} R_{pq;rs} + 153 R^{pq} R^{rs;t} R_{rtsp;q} + 6 R^{;pq} \Box R_{pq} 
\nonumber\\
&+ \frac{33}{4} R^{pqrs;tu} R_{pqrs;tu} + 6 R^{;pq} \tensor{R}{^{rst}_q} R_{rstp} - \frac{21}{2} \Box R^{pq} \tensor{R}{^{rst}_p} R_{rstq} + 36 R^{pq;rs} \tensor{R}{^t_p^u_r} R_{tqus} 
\nonumber\\
&+ 183 R^{pqrs} \tensor{R}{_p^{tuv}} R_{qtru;sv} - 51 R^{pq} \tensor{R}{^{rs}_{;p}} R_{rs;q} - \frac{177}{2} R^{;p} R^{qr;s} R_{pqrs} - 81 R^{pq;r} \tensor{R}{_p^{s;t}} R_{qrst} 
\nonumber\\
&- 93 R^{pq;r} \tensor{R}{_p^{s;t}} R_{qsrt} - 60 R^{pq;r} \tensor{R}{^{st}_p^u} R_{stqr;u} + 33 R^{pqrs} \tensor{R}{^{tuv}_{p;q}} R_{tuvr;s} \nonumber\\
& - 27 R^{pqrs} \tensor{R}{^{tuv}_{p;r}} R_{tuvq;s} 
- 63 R^{pqrs} \tensor{R}{^t_p^u_r^{;v}} R_{tqus;v} + 6 R^{;pq} \Box R_{pq}\, ,
\\
\mathcal{L}_{(8,4)}^{(4)} =&+ 52 R^{p q} R^{rs;t} R_{rtsp;q} + 8 R^{pq;rst} R_{prqs;t}  -4 \Box R^{pq} \Box R_{pq} -20 R^{pq;r} \tensor{R}{_p^{s;t}} R_{qsrt} \nonumber\\
&- 24 R^{pq;r} \tensor{R}{^{st}_p^u} R_{stqr;u} 
-4 R^{pq;r} \tensor{R}{^s_p^{tu}} R_{sqtr;u} + 12 R^{pqrs}\tensor{R}{^{tuv}_{p;q}} R_{tuvr;s} \nonumber\\
& - 10 R^{pqrs} \tensor{R}{^{tuv}_{p;r}} R_{tuvq;s} -20 R^{pqrs} \tensor{R}{^t_p^u_r^{;v}} R_{tqus;v} 
+ 2 R^{;pq} \Box R_{pq} + 3R^{pqrs;tu} R_{pqrs;tu} \nonumber\\
&+ 72 R^{pqrs} \tensor{R}{_p^{tuv}} R_{qtru;sv} - 8R^{pq} \tensor{R}{^{rs}_{;p}} R_{rs;q} -22 R^{;p} R^{qr;s} R_{pqrs} - 36 R^{pq;r} \tensor{R}{_p^{s;t}} R_{qrst}\, ,
\\
\mathcal{L}_{(8,4)}^{(5)} =&+ 1178 R^{pq} R^{rs;t} R_{rtsp;q} + 171 R^{pq;rst} R_{prqs;t} -95 \Box R^{pq} \Box R_{pq} -76 R^{pq} R^{rs} R_{pq;rs} \nonumber\\
&- 646 R^{pq;r} \tensor{R}{_p^{s;t}} R_{qsrt} 
-475 R^{pq;r} \tensor{R}{^{st}_p^u} R_{stqr;u} + 228 R^{pq;r} \tensor{R}{^s_p^{tu}} R_{sqtr;u} \nonumber\\
&+ 266 R^{pqrs} \tensor{R}{^{tuv}_{p;q}} R_{tuvr;s} - 209 R^{pqrs} \tensor{R}{^{tuv}_{p;r}} R_{tuvq;s} 
-494 R^{pqrs} \tensor{R}{^t_p^u_r^{;v}} R_{tqus;v} \nonumber\\
&+ \frac{95}{2} R^{;pq} \Box R_{pq} + \frac{133}{2} R^{pqrs;tu}R_{pqrs;tu} + 38 R^{;pq} \tensor{R}{^{rst}_q} R_{rstp} 
+ 228  R^{pq;rs} \tensor{R}{^t_p^u_r} R_{tqus}\nonumber\\
& + 1520 R^{pqrs} \tensor{R}{_p^{tuv}} R_{qtru;sv} - 342 R^{pq} \tensor{R}{^{rs}_{;p}} R_{rs;q} - 646  R^{;p} R^{qr;s} R_{pqrs}
 \nonumber\\
&- 646 R^{pq;r} \tensor{R}{_p^{s;t}} R_{qrst}\, .
\end{align}
The integrated equations for each of these densities read, respectively,
\begin{align}
\mathcal{F}_{(8,0)}^{(1)} &= - \frac{12 f'}{5r^3} \bigg[ \frac{r f f''}{2} \left(r f' + 2 (k-f) \right) - \frac{f'}{3} \left(\frac{3 r^2 f'^2}{8} +  \frac{r f'}{2} (f + 2k) + 3 f (k-f) \right) \bigg]\, ,
\\
\mathcal{F}_{(8,2)}^{(2)} &= -\frac{4 f^2}{r^5} \alpha^2 \, ,
\\
\mathcal{F}_{(8,4)}^{(3)} &=+ \frac{3 f^2}{2 r^5} (5 \alpha^2-2r  \alpha \alpha'+r^2 \alpha'^2) \, ,
\\
\mathcal{F}_{(8,4)}^{(4)} &= - \frac{2 f^2}{r^5} \alpha (4 \alpha-\alpha'' r^2) \, ,
\\
\mathcal{F}_{(8,4)}^{(5)} &= +\frac{19 f^2}{4r^5} (4(\alpha-\alpha' r )\beta+\alpha'' r^2 (\alpha+\beta))\,,
\end{align}
where we defined the functions\footnote{The functions $\alpha(r)$ and $\beta(r)$ are directly proportional to the non-trivial components of the traceless Ricci tensor and Weyl tensor for the single-function static, spherically symmetric background, respectively.}
\begin{equation}
\alpha(r)\equiv 2(k-f(r))+r^2 f''(r)\, , \quad \beta(r)\equiv 2(k-f(r))+2rf'(r)-r^2 f''(r)\, .
\end{equation}
From the densities involving covariant derivatives, while the first three exclusively depend on $\alpha(r)$ and its derivatives, the fourth one also includes a dependence on $\beta(r)$ ---which can not be expressed in terms of $\alpha(r)$ and its derivatives.

Observe that $\alpha(r)$ and $\beta(r)$ identically vanish when evaluated for a maximally symmetric background. Namely, if we set 
\begin{equation}\label{ads}
f(r)|_{\rm (A)dS}\equiv \frac{r^2}{L_{\star}^2}+k \quad  \Rightarrow \quad \alpha(r)|_{\rm (A)dS}=\beta(r)|_{\rm (A)dS}=0\, ,
\end{equation}
 and therefore
\begin{equation}
\mathcal{F}_{(8,2)}^{(2)}|_{\rm (A)dS}=\mathcal{F}_{(8,4)}^{(3,4,5)}|_{\rm (A)dS}=0\, ,
\end{equation}
or, in other words, the equations of motion of the new GQTs identically vanish for maximally symmetric backgrounds. Furthermore, it is easy to see that the usual Schwarzschild-(A)dS solution satisfies the equations of the new densities. This follows from the fact that 
\begin{equation}
\alpha(r)|_{ \text{Sch-(A)dS}}=0\, , \quad \beta(r)|_{ \text{Sch-(A)dS}}=\frac{12M}{r}\, ,
\end{equation}
where
\begin{equation}
f(r)|_{\text{Sch-(A)dS}}\equiv \frac{r^2}{L_{\star}^2}+k-\frac{2M}{r}\, .
\end{equation}
Since all terms appearing in $\mathcal{F}_{8}^{(i=2,3,4,5)}$ are proportional to $\alpha(r)$ or its derivatives, it follows that 
\begin{equation}
\mathcal{F}_{(8,2)}^{(2)}|_{\rm Sch-(A)dS}=\mathcal{F}_{(8,4)}^{(3,4,5)}|_{\rm Sch-(A)dS}=0\, .
\end{equation}
This implies that if we couple the new densities to Einstein gravity, the Schwarzschild solution will not receive corrections from such terms. 

\subsubsection{Ten-derivative actions}

\noindent To the best of our knowledge, a full classification of higher-derivative invariants at ten-derivative order has not been undertaken. Therefore, our analysis in this section is necessarily incomplete but, as we shall see, interesting.

To study ten-derivative actions we do the following. We construct all possible combinations of ten-derivative actions built from lower-order densities ---for example, by multiplying all six-derivatives densities by the four-derivative ones, and so on. In addition to this, we include 20 additional terms that are explicitly order ten in derivatives. We list the ones used for this purpose in appendix \ref{chp:App-basis}. However, it is particularly relevant that our set contains the contraction of five Weyl tensors and the following density,
\begin{equation} 
C_{abcd} C^{abcd} C_{ef rs ;u} C^{ef rs ;u} \, .
\end{equation}
As discussed in~\cite{Ruhdorfer:2019qmk,Li:2023wdz}, in four space-time dimensions there are four non-trivial parity-preserving contributions to the effective field theory of gravity at the ten-derivative level. Two of them involve the square of a dual Riemann tensor and hence they vanish identically on spherically symmetric spacetimes. We thus are left with two contributions that modify spherically symmetric solutions. The first contribution can be taken, as usual, to be a contraction of five Weyl tensors. The density appearing above is a particular choice for the second non-trivial contribution. 

The ten-derivative action is the first instance where more than one non-trivial contribution to the EFT appears. Moreover, it is the first instance where terms involving covariant derivatives of the metric play an essential role ---\ie cannot be removed by field redefinitions. For these reasons, we expected to find novel GQT theories at this order that explicitly modify the solutions to vacuum Einstein gravity, corresponding to the two possible non-trivial effective field theory contributions. This expectation will be borne out.  

From the entire set of ten-derivative invariants that we construct, there turn out to be 21 independent contributions. This represents notable growth compared to the eight-derivative case where there were five independent contributions. Of the 21 independent ten-derivative GQT theories, only two of these are non-trivial when evaluated on the Schwarzschild solution ---corresponding to $\mathcal{F}_{(10, 0)}^{(1)}$ and $\mathcal{F}_{(10, 4)}^{(9)}$ below--- and hence will give raise to deformations of the Schwarzschild solution \cite{Aguilar-Gutierrez:2023kfn}. Of the 21 theories, 5 have second-order integrated equations, 7 have third-order, 6 have forth-order, 2 have fifth-order, and 1 has sixth-order. As we have not included all possible 10 derivative densities in our starting action, these numbers are likely to be incomplete. However, we expect that any additional GQTs, should they exist, will not correct the solutions of vacuum general relativity. The list of 21 inequivalent integrated equations reads
\begin{align}
\mathcal{F}_{(10, 0)}^{(1)} =&+ \frac{f'^2}{r^2} \bigg[\frac{f'^3}{5} + \frac{2(f+k) f'^2}{4 r} - \frac{2f(f-k)f'}{r^2} - \frac{f f''}{r} \left( r f' + 2(k-f) \right) \bigg] \, ,
\\
\mathcal{F}_{(10, 4)}^{(2)} =&+ \frac{f^2 \alpha^2 (f-k)}{r^7} \, , 
\\
\mathcal{F}_{(10, 4)}^{(3)} =&+ \frac{f^2 \alpha (\alpha + \beta)^2}{r^7} \, ,
\\
\mathcal{F}_{(10, 4)}^{(4)} =&+ \frac{f^2 \alpha \left( 8 \alpha (f-k)  + \alpha^2 - \beta^2 \right)}{r^7} \, ,
\\
\mathcal{F}_{(10, 4)}^{(5)} =&+ \frac{f^2 \alpha^2 \left(6(k-f) - \alpha \right)}{r^7} \, ,
\\
\mathcal{F}_{(10, 4)}^{(6)} =&+ \frac{f^2 (\alpha + \beta)^2 \left(r \alpha' - 2 \alpha  \right)}{r^7}\,, 
\\ 
\mathcal{F}_{(10, 4)}^{(7)} =&+ \frac{f^2 \alpha (\alpha + \beta) \left( r \alpha' - \alpha + \beta \right)}{r^7} \, ,
\\
\mathcal{F}_{(10, 4)}^{(8)} =& -\frac{f^2 \alpha \beta \left(r \alpha' - \alpha + \beta \right)}{r^7}  \, , 
\\
\mathcal{F}_{(10, 4)}^{(9)} =&+ \frac{1}{r^7} \bigg[ 48 f^2 r^3 \left(-r^2 f'' + 2 r f' + 2 (k-f) \right) (k-f) f''' 
\nonumber
\\
&+  f \left(2 \left( 65 f + 16 k \right) r f' 
+ 4 \left(2 k - 65 f \right) (k-f) \right) \left(\frac{r f'}{2} + k - f \right) r^2 f'' - 4 k r^4 f'^4 
\nonumber
\\
&- 3 \left(3 k^2 + 4 k f + 121 f^2 \right) r^3 f'^3 - 2 \left(2 k^2 + 38 k f + 1271 f^2 \right)(k-f) r^2 f'^2 
\nonumber
\\
&- 40 r f (k + 122 f) \left(k-f \right)^2 f' - 3448 f^2 (k-f)^3 \bigg] \, ,
\\
\mathcal{F}_{(10, 4)}^{(10)} =&+ \frac{f^2}{r^7} \bigg[4 r^2 (k-f) \alpha'^2 - 2 r \left( 4(k-f) + \alpha + \beta \right) \alpha \alpha'  - \alpha (\alpha + \beta)(\beta - 3 \alpha) \bigg] \, , 
\\
\mathcal{F}_{(10, 4)}^{(11)} =&+ \frac{f^2}{r^7} \bigg[ r^2 \left(4(k-f) - \alpha -\beta \right) \alpha'^2 
\nonumber
\\
&- 2 r \left(\left(4(k-f) + \beta \right) \alpha + \beta^2 \right) \alpha'  + 3 \alpha (\alpha + \beta)^2 \bigg] \, , 
\\
\mathcal{F}_{(10, 4)}^{(12)} =&+\frac{f^2}{r^7} \bigg[2r^2 \left(2(k-f) - \alpha \right) \alpha'^2 - 2 r \left(4(k-f) + 3 \beta - \alpha \right) \alpha \alpha' 
\\
&+ \alpha \left(\alpha^2  + 6 \alpha \beta  -3 \beta^2 \right) \bigg] \, , 
\\
\mathcal{F}_{(10, 4)}^{(13)} =&+ \frac{f^2}{r^7} \bigg[ r^2 (k-f) (\alpha +\beta) \alpha'' - r \left(\alpha^2 + \alpha \beta + 4 \beta(k-f) \right) \alpha'  
\nonumber
\\
&+ \left(2 \alpha^2  - \alpha \left(4(k-f) - 2 \beta \right) + 4 \beta(k-f) \right) \alpha  \bigg] \, ,
\\ 
\mathcal{F}_{(10, 4)}^{(14)} =&+ \frac{f^2}{r^7} \bigg[ r^2 (\alpha + \beta) \left(8(k-f) - \alpha - \beta \right) \alpha'' - 4 r \left(\alpha^2 + \beta \left( 8 (k-f) - \beta \right) \right) \alpha' 
\nonumber
\\
&+ 8 (\beta-\alpha) \left(4(k-f) - \alpha -\beta \right) \alpha \bigg]\,,
\\ 
\mathcal{F}_{(10, 4)}^{(15)} =&+ \frac{f^2 \alpha }{r^7} \bigg[ r^2(k-f) \alpha'' - (\alpha + \beta) \left(r \alpha' - \alpha + \beta \right) \bigg] \, ,
\\ 
\mathcal{F}_{(10, 4)}^{(16)} =&+ \frac{f^2  }{r^7} \bigg[r^2 \left(-\alpha^2 + \left(6(k-f) - \beta \right) \alpha + 2 \beta (k-f) \right) \alpha'' - 2 r \left(\alpha^2 + \alpha \beta + \beta(k-f)  \right) \alpha' 
\nonumber\\
&+ 4 \left(\alpha^2 - \left(2(k-f) - \beta \right) \alpha + 2 \beta (k - f) \right) \alpha \bigg] \, ,
\\
\mathcal{F}_{(10, 4)}^{(17)} =&+  \frac{f^2 \alpha }{r^7} \bigg[ r^2 \left( 4(k-f) - \alpha \right) \alpha'' - 4 \left(r \alpha' - \alpha + \beta \right) \beta \bigg]\,,
\\ 
\mathcal{F}_{(10, 4)}^{(18)} =&+ \frac{f^2  }{r^7} \bigg[ 
+ 16 r^2 \left(  3 \alpha + 3 \alpha + 4 f - 9 k \right) \alpha'^2
\nonumber
\\
& r^2 \left( -4 r \left(3 \alpha + 3 \beta + 4 f - 12 k \right) \alpha' + 4 \left(3 \alpha^2 + 4 f \alpha - 12 k \alpha - 3 \beta^2 - 4 f \beta + 4 k \beta \right)  \right) \alpha''
\nonumber
\\
&- 2 r \left(47 \alpha^2 - \alpha \beta + 92 f \alpha - 28 k \alpha - 48 \beta^2 - 32 f \beta + 32 k \beta \right) \alpha' 
\nonumber
\\
&+ \left( 5 \alpha^2 - 178 \alpha \beta + 176 f \alpha + 464 k \alpha - 183 \beta^2 - 64 f \beta + 64 k \beta \right) \alpha \bigg) \bigg]\,,
\\
\mathcal{F}_{(10, 4)}^{(19)} =&+ \frac{f^2  }{r^7} \bigg[96 f r^3 \alpha \alpha''' - 32 r^2 \left(-r f \alpha' + (19f - 9k) \alpha + \beta (k-f) \right) \alpha'' +  64 r^2 \left(3k-2f \right) \alpha'^2 
\nonumber
\\
&+  4 r \left(-163 \alpha^2 + \left(-148 k - 76 f - 163 \beta \right) \alpha + 32 \beta (k-f) \right) \alpha' 
\nonumber
\\
&- 2  \left( - 722 \alpha^2 - 2 \left( 35 \beta + 248 f + 392 k \right)  \alpha  + \beta \left(64(k-f) + 291 \beta \right) \right) \alpha\bigg] \, ,
\\
\mathcal{F}_{(10, 4)}^{(20)} =&+ \frac{f^2}{r^7} \bigg[ 6 r^3 f \left( r \alpha' - 2 \alpha \right) \alpha''' + 6 f r^4 \alpha''^2
\nonumber
\\
&- 4 \left( 4 r f \alpha'  + \left( 18 k - 13 f \right) \alpha + 5 \beta(k-f) \right)r^2 \alpha'' 
\nonumber
\\
&- 4 \left(3 k + 14 f \right) r^2 \alpha'^2 + 80 \left( \frac{\alpha^2}{16} + \left( \frac{23 k}{20} + \frac{41 f}{20} + \frac{\beta}{16} \right) \alpha  + \beta(k-f) \right) r \alpha' 
\nonumber
\\
&- 80 \left( \frac{7 \alpha^2}{32} + \left(-\frac{13 k}{10} +  \frac{23 f}{10} + \frac{5 \beta}{16} \right)\alpha + \beta \left( k -f  + \frac{3 \beta}{32} \right) \right) \alpha \bigg]\,,
\\
\mathcal{F}_{(10, 4)}^{(21)} =&-\frac{3f^2}{r^7}\bigg[-\frac{1}{3} \left( -\alpha' f r -\frac{9 \alpha^ 2}{2}+\left(4k+2f-\frac{9\beta}{2}\right)\alpha + \beta (k-f)\right)r^2\alpha'' \nonumber \\
& -\frac{r^4 f \alpha \alpha'''' }{6} +r^3  \left( -\frac{\beta}{4}+k+\frac{f}{3}-\frac{\alpha}{4}\right) \alpha \alpha''' +2\left(k-\frac{2f}{3}\right)r^2\alpha''^2 \nonumber \\
&+\frac{4}{3}\left(-\frac{163\alpha^2}{32}-\left(\frac{77k}{8}+\frac{21 f}{8}+\frac{163\beta}{32} \right)\alpha+\beta(k-f) \right)r \alpha' \nonumber\\
&-\frac{4\alpha}{3} \left(\frac{\beta' f r}{4}-\frac{361 \alpha^2}{64}-\left(\frac{69k}{4}+8f+\frac{35\beta}{32} \right)\alpha+\beta \left(k-\frac{3f}{4}+\frac{291\beta}{64} \right) \right)\bigg]\, .
\end{align}
As we can see, all densities but $\mathcal{F}_{(10, 0)}^{(1)}$ and $\mathcal{F}_{(10, 4)}^{(9)}$ involve linear combinations of terms proportional to either $\alpha(r)$, or $\beta(r)$, or their derivatives. Hence, for all those, the Schwarzschild metric solves the corresponding equations of motion. The explicit form of the covariant densities is rather complicated in general, so we have preferred not to include the full list here. The corresponding expressions for  $\mathcal{F}_{(10, 0)}^{(1)}$ and $\mathcal{F}_{(10, 4)}^{(9)}$ read, respectively,
\begin{align}
\mathcal{L}^{(1)}_{10, 0} =&+\frac{1}{2160} \bigg[5 R^5 + 132 R  \left( R_{ab} R^{ab} \right)^2 + 18 R \left(R_{abcd} R^{abcd} \right)^2 - 272 R^2 \tensor{R}{_a^b_c^d}\tensor{R}{_b^e_d^f} \tensor{R}{_e^a_f^c} 
\nonumber
\\
&+ 10 R^2 \tensor{R}{_{ab}^{cd}} \tensor{R}{_{cd}^{ef}} \tensor{R}{_{ef}^{ab}} - 30 R^3 R_{ab}R^{ab} - 102 R R_{ab} R^{ab} R_{cdef} R^{cdef} 
\nonumber
\\
&+ 552 R_{ij} R^{ij} \tensor{R}{_a^b_c^d}\tensor{R}{_b^e_d^f} \tensor{R}{_e^a_f^c} - 156 R_{ijkl} R^{ijkl} \tensor{R}{_a^b_c^d}\tensor{R}{_b^e_d^f} \tensor{R}{_e^a_f^c}\bigg]\,,
\\
\mathcal{L}^{(9)}_{10, 4} =& -\frac{1113943}{20864} \tensor{C}{_{abcd}} \tensor{C}{^{abcd}}  \tensor{C}{^{efgh;i}} \tensor{C}{_{efgh;i}}
+ \frac{19309071}{39446} R_b^a R_c ^b \tensor{R}{_{ae}^{cd}} \tensor{R}{_{gh}^{ef}} \tensor{R}{_{df}^{gh}}  
\nonumber
\\
&+ \frac{2168502179}{4733520} R_c^a R_d ^b \tensor{R}{_{ef}^{cd}}\tensor{R}{_{gh}^{ef}} \tensor{R}{_{ab}^{gh}} - \frac{23092199}{10758} R_b^a \tensor{R}{_{ad}^{bc}} \tensor{R}{_{fh}^{de}}\tensor{R}{_{c i}^{fg}} \tensor{R}{_{eg}^{hi}}  
\nonumber
\\
&+ \frac{7605694303}{4733520} R_b^a \tensor{R}{_{de}^{bc}}\tensor{R}{_{cf}^{de}} \tensor{R}{_{hi}^{fg}} \tensor{R}{_{ag}^{hi}}  + \frac{2051116779}{788920} \tensor{R}{_{cd}^{ab}} \tensor{R}{_{eg}^{cd}} \tensor{R}{_{ai}^{ef}} \tensor{R}{_{fj}^{gh}} \tensor{R}{_{bh}^{ij}} 
\nonumber
\\
&- \frac{6886022969}{2366760}  \tensor{R}{_{ce}^{ab}} \tensor{R}{_{af}^{cd}} \tensor{R}{_{gi}^{ef}} \tensor{R}{_{bj}^{gh}} \tensor{R}{_{dh}^{ij}}  + \frac{176696887}{215160} \tensor{R}{_{ce}^{ab}} \tensor{R}{_{fg}^{cd}} \tensor{R}{_{hi}^{ef}} \tensor{R}{_{aj}^{gh}} \tensor{R}{_{bd}^{ij}} 
\nonumber
\\
&+ \frac{237411}{7172} R_{ab}R^{ab} \tensor{R}{^{pqrs;t}} \tensor{R}{_{pqrs;t}} - \frac{388530029}{11360448} R_{abcd} R^{abcd} R^{;p}R_{;p}  + \frac{649013}{7824} R_{abcd} R^{abcd} R^{pq;r}R_{pq;r} 
\nonumber
\\
&+ \frac{472435}{43032} R_{abcd} R^{abcd} R^{pq;r} R_{pr;q} - \frac{1802455}{57376} R_{abcd} R^{abcd} R^{pq;rs} R_{p r q s} 
\nonumber
\\
&- \frac{1065937721}{2840112} R_{pqrs} R^{pqrs} \tensor{R}{_a^c_b^d} \tensor{R}{_c^e_d^f}\tensor{R}{_e^a_f^b} + \bigg[ \frac{1535482063}{22720896}R^{;a} R_{;a}  - \frac{34589843}{631136}  R^{pq;r} R_{pq;r}  
\nonumber
\\
& - \frac{90238183}{946704} R^{pq;r} R_{pr;q}    + \frac{21168615}{2524544} R^{pqrs;t} R_{pqrs;t}   + \frac{315857975}{1893408} R^{;pq} R_{pq}   - \frac{6185788187}{17040672} R_a^b R_b^c R_c^a 
\nonumber
\\
&- \frac{939340}{177507} \tensor{R}{_a^c_b^d} \tensor{R}{_c^e_d^f}\tensor{R}{_e^a_f^b}  - \frac{26416471}{516384} \tensor{R}{_{ab}^{cd}} \tensor{R}{_{cd}^{ef}} \tensor{R}{_{ef}^{ab}}\bigg] \Box R 
\nonumber
\\
&+ \bigg[ \frac{1612029697}{1893408}  R^{pq}R_p^r \Box R_{qr} + \frac{74535679}{118338} R^{p q} R^{r s} R_{pq ; rs} + \frac{48934355}{236676} R^{p q ; rs} \tensor{R}{^t_p^u_r} R_{t q u s}
\nonumber
\\
&+\frac{23734313}{59169} R^{pq} R^{rs;qt} R_{p r s t}+\frac{35456237}{946704} R^{pq} R^{r s t u} R_{r s t u ;p q}-\frac{44597992 }{59169} R^{pq} R^{rs} R_{p r ; q s} 
\nonumber
\\
&-\frac{619200179 }{1420056} R^{;p q} R^{r s} R_{p r q s}-\frac{315857975 }{1893408} R^{p q} \nabla_q \nabla_p \Box R +\frac{90238183 }{473352} R^{p q ; r} \nabla_q \Box R_{p r}
\nonumber
\\
&-\frac{21168615}{315568} R^{p q; rst}R_{p r q s ; t}+\frac{293954069 }{2840112}R^{;pq} \Box R_{p q} +\frac{1588811801 }{5680224} R R^{p q ; r s} R_{p r q s} 
\nonumber
\\
&+\frac{15025369}{19723} R^{pq} \Box R^{r s} R_{p r q s}+\frac{1535482063 }{11360448} R^{;pq} R_{;pq} +\frac{90238183 }{473352} R^{p q ; r s} R_{p r ; q s}
\nonumber
\\
&+\frac{26525693 }{258192} R^{;p q} R_p^r R_{q r}+\frac{34589843 }{315568} R^{p q; r s} R_{p q; rs}-\frac{681365365 }{946704}R R^{pq} \Box R_{pq}
\nonumber
\\
&+\frac{34589843 }{315568} R^{p q; r} \nabla_r \Box R_{p q}+\frac{293954069 }{1420056} R^{;p q r} R_{pq;r} +\frac{98790361 }{473352} R^{pq;r} \tensor{R}{_p^{s;t}} R_{q r s t}
\nonumber
\\
&-\frac{333105233 }{315568} R^{p q; r} \tensor{R}{_p^{s;t}} R_{q s r t}+\frac{17765777 }{86064} R^{pq;r} \tensor{R}{^{st}_p^u} R_{s t q r ;u}
-\frac{425439281 }{946704} R^{p q ;r} \tensor{R}{^s_p^{tu}} R_{s q t r ;u} 
\nonumber
\\
&+\frac{16960493 }{187776} R^2 \Box R +\frac{421946281}{315568} R^{p q r s} \tensor{R}{^{tuv}_{p;q}} R_{tuvr;s} -\frac{21168615 }{1262272} R^{p q r s ; t u} R_{p q r s ; t u} 
\nonumber
\\
&+\frac{238362363 }{631136} R^{p q r s} \tensor{R}{^{tuv}_{p;r}} R_{tuvq;s}-\frac{9210385 }{315568} R^{p q r s} \tensor{R}{^t_p^u_r^{;v}} R_{t q u s ;v} 
\nonumber
\\
&+\frac{298907053 }{1893408} \tensor{R}{^{m n r s}} \tensor{R}{_m^d_r^g} \tensor{R}{_d^c_g^i} R_{n c s i}-\frac{298907053}{7573632} \tensor{R}{^{m n r s}} \tensor{R}{_{m n} ^{ d g}} \tensor{R}{_{d g}^{c i}} \tensor{R}{_{r s c i}}\bigg] R\, .\label{L1049}
\end{align}


\section{Conclusions}\label{conclu}
A summary of the main findings of this chapter can be found in the introduction. Let us close with some remarks regarding open questions and future work. 

In this chapter, we have initiated the study of GQTs with covariant derivatives. Our analysis has been restricted to four dimensions and to the first few curvature orders. It would be interesting to pursue a full classification of GQTs with covariant derivatives in general dimensions as well as for arbitrary curvature orders, similar to the one achieved for polynomial GQTs in \cite{Bueno:2022res,Moreno:2023rfl}.

Additionally, it would be interesting to prove that any gravitational effective action can be mapped to a GQT. This is established for general polynomial densities \cite{Bueno:2019ltp}, but the proof for terms involving covariant derivatives is thus far limited to theories with up to eight derivatives  of  the metric and also for theories with any number of Riemann tensors and two covariant derivatives.

On a different front, it would be interesting to characterise the generalized symmetries of general linearized higher-derivative gravities with covariant derivatives along the lines of \cite{Benedetti:2023ipt}, where such analysis was performed for $\mathcal{L}(g^{ab},R_{abcd})$ theories.


\cleardoublepage
\part{Brane-Worlds}\label{part:BWs}
\lhead{Chapter 3}
\rhead{Gravity on Brane-Worlds}

\chapter{Gravity on Brane-Worlds}
\label{chp:ReviewBWs}

\section{Introduction}
\label{sec:IntroBWs}

Even though the possibility that we ---the Standard Model fields--- live in a four-dimensional subspace of a higher-dimensional spacetime had been around since the eighties \cite{Rubakov:1983bb, Lukas:1998yy}, it was not until Randall and Sundrum proposed brane-worlds in an AdS bulk \cite{Randall:1999ee, Randall:1999vf} that the field gained momentum.
Before their work, it was widely believed that the only way to obtain an effective four-dimensional description of our Universe from higher-dimensional theories of quantum gravity, such as 10-dimensional superstring theories or 11-dimensional M theory, was by making the extra dimensions compact and sufficiently small \cite{Antoniadis:1990ew}.
This is because one can confine matter fields to a lower-dimensional subspace, but gravity \textit{is} geometry, and so it must necessarily feel all dimensions.

Indeed, in general, spacetimes with $D$ large non-compact dimensions present a potential that behaves as $1/r^{D-3}$ around point-like masses. By compactifying and shrinking the extra dimensions to microscopic size, gravity will only feel them at such small scales, thus recovering the usual Newtonian $1/r$ potential at large distances.

Following this reasoning, Arkani-Hamed, Dimopoulous, and Dvali had previously proposed a model  \cite{Arkani-Hamed:1998jmv} in which, by confining the Standard Model fields to a four-dimensional brane, the compact dimensions could reach the millimetre scale, but that was as big as one could go to avoid conflicts with known results from high-energy collider experiments.
Instead, the solution proposed by Randall and Sundrum allowed for a large, non-compact, extra dimension. 
In particular, it consisted of embedding a four-dimensional brane with flat geometry into a five-dimensional AdS bulk spacetime.

Due to the curvature of the bulk, a massless graviton mode becomes localized on the brane. 
Although there is also a continuum of
Kaluza-Klein modes with arbitrarily small mass, their wave function is suppressed on the brane.
Thus, gravity on the brane becomes effectively four-dimensional at low enough energies, and indeed it displays a Newtonian potential at large scales \cite{Garriga:1999yh}.
However, at high energies, that is, at distances smaller than the curvature radius of the AdS bulk, gravity feels the true high-dimensional nature of the bulk.

Through the holographic duality \cite{Maldacena:1997re}, it is possible to reinterpret the effects of this large extra dimension as CFT radiation on the brane \cite{Gubser:1999vj, Verlinde:1999fy, Hawking:2000kj,deHaro:2000wj}. Alternatively, since the brane is at some finite distance from the asymptotic boundary, we can think of this model as imposing a UV cut-off to the dual CFT, and then coupling it to dynamical gravity on the brane. This gravitational theory is induced by integrating out the CFT degrees of freedom above the cut-off, in the same spirit as in Sakharov's induced gravity proposal \cite{Sakharov:1967pk}.

In order to force the brane to have a flat geometry, Randall and Sundrum had to fine-tune its tension to a critical value.
This requirement was relaxed by Karch and Randall \cite{Karch:2000ct} soon afterwards.
A brane tension smaller than the critical value allows the brane to have an AdS geometry, while the brane gets a dS geometry if its tension is larger than critical.
Nevertheless, if the value of the tension is close enough to criticality, then the brane sits close to the asymptotic boundary of the AdS bulk and one recovers almost-flat four-dimensional gravity on the brane.

\begin{figure}[t]
    \centering
    \includegraphics[scale=0.7]{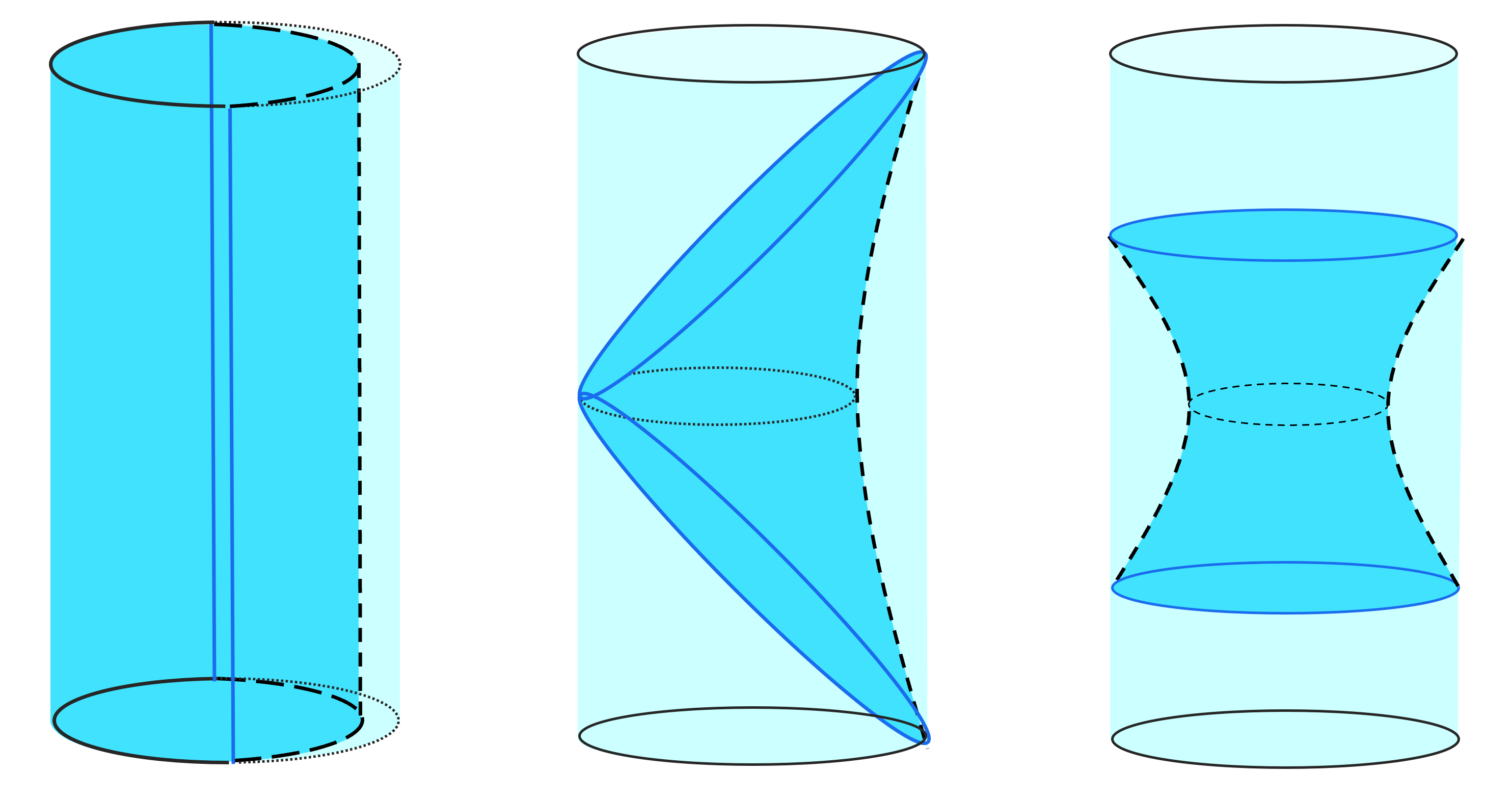}
    \caption{\small{The three possible maximally symmetric brane-world models, in global AdS. The branes are denoted with a black dashed line in the bulk, or with a dark blue line where they meet the asymptotic boundary. The part of the bulk spacetime that is in lighter blue is removed from the set-up. 
    Left: (Karch-Randall) AdS brane. Center: Randall-Sundrum flat brane. Right: dS brane.
    }}
    \label{fig:Branes}
\end{figure}

Even though non-flat geometries had already been considered, either in the context of cosmology \cite{Ida:1999ui, Kaloper:1999sm} or black hole physics \cite{Emparan:1999fd}, Karch and Randall were the first to show that gravity still localizes on the brane without the need of a second IR brane or a critical tension.
Moreover, they showed that the effective gravitational theory on AdS branes is not Einstein gravity but massive gravity.

As in the original Randall-Sundrum model, it is also possible to reinterpret these general Karch-Randall brane-world models through the holographic duality, as an effective four-dimensional gravitational theory coupled to cut-off CFT radiation dual to the five-dimensional AdS bulk \cite{Porrati:2001gx}.
Again, this effective brane picture is clearest when the brane is close to the asymptotic boundary, since the cut-off scale of the CFT is related to the distance in the bulk between the brane and the asymptotic boundary. However, we will argue that this interpretation can be extended beyond the cut-off, since the bulk picture provides a (partial) UV-completion of the system, that is, up to the scale of the brane thickness, or the bulk string or Planck scales.

It should also be mentioned here that the effective gravitational theory that is holographically induced on the brane is not simply Einstein gravity,
but a higher-derivative theory of gravity \cite{deHaro:2000wj, Neuenfeld:2021wbl}. 
As we will show in Chapter \ref{chp:Alg} of this thesis, the terms in the expansion can be computed algorithmically from the bulk theory \cite{Bueno:2022log}, following the same standard procedure that allows one to find the counterterms for holographic renormalization \cite{deHaro:2000wj, Kraus:1999di, Skenderis:2002wp, Emparan:1999pm}.
Again, the expansion parameter that controls the higher-curvature operators in the gravitational effective action is related to the distance from the brane to the asymptotic boundary.
We will study some properties of this higher-derivative gravity on the brane in Part III of this thesis.


Brane-world proposals were met with great interest, especially by cosmologists and phenomenologists, since they were not directly ruled out by existing observational or experimental data but nonetheless offered a way to test novel predictions to general relativity loosely inspired by M theory (see \eg \cite{Maartens:2010ar, Brax:2003fv, Kubyshin:2001mc} and references therein). 
For example, brane-worlds offered new ways of looking at the cosmological constant problem \cite{Arkani-Hamed:2000hpr}. Instead of wondering why our Universe seems to have such a small vacuum energy density on large scales, if we were to live on a Karch-Randall brane-world, we should ask why the brane tension is fine-tuned to being almost critical.

One should be careful, however, when considering predictions from brane-world models as robust outcomes of string theory. 
Although there exist some top-down constructions of brane-world set-ups (see  \eg \cite{Verlinde:1999fy, Chiodaroli:2012vc, Gutperle:2020gez, Uhlemann:2021nhu, Karch:2022rvr}), most brane-world constructions are bottom-up models in which the brane is infinitely thin and purely tensional, with no other charges.
From an EFT perspective, the brane tension is only the first term of a series expansion describing the brane action, so we could and should consider adding higher-order operators to it, as we will do in the following chapter.


Our interest in brane-world models, however, does not stem from cosmology or phenomenology, but from their applications to holography and black hole physics.
As we have mentioned before, from the brane perspective, brane-worlds describe dynamical gravity coupled to a cut-off CFT. 
Moreover, this description is under control, since the effective theory on the brane is regulated by the distance of the brane to the boundary.
Consequently, they can be the perfect theoretical laboratory in which to study black holes coupled to quantum matter or test the black hole information paradox.

Recently, brane-worlds have been used to show that one can indeed recover a unitary Page curve for evaporating large AdS black holes \cite{Almheiri:2020cfm}. These models suggest that the correct way of computing entropies in semi-classical gravity is by using quantum extremal surfaces \cite{Penington:2019npb, Almheiri:2019hni, Almheiri:2019yqk, Rozali:2019day, Chen:2020hmv, Chen:2020uac}, in what has become known as the island formula. In brane-worlds, this recipe corresponds to simply using the usual Ryu-Takayanagi prescription \cite{Ryu:2006bv} in the dual bulk picture, while allowing the RT surface to end on the brane. 

Brane-worlds have also been used to study (quantum) black holes interacting with strongly coupled CFTs, thanks to the use of the many versions of the C-metric \cite{Emparan:1999fd, Emparan:1999wa, Emparan:2000rs,Emparan:2020znc, Emparan:2022ijy, Emparan:2023dxm, Feng:2024uia, Climent:2024nuj}.

\begin{figure}[t]
    \centering
    \includegraphics[scale=0.5]{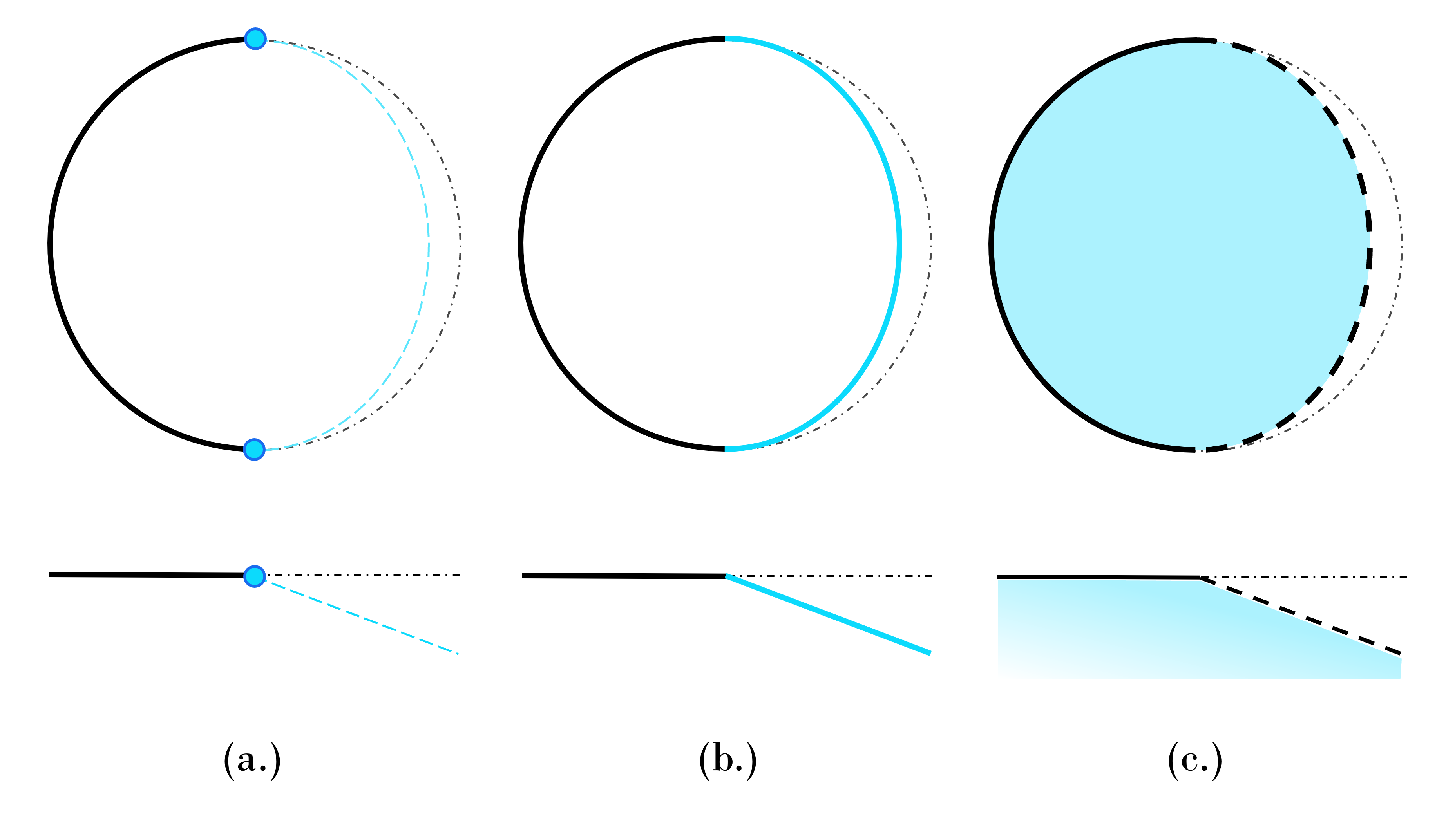}
    \caption{\small{The three different perspectives of brane-world holography in $d+1$ dimensions with an AdS$_d$ brane. 
    Top pictures are of a horizontal time slice of the system, while the bottom ones are in Poincaré-like coordinates, zooming in where the brane meets the asymptotic boundary. 
    From left to right: (a.) Boundary perspective: BCFT$_d$. A CFT$_d$ living on a $d$-dimensional fixed spacetime (black) with a boundary (blue dots), where it couples to a CFT$_{d-1}$.
    (b.) Brane perspective: It consists of a CFT$_d$ living on the fixed geometry of the asymptotic boundary (black), plus an effective CFT$_d$ with a UV cut-off coupled to (massive) gravity (with higher-curvature corrections) on the brane geometry (blue). There are transparent boundary conditions between the two CFTs at the defect
    where the brane meets the asymptotic boundary.
    (c.) Bulk perspective: Einstein gravity in an AdS$_{d+1}$ spacetime (blue), containing an AdS$_d$ Karch-Randall brane as an end-of-the-world brane (dotted line).
    }}
    \label{fig:KRPoinc}
\end{figure}

\begin{figure}[ht]
    \centering
    \includegraphics[scale=0.5]{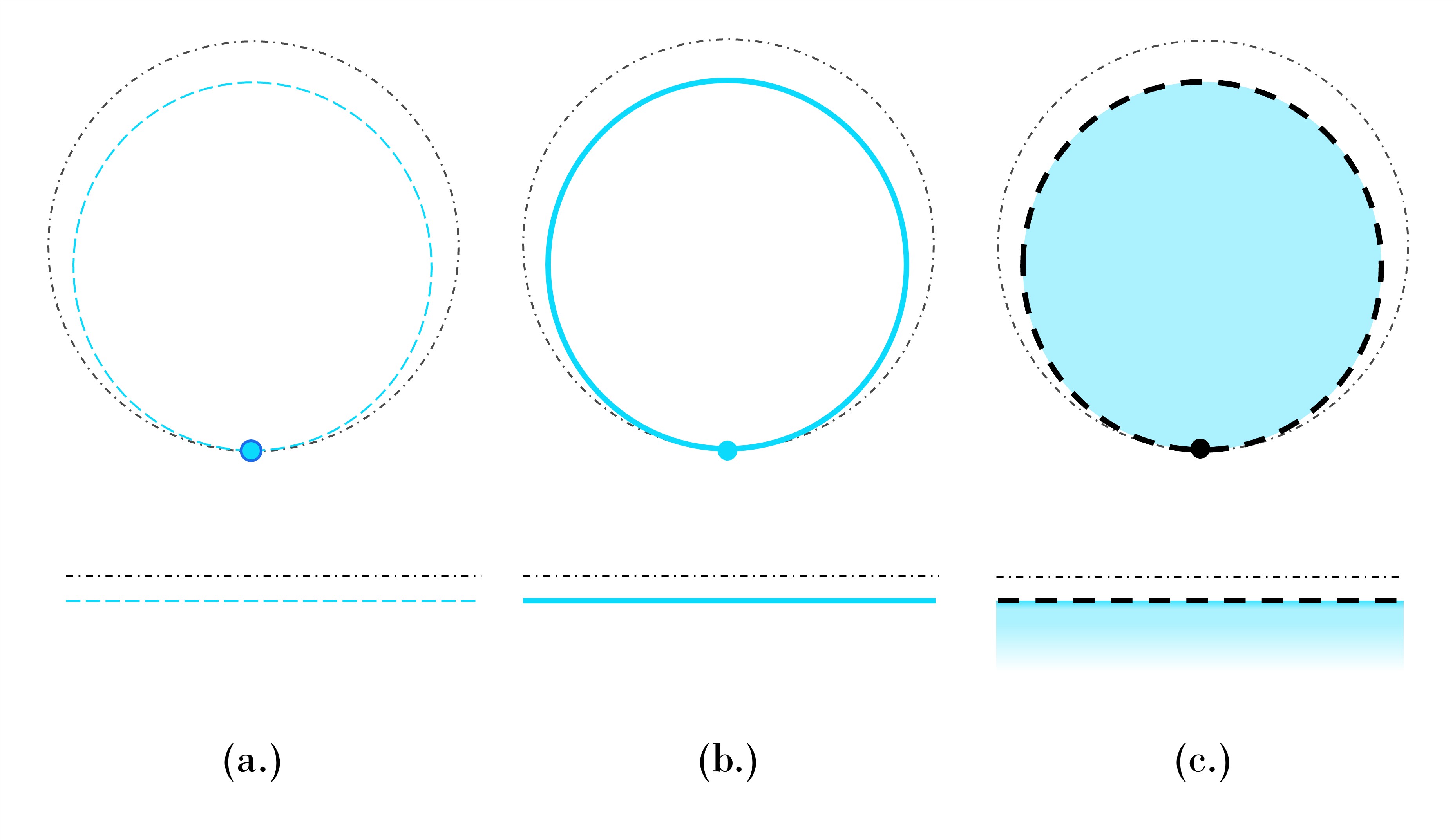}
    \caption{\small{The different perspectives of brane-world holography with a Minkowski brane. Top pictures are of a horizontal time slice of the system, while the bottom ones are in Poincaré-like coordinates, zooming in on the region far from where the brane meets the asymptotic boundary. From left to right: (a.) Boundary perspective: Unclear. Might be interpreted as a null defect at the ($d-1$)-dimensional null hypersurface where the bulk brane meets the asymptotic boundary, but it is not known what kind of theory lives on the defect.
    (b.) Brane perspective: Effective cut-off CFT$_d$ coupled to gravity (with higher-curvature corrections) on the brane geometry (blue). This view is only valid far from the boundary null defect. (c.) Bulk perspective: Einstein gravity in a subregion of an AdS$_{d+1}$ spacetime (blue), containing a $d$-dimensional Randall-Sundrum flat brane as an end-of-the-world brane (dotted line).
    }}
    \label{fig:RSBW}
\end{figure}

\begin{figure}[ht]
    \centering
    \includegraphics[scale=0.5]{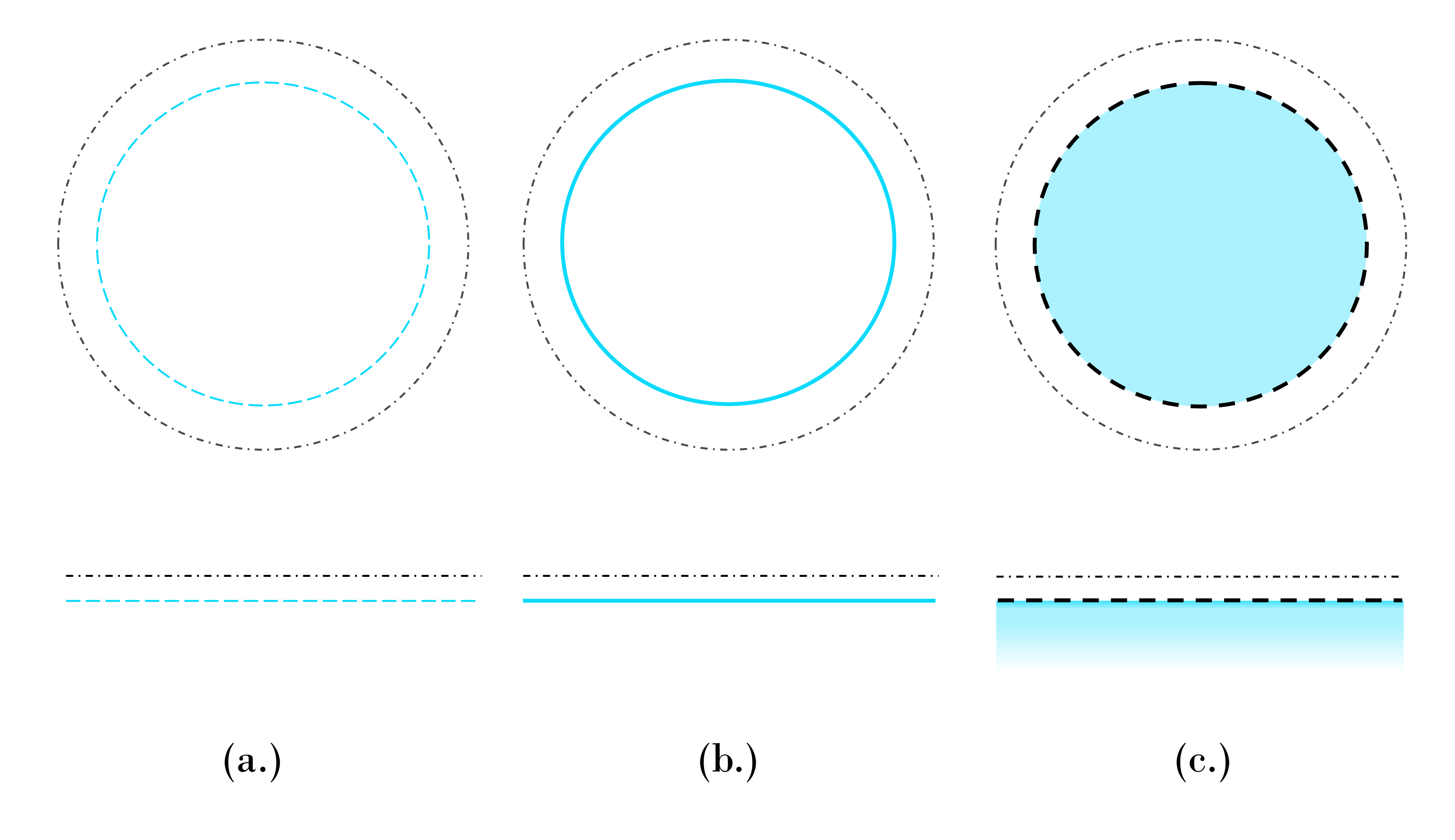}
    \caption{\small{The different perspectives of brane-world holography with a dS brane. Top pictures are of a horizontal time slice of the system, while the bottom ones are in Poincaré-like coordinates, zooming in on the brane. From left to right: (a.) Boundary perspective (not seen in the picture, see Fig. \ref{fig:Branes}): Unclear. Might be interpreted as spatial defects at the two ($d-1$)-dimensional spacelike hypersurfaces where the bulk brane meets the asymptotic boundary. The brane is created at some global time $T$ on one of these defects, and ceases to exist on the other one, some period later, at $T+\pi$ \cite{Emparan:2022ijy}.
    It is not known what kind of theory lives on the defects, or if the boundary picture corresponds to some non-local operator insertions at those two instances of time. 
    (b.) Brane perspective: Effective cut-off CFT$_d$ coupled to gravity (with higher-curvature corrections) on the brane geometry (blue). 
    (c.) Bulk perspective: Einstein gravity in a subregion of an AdS$_{d+1}$ spacetime (blue), containing an $d$-dimensional dS brane as an end-of-the-world brane (dotted line).
    }}
    \label{fig:dSBW}
\end{figure}

Although the C-metric allows for BHs in flat and even dS brane-worlds, most of these
works rely on the case of AdS Karch-Randall brane-worlds, which are qualitatively different from their flat or dS counterparts.
The main reason behind these dissimilarities is the fact that, for AdS branes, the brane only cuts off part of the asymptotic boundary, as opposed to the other two cases, in which the asymptotic boundary is completely removed from the set-up. This is shown in Figures \ref{fig:Branes}, \ref{fig:KRPoinc}, \ref{fig:RSBW}, and \ref{fig:dSBW}.

In dual terms, this means that AdS brane-worlds are not only dual to a cut-off CFT living on the dynamically gravitating brane geometry, but that this CFT is also coupled to a CFT living on the fixed geometry of the remaining asymptotic boundary. Both CFTs are connected through transparent boundary conditions at the defect where the brane reaches the asymptotic boundary.

That is the main reason why AdS brane-worlds are often used in the context of the BH information problem: they provide a model of a BH coupled connected to a non-gravitating bath, and so they are useful as models of BH evaporation, since they allow for a clean separation between the BH and radiation degrees of freedom.

The first consequence of having brane-world models with an AdS brane is that the bulk graviton mode that localizes on the brane is not massless but massive \cite{Karch:2000ct, Miemiec:2000eq, Schwartz:2000ip}.
From the bulk perspective, the graviton mode localized on the brane gets a mass because it is a mixture of a normalizable and a non-normalizable mode\footnote{Both modes are normalizable, since the brane sits at a finite distance from the boundary. However, if there were no brane, one of the modes would diverge, while the other would not, and so we still use this distinction.} \cite{Neuenfeld:2021wbl}, as opposed to the flat and dS brane cases, in which the brane graviton is massless since it comes from a purely non-normalizable bulk mode.
From the brane perspective, the graviton on the brane acquires a mass due to its interaction with the CFT radiation, which has transparent boundary conditions at infinity, where the brane meets the boundary.
This is, in fact, a general feature of gravity coupled to matter in AdS with transparent boundary conditions at infinity. If CFT radiation is allowed to leak out of an AdS spacetime (the brane, in our case), then the graviton acquires a mass through a Higgs-like mechanism \cite{Porrati:2001db, Porrati:2003sa}.

Secondly, in the case of AdS brane-worlds, the Kaluza-Klein modes do not form a continuum as in the flat or dS cases, but a discrete spectrum \cite{Karch:2000ct}.
This is due to bulk modes being sensitive both to the Dirichlet boundary conditions on the boundary and the Neumann-like boundary conditions on the brane, and so they feel as if they were trapped in a potential well.

Finally, there is a third unique way to interpret AdS Karch-Randall brane-worlds, obtained by dualizing the whole bulk, brane included.
The fully dual picture is that of a BCFT, that is, a CFT living on a $d$-dimensional geometry with a boundary, where it couples to a CFT$_{d-1}$.
The nature of this third boundary interpretation for flat or dS brane-worlds is not clear.
The BCFT picture for AdS Karch-Randall brane-worlds was already proposed by the original authors themselves in \cite{Karch:2000gx}, and later refined by \cite{DeWolfe:2001pq}. 
The fact that the brane graviton is massive translates, in the BCFT perspective, to the fact that, even though the stress-tensor of the full BCFT system is conserved, the stress-energy tensor of the CFT$_{d-1}$ defect is not, and so it gets an anomalous dimension \cite{Aharony:2003qf}.

This AdS/BCFT correspondence was later independently proposed by Takayanagi et al. \cite{Takayanagi:2011zk, Fujita:2011fp}, with a different philosophy, but technically similar methods.
Besides proposing the dictionary between the AdS bulk and the BCFT description, one of the key results of Takayanagi was proving a holographic $g$-theorem.
As we mentioned before, tuning the brane tension amounts to moving the position of the brane in the bulk.
As we move the brane deeper into the bulk by decreasing its tension, its curvature also decreases. In CFT terms, this translates to the fact that flowing to the IR reduces the number of degrees of freedom that are dual to the brane. 
Alternatively, the closer the brane is to the boundary, the more degrees of freedom has its dual defect CFT.

From this result, one might be tempted to say that the brane, containing gravity plus a cut-off CFT, in the intermediate picture, is exclusively dual to the defect CFT$_{d-1}$ in the boundary description.
Consequently, AdS Karch-Randall brane-world models have also become known as doubly-holographic models, since one may naively think that by dualizing once, we can go from the AdS bulk description to the intermediate brane picture, and then we can dualize again the AdS brane to get the BCFT picture.
But that is not exactly the case, as some top-down constructions show \cite{Karch:2022rvr, Uhlemann:2021nhu} that we should be careful when making such identifications.
The AdS/BCFT dictionary is well-described \cite{Fujita:2011fp, Park:2024pkt}, as well as an approximate dictionary between the bulk and the brane pictures \cite{Neuenfeld:2021wbl}, but, so far, the dictionary between the intermediate brane perspective and the BCFT description is far from clear, although some progress has been made in low-dimensional models \cite{Neuenfeld:2024gta}.
Even more so, it has been shown that BCFTs which have a holographic dual with a localized gravitating end-of-the-world brane are, in fact, not generic \cite{Reeves:2021sab}.
We can put this discussion aside, however, since we will not explore the BCFT interpretation in detail anywhere in this thesis.


In this chapter, we will redo and expand the original works of Randall, Sundrum, and Karch \cite{Randall:1999vf, Karch:2000ct}.
We will present the computations in a simplified way, generalizing them to any number of spacetime dimensions, while also dealing simultaneously with all three possible maximally symmetric brane geometries. 
We will refine the formula of the graviton mass as a function of the brane location for the case of AdS branes, and give new expressions for the mass of the higher harmonics.
We will also discuss how to reinterpret these results from the brane perspective \cite{Neuenfeld:2021wbl}. 
Even though we will study all three brane cases, we will only give a detailed discussion of the AdS Karch-Randall case, since it is the one that is more relevant for holography.

First, we will introduce the basic ingredients of our set-up in section \ref{sec:SetUp}, namely a $d$-dimensional brane with maximally symmetric geometry embedded in a ($d+1$)-dimensional AdS bulk. 
Then, in section \ref{sec:LLG}, we will linearly perturb the bulk and show how a bulk mode localizes on the brane, effectively describing $d$-dimensional gravity on the brane.
Finally, in section \ref{sec:BranePOV}, we will describe the effective brane picture, by integrating the bulk into the aforementioned cut-off CFT radiation. 
We will comment on the dictionary that relates both perspectives \cite{Neuenfeld:2021wbl}, but we will leave the detailed study of the higher-derivative gravitational theory on the brane for the following chapters.


\section{Set-up}\label{sec:SetUp}
Our starting point is the action
\begin{equation}\label{ActionBulk}
    I = I_\text{bulk} + I_\text{brane}\,,
\end{equation}
consisting simply of Einstein-Hilbert gravity in a $(d+1)$-dimensional asymptotically AdS bulk $\mathcal{M}$ which ends on a co-dimension one brane with tension $\tau$,
\begin{align}\label{IBulk}
    I_{\text{bulk}} & = \frac{1}{16 \pi G_N} \left[ \int_{\mathcal{M}} \df^{d+1} x \sqrt{-G} \left(  R[G] - 2\Lambda \right) + 2 \int_{\partial\mathcal{M}} \df^dx \sqrt{-g} \ K \right]\,,\\
    I_\text{brane} & = - \int_{\partial\mathcal{M}_b} \df^dx \sqrt{-g} \ \tau\,,
    \label{IBrane}
\end{align}
where $G$ denotes the bulk metric, $g$ is the induced metric on the brane, $G_N$ is the bulk Newton's constant, and $\Lambda$ is the bulk AdS$_{d+1}$ cosmological constant, with curvature radius $L$. 
The boundary of $\mathcal{M}$ is $\partial\mathcal{M} \supseteq \partial\mathcal{M}_b$, where we have included the Gibbons-Hawking-York boundary term explicitly to correctly get the desired boundary conditions for our set-up. 

Since the bulk spacetime ends at the brane, we will sometimes refer to it as an end-of-the-world (EOW) brane. From now on, when we speak about the boundary, we will generally only mean the asymptotic boundary at infinity, and not the EOW brane. 
In the original brane-world articles \cite{Randall:1999vf,Karch:2000ct}, as well as in many others, they did not consider the brane to be an EOW brane and instead imposed bulk $\mathbb{Z}_2$ symmetry, with the brane being the axis of symmetry.
This symmetry only introduces an extra factor of 2 in some equations that is not relevant to our discussion.


\subsection{EOMs and Junction Condition}

Varying the action \eqref{ActionBulk} with respect to the bulk metric $G_{MN}$, while imposing Dirichlet boundary conditions on the boundary and Neumann-like boundary conditions on the brane, we obtain the usual bulk AdS$_{d+1}$ Einstein equations,
\begin{equation}\label{EEsBulk}
    R_{MN} [G] - \frac{1}{2} G_{MN} R[G] + \Lambda G_{MN} = 0\,,
\end{equation}
plus the Israel junction condition on the brane \cite{Israel:1966rt}
\begin{equation}
    K_{ab} - K g_{ab} = - 8 \pi G_N \tau g_{ab}\,,
    \label{IJC}
\end{equation}
where $K_{ab}$ is the extrinsic curvature on the brane, defined as the Lie derivative of the metric in the direction normal to the brane,
\begin{equation}
    K_{ab} = \frac{1}{2} \mathcal{L}_n g_{ab}\,.
\end{equation}
and $K = g^{ab}K_{ab}$ is its trace.

Taking the trace with respect to the induced metric $g^{ab}$, we see that we must place the brane so that the trace of its extrinsic curvature is constant and proportional to its tension,
\begin{equation}\label{Kct}
    K = 8 \pi G_N \tau \frac{d}{d-1}\,.
\end{equation}
Plugging this result back into eq. \eqref{IJC}, the junction condition now reads
\begin{equation}\label{KabTU}
    K_{ab} = \frac{8 \pi G_N \tau}{d-1} g_{ab}\,.
\end{equation}
This greatly restricts the geometry of the brane and its location ---the brane is forced to sit on a totally umbilic hypersurface.


\subsection{Background Solution}

Our ansatz for the bulk metric in Poincaré-like coordinates takes the form
\begin{equation}\label{SlicingMetric}
    d{s}^2_{d+1} = G_{\mu \nu} (x,z) dy^\mu dy^\nu = \frac{L^2}{\left(f(z)\right)^2} \left[dz^2 + \hat{g}_{ij}(x)dx^idx^j \right]\,,
\end{equation}
where the $d$-dimensional metric $\hat{g}_{ij}(x)$ is either flat, or an (A)dS$_d$ metric with unit curvature radius.
This metric slices the bulk in slices of constant $z$ where the geometry is that of a $d$-dimensional maximally symmetric spacetime. See also Appendix \ref{chp:App-SlicingMetric} for more information on this metric and its curvature tensors.

\begin{figure}
    \centering
    \includegraphics[scale=0.55]{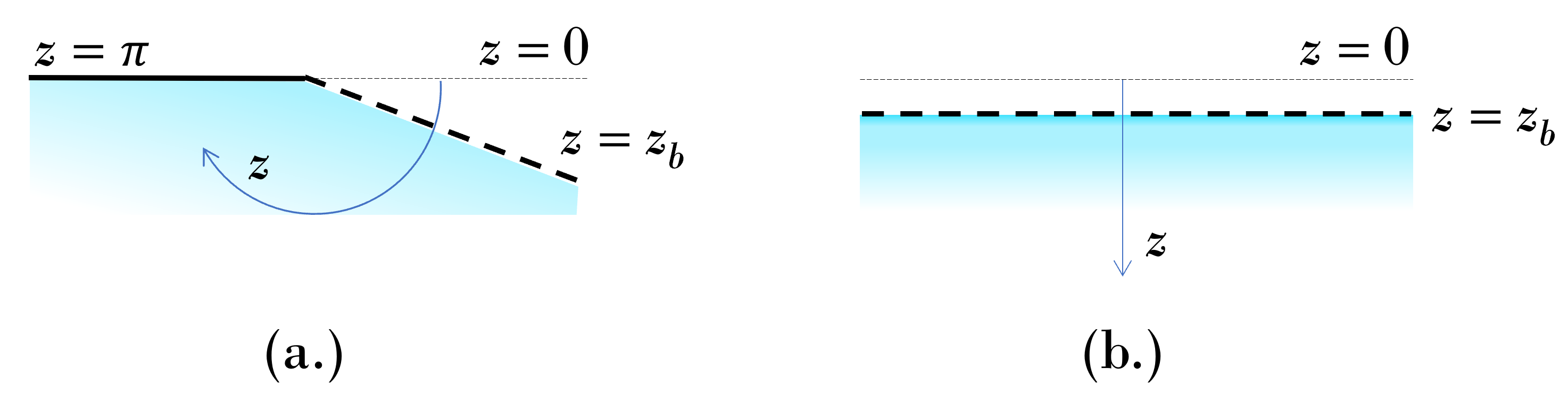}
    \caption{\small{(a.): For AdS branes, the $z$ coordinate goes from $z = z_b$ to $z = \pi$. (b.): For dS and Flat branes, the $z$ coordinate goes from $z = z_b$ to the horizon at $z = \infty$. In both cases, we will work in the limit where the brane is close to the boundary, $z_b \to 0$.}}
    \label{fig:zcoords}
\end{figure}

The radial coordinate $z$ starts at the position of the brane at $z = z_b$ (see Fig. \ref{fig:zcoords}). 
We will usually work in the limit of small $z_b$, that is, in the limit in which the brane is close to the boundary at $z = 0$. 
For AdS branes, the $z$ coordinate goes up to $z = \pi$, while for dS and flat branes, the $z$ coordinate goes all the way to the horizon at $z = \infty$.
In the case of flat branes, this horizon is the usual Poincaré horizon of the bulk. 
For dS branes, however, the horizon is a Rindler horizon. This horizon is present in our set-up because the brane must be accelerated in order not to fall deep into the AdS bulk.

The function $f(z)$ is a function which behaves as $f(z) \sim z$ for small values of $z$, and whose specific form depends on the brane geometry,
\begin{equation}\label{f(z)Def}
    f(z) \ = \  
    \begin{cases}
        \sin(z) \quad \ \text{for AdS branes,}\\
         \ \ \ z \quad \quad \ \text{for flat branes,}\\
        \sinh(z) \quad \text{for dS branes.}
    \end{cases}
\end{equation}
It is easy to check that these metrics indeed fulfil the bulk Einstein Equations \eqref{EEsBulk}.

Notice that the induced metric on constant $z$ slices is 
\begin{equation}
    g_{ij}(x,z) = \frac{L^2}{\left(f(z)\right)^2} \hat{g}_{ij}(x)\,.
\end{equation}
Therefore, for (A)dS branes, the actual curvature radius of the induced brane geometry is 
\begin{equation}\label{lbrane}
    l^2 = \frac{L^2}{\left(f(z_b)\right)^2}\,.
\end{equation}

Since the metric is block diagonal, the unit normal vector to the brane at $z=z_b$ is simply given by
\begin{equation}
    \partial_n = - \frac{f(z)}{L} \partial_z \,,
\end{equation}
where the minus sign comes from the fact that we want to pick our unit normal vector to be \textit{outward} directed \cite{Chen:2020uac}. In our case, since we are excising the part of the spacetime behind the brane, with $0 < z < z_b$, we need the minus sign so that $n$ points towards decreasing values of $z$.

The extrinsic curvature for constant $z$ slices is
\begin{align}
    K_{ij} & = \frac{f'(z)}{L} g_{ij}\,,
\end{align}
where the prime denotes the derivative with respect to the holographic coordinate $z$, \ie $' = \partial_z$.

Now, it is easy to see that the Israel junction condition \eqref{IJC} for the brane at $z = z_b$ is fulfilled provided that
\begin{equation}\label{IJCTau}
    \tau = \tau_c f'(z_b) \,,
\end{equation}
where we have defined
\begin{equation}\label{TauCrit}
    \tau_c = \frac{d-1}{8 \pi G_N L}\,.
\end{equation}
Since $\cos(z_b) < 1$, we must have $\tau < \tau_c$ for AdS branes, while $\tau > \tau_c$ corresponds to dS branes, with $\cosh(z_b) > 1$.
For flat branes, the tension of the brane becomes critical $\tau = \tau_c$, and we can place it on any value of the holographic coordinate $z$.

It will be useful for later to rewrite the above metrics in a Fefferman-Graham fashion,
\begin{equation}\label{FGbulk}
    ds^2_{d+1} = \frac{L^2}{4\rho^2} d\rho^2 + \frac{L^2}{\rho} \tilde{g}_{ij}(\rho,x)dx^i dx^j\,,
\end{equation}
where
\begin{equation}
    \tilde{g}_{ij}(\rho,x) = F(\rho) \hat{g}_{ij}(x)\,,
\end{equation}
with
\begin{equation}\label{f(rho)Def}
    F(\rho) \ = \  
    \begin{cases}
        \left( \frac{1+\rho}{2} \right)^2 \quad \text{for AdS branes,}\\
        \quad \ 1 \quad \quad \ \ \text{for flat branes,}\\
        \left( \frac{1-\rho}{2} \right)^2 \quad \text{for dS branes.}
    \end{cases}
\end{equation}
The necessary changes of variables can be found by solving the following ODE
\begin{equation}
    \frac{d \rho}{2 \rho} = \frac{dz }{f(z)}\,,
\end{equation}
which gives
\begin{equation}\label{CoV-rhotoz}
    \rho \ = \  
    \begin{cases}
        \ \tan^2 \left( \frac{z}{2} \right) \quad \text{for AdS branes,}\\
        \quad \ \ \ z^2 \quad \quad \ \text{for flat branes,}\\
        \tanh^2 \left( \frac{z}{2} \right) \quad \text{for dS branes.}
    \end{cases}
\end{equation}
Alternatively,
\begin{equation}\label{fCoV-ztorho}
    f(z) \ = \  
    \begin{cases}
        \ \sin{z} = \frac{2\sqrt{\rho}}{1+\rho} \quad \text{for AdS branes,}\\
        \quad \ z \ = \sqrt{\rho} \quad \ \text{for flat branes\,,}\\
        \sinh z = \frac{2\sqrt{\rho}}{1-\rho} \quad \text{for dS branes.}
    \end{cases}
\end{equation}

Notice that the coordinate $\rho$ starts at a value $\rho_b$ close to the boundary at $\rho = 0$.
For the AdS case, we find the other side of the asymptotic boundary at $\rho = \infty$ corresponding to $z = \pi$, while we find the horizon at $\rho \to \infty$ and $\rho = 1$ for the flat and dS cases, respectively.


\section{Locally Localized Gravity - The Bulk Perspective}\label{sec:LLG}

Now, let us perturb our previous metric \eqref{SlicingMetric} with a linear axial transverse and traceless perturbation ($\delta G_{\mu z} = 0$, $\hat{g}^{ij} \delta \hat{g}_{ij} = 0$, $\hat{\nabla}^i \delta \hat{g}_{ij} = 0$),
\begin{equation}\label{PertSlicingMetric}
    d{s}^2_{d+1} = \frac{L^2}{\left(f(z)\right)^2} \left[ dz^2 + \left(\hat{g}_{ij}(x) + \delta \hat{g}_{ij}(x,z)\right) dx^i dx^j \right]\,,
\end{equation}
We are interested in transverse and traceless perturbations since they are the ones that look like gravitons from the brane perspective. Moreover, the other modes can be shown to be non-dynamical \cite{Karch:2001jb}.
And we can always choose an axial gauge.

Taking this metric \eqref{PertSlicingMetric} and plugging it into the AdS$_{d+1}$ bulk Einstein Equations \eqref{EEsBulk}, and using the fact that the brane itself is a $d$-dimensional maximally symmetric spacetime, 
one obtains the following equation
\begin{equation}\label{EEPert}
    \left[ \partial^2_z - (d-1) \frac{f'(z)}{f(z)} \ \partial_z + \left( \hat{\square} - 2 \sigma \right) \right]\delta \hat{g}_{ij}(x,z)  = 0,
\end{equation} 
where $\hat{\Box} = \hat{\nabla_i}\hat{\nabla}^i$, with $\hat{\nabla}$ being the Levi-Citiva connection of the unperturbed $\hat{g}_{ij}$ brane metric, and we have defined
\begin{equation}\label{sigma}
    \sigma \ = \  
    \begin{cases}
        -1 \quad \text{for AdS branes,}\\
         \ 0 \quad \ \text{for flat branes,}\\
        +1 \quad \text{for dS branes.}
    \end{cases}
\end{equation}
We can also substitute our perturbed metric \eqref{PertSlicingMetric} into the Israel junction condition \eqref{IJC} to find its boundary condition on the brane, with $K_{ij}$ and $K$ being, to linear order in perturbation,
\begin{align}
    \delta K_{ij} & = 
    \frac{L f'}{f^2} \delta \hat{g}_{ij} - \frac{f}{2L} \delta \hat{g}'_{ij}\,, &
    \delta K & = 0\,.
\end{align}
Then, using the unperturbed junction condition \eqref{IJCTau} to simplify our calculations, we are simply left with the Neumann boundary condition
\begin{equation}\label{BCPert}
    \left[ \partial_z \delta \hat{g}_{ij}(x,z) \right]_{z_b} = 0\,.
\end{equation}
Notice that indeed the trace of the extrinsic curvature has remained constant, and that the brane still lies on a totally umbilic hypersurface, as equations \eqref{Kct} and \eqref{KabTU} dictate.

Finally, assuming that the perturbation fulfils the following separation ansatz,
\begin{equation}\label{separPert}
   \delta \hat{g}_{ij} (x,z) = H(z) h_{ij}(x)\,,
\end{equation}
and introducing the separation constant $E^2$, equation \eqref{EEPert} translates into
\begin{align}\label{EqRad}
     \left[ \partial^2_z - (d-1) \frac{f'(z)}{f(z)} \ \partial_z \right] H(z) & = - E^2 H(z)\,, \\
    \left( \hat{\square} - 2 \sigma \right)h_{ij}(x) & = E^2 h_{ij}(x)\,, \label{EqBrane}
\end{align}
and the boundary condition \eqref{BCPert} simply becomes
\begin{equation}\label{BCBrane}
    H'(z_b)=0\,.
\end{equation}
Details for these calculations can be found in Appendix \ref{chp:App-SlicingMetric}.


\subsection{The Brane Equation}
Let us first have a look at eq. \eqref{EqBrane}. This equation describes the behaviour of $\delta \hat{g}_{ij}(x,z)$ on the hypersurfaces of constant $z$.
The separation constant $E^2$ is an eigenvalue of the Lichnerowicz operator $(\hat{\Box} - 2\sigma)$.
Now, at $z=z_b$, we can rescale eq. \eqref{EqBrane}  to see that, from the point of view of the induced brane metric, this equation is
\begin{equation}\label{EqBrane2}
    \left( \square - \frac{2 \sigma}{l^2} \right)h_{ij}(x) = m^2 h_{ij}(x)\,,
\end{equation}
where 
\begin{equation}\label{m2toE2}
    m^2 = \frac{1}{l^2} E^2\,,
\end{equation}
and $\Box = \nabla_i \nabla^i$, with $\nabla$ being the Levi-Citiva connection of the unperturbed induced metric on the brane $g_{ij}$, and recall that the radius $l$ is the actual curvature radius of the induced brane geometry defined at eq. \eqref{lbrane}, $l^2 = L^2 / \left(f(z_b)\right)^2$.

As we saw in Part \ref{part:HDGs} of this thesis ---see \eg eq. \eqref{spin2m}---, equation \eqref{EqBrane2} describes a spin-2 massive mode in a maximally symmetric spacetime of radius $l$.
Therefore, from the perspective of the brane metric, indeed these $h_{ij}(x)$ perturbations look like massive spin-2 gravitons with mass $m^2$. 
We will be able to find the allowed masses $m^2$ by studying the radial equation \eqref{EqRad}.


\subsection{The Radial Equation}\label{sec:EqRad}

Let us now study the radial equation \eqref{EqRad} with boundary condition \eqref{BCBrane} on the brane. This equation describes the behaviour of $\delta \hat{g}_{ij}(x,z)$ along the holographic direction $z$.

We will further impose Dirichlet boundary conditions on the asymptotic boundary on the other side of the spacetime
\begin{equation}
    H(z = \pi) = 0\,,
\end{equation}
for the AdS brane case, and regularity at the horizon at $z \to \infty$ 
for dS and flat branes..

There are many ways to solve this equation. We shall illustrate a couple of them here, and leave the rest for Appendix \ref{chp:App-BWresults}.


\subsubsection*{The Volcano Potential}

To gain some intuition before crunching the numbers, let us have a look at the original way \cite{Randall:1999vf,Karch:2000ct} of solving equation \eqref{EqRad}. It consists in redefining the radial function $H(z)$ as
\begin{equation}
   \tilde{H}(z) = \left( \frac{L}{f(z)} \right)^{\frac{d-1}{2}}H(z)\,,
\end{equation} 
to obtain a classical time-independent Schrödinger equation 
\begin{equation}\label{HRadSch}
    \left[- \partial_z^2 + V(z) \right] \tilde{H}(z) = E^2 \tilde{H}(z)\,,
\end{equation}
with potential
$$
    V(z)  =  \frac{d^2-1}{4}\frac{1}{\left(f(z)\right)^2}  + \sigma \frac{(d-1)^2}{4}\,.
$$

\begin{figure}[ht]
    \centering
    \includegraphics[scale=0.6]{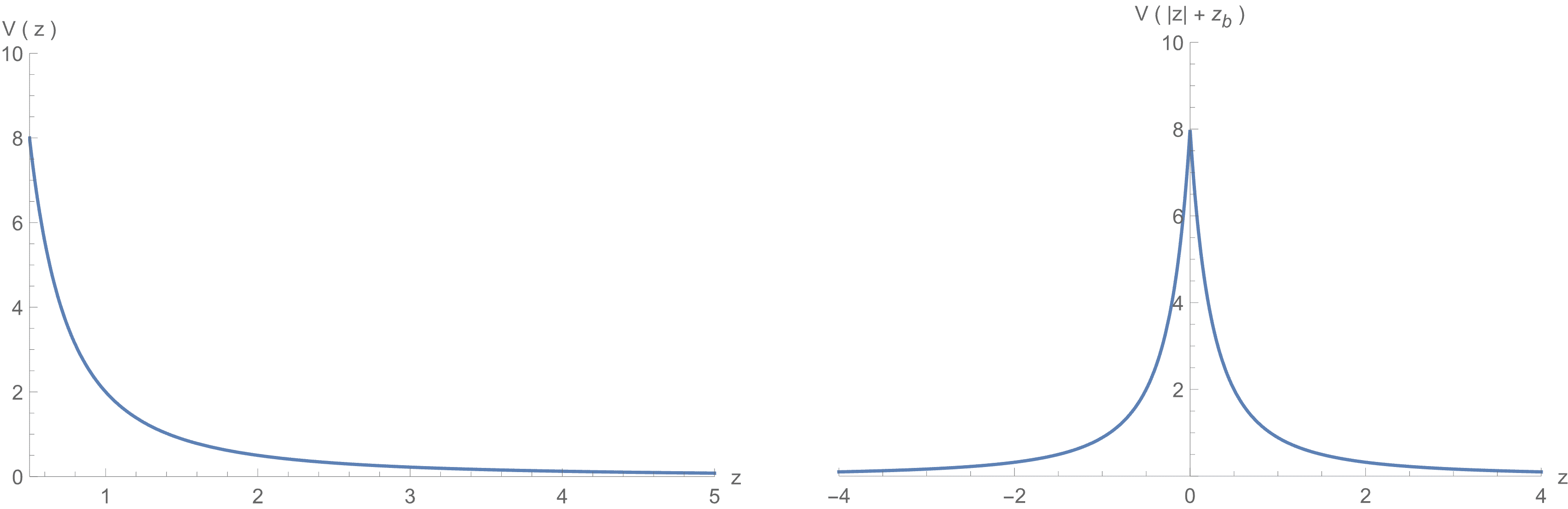}
    \caption{\small{Left: Flat-brane potential $V(z)$ for $d = 3$ and $z_b = 1/2$. Right: $\mathbb{Z}_2$ symmetric flat-brane potential $V(z)$ for the same values.}}
    \label{VMink}
\end{figure}

\begin{figure}[ht]
    \centering
    \includegraphics[scale=0.6]{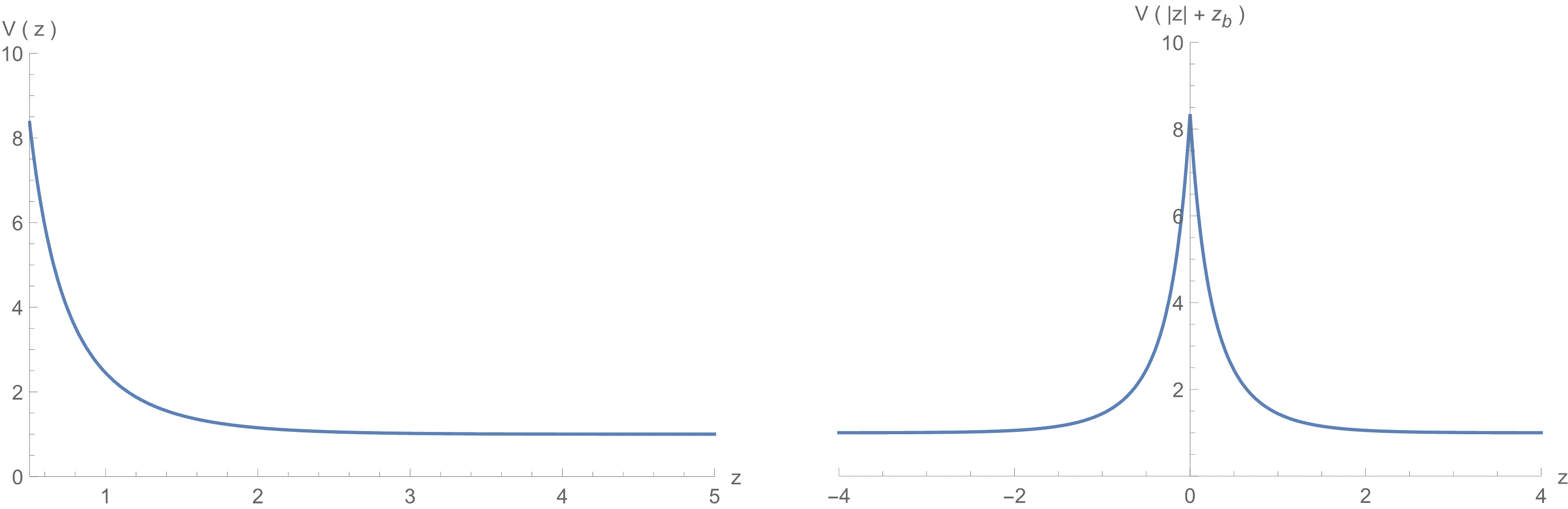}
    \caption{\small{Left: dS-brane potential $V(z)$ for $d = 3$ and $z_b = 1/2$. Right: $\mathbb{Z}_2$ symmetric dS-brane potential $V(z)$ for the same values.}}
    \label{VdS}
\end{figure}

\begin{figure}[ht]
    \centering
    \includegraphics[scale=0.6]{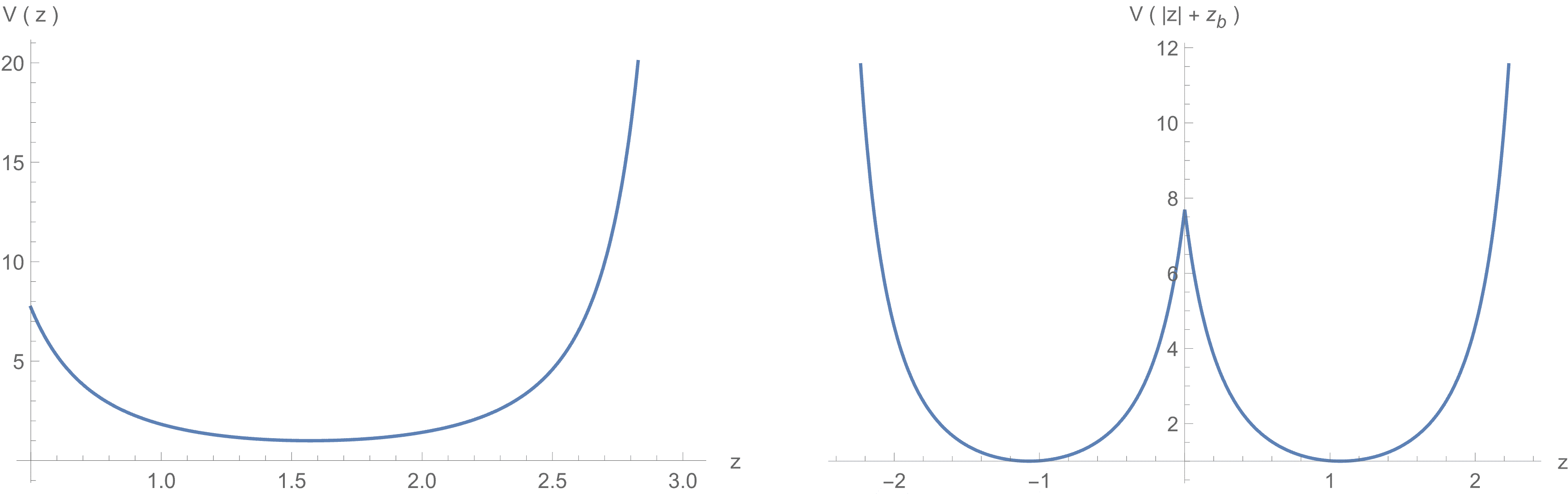}
    \caption{\small{Left: AdS potential $V(z)$ for $d = 3$ and $z_b = \pi/8$. Right: Same potential with $\mathbb{Z}_2$ symmetry, to illustrate the name ``volcano potential''. Notice how the potentials diverge as we approach the asymptotic boundary, at $ z + z_b = \pm \pi$.}}
    \label{VAdS}
\end{figure}

Borrowing insight from undergraduate quantum mechanics, it is easy to see from the shape of the potentials $V(z)$ that the spectrum of eigenvalues will be continuous in the case of flat and dS branes, since their potentials fall off at infinity (Figures \ref{VMink} and \ref{VdS}). On the other hand, the spectrum will be discrete in the case of AdS branes, since the potential looks like a well (Figure \ref{VAdS}). This is due to the bulk modes being sensitive to the Dirichlet boundary condition at the asymptotic boundary at $z = \pi$.

Moreover, after this redefinition, the boundary condition on the brane has become
\begin{equation}
    \tilde{H}'(z_b) + \frac{d-1}{2}\frac{f'(z_b)}{f(z_b)}\tilde{H}(z_b) = 0\,,
\end{equation}
which acts as a delta function pointing downwards on the potential at $z = z_b$. This ensures that there exists a lowest-lying mode whose wavefunction is localized on the brane. 

For flat and dS branes, the lowest-lying eigenvalue is exactly massless, that is, there exists a solution to equation \eqref{HRadSch} with $E^2=0$. We will show this explicitly, as well as the localized radial profile of these modes, in the next subsection.
Moreover, we can see that the potential falls off to zero at infinity, $V(z \to \infty) \to 0$, for flat branes, while $V(z \to \infty) \to (d-1)^2/4$ for dS branes. That means that there's no mass gap between the lowest-lying eigenvalue and the continuum modes for flat branes, but that there's a mass gap for dS branes, since the first excited state needs a minimum energy of $E^2_{\text{gap}} = (d-1)^2/4$.
In any case, we will see that all these excited modes are not localized on the brane.

For AdS branes, one can find the solutions to the Schrödinger equation \eqref{HRadSch} using standard techniques from quantum mechanics. The lowest-lying mode is almost massless \cite{Karch:2000ct, Miemiec:2000eq, Schwartz:2000ip, Neuenfeld:2021wbl}, with its mass going to zero as the brane gets closer to the boundary as
\begin{equation}
    E^2_0 \simeq \frac{d-2}{2^{d-1}} \frac{\Gamma(d)}{\left(\Gamma(d/2)\right)^2} z_b^{d-2}\,,
\end{equation}
while the excited modes have a mass of
\begin{equation}
    E^2_n = n(n+d-1) + \mathcal{O}(z_b^{d-2})\,.
\end{equation}
We will provide the details on how to compute the whole spectrum analytically and numerically in the next subsection, where we solve the original equation \eqref{EqRad} directly.


\subsubsection*{Full Solution}

Now that we have an idea of the problem at hand, let us try to solve it directly. We will first study the case of flat and dS branes together, since they are qualitatively very similar, and leave the (more interesting) AdS case for later.

\begin{figure}[ht]
    \centering
    \includegraphics[scale=0.6]{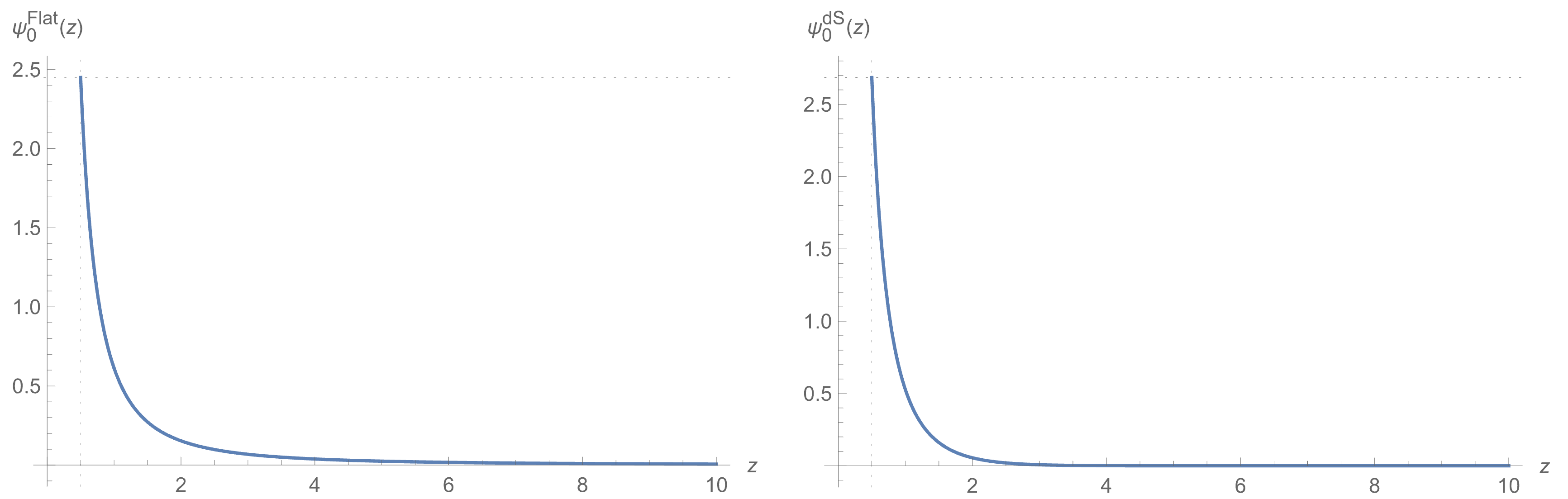}
    \caption{\small{Left: Normalized radial profile for the massless mode on a flat brane at $z_b = 1/2$, with $d = 3$. Right: Normalized radial profile for the massless mode on a dS brane at $z_b = 1/2$, with $d = 3$.}}
    \label{fig:ZeroModesFdS}
\end{figure}

First, we will look for the massless mode on flat and dS branes. 
Solving the radial equation \eqref{EqRad} for $H(z)$ while imposing $E_0^2 = 0$ and the boundary condition \eqref{BCBrane} gives a constant value for $H(z)$, both for the dS and flat brane cases.
Recall however, from our definitions of the bulk perturbations in eqs. \ref{PertSlicingMetric} and \ref{separPert}, that once we take the warp factor into account, the actual radial profile of the bulk modes is 
\begin{equation}
	\psi(z) = \frac{L^2}{\left( f(z) \right)^2} H(z)\,.
\end{equation}
Therefore, the radial profile of the zero mode goes as $\psi_0(z) \sim L^2/z^2$ for flat branes, and as $\psi_0(z) \sim L^2/\sinh{z}^2$ for dS branes, 
and so indeed the massless modes are localized on the brane (see Fig. \ref{fig:ZeroModesFdS}).

For the flat brane case, we can solve equation \eqref{EqRad} for the excited states of energy $E^2$ to find
\begin{equation}\label{HBessel}
    H(z) = c_1 z^{d/2} J_{d/2}(E z) + c_2 z^{d/2} Y_{d/2}(E z)\,,
\end{equation}
where the functions $J$ and $Y$ are the usual Bessel functions of the first and second kind, respectively, and $c_1$ and $c_2$ are arbitrary constants. 

For the dS brane case, equation \eqref{EqRad} can be similarly solved, and we find that the radial function of the excited states can be written as a linear combination of two hypergeometric functions which have the hyperbolic tangent as their argument,
\begin{align}\label{HHyper}
    H(z) = \ & c_1 \frac{\left(\sinh(z)\right)^d}{\left(\cosh(z)\right)^{1+\nu_+}} \prescript{}{2}{F_1}\left(\frac{1+\nu_+}{2},1+\frac{\nu_+}{2};1+\frac{d}{2};\tanh^2(z)\right) \\
    & + c_2 \frac{\left(\sinh(z)\right)^d}{\left(\cosh(z)\right)^{1+\nu_-}} \prescript{}{2}{F_1}\left(\frac{1+\nu_-}{2},1+\frac{\nu_-}{2};1-\frac{d}{2};\tanh^2(z)\right)\,.
\end{align}
where again $c_1$ and $c_2$ are some arbitrary complex numbers, and the $\nu_\pm$ are defined as
\begin{equation}
    \nu_\pm = \frac{\pm d - 1 + \sqrt{(d-1)^2-4E^2}}{2}\,.
\end{equation}

In both flat and dS cases, imposing the boundary condition \eqref{BCBrane} on the brane fixes the ratio of the constants $c_1/c_2$ as a function of the position of the brane $z_b$, \ie the boundary condition simply chooses a specific combination of the two independent solutions.
\begin{figure}[ht]
    \centering
    \includegraphics[scale=0.6]{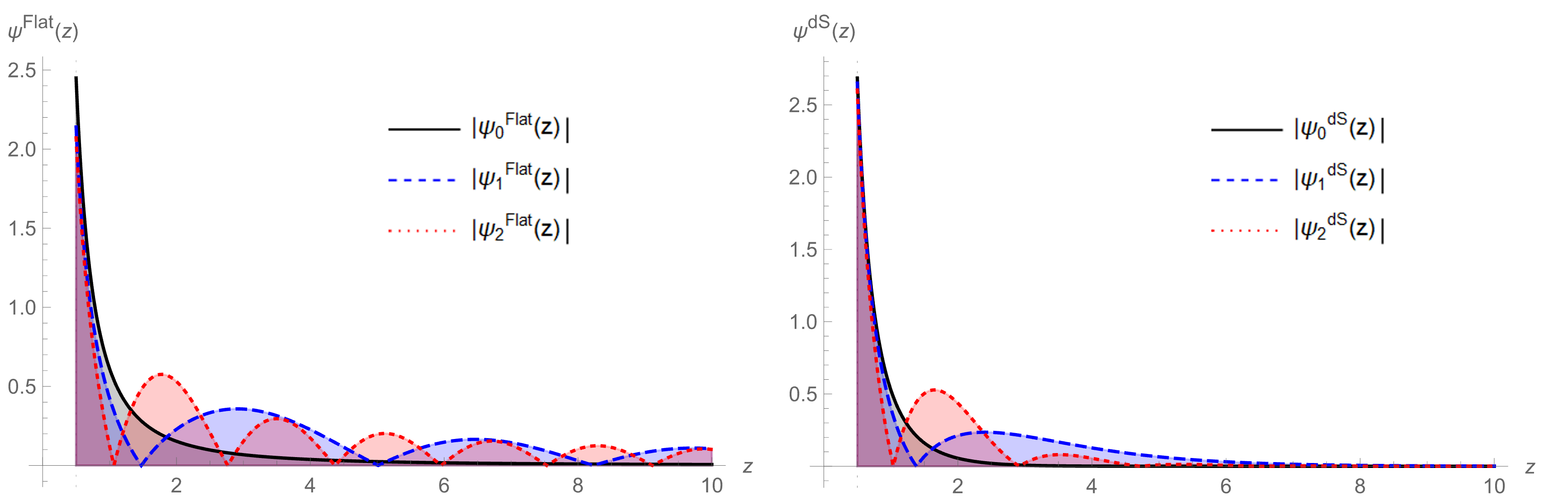}
    \caption{\small{Left: Normalized radial profiles on a flat brane at $z_b = 1/2$, with $d = 3$, for $E^2 = 0$, $E^2 = 1$, and $E^2 = 4$. Right: Normalized radial profiles on a dS brane at $z_b = 1/2$, with $d = 3$, for $E^2 = 0$, $E^2 = 1$, and $E^2 = 4$.}}
    \label{KKModes}
\end{figure}

The solution to equation \eqref{EqRad} for AdS branes is
\begin{equation}\label{HLegendrec1c2}
H(z) = c_1 (\sin z)^{\frac{d}{2}}P^{d/2}_\nu(\cos z) + c_2 (\sin z)^{\frac{d}{2}}Q^{d/2}_\nu(\cos z)\,,
\end{equation}
where $P^{d/2}_\nu$ and $Q^{d/2}_\nu$ are associated Legendre polynomials, $\nu$ is defined as
\begin{equation}\label{Defnu}
    \nu = \frac{-1 + \sqrt{(d-1)^2+4E^2}}{2}\,,
\end{equation}
and $c_1$ and $c_2$ are complex arbitrary constants \cite{Neuenfeld:2021wbl}.
Imposing the Dirichlet boundary condition at the asymptotic boundary, $H(z = \pi) = 0$, we find that the ratio between $c_1$ and $c_2$ must be
\begin{equation}\label{c2c1}
    \frac{c_2}{c_1} = \frac{2}{\pi} \cot \left( \pi \nu + \frac{\pi}{2} \right)\,,
\end{equation}
so we can write
\begin{equation}\label{HLegendreFixed}
    H(z) = (\sin z)^{\frac{d}{2}} \left[P^{d/2}_\nu(\cos z) + \frac{2}{\pi}\cot \left( \pi \nu + \frac{\pi}{2} \right) Q^{d/2}_\nu(\cos z) \right]\,.
\end{equation}
Further imposing the boundary condition \eqref{BCBrane} on the brane discretizes the spectrum. This can be done numerically, or analytically in the limit $z_b \to 0$. In any case, it is easier to do so after switching from the coordinate $z$ to the Fefferman-Graham $\rho$, using the change of variables \eqref{fCoV-ztorho} from the previous subsection. In these new coordinates, the boundary condition on the brane also reads
\begin{equation}\label{BCBraneRho}
    \left[ \partial_\rho H(\rho) \right]_{\rho_b} = 0\,.
\end{equation}
For convenience, we will study the eigenvalues $E^2$ as a function of the position of the brane $\rho_b$, and not its tension $\tau < \tau_c$. Recall that both quantities are directly related to one another by eq. \eqref{IJCTau}, and that tuning the tension close to its critical value, $\tau \to \tau_c$, brings the brane closer to the boundary, $\rho_b \to 0$.

Analytically, in the limit $\rho_b \to 0$, we find
\begin{equation}\label{E2nd}
    E^2_{(n,d)} \simeq n(n+d-1) + \frac{1}{2}(d-2)(2n+d-1) \frac{\Gamma(n+d-1)}{\Gamma(n+1)(\Gamma(d/2))^2} \rho_b^{d/2-1}\,,
\end{equation}
where $n = 0,1,2,\ldots$ . We can see that as the brane is sent to the boundary, that is, as $\rho_b \to 0$, we recover the usual eigenvalues for the graviton modes of global AdS$_{d+1}$,
\begin{equation}\label{E2AdS}
    E^2_{(n,d)} (\rho_b = 0) = n(n+d-1)\,.
\end{equation}
This fact is due to the brane becoming stiffer as it is sent to the boundary, so the Neumann-like boundary condition \eqref{BCBrane} for the radial equation becomes a Dirichlet boundary condition at infinity.
Physically, we can think of it this way: the brane is allowed to fluctuate, but as we send it to the boundary by increasing its tension, it becomes stiffer, and so it's harder to wiggle it around. In the limit where the brane is sent all the way to the boundary, it is as if the brane were infinitely stiff, and so we recover the usual Dirichlet boundary conditions used in standard AdS/CFT.
Mathematically, this result is due to the behaviour of the associated Legendre polynomials in eq. \eqref{HLegendreFixed} close to $z = 0$ with (half-)integer $\nu$, since
\begin{equation}
    \nu (\rho_b = 0) = \frac{d}{2} + n - 1\,,
\end{equation}
and imposing both $H(\pi)=0$ and $\partial_z H(0)=0$ also implies $H(0)=0$, since then, $P^{d/2}_\nu (\cos z)$ goes to zero while the cotangent multiplying $Q^{d/2}_\nu (\cos z)$ also vanishes, or vice versa, depending on the parity of the number of brane dimensions $d$.

Notice also that, in the strict limit $\rho_b = 0$, equation \eqref{E2AdS} is only valid for $n = 1,2,\ldots$, and not $n=0$. Having Dirichlet boundary conditions on both sides of the spacetime kills this zero-mode, since it is now non-normalizable and must be removed from the spectrum.

To find the analytical expansion \eqref{E2nd} above, we derived $H(\rho)$ with respect to $\rho$, and then we expanded it in terms of $u = \sqrt{\rho}$ around $u = 0$ to ($d-2$) order. Afterwards, we expanded the expression again, now in terms of $E^2_{(n,d)}$ around $n(n+d-1)$ to linear order, and only then did we impose the boundary condition \eqref{BCBraneRho} to find an analytic approximation for $E^2_{(n,d)}$.

With this method, we have been able to find a better approximation for the $d=3$ eigenvalues, namely
\begin{equation}\label{E2nd3}
    E^2_{(n, d=3)} \simeq \frac{n(n+2)\pi + (n^2+2n+4)\sqrt{\rho_b}}{\pi - 3\sqrt{\rho_b}}\,.
\end{equation}
We have also constructed similar expressions for $d=5$ and $d=7$ case by case in $n$, but we have not been able to find an improved general formula for all $(n,d)$. These results can be found in Appendix \ref{chp:App-BWresults}.

\begin{figure}[ht]
    \centering
    \includegraphics[scale=0.47]{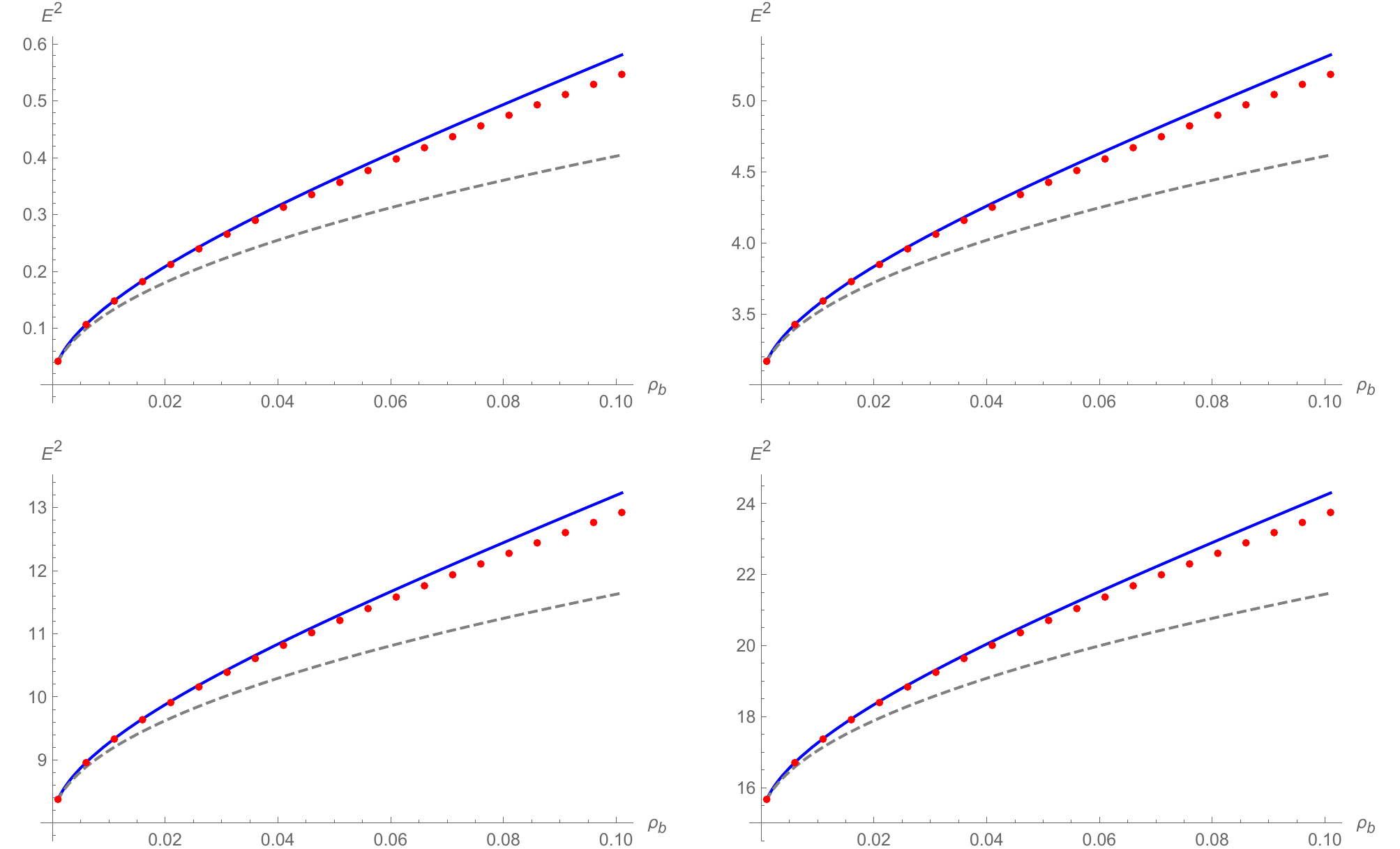}
    \caption{\small{Lowest-lying eigenvalues $E^2_n$ for $d=3$ as a function of the position of the brane $\rho_b$. The red dots are numerical results, the blue line corresponds to the analytical approximation shown in \eqref{E2nd3}, while the gray dashed line corresponds to \eqref{E2nd}. Top-left: almost-massless mode, with $n = 0$. Top-right: first excited mode, with $n = 1$. Bottom-left: second excited mode, with $n = 2$. Bottom-right: third excited mode, with $n = 3$.
    }}
    \label{Eigen}
\end{figure}

Now, in the limit $\rho_b \to 0$, following \eqref{fCoV-ztorho}, we have
$2\sqrt{\rho_b} \simeq z_b $. Therefore, as seen from the brane, the graviton masses \eqref{m2toE2} are
\begin{equation}
    m^2_{(n,d)} \simeq n(n+d-1)\frac{z_b^{2}}{L^2} + \frac{(d-2)(2n+d-1)}{2^{d-1}} \frac{\Gamma(n+d-1)}{\Gamma(n+1)(\Gamma(d/2))^2} \frac{z_b^{d}}{L^2}\,.
\end{equation}
In particular, the mass of the lowest-lying mode, with $n=0$, is 
\begin{equation}\label{GravMass}
    m^2_{(0,d)} \simeq \frac{(d-2)}{2^{d-1}} \frac{\Gamma(d)}{(\Gamma(d/2))^2} \frac{z_b^{d}}{L^2}\,,
\end{equation}
which agrees with the results found in \cite{Neuenfeld:2021wbl}.

\begin{figure}[ht]
    \centering
    \includegraphics[scale=0.67]{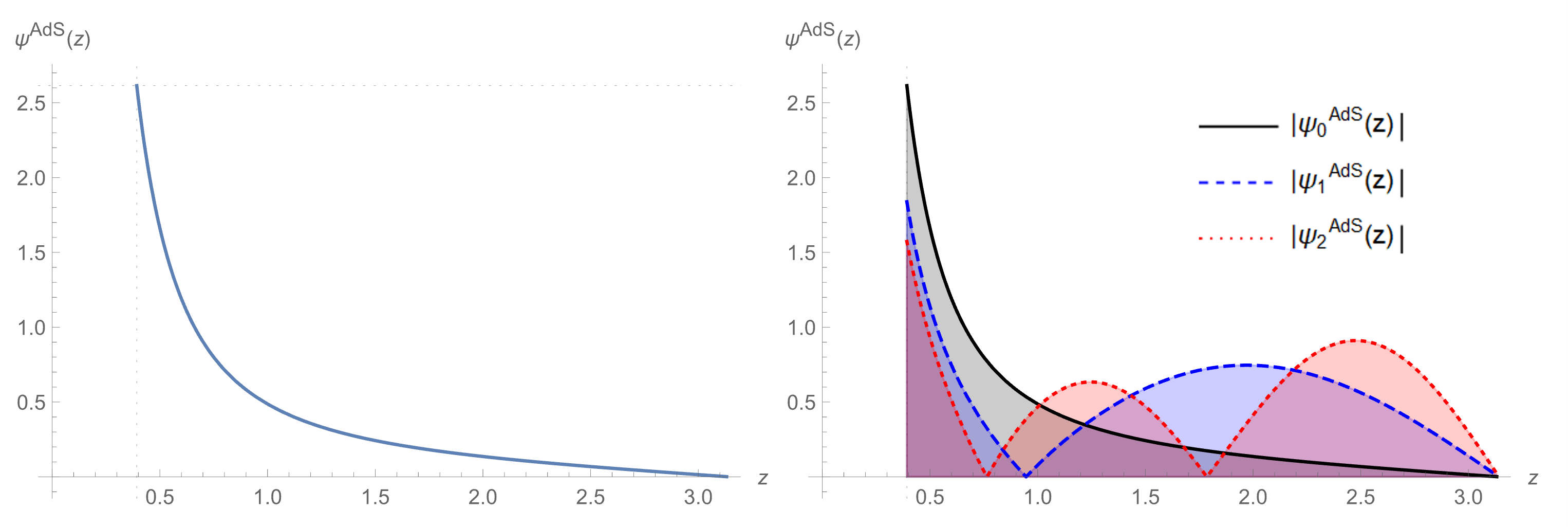}
    \caption{\small{Left: Normalized radial profile for the almost-massless mode on an AdS brane at $z_b = \pi/8$, with $d = 3$. Right: Normalized radial profiles on an AdS brane at $z_b = \pi/8$, with $d = 3$, for $n = 0$, $n = 1$, and $n = 2$.}}
    \label{ZeroModesAdS}
\end{figure}

Again, one can easily check that the lowest-lying mode is localized on the brane, since its radial behaviour goes mainly as $\psi_0(z) \sim 1/\sin^2(z)$ for $|z - z_b| \ll 1$ and when the brane is close to the boundary, $z_b \ll 1$ (see Fig. \ref{ZeroModesAdS}).
This is due to the mode being a mixture of a non-normalizable ---would-be divergent if there were no brane--- mode, behaving as $ \psi_0^\text{div}(z) \sim 1/\sin^2(z)$ for $z$ close to the brane, and a normalizable mode, behaving as $\psi_0^\text{norm}(z) \sim \sin^{d-2}(z)$.
When $d$ is odd, the non-normalizable term $\psi^\text{div}_0(z)$ corresponds to the one proportional to the Legendre polynomial $P^{d/2}_\nu(\cos z)$ in equation \eqref{HLegendreFixed}, and the normalizable term $\psi^\text{norm}_0(z)$ is proportional to $Q^{d/2}_\nu(\cos z)$. 
As the brane is sent to the boundary, $z_b \to 0$, the lowest-lying eigenvalue goes to zero, and so the ratio between $c_2$ and $c_1$ goes to zero following \eqref{c2c1} and we are left with the non-normalizable 
term $\psi^\text{div}_0(z)$ dominating the almost-massless mode.
An analogous discussion follows for the case of even $d$, where now the divergent polynomial is $Q^{d/2}_\nu(\cos z)$, while $P^{d/2}_\nu(\cos z)$ is perfectly normalizable. This discussion will become relevant in the following Section \ref{sec:BranePOV}, in which we will reinterpret these results from the dual brane picture.

\subsection{Gravity on the Brane}

So far, we have shown that we can recover an effective $d$-dimensional description of \textit{linearized} gravity on the brane.
But one needs to do some more work to show that indeed we recover $d$-dimensional gravity on the brane at the non-linear level.
A first step is reproducing the desired Newtonian potential in non-relativistic scenarios, by studying the behaviour of the graviton modes on the brane.

For the flat brane case and $d = 4$, Randall and Sundrum already showed that the contribution of the massless mode to the gravitational potential, $V_N \sim h_{00}/2$,
gave the correct four-dimensional result $V_N(r) \propto r^{-1}$, at a distance $r$ on the brane away from a point-like mass.

Then, taking the effect of the continuum of Kaluza-Klein modes into account, Garriga and Tanaka \cite{Garriga:1999yh} proved that indeed, on four-dimensional Randall-Sundrum brane-worlds, these models reproduce a Newtonian potential at a distance $r$ around a point-like mass $M$ of the form
\begin{equation}\label{NewtVBrane}
    V(r) = \frac{G_{N,\text{eff}} M}{r} \left( 1 + \frac{2 L^2}{3 r^2} \right)\,,
\end{equation}
for $r \gg L$, and where 
\begin{equation}
    G_{N,\text{eff}} = \frac{2}{L}G_{N}\,.
\end{equation}
Again, the factor of 2 missing between their definition of $G_{N, \text{eff}}$ and ours is simply because we are not orbifolding our space-time along the brane with a $\mathbb{Z}_2$ symmetry.

Therefore, at sufficiently low energies, gravity on the brane becomes four-dimensional, exhibiting the desired Newtonian potential at large scales. However, gravity experiences the higher-dimensional character of the bulk at high energies, \ie at distances shorter than the AdS bulk's curvature radius $L$, through the presence of the Kaluza-Klein modes.

Later on, by decomposing and rearranging the bulk five-dimensional Einstein Equations, Shiromizu, Maeda, and Sasaki \cite{Shiromizu:1999wj} proved that one can covariantly obtain the full four-dimensional Einstein Equations on the brane.
Their approach, however, might be misleading if not interpreted with care.
There is a term that appears on the matter side of the four-dimensional Einstein Equations, which corresponds to a projection of the bulk Weyl tensor on the brane and cannot be neglected. It captures the corrections from the effects of five-dimensional gravity, and it is the non-linear generalisation of the Kaluza-Klein modes we have found in our linearized study.
From the perspective of an observer on the brane, these KK effects are non-local, since they come from the full five-dimensional bulk, and therefore cannot be determined purely from data on the brane.
Using the holographic duality, we will reinterpret these non-local terms on the matter side of the four-dimensional Einstein Equations as the CFT radiation dual to the AdS bulk.

Finally, let us just add that these results can be easily generalized to any number of spacetime dimensions, now taking
\begin{equation}
    G_{N,\text{eff}} = \frac{d-2}{L}G_{N}\,,
\end{equation}
and to the case of AdS and dS branes, provided that they sit close enough to the asymptotic boundary so that locally the geometry on the brane looks almost flat.
This generalization to (A)dS branes relies on the fact that the behaviour of the linear graviton modes is dominated by the warp factor $1/\left(f(z)\right)^2$ for $z \ll 1$, and that $f(z)$ behaves as $f(z) \sim z$ for small values of $z$ regardless of the brane geometry.


\section{Induced Gravity on the Brane - The Brane Perspective}\label{sec:BranePOV}

Up to this point, we have been exploring brane-worlds from the bulk perspective. 
We will now make use of the holographic duality to get an understanding of the brane physics solely in terms of brane quantities. 
However, as we learned in the previous subsection, we will have to give up locality if we only want to use brane variables \cite{Shiromizu:1999wj} ---from the point of view of the brane, bulk effects can be non-local.

We will use the holographic duality to encode the bulk as the stress-energy tensor of a strongly coupled CFT on the brane \cite{deHaro:2000vlm, Neuenfeld:2021wbl}. 
As in standard AdS/CFT, this quantity is proportional to a particular coefficient of the bulk metric written in a Fefferman-Graham expansion \cite{Skenderis:2002wp}.
Therefore, we will only be able to compute it directly if the bulk metric is known.
Alternatively, given a metric on the brane, we can define the CFT stress-energy tensor as the right-hand-side of the Einstein's Equations on the brane.
But this can only be made consistently if one then solves the bulk boundary problem to show that indeed there is a bulk metric fulfilling the bulk Einstein Equations with the required boundary conditions on the brane. 
Realistically, this can only be done if the bulk metric is known beforehand, as it is done in the C-metric papers describing quantum black holes on branes (see \eg \cite{Emparan:2020znc, Emparan:2022ijy, Emparan:2023dxm, Feng:2024uia, Climent:2024nuj}).


\subsection{The Brane Effective Action}
\label{subsec:IeffBrane}

Let us now proceed with this reinterpretation.
As we will explain in detail in Chapter \ref{chp:Alg}, we can integrate the bulk following a ``finite'' holographic renormalization prescription \cite{Neuenfeld:2021wbl, deHaro:2000wj, Skenderis:2002wp, Bueno:2022log} to obtain an effective description of the brane dynamics written purely in brane variables. The result is the following effective action on the brane,
\begin{equation}\label{Ieff}
    I_{\text{eff}} = I_\text{bgrav} + I^{UV}_\text{CFT}\,.
\end{equation}
On the one hand, the term $I^{UV}_\text{CFT}$ describes a holographic CFT with a UV cut-off, and it is the dual of the AdS$_{d+1}$ bulk. 
The fact that this holographic CFT has a UV cut-off is directly related to the fact that the bulk ends at the EOW brane, at some finite distance from the boundary.
On the other hand, the term $I_\text{bgrav}$ is an effective higher-derivative theory of gravity on the brane,
\begin{equation}\label{Ibgrav}
     I_\text{bgrav} = \frac{1}{16 \pi G_{N,\text{eff}} } \int \df^d x \sqrt{-g} \left[R-2\Lambda_{\text{eff}} + \frac{L^2}{(d-4)(d-2)}( R^{ab}R_{ab} - \frac{d}{4(d-1)} R^2) + \cdots \right]\,,
\end{equation}
where all curvature tensors are built from the induced metric on the brane, and
\begin{align}\label{GdLd}
    G_{N,\text{eff}} & = \frac{d-2}{L}G_{N}\,, & \Lambda_{\text{eff}} & = - \frac{(d-1)(d-2)}{L^2} \left( 1 - \frac{\tau}{\tau_c} \right)\,.    
\end{align}
Notice how indeed the effective Newton's constant $G_{N,\text{eff}}$ on the brane coincides with the one that appears in equation \eqref{NewtVBrane} for the brane Newtonian potential \cite{Garriga:1999yh}.

This term $I_\text{bgrav}$ is generated when the bulk Einstein equations are solved in the near-boundary region excluded by the EOW brane.
In dual terms, this translates to the fact that integrating out the ultraviolet degrees of freedom of the CFT above the cut-off induces gravitational dynamics on the brane.

Let's make this statement more precise. 
How does the full effective action \eqref{Ieff} arise?
To answer this question, we need to review the standard holographic renormalization procedure, with no EOW brane present. 
In Chapter \ref{chp:Alg}, we will present it explicitly, but for now, we only need a qualitative understanding of it.

In conventional AdS/CFT, the bulk partition function diverges,
since the AdS$_{d+1}$ asymptotics dictate that bulk distances and volumes diverge near the boundary. 
These divergences are long-distance (IR) divergences from the perspective of the bulk, but they correspond to UV divergences on the CFT side of the duality.
In order to remove these divergences and obtain a useful, finite, partition function which we can equate to the CFT partition function, one must add counterterms,
\begin{equation}\label{Ifinbulk}
    I_\text{bulk}^\text{fin} = I_\text{bulk} + I_{\text{ct}}\,.
\end{equation}
These counterterms can be written in terms of local curvature tensors of the boundary metric \cite{Skenderis:2002wp, Kraus:1999di, deHaro:2000vlm, Emparan:1999pm}, and we will show how to compute them in detail in Chapter \ref{chp:Alg}.

Now, since our spacetime ends on the brane at some finite distance $z_b$ from the boundary, the on-shell action no longer diverges.\footnote{
In the case of AdS branes, the bulk reaches the asymptotic boundary on the side of the spacetime far from the brane, at $z = \pi$. There the on-shell action diverges, and we must add the usual counterterms.}
If the brane is sufficiently close to the boundary, however, the dependence of $I_\text{bulk}$ with $z_b$ has the same structure as the counterterms $I_\text{ct}$, since they are just a reflection of the AdS asymptotics.
We collect these structured finite ``counterterms'' under $I^{z_b}_{\text{ct}}$.
The terms at order $n < d$ in derivatives of the metric would diverge as the brane is sent to the boundary, 
while the terms at order $n > d$ in derivatives would vanish as $z_b \to 0$.
The term at order $n = d$ would give rise to the trace anomaly in standard AdS/CFT \cite{Henningson:1998gx} for spacetimes with an even number of dimensions.
However, since the brane sits at a finite distance from the boundary, all terms are finite, so $I^{z_b}_{\text{ct}}$ contains an infinite tower of higher-derivative terms.
It is then useful to add and subtract these finite ``counterterms'' $I^{z_b}_\text{ct}$, to write our initial action \eqref{ActionBulk} as \cite{Neuenfeld:2021wbl}
\begin{equation}
    I_\text{bulk} + I_\text{brane} = \left( I_\text{bulk} + I^{z_b}_{\text{ct}}\right) + \left(I_\text{brane} - I^{z_b}_{\text{ct}}\right) \,.
\end{equation}
We can now identify the first term as $I^{UV}_\text{CFT}$,
\begin{equation}\label{IUVCFT}
    I^{UV}_\text{CFT} = I_\text{bulk} + I^{z_b}_{\text{ct}}\,,
\end{equation}
a CFT with a UV cut-off given by the distance of the brane to the boundary. Indeed, if we were to push the brane all the way to the boundary, this first term would become $I_\text{bulk}^\text{fin}$, as seen in eq. \eqref{Ifinbulk}, which we would translate into the standard CFT partition function using the holographic dictionary.

The remaining term then becomes the effective gravitational action for the brane dynamics,
\begin{equation}
    I_\text{bgrav} = I_\text{brane} - I^{z_b}_{\text{ct}}\,,
\end{equation}
with its explicit expression given in \eqref{Ibgrav} above. 
$I^{z_b}_{\text{ct}}$ contains the full tower of higher-derivative operators seen in eq. \eqref{Ibgrav}, which have the structure of the standard counterterms, while $I_\text{brane}$ is simply the tension on the brane \eqref{IBrane}, which tunes the cosmological constant $\Lambda_\text{eff}$ to the  value shown in eq. \eqref{GdLd}.
Notice that $I_{\text{bgrav}}$ is large when $z_b$ is small, since it is mostly $-I_\text{ct}^{z_b}$, whose first terms would diverge as $z_b \to 0$. 
This shows the strong localization of gravity on the brane.

If $\tau < \tau_c$, the cosmological constant on the effective action is negative, so we will have AdS asymptotics on the brane, while if $\tau > \tau_c$, then the effective cosmological constant is positive, and so the brane will have dS asymptotics. 
As excepted, the cosmological constant becomes zero when $\tau = \tau_c$.
This is in accordance with our results from the previous section.

We can rewrite the brane effective cosmological constant given in eq. \eqref{GdLd} as 
\begin{equation}
    \Lambda_{\text{eff}} = \sigma \frac{(d-1)(d-2)}{2 \ell^2}\,,
\end{equation}
where $\sigma$ again denotes the sign of the cosmological constant, and we have defined the effective curvature radius $\ell$ as
\begin{equation}\label{ellbrane}
    \ell^2 = \frac{L^2}{2} \left| 1 - \frac{\tau}{\tau_{c}} \right|^{-1}\,.
\end{equation}
For (A)dS branes, in the limit $\tau \to \tau_c$, if we define the small parameter $\varepsilon$ as
\begin{equation}
    \varepsilon = \frac{1}{2}\left| 1 - \frac{\tau}{\tau_c} \right|,
\end{equation}
we have
\begin{equation}
    \frac{L^2}{\ell^2} = 2 \left|1 - \frac{\tau}{\tau_c} \right| = 4\varepsilon\,.
\end{equation}
This makes it clearer that indeed the action \eqref{Ieff} is an effective action, with each higher-curvature term parametrically smaller than the previous one \cite{Emparan:2020znc, Emparan:2022ijy}.

Although $\ell$ looks very similar to the brane curvature radius $l$ defined in the previous sections, they only match to linear order in $\varepsilon$. Indeed, for (A)dS branes, we can use eq. \eqref{IJCTau} and the Pythagorean trigonometric identity for $f(z)$, to write the curvature radius $l$ defined in eq. \eqref{lbrane} as
\begin{equation}
    l^2
    = \frac{L^2}{\left|1 - \left(\frac{\tau}{\tau_c}\right)^2\right|}\,,
\end{equation}
which in the limit $\tau \to \tau_c$ exactly reads
\begin{equation}
    \frac{L^2}{l^2} = 4\varepsilon + 4\varepsilon^2\,.
\end{equation}

Making use of equations \eqref{IJCTau} and \eqref{fCoV-ztorho}, it is easy to see that the position of the brane is also controlled by this same parameter $\varepsilon$. In Poincaré-like coordinates \eqref{SlicingMetric}, it is
\begin{equation}
    z_b = (f')^{-1}\left(\tau/\tau_c\right) \simeq 2\sqrt{\varepsilon}\ + \mathcal{O}(\varepsilon)\,,
\end{equation}
while in Fefferman-Graham coordinates \eqref{FGbulk}, we have
\begin{equation}
    \rho_b = \left| \frac{\tau_c-\tau}{\tau_c+\tau} \right| \simeq \varepsilon\ + \mathcal{O}(\varepsilon^2)\,.
\end{equation}


\subsection{Graviton Mass from the Brane Perspective}
\label{subsec:GravMassBranePOV}

We will now reinterpret our results from the previous section from the brane perspective. In particular, we will explain in detail how the graviton on AdS branes acquires its mass, which may not be clear from the point of view of the effective action on the brane \eqref{Ieff}.
One might be tempted to say that the mass of the graviton, and similarly, the mass of all the Kaluza-Klein modes, comes from the higher-derivative operators on the brane.
After all, we saw in Part \ref{part:HDGs} of this thesis how higher-derivative terms induce massive modes in the spectrum, when linearizing around a maximally symmetric spacetime.
Moreover, the first higher-curvature correction in equation \eqref{Ieff}, at quadratic order, is precisely a term known for generating theories of massive gravity ---in $d=3$, it precisely coincides with the term in ``New Massive Gravity'' \cite{Bergshoeff:2009hq, Bergshoeff:2009aq}.

We will now see, however, that the mass of the lowest-lying graviton on AdS branes does not come from the higher-derivative terms,
but from the interaction between the gravity and CFT on the brane.
To do so, we will linearize the equations of motion of the full effective action on the brane \eqref{Ieff}, and then relate each side of the equations to bulk quantities, to make use of our results from the previous section \ref{sec:LLG}.
That is, we will not find a \textit{new} way to compute the mass of the graviton, but a way to reinterpret the bulk results from the brane perspective.
We will closely follow \cite{Neuenfeld:2021wbl} in this subsection.

Varying the brane effective action \eqref{Ieff}, we obtain
\begin{equation}\label{BraneEEs}
    R_{ab} - \frac{1}{2} R g_{ab} + \Lambda_\text{eff} \ g_{ab} + \cdots = 8 \pi G_{N,\text{eff}} T^{\text{CFT}}_{ab} \,,
\end{equation}
where $T^{\text{CFT}}_{ab}$ is the stress-energy tensor obtained from varying $I^{UV}_\text{CFT}$ with respect to the induced brane metric $g_{ab}$, and the ellipses denote the equations of motion of the higher-derivative terms in $I_\text{bgrav}$.

Since we know that $I^{UV}_\text{CFT}$ is given by equation \eqref{IUVCFT}, we can adapt our knowledge of standard AdS/CFT to this new set-up to argue that, to leading order, the CFT stress-energy tensor on the brane is \cite{Neuenfeld:2021wbl}
\begin{equation}\label{TijCFT}
    \langle T^{\text{CFT}}_{ij} \rangle = \varepsilon^{d/2-1} \left( \frac{d L}{16 \pi G_N} \tilde{g}^{(d)}_{ij} + X^{(d)}_{ij} \left[ \tilde{g}^{(0)} \right] \right)\,,
\end{equation}
where $\tilde{g}^{(0)}_{ij}$ and $\tilde{g}^{(d)}_{ij}$ are the terms that appear, respectively, at order $\rho^0$ and $\rho^{d/2}$ in the Fefferman-Graham expansion of the bulk metric on the brane,
\begin{equation}\label{FGg}
    G_{ij}(\rho,x) = \frac{L^2}{\rho} \left( \tilde{g}^{(0)}_{ij}(x) + \tilde{g}^{(2)}_{ij}(x) \rho + \cdots + \tilde{g}^{(d)}_{ij}(x) \rho^{d/2} + \tilde{h}^{(d)}_{ij}(x)  \rho^{d/2} \log (\rho) + \mathcal{O}(\rho^{d/2+1}) \right) \,.
\end{equation}
The term $X^{(d)}_{ij}$ in eq. \eqref{TijCFT} and the coefficient $\tilde{h}^{(d)}_{ij}(x)$ above only appear when $d$ is even \cite{Fefferman:2007rka}.
Through the bulk Einstein Equations, they are fixed in terms of $\tilde{g}^{(0)}_{ij}$, and give raise to the trace anomaly of $\langle T^{\text{CFT}}_{ij} \rangle$ \cite{Henningson:1998gx}.
Again, we will give more details on this computation in Chapter \ref{chp:Alg}.

In the limit $\tau \to \tau_c$, one can check that the brane Einstein Equations \eqref{BraneEEs}, including the first few order-$n$ curvature terms, are solved by a vacuum AdS$_d$ metric with curvature radius $l$ up to order $\varepsilon^{n+1}$, even though we saw that the curvature radii $\ell$ and $l$ only coincide to order $\varepsilon$. This is because, in maximally symmetric geometries, the higher curvature terms act as extra contributions to the cosmological constant, thus bringing $\ell$ and $l$ together.

Upon perturbing the brane metric $g_{ij} \to g_{ij} + \delta g_{ij}$ around this AdS$_d$ spacetime of radius $l$, we obtain
\begin{equation}\label{LinEqBrane}
    \left( \Box + \frac{2}{l^2} + \cdots \right) \delta g_{ij} = -16\pi G_{N,\text{eff}} \ \delta T^\text{CFT}_{ij}.
\end{equation}
But this CFT stress-energy tensor is not an arbitrary stress-energy tensor: it is the one that comes from integrating the bulk, as explained before. In particular, it is proportional to the $\tilde{g}^{(d)}_{ij}$ coefficient of the bulk metric expressed in Fefferman-Graham coordinates. Similarly, the induced metric on the brane, to leading order, is proportional to the $\tilde{g}^{(0)}_{ij}$ term of the bulk metric in the FG expansion. 
Therefore, we can relate $\delta g_{ij}$ and $\delta T^\text{CFT}_{ij}$, making use of the expansion \eqref{FGg} and the following expressions for the brane metric perturbations
\begin{equation}
    \delta g_{ij}(x) = \delta G_{ij} (\rho_b,x) = \frac{L^2}{\rho_b} H(\rho_b) h_{ij}(x)\,,
\end{equation}
to write \cite{Neuenfeld:2021wbl},
\begin{equation}
    \left.\begin{cases}
        \ \tilde{g}^{(d)}_{ij} (x) = B_0 h_{ij} (x) \\
        \ \delta g_{ij}(x) = \frac{L^2}{\rho_b} A_0 h_{ij} (x) + \mathcal{O}(\rho_b)
    \end{cases} \right\}
    \implies \tilde{g}^{(d)}_{ij} (x)
    \simeq \frac{\rho_b}{L^2} \frac{B_0}{A_0} \delta g_{ij} (x)\,,
\end{equation}
where $B_0$ is the coefficient of the term $\propto \rho^{d/2}$ and $A_0$ is the coefficient of the term $\propto \rho^{0}$ of $H(\rho)$ when expanded close to $\rho \to 0$.
Finally, substituting all these results into eq. \eqref{LinEqBrane}, and relating again the bulk and brane Newton's constant through eq. \eqref{GdLd}, we obtain \cite{Neuenfeld:2021wbl}  
\begin{equation}\label{LinEqBrane2}
    \left( \Box + \frac{2}{l^2} + \cdots \right) \delta g_{ij} = -d(d-2) \frac{B_0}{A_0}\frac{\varepsilon^{d/2}}{L^2} \delta g_{ij} + \cdots\,.
\end{equation}

Ignoring the higher-curvature terms, we see that this equation above is an equation for a massive graviton, and is, in fact, a rescaled version of the brane equation \eqref{EqBrane2}, if we identify
\begin{equation}\label{GravMassBrane}
    m^2_0 = -d(d-2)\frac{B_0}{A_0}\frac{\varepsilon^{d/2}}{L^2}\,.
\end{equation}
Therefore, we will be able to find the mass of the graviton on AdS branes if we can expand the bulk metric à la Fefferman-Graham and find the coefficients $A_0$ and $B_0$.

Indeed, in the previous section, we found that the radial profile of bulk perturbations $\psi(\rho)$
could be written as a superposition of associated Legendre polynomials,
\begin{equation}
    \psi(\rho) = \sum_{n} \frac{L^2}{\rho}H_n(\rho)\,.
\end{equation}
In the FG coordinates given in eq. \eqref{FGbulk}, near the brane at $\varepsilon \to 0$, these Legendre polynomials can be expanded into \cite{NIST:DLMF}, 
\begin{equation}\label{HlimRhob}
    H_n(\rho) \ \simeq
    \begin{cases}
        -2^{d/2} \Bigg[(-1)^{\frac{d-1}{2}}\frac{\pi \cos(\nu\pi)}{2\Gamma(1-d/2)} +\frac{\pi \sin(\nu\pi) \Gamma(\nu+d/2+1)}{2\Gamma(d/2+1)\Gamma(\nu-d/2+1)} \rho^{d/2} \Bigg] + \cdots \ \ \quad \quad \quad \quad \quad \text{for odd $d$,}\\
        - \frac{\sin(\pi \nu)}{\pi} 2^{d/2} \left[ \Gamma(d/2) + (-1)^{d/2} \cos(\nu\pi) \frac{\Gamma(d/2-\nu)\Gamma(d/2+\nu+1)}{\Gamma(d/2+1)} \rho^{d/2} \right] + \cdots  \quad \text{for even $d$,}
    \end{cases}
\end{equation}
where $\nu$ is defined in eq. \eqref{Defnu}. 

We can now read $A_0$ and $B_0$ from this equation \eqref{HlimRhob}. We see,
\begin{equation}
    \frac{B_n}{A_n} =
    \begin{cases}
         (-1)^{\frac{d-1}{2}} \sin(\nu\pi) \frac{\Gamma(d/2-\nu)\Gamma(d/2+\nu+1)}{\Gamma(d/2)\Gamma(d/2+1)}\,,  \quad \text{for odd $d$,}\\
         (-1)^{d/2} \cos(\nu\pi) \frac{\Gamma(d/2-\nu)\Gamma(d/2+\nu+1)}{\Gamma(d/2)\Gamma(d/2+1)}\,,\quad \text{for even $d$,}
    \end{cases}
\end{equation}
where we have used the relation \cite{NIST:DLMF}
\begin{equation}
    \Gamma(\omega) \Gamma(1-\omega) = \frac{\pi}{\sin(\pi \omega)}\,.
\end{equation}
Now, since the lowest-lying eigenvalue behaves as $E_0^2 \to 0$ as $\varepsilon \to 0$, we have, from equation \eqref{Defnu}, that $\nu \to \frac{d}{2} - 1$,
and so we see that, regardless of the parity of our number of brane dimensions $d$,
\begin{equation}\label{BoverA}
    \frac{B_0}{A_0} \simeq -\frac{2}{d}\frac{\Gamma(d)}{(\Gamma(\frac{d}{2}))^2}\,.
\end{equation}
Plugging this into equation \eqref{GravMassBrane} for $m_0^2$, we finally obtain
\begin{equation}
    m^2_0 = \frac{2(d-2)\Gamma(d)}{(\Gamma(\frac{d}{2}))^2}\frac{\varepsilon^{d/2}}{L^2}\,,
\end{equation}
which coincides with our results from the previous section, as seen in equation \eqref{GravMass} with $2 \sqrt{z_b} \simeq \varepsilon$. This argument is not perfect but slightly circular, since we are \textit{assuming} that $E_0^2$ is small and goes to zero as the brane is sent to the boundary to expand the quotient \eqref{BoverA} around $\nu \simeq \frac{d}{2}-1 + \mathcal{O}(E^2_0)$, and then finding that this is indeed the case.
However, we may argue that we already knew that $E^2_0$ went to zero with $\varepsilon$ from our discussion of the volcano potential or our numerical studies, and then this approximation is justified in order to get an analytical formula.

Finally, a word on why the graviton of the flat and dS brane cases is massless. Recall that the zero mode for the flat and dS branes had constant $H(z)$, and so it had radial profile,
\begin{equation}
    \psi(\rho) = \frac{1}{\rho} H(\rho) \sim \frac{1}{\rho}\,,
\end{equation}
which means that the zero mode contains only a non-normalizable piece. Therefore, it does not contribute to the brane CFT stress-energy tensor, agreeing with the fact that it is massless, following an argument along the same lines as the one above for the AdS case.


\section{Conclusions}

In this chapter, we have reviewed and generalized the original brane-world constructions of Randall, Sundrum and Karch \cite{Randall:1999vf, Karch:2000ct}.
We have presented detailed calculations showing the localization of (linearized) gravity on the brane from the bulk perspective. In particular, we have shown how to compute all eigenvalues for bulk graviton modes, both analytically and numerically, and shown their radial profile.

We have then used the holographic duality to reinterpret brane-world models from the brane perspective. We have shown how the bulk can be dualized into a UV-cut-off CFT, and how the mass of the AdS brane graviton can be understood as coming from the interaction between gravity and CFT on the brane.
It would be interesting to see if one can also reinterpret the mass of the excited Kaluza-Klein modes from the brane perspective.

In the following chapter, we will redo and expand the analysis on this chapter, but now with an explicit Einstein-Hilbert term on the brane.
\cleardoublepage
\lhead{Chapter 4}
\rhead{Brane-Worlds with DGP}

\chapter{Brane-Worlds with DGP Terms}
\label{chp:BWsWithDGP}


In this chapter, we will study Karch-Randall brane-worlds with a DGP term ---an explicit Einstein-Hilbert term on the brane action---, constraining the allowed range of the DGP coupling.
It is based entirely on new, unpublished results.

\section{Introduction}
\label{sec:IntroDGP}

Soon after the original brane-world papers, Dvali, Gabadadze and Porrati \cite{Dvali:2000hr} presented an alternative way to obtain a four-dimensional description of gravity on a brane lying in a five-dimensional spacetime.
They wanted to recover 4-dimensional gravity from a 5-dimensional flat and infinite bulk, so instead of considering a purely tensional brane in an AdS bulk, they considered a flat brane with an explicit Einstein-Hilbert term sitting on a Minkowski bulk,
\begin{equation}
    I_{\text{DGP}} = \frac{1}{16 \pi G_N} \left( \int \df^5 x \sqrt{-G} \ R[G] + \lambda \int \df^4x \sqrt{-g} \ R \right)\,,
\end{equation}
where $\lambda$ is some length scale.

Unlike the original Randall-Sundrum and Karch-Randall models, however, their work recovers four-dimensional gravity at \textit{short} scales but not at distances larger than the scale $\lambda$ \cite{Luty:2003vm} ---gravity leaks off the brane into the bulk at large scales. Then, generalizing DGP models to allow for a FLRW geometry on the brane \cite{Deffayet:2000uy}, this weakening of gravity at low energies induces an accelerated expansion of the four-dimensional brane \cite{Maartens:2010ar}.
This attracted much attention from cosmologists, since 
supernovae observations had just reported an accelerated expansion of our Universe \cite{SupernovaCosmologyProject:1998vns, SupernovaSearchTeam:1998fmf}, and this model offered an alternative to the cosmological constant.

However, as cosmological measurements improved, along with a deeper understanding of DGP models, it soon became clear that observations were in tension with the model's predictions \cite{Maartens:2010ar}.
To make things worse, \cite{Charmousis:2006pn, Koyama:2007za} later discovered that the DGP model with dS branes is theoretically unstable, since the scalar sector of gravitational perturbations contains an infrared ghost.
It seems unlikely that an infrared issue can be resolved by a UV completion of these models within string theory, so the original DGP construction was ruled out as a model for our Universe \cite{Maartens:2010ar}.
Nevertheless, DGP brane-worlds have remained a useful playground for testing modified gravity models, and we will draw inspiration from them to study the physics of new brane-world models.


Again, however, we are interested in holography, not cosmology or phenomenology, so we will consider DGP Karch-Randall branes sitting on an AdS bulk and not Minkowski space.
From an EFT point of view, after the brane tension, the DGP term is the next natural term in an effective expansion of the brane action, and so this model is a logical generalisation of the brane-world models studied in the previous chapter.
Nevertheless, we will depart from an EFT perspective, sometimes considering unnaturally large DGP couplings in order to characterise their possible effects.

Set-ups similar to this one have already been studied, but with an emphasis on other topics.
For example, \cite{Tanaka:2003zb} explored how to recover a four-dimensional Newtonian potential around static, spherically symmetric masses on the brane. 
Their model allowed for a non-zero cosmological constant in the bulk, but the brane tension was tuned to the critical value, fixing the brane geometry to be flat.
Indeed, they showed that at very large scales, the model presented a four-dimensional Newtonian potential due to the Randall-Sundrum mechanism, while it also displayed four-dimensional gravity at short scales due to the DGP term.
At intermediate scales, however, gravity remained five-dimensional.

From a more modern perspective, Karch-Randall brane-world models with an explicit DGP term have been used to explore the black hole information problem \cite{Chen:2020hmv}.
From the bulk perspective, the DGP term on the brane simply alters the boundary conditions for the bulk equations.
When using the bulk Ryu-Takayanagi prescription to compute the generalized holographic entanglement entropy of brane subregions, this change in the boundary conditions translates into changes in the way RT surfaces attach to the brane.
In the standard case, RT surfaces attach to the brane at a ninety-degrees angle, as they are extremal. Adding a DGP term on the brane changes this angle \cite{Perez-Pardavila:2023rdz}.
Moreover, imposing well-known properties of holographic entanglement entropies, it is possible to constrain the allowed value of the DGP coupling, as was done in \cite{Geng:2023iqd}.
Their results qualitatively match ours, although they may not directly apply to our set-up; they considered a model of wedge holography \cite{Akal:2020wfl}, with two AdS branes cutting out the entirety of the bulk asymptotic boundary.


In this chapter, we will consider an AdS bulk containing a Karch-Randall brane with a DGP term, in addition to the brane tension.
We will allow for all three possible brane maximally symmetric geometries, and study the brane location as a function of the DGP coupling.
Then, we will follow the steps of the previous chapter to explore how the localization of gravity on the brane changes due to the presence of the DGP term.\footnote{
Recently, \cite{Miao:2023mui} explored the issue of localized gravity on AdS Karch-Randall brane-worlds, as we will do in this chapter.
However, their procedure and results are unclear to us, and so we will proceed independently, without comparing our results to theirs.
}
In particular, we will look for the presence of inconsistencies or pathologies in the theory, which will put a bound on the allowed values for the DGP coupling.
We will see that positive values for the DGP coupling are always allowed, as well as a small enough negative coupling.
However, we cannot have large negative DGP couplings, or the whole construction breaks down.
In Section \ref{sec:BranePOV} we will again reinterpret these results from the perspective of the dual brane picture.
Throughout the whole chapter, our emphasis will be on the AdS brane case, since it is the one most relevant for holographic studies.
We will end this chapter peeking into the possibility of adding the next natural higher-derivative operators on the brane.


\section{Set-up}
\label{sec:SetUpDGP}

Our set-up is the same as in the previous chapter, namely, a $(d+1)$-dimensional AdS bulk with radius $L$ ending on a co-dimension one brane, as described in the action \eqref{IBulk}, except that we now add an explicit Einstein-Hilbert term on the brane. The brane action reads
\begin{equation}
    I_{\text{brane}} = \int_{\partial\mathcal{M}_b} \df^d x \sqrt{-g} \left( -\tau + A \frac{L}{8 \pi G_N (d-2)} R \right)\,,
    \label{IbrDGP}
\end{equation}
where $R$ is the Ricci scalar built from the brane induced metric $g_{ab}$, and we have chosen to normalize the DGP coupling constant $A$ in this way for simplicity in future calculations.

Since the DGP term is only present on the brane, the bulk Einstein equations \eqref{EEsBulk} remain unchanged. However, the Israel junction condition on the brane now reads
\begin{equation}\label{IJCDGP}
    K_{ab} - K g_{ab} = - 8 \pi G_N \tau g_{ab} + A\frac{L}{d-2} \left( R g_{ab} - 2R_{ab} \right)\,,
\end{equation}
where all tensors are built from the induced metric on the brane $g_{ab}$.


Again, our ansatz for the background solution is an AdS$_{d+1}$ metric written in slicing coordinates, as in eq. \eqref{SlicingMetric}. 
We can substitute into eq. \eqref{IJCDGP} our results from Appendix \ref{chp:App-SlicingMetric} to find that the Israel junction condition now reads
\begin{equation}\label{IJCDGPf}
    -\frac{d-1}{L}f'(z_b) = -8 \pi G_N \tau + \sigma A  \frac{(d-1)}{L} \left(f(z_b)\right)^2\,,
\end{equation}
where again $\sigma$ denotes the sign of our spacetime, as defined in eq. \eqref{sigma}.
Therefore, the position of the brane will now depend not only on the brane tension $\tau$ but also on the value of the DGP coupling $A$.
Notice how the brane position is unaffected in the flat case, where $\sigma=0$, so $\tau=\tau_c$ and the position of the brane remains free. 

\begin{figure}[ht]
    \centering
    \includegraphics[scale=0.54]{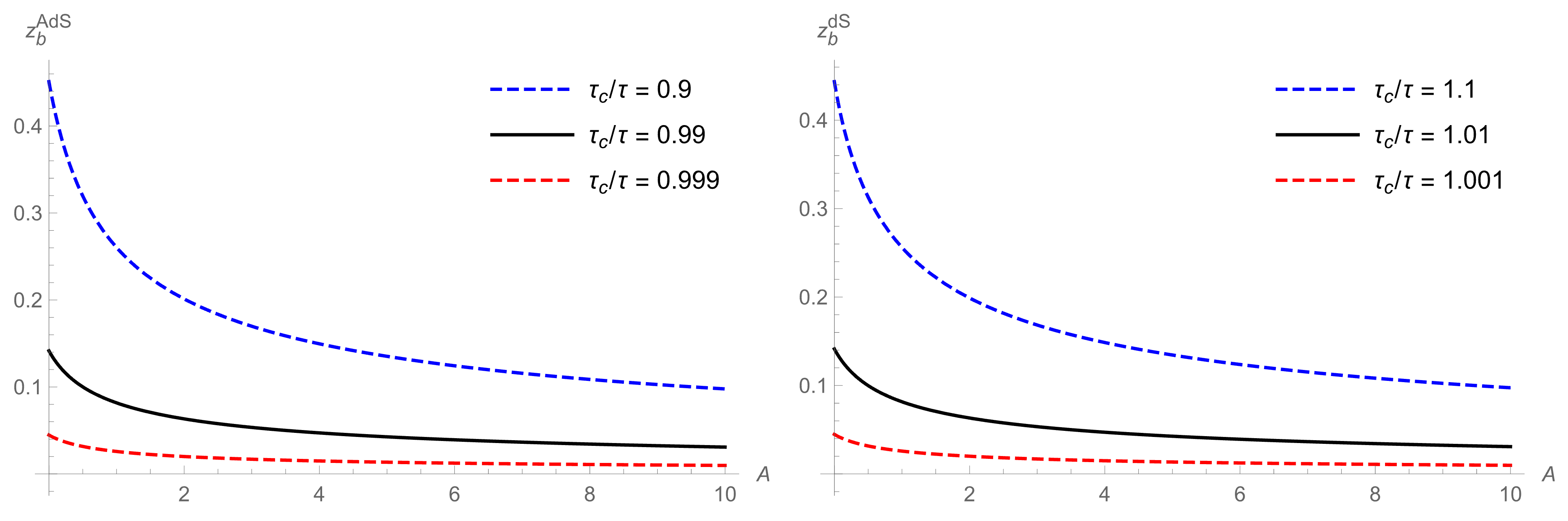}
    \caption{\small{Brane position $z_b$ as a function of $A$, for positive values of $A$. This shows that a positive DGP coupling always moves the brane closer to the boundary. Left: $z_b$ for an AdS brane with $\tau/\tau_c = 0.9$ (blue dashed), $\tau/\tau_c = 0.99$ (black), $\tau/\tau_c = 0.999$ (red dashed). Right: $z_b$ for a dS brane with $\tau/\tau_c = 1.1$ (blue dashed), $\tau/\tau_c = 1.01$ (black), $\tau/\tau_c = 1.001$ (red dashed).}}
    \label{fig:zBDGP}
\end{figure}

For the (A)dS cases, we can relate $f'(z_b)$ and $f(z_b)$ through
\begin{equation}\label{Pyth}
    \left(f(z_b)\right)^2 = \sigma \left( (f'(z_b))^2-1 \right)\,,
\end{equation}
since the functions $f(z)$ are trigonometric functions. Substituting this Pythagorean relation into \eqref{IJCDGPf}, 
we can write the following equation, which explicitly relates the position of the brane $z_b$ with its tension and the DGP coupling $A$,
\begin{equation}
    A (f'(z_b))^2 + f'(z_b) - \left( A + \frac{\tau}{\tau_c}\right)\ = 0\,.
\end{equation}
Solving for $f'(z_b)$, we find
\begin{equation}\label{zDGP}
    f'(z_b) = \frac{-1+\sqrt{1+4 A (A + \tau/\tau_c)}}{2A}\,,
\end{equation}
where we must choose the plus sign in the quadratic formula, since we want $z_b$ to be close to the asymptotic boundary at $z = 0$ for $\tau \to \tau_c$ and small values of $A$, in order to match with our results from the previous chapter.
Expanding for small values of $A$, we obtain
\begin{equation}
    f'(z_b) \simeq \frac{\tau}{\tau_c} + \left(1-\frac{\tau^2}{\tau^2_c} \right) A\, + \mathcal{O}(A^2).
\end{equation}
From this equation, we can see that turning on a positive DGP coupling will bring the brane closer to the boundary, both for the AdS and dS cases, as shown in Fig. \eqref{fig:zBDGP}.
This behaviour remains true for larger values of $A$, since one can easily see that the full equation \eqref{zDGP} is monotonous in $A$.
Physically, this is due to the brane having a maximally symmetric geometry, and so the Ricci scalar $R$ is constant and acts as an extra tension term on the brane.

\begin{figure}[ht]
    \centering
    \includegraphics[scale=0.53]{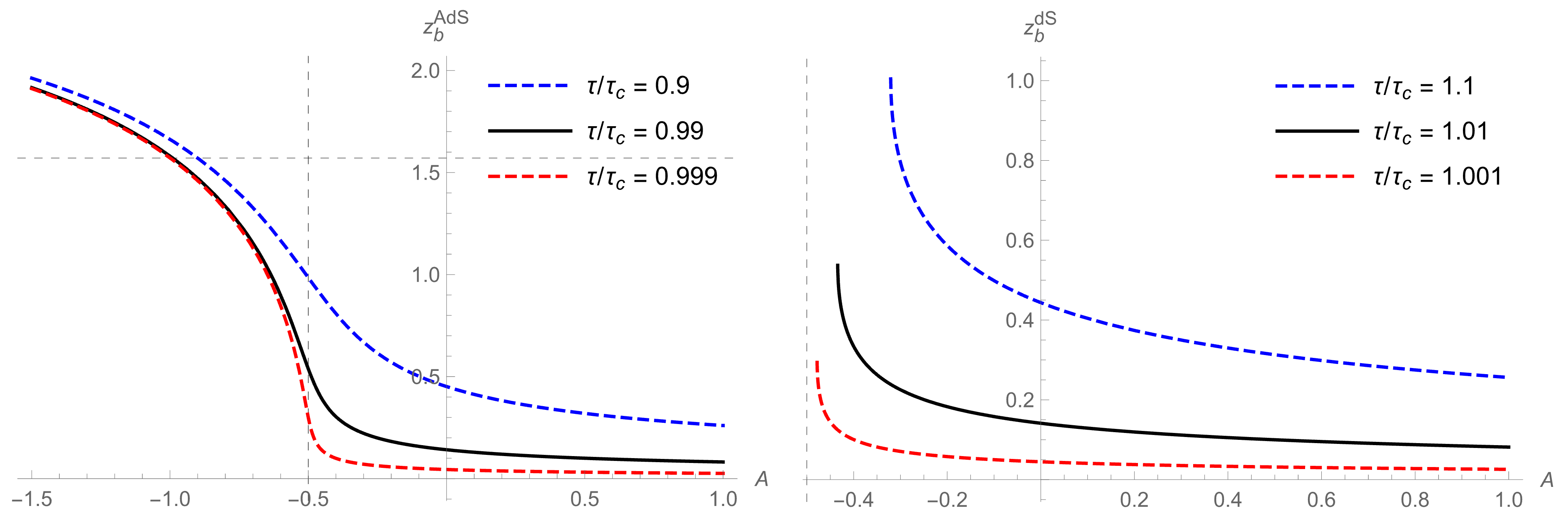}
    \caption{\small{Brane position $z_b$ as a function of $A$, for negative values of $A$. The vertical line corresponds to $A = -1/2$ on both graphs. Left: $z_b$ for an AdS brane with $\tau/\tau_c = 0.9$ (blue dashed), $\tau/\tau_c = 0.99$ (black), $\tau/\tau_c = 0.999$ (red dashed). Notice how the brane moves to the other side of the spacetime at $A \simeq -1/2$. Right: $z_b$ for a dS brane with $\tau/\tau_c = 1.1$ (blue dashed), $\tau/\tau_c = 1.01$ (black), $\tau/\tau_c = 1.001$ (red dashed). The brane position has no solution for $A \lesssim -1/2$.}}
    \label{fig:zBDGPneg}
\end{figure}

If we choose a negative DGP coupling, however, things can change drastically. 
For a small negative DGP coupling, the brane simply moves slightly away from the boundary, up to approximately $A \sim -1/2$.
This behaviour is true for both (A)dS cases, as shown in Fig. \ref{fig:zBDGPneg}.

Now, what happens at larger negatives values of $A$?
For AdS branes with fixed $\tau$, the position of the brane as a function of $A$ is continuous, from $z_b \to \pi$ as $A \to -\infty$, to $z_b \to 0$ as $A \to +\infty$. For values of $\tau$ close to $\tau_c$, the brane remains close to the boundary up to some value $A_\text{min} \gtrsim -1/2$, when it rapidly jumps to the other side of the bulk spacetime, with $z_b > \pi/2$.
For dS branes, there is a limit on how negative $A$ can get before the position of the brane turns complex,
\begin{equation}
    A_{\text{min}} = -\frac{1}{2}\left(\frac{\tau}{\tau_c} - \sqrt{\frac{\tau^2}{\tau_c^2}-1} \right) \geq -\frac{1}{2}\,.
\end{equation}
Therefore, it makes no sense to consider DGP couplings with $A < -1/2$ in either case, since then, the position of the brane is either far from the asymptotic boundary at $z=0$ or not even well-defined.

\begin{figure}[ht]
    \centering
    \includegraphics[scale=0.54]{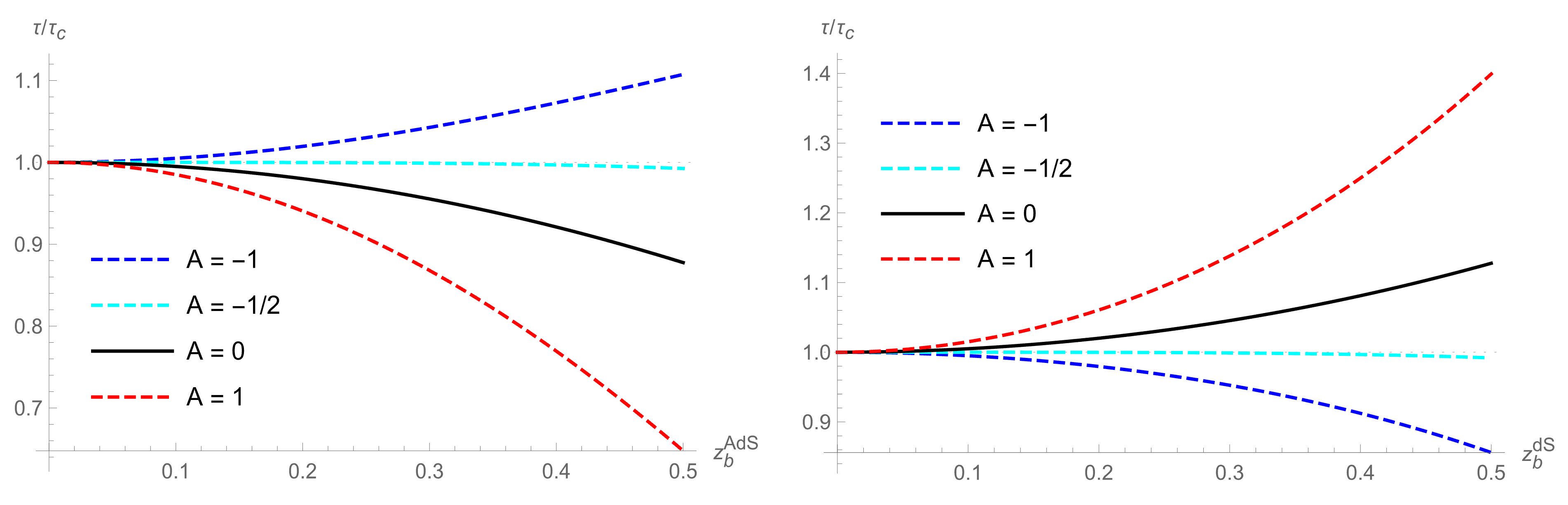}
    \caption{\small{Normalized brane tension $\tau/\tau_c$ as a function of the brane position $z_b$, following eq. \eqref{TauDGPEq}, for four different values of $A$. The horizontal line corresponds to the critical tension $\tau = \tau_c$ on both graphs. Left: $\tau/\tau_c$ for an AdS brane with $A = -1$ (blue dashed), $A = -1/2$ (cyan dashed), $A = 0$ (black), and $A = 1$ (red dashed). Notice how we need supercritical tensions to have AdS branes close to the boundary with $A < -1/2$. Right: $\tau/\tau_c$ for a dS brane with $A = -1$ (blue dashed), $A = -1/2$ (cyan dashed), $A = 0$ (black), and $A = 1$ (red dashed). Notice how we need subcritical tensions to have dS branes close to the boundary with $A < -1/2$.}}
    \label{fig:TauDGP}
\end{figure}

In the following section, however, we will usually work with the position of the brane $z_b$ and the DGP coupling $A$ as free parameters, and instead tune the brane tension to be
\begin{equation}\label{TauDGPEq}
    \tau = \tau_c \left[ f'(z_b) + \sigma A \left(f(z_b)\right)^2 \right]\,.
\end{equation}
As seen in Figure \ref{fig:TauDGP}, it is possible to have AdS (dS) branes close to the asymptotic boundary with $A < -1/2$ if we allow for a supercritical (subcritical) brane tension.
Nevertheless, in the following section, we will see that these branes will also show pathological behaviour.

\section{Locally Localized Gravity with DGP}

Let us now perturb the bulk metric with axial transverse and traceless perturbations, as defined in \eqref{PertSlicingMetric}, following the same procedure as in the previous chapter.
Again, we will assume that the perturbed equations \eqref{EEPert} separate through some separation constant $E^2$ by writing $\delta \hat{g}_{ij}(x,z) = H(z) h_{ij}(x)$.

Since the bulk Einstein equations are unchanged, we obviously obtain the same equations for $h_{ij}(x)$ and $H(z)$ as before, eqs. \eqref{EqBrane} and \eqref{EqRad}, respectively.
Now, however, the boundary condition on the brane will have changed.

Substituting our results from Appendix \ref{chp:App-SlicingMetric} for the perturbed metric \eqref{PertSlicingMetric} into the new Israel junction condition \eqref{IJCDGP}, we find
\begin{equation}
    H'(z_b)h_{ij}(x) + \frac{2 A f(z_b)}{d-2} H(z_b) \left( \hat\Box + 2\sigma \right) h_{ij}(x) = 0\,.
\end{equation}
We can now factor out $h_{ij}(x)$ using the brane equation \eqref{EqBrane}, to trade the Lichnerowicz operator for the eigenvalue $E^2$. Then, the boundary condition for the radial equation reads
\begin{equation}\label{BCBraneDGP}
    H'(z_b) + \frac{2 A f(z_b)}{d-2} E^2 H(z_b) = 0\,.
\end{equation}
How will the spectrum of the eigenvalues $E^2$ change after this change of boundary conditions?


On the one hand, the spectrum will not change much for flat and dS branes.
First, notice that the massless mode remains totally unchanged in both cases, since we recover the previous $H'(z_b) = 0$ boundary condition for $E^2 = 0$.

Then, one could easily argue, following the volcano potential argument in Subsection \ref{sec:EqRad}, that the continuum of excited eigenvalues will qualitatively have the same properties as if the DGP term were not there. 
That is, there will be no mass gap for flat branes, while there will be a mass gap of $E^2_\text{gap} = (d-1)^2/4$ for dS branes.

Moreover, since bulk equations remain unchanged, the solutions to \eqref{EqRad} are still the same linear combination of Bessel functions \eqref{HBessel} for the case of flat branes, and the same linear combination of hypergeometric functions \eqref{HHyper} for dS branes.
Upon imposing the new boundary condition \eqref{BCBraneDGP}, the only thing that will change is the ratio between the two constants $c_1/c_2$ appearing in the solutions, which now not only depends on the position of the brane $z_b$, but also on the DGP coupling $A$.


On the other hand, for AdS branes, the discrete spectrum of eigenvalues $E^2_{(n,d)}$ changes. As advertised before, we will study it as a function of the DGP parameter $A$ and the position of the brane $z_b$ (or equivalently $\rho_b$, given by the change of variables in eq. \eqref{fCoV-ztorho}), and not the tension $\tau$, which we will assume to be given by eq. \eqref{TauDGPEq}.

First, it is easy to see that the solution to equation \eqref{EqRad} for AdS branes still must be \eqref{HLegendrec1c2}. We are still imposing Dirichlet boundary conditions on the asymptotic boundary $H(z=\pi) = 0$, which fixes the linear combination of the two independent solutions to be the one shown in \eqref{HLegendreFixed}. Further imposing the boundary condition \eqref{BCBraneDGP} on the brane discretizes the spectrum, but this discretization now depends on the DGP coupling.

Analytically, following the same procedure as in the previous chapter, in the limit $\rho_b \to 0$, we find
\begin{equation}\label{EndDGP}
    E^2_{(n,d)}(A) \simeq n(n+d-1) + \frac{1}{2}(d-2)(2n+d-1) \frac{\Gamma(n+d-1)}{(\Gamma(d/2))^2\Gamma(n+1)} \frac{\rho_b^{d/2-1}}{(1+2A)}\,.
\end{equation}
Notice how the only change comes from a $(1+2A)$ factor on the denominator of the $\rho_b^{d/2-1}$ term.
We were also able to find an improved expression for the case of $d=3$,
\begin{equation}\label{En3DGP}
    E^2_{(n,d=3)}(A) \simeq \frac{n(n+2)(1+2A)\pi + (n^2+2n+4)\sqrt{\rho_B}}{(1+2A)\pi - 3\sqrt{\rho_B}}\,.
\end{equation}

\begin{figure}[ht]
    \centering
    \includegraphics[scale=0.75]{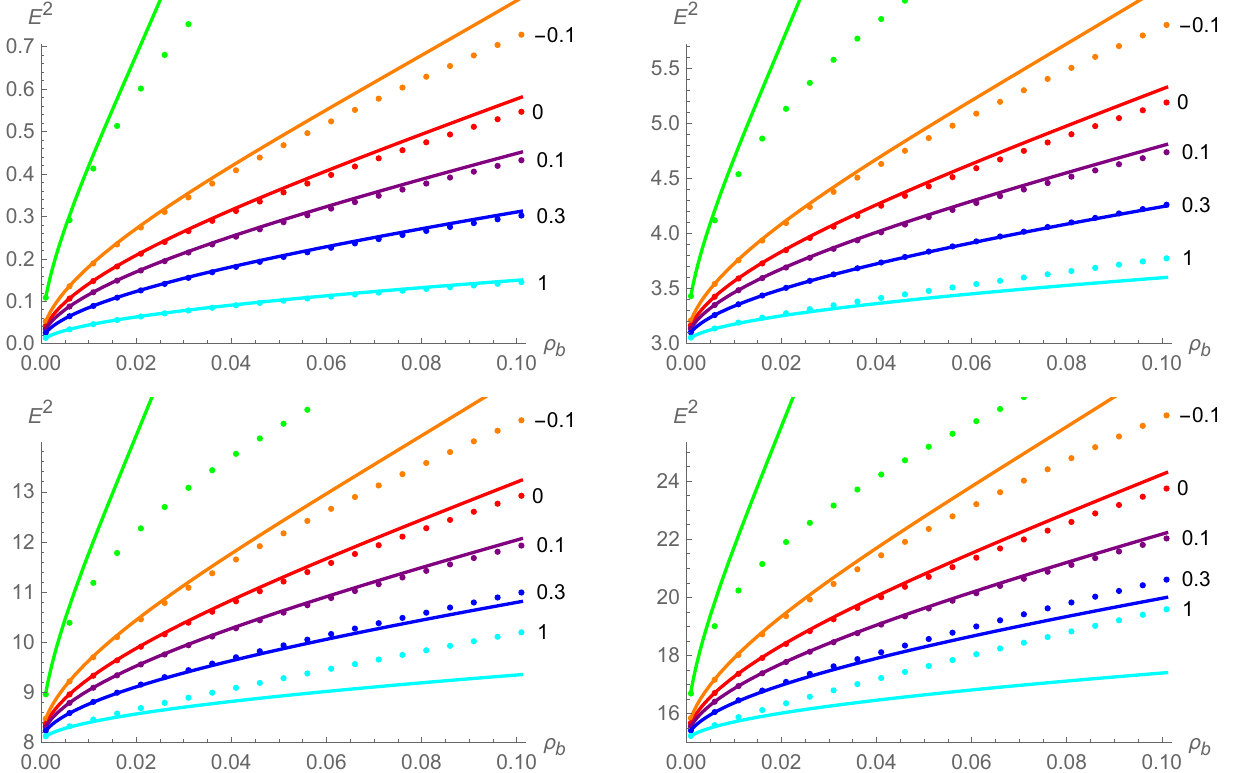}
    \caption{\small{Lowest-lying eigenvalues $E^2_n$ for $d=3$ as a function of the position of the brane $\rho_b$, for different values of $A$. The dots show the numerical results, and the lines correspond to the improved approximation shown in equation \ref{En3DGP}. Green colour corresponds to $A = -0.3$, while the value of $A$ for the other colours is shown next to the corresponding data. 
    Top-left: almost-massless mode, with $n = 0$. Top-right: first excited mode, with $n = 1$. Bottom-left: second excited mode, with $n = 2$. Bottom-right: third excited mode, with $n = 3$.
    }}
    \label{EigenDGP}
\end{figure}

\begin{figure}[ht]
    \centering
    \includegraphics[scale=0.75]{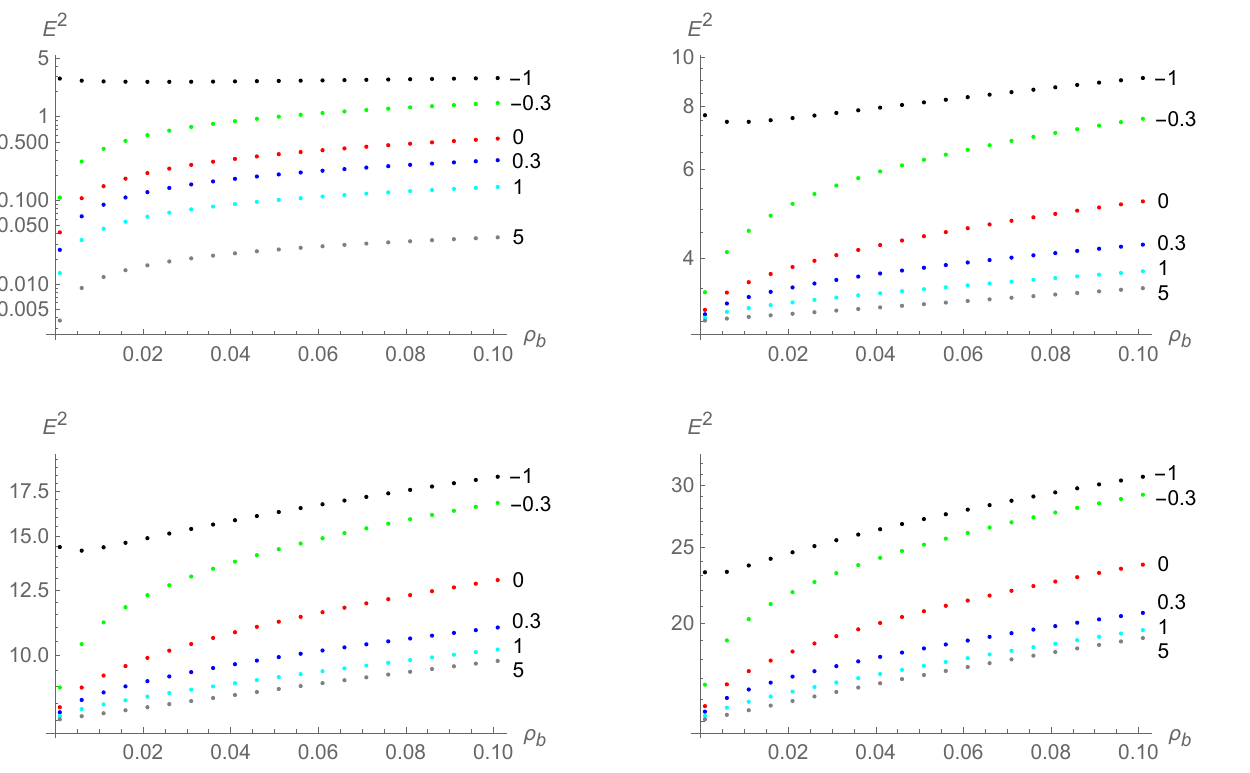}
    \caption{\small{Numerical results in log-scale of the lowest-lying eigenvalues $E^2_n$ for $d=3$ as a function of the position of the brane $\rho_b$, for different values of $A$. The value of $A$ is shown next to the corresponding data. 
    Notice how for $A \lesssim -1/2$ the value of the $n$-th eigenvalue jumps closer to the next one, $(n+1)(n+3)$, from below. Top-left: almost-massless mode, with $n = 0$. Top-right: first excited mode, with $n = 1$. Bottom-left: second excited mode, with $n = 2$. Bottom-right: third excited mode, with $n = 3$.
    }}
    \label{EigenLog}
\end{figure}

As shown in Fig. \ref{EigenDGP}, we see that for positive values of $A$, the term proportional to $\rho_b^{d/2-1}$ becomes smaller, so $E^2_{(n,d)}$ moves closer to $n(n+d-1)$, while we get the opposite effect for negative values of $A$ up to $A \sim -1/2$, when the expression blows up. 
Numerically, we observe this same behaviour. 
Moreover, we can see that for $A \lesssim -1/2$, the eigenvalues $E^2_{(n,d)}$ jump from being slightly larger than $n(n+d-1)$ to approaching the next level $(n+1)(n+d)$ from below, as shown in Fig. \ref{EigenLog}.

These results match our previous findings, even though we are now tuning the tension as we change $A$ to keep the position of the brane fixed. 
Turning on a positive DGP coupling, the eigenvalues $E^2_{(n,d)}$ become closer to the ones of empty global AdS.
One can also check that the almost-zero mode becomes more strongly localized on the brane as we turn up $A$, while the opposite is true for the higher overtones.

Turning on a small, negative DGP coupling has a small effect on the eigenvalues, moving them a bit further away from the ones of empty global AdS.
Therefore, we see that a positive or small enough negative DGP couplings are allowed, and we still obtain $d$-dimensional gravity localized on the brane.

However, if the negative DGP is large enough, $A \lesssim -1/2$, eigenvalues jump away from their value close to $n(n+d-1)$ and we lose the almost-massless mode.
Therefore, gravity no longer localizes on the brane for large negative DGP couplings.

Moreover, numerically and for negative values of the DGP coupling $A$, we have found a mode with negative mass $E^2_t < 0$ in the spectrum.
For small values of negative $A$, this eigenvalue is complex, with a large (negative) mass, of order $\mathcal{O}(z_b^{-1})$ and a tiny imaginary part.
From an EFT perspective, it is way beyond the energy scale at which one should trust the theory, and so we believe that this does not compromise the validity of the theory at small values of negative $A$.
But again, for $A \lesssim -1/2$, the mass of this tachyonic mode becomes real and of $\mathcal{O}(1)$, signalling at an instability of the theory, which agrees with all our previous discussions. 

\begin{figure}[ht]
    \centering    \includegraphics[scale=0.55]{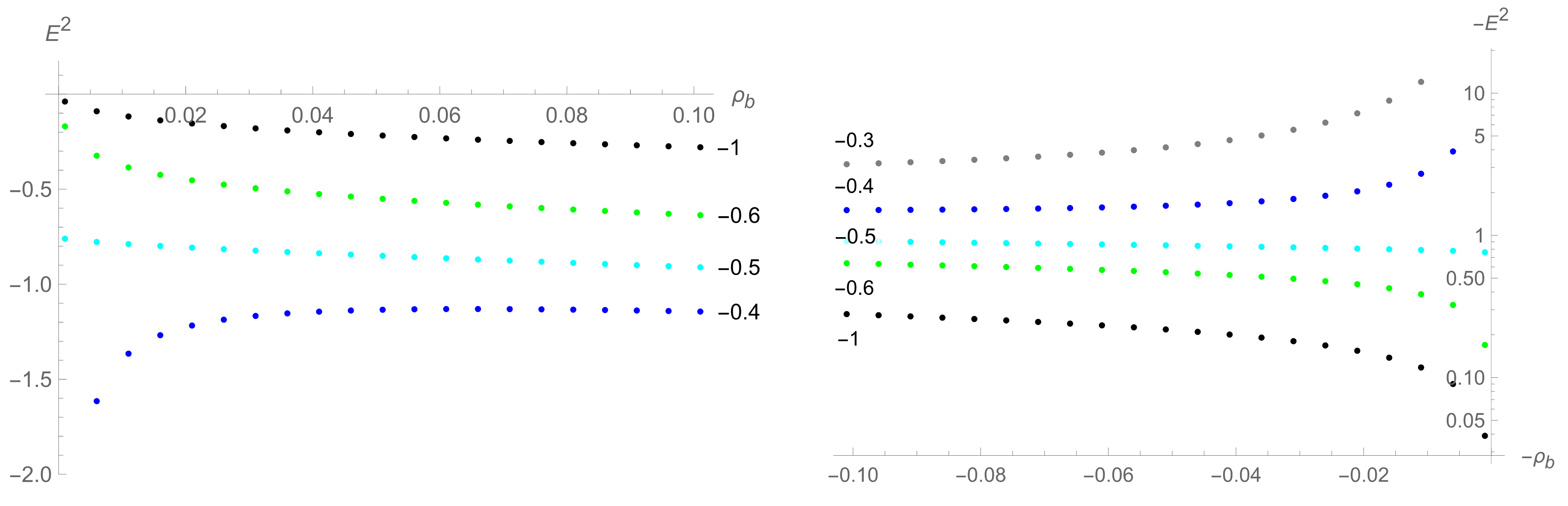}
    \caption{\small{Tachyon mode mass, in $d=3$, as a function of $\rho_b$ for different values of $A$, labelled next to the respective data. These results have been computed numerically. Left: Linear plot. Right: Log scale on the vertical axis. Notice how we have flipped the signs of both axes, to avoid problems with the logarithm.}}
    \label{fig:TachDGP}
\end{figure}


\section{Induced Gravity on the Brane with DGP}\label{sec:BranePOVDGP}

Let us now dualize the bulk to reinterpret these results from the brane perspective. 
Following the same procedure for ``finite'' holographic renormalization described in \ref{sec:BranePOV}, the brane effective action now reads
\begin{equation}
    I_{\text{eff}} = \frac{1}{16 \pi G_{N,\text{eff}} } \int d^dx \sqrt{-\gamma} \left[-2\Lambda_{\text{eff}} + \left(1+2A  \right)R + \cdots \right] + I^{UV}_\text{CFT},
\end{equation}
where, as before,
\begin{equation}
    G_{N,\text{eff}} = \frac{d-2}{L}G_{N}, \quad \quad \quad \Lambda_{\text{eff}} = - \frac{(d-1)(d-2)}{L^2} \left( 1 - \frac{\tau}{\tau_c} \right)\,.
\end{equation}

The only change with respect to the effective action written in eq. \eqref{Ieff} is the coefficient in front of the Einstein-Hilbert term.

It is interesting to see how for $A < -1/2$ the Einstein-Hilbert term picks up a minus sign, which again signals to an instability of the theory, since, upon linearization, the graviton picks up the wrong kinetic sign and becomes a ghost. 
This once more agrees with our discussion in previous sections, where we saw that for $A < -1/2$ the brane moves far away from the boundary at $z = 0$ for the AdS case, while it simply ceases to have a well-defined position for the dS case.


\section{Conclusions}\label{DGPConclusions}

In this chapter, we have studied Karch-Randall branes with a DGP term, exploring all three possible maximally symmetric brane geometries.

We have investigated how the existence of the DGP term affects the localization of gravity on the brane, searching for pathologies or inconsistencies of the theory. 
This has allowed us to put bounds on the values of the DGP coupling on the brane.
Positive couplings are always permitted, and sufficiently small negative couplings are allowed too.
Nevertheless, we have discovered that the model is unstable for large negative DGP couplings.

If we keep the brane tension fixed while we change $A$, we have observed that, as the DGP coupling $A$ becomes more negative than $-1/2$, the position of the brane ceases to be well-defined.
For AdS branes, it jumps to the other side of the bulk spacetime, close to $z = \pi$.
For dS branes, its position turns complex, meaning that there cannot be dS branes with $A < -1/2$, since we cannot find a totally umbilic hypersurface in the AdS bulk which can support them.

If instead we keep the position of the brane fixed while changing $A$, we have found that gravity no longer localizes on the brane for AdS branes with $A < -1/2$.
The almost-zero mode is lost, and instead we get a tachyon in the spectrum with a negative small mass of $\mathcal{O}(1)$.

These results are consistent with the effective action of the dual brane picture, in which the Einstein-Hilbert term picks up the wrong sign if $A < -1/2$.

We have also been exploring whether one can use these brane-world models with a DGP term to study models of holography with dynamical gravity at the boundary \cite{Compere:2008us,Ishibashi:2023luz,Ghosh:2023gvc}.
The idea would consist in sending the DGP Karch-Randall brane to the boundary while rescaling its tension and DGP coupling so that these operators remain finite at the boundary.
Now, however, since we are going all the way to the boundary, we need to add counterterms so that $I_\text{CFT}$ does not diverge.
Usually, counterterms are added on a regulating hypersurface at some finite $\varepsilon$ distance from the boundary, and then the limit $\varepsilon \to 0$ is taken.
It seems natural, then, to add these counterterms on the brane, which is itself a regulating surface at some distance $\rho_b \to 0$.
The problem is that, seen as operators on the brane, the first few counterterms are precisely a critical tension $\tau_c$ and a DGP term with coupling with $A = -1/2$, where we saw that the DGP Karch-Randall construction breaks down.
This could either mean that holographic models with dynamical boundary are pathological, or that DGP Karch-Randall brane-worlds cannot be used as a way to study them, or simply that we have not been able to take the limit $z_b \to 0$ properly.
Therefore, we have been unable to find conclusive results on this topic, and we have thus decided not to include it in this thesis.

\section{Adding Higher-Derivative Terms on the Brane}\label{sec:further}

From an EFT point of view, the brane tension and the DGP term are only the first of a series expansion of the brane action.
The next natural terms would be terms quadratic in curvature.

Let us now consider a Karch-Randall model with bulk action \eqref{IBulk}, and brane action
\begin{equation}
    I_{\text{brane}} = \int_{\partial\mathcal{M}_b} \df^d x \sqrt{-g} \left[ -\tau + A\frac{L}{8 \pi G_N (d-2)} R + \frac{1}{8 \pi G_N} \left( \beta_1 R^2 + \beta_2 R_{ab}R^{ab} + \beta_3 R_{abcd}R^{abcd} \right) \right]\,.
    \label{Ibrbeyond}
\end{equation}
Varying it, we find the usual AdS bulk Einstein Equations, and that the Israel junction condition on the brane now reads
\begin{equation}\label{IJCbeyond}
    K_{ab} - K g_{ab} = - 8 \pi G_N \tau g_{ab} + A\frac{L}{d-2} \left( R g_{ab} - 2R_{ab} \right) + \beta_1 E_{ab}^{(1)} + \beta_2 E_{ab}^{(2)} + \beta_3 E_{ab}^{(3)}\,,
\end{equation}
where the $E_{ab}^{(k)}$ are simply the equations of motion of these curvature-squared terms, times (-2),
\begin{align}
   E_{ab}^{(1)} & = g_{ab}R^2 - 4 R_{ab} R + 4 \nabla_a \nabla_b R - 4 g_{ab} \Box R \,, \\
   E_{ab}^{(2)} & = g_{ab}R_{cd}R^{cd} - 4 R^{cd} R_{acbd} + 2 \nabla_a \nabla_b R - 2 \Box R_{ab} - g_{ab} \Box R \,, \\
   E_{ab}^{(3)} & = g_{ab} R_{cdef}R^{cdef} - 4 R_{a}{}^{cde}R_{bcde} - 8 R^{cd}R_{acbd} + 8 R_{a}{}^c R_{bc}  + 4 \nabla_a \nabla_b R - 4 \Box R_{ab}\,.
\end{align}

Again, considering a maximally symmetric ansatz written in slicing coordinates \eqref{SlicingMetric}, the junction condition becomes
\begin{align}
    -\frac{d-1}{L}f'(z_b) = & -8 \pi G_N \tau + \sigma A  \frac{(d-1)}{L} \left(f(z_b)\right)^2 \nonumber \\
    & + \left[ d(d-1)\beta_1 + (d-1)\beta_2 + 2 \beta_3 \right] \frac{(d-1)(d-4)}{L^4} \left(f(z_b)\right)^4  \,.
\end{align}
Through the Pythagorean identity \eqref{Pyth}, this would become an algebraic equation of fourth order for $f'(z_b)$. 
Out of the four possible solutions, we would need to choose the one that coincides with eq. \eqref{zDGP} when $\beta_k \to 0$ for all $k = 1,2,3.$

One might be worried that upon linearization, due to the complicated appearance of the $E^{(k)}_{ab}$, we cannot longer factor out $h_{ij}(x)$ from the boundary condition on the brane.
However, since we are only considering axial transverse traceless perturbations, the linearized junction condition can be all written in terms of powers of the box operator acting on $h_{ij}(x)$.
Therefore, we can still use the brane equation \eqref{EqBrane} to get a boundary condition for the bulk radial equation that only depends on $H(z)$ and its first derivative, evaluated on the brane.
The resulting boundary condition for eq. \eqref{EqRad} is
\begin{align}
    H'(z_b) + \frac{2 A f(z_b)}{d-2} & E^2 H(z_b) \nonumber \\ & + \left[ \beta_1 C_1(E^2) + \beta_2 C_2(E^2) + \beta_3 C_3(E^2) \right] \frac{f(z_b)^3}{L^3} H(z_b) = 0\,,
\end{align}
where
\begin{align}
    C_1(E^2) & = d(d-1)\left(4+d(d-5)-E^2\right)\,,\\
    C_2(E^2) & = (d-4)(d-1)^2-2(d-1)-2E^2+E^4\,,\\
    C_3(E^2) & = 2 \left((d-1)(d-4) + 2(d-4)E^2 + 2E^4\right) \,.
\end{align}
We postpone the exploration of the allowed values of these couplings for future work ---the important conclusion from this last section is that it is indeed possible to study the range of validity of higher-derivative operators on the brane action.
Even if the Israel junction condition will get higher-derivative terms, it will always be possible to factor out $h_{ij}(x)$ using \eqref{EqBrane} to get a boundary condition for the radial equation in terms only of $H(z_b)$ and $H'(z_b)$.

\cleardoublepage
\part{Higher-Derivative Gravities from Brane-Worlds}\label{part:HDGsBWs}
\lhead{Chapter 5}
\rhead{Computing Counterterms}

\chapter{Computing Counterterms}
\label{chp:Alg}


\section{Introduction}

The quantum fluctuations of a field in a curved spacetime give rise to ultraviolet divergences that take the form of invariants of the metric and curvature in the quantum effective action. For holographic conformal field theories dual to Anti-de~Sitter spacetime in $d+1$ dimensions with radius $L$, the form of this action is \cite{deHaro:2000vlm,Balasubramanian:1999re,Emparan:1999pm}
\begin{align}\label{divact}
    I_{\text{div}}=\frac{L}{16\pi G_N(d-2)}\int_{\partial\mathcal{M}} \df^d x\sqrt{-g}\biggl[
    &\frac{2(d-1)(d-2)}{L^2}
    +R\nonumber\\
    &+\frac{L^2}{(d-2)(d-4)}\left(R_{ab}R^{ab}-\frac{d}{4(d-1)}R^2\right)
    +\dots
    \biggr].
\end{align}
Here $g_{ab}$ is the metric induced near the AdS boundary $\partial\mathcal{M}$, and the divergences arise because $g_{ab}$ grows infinitely large as the asymptotic boundary is approached. After regularization, counterterms are added with the same structure as \eqref{divact} in order to renormalize the theory.

The effective action expansion in \eqref{divact} can be systematically derived from the bulk Einstein equations in asymptotically AdS spacetimes \cite{Kraus:1999di,Papadimitriou:2004ap,Papadimitriou:2010as, Papadimitriou:2016yit,Elvang:2016tzz}.
In this chapter, we will review this procedure, known as holographic renormalization, for asymptotically AdS spacetimes with no matter content. 
We will present two different ways to perform this computation: the original one devised by Skenderis et al. \cite{Henningson:1998gx,deHaro:2000vlm, Skenderis:2002wp}, based on the Fefferman-Graham expansion of the bulk metric \cite{Fefferman:2007rka}, and the algorithm of Kraus, Larsen, and Siebelink \cite{Kraus:1999di}, based on iteratively integrating the Gauss constraint on radial hypersurfaces.
Both methods are conceptually very similar, and have in fact be shown to be equivalent \cite{Papadimitriou:2004ap, Papadimitriou:2010as, Papadimitriou:2016yit}.
We will expand their work, and give the explicit curvature invariants of \eqref{divact} up to quintic order for general dimension $d$, and to sextic order for $d=3$.

The coefficients of each of the individual curvature invariants reflect the ultraviolet structure of holographic CFTs,\footnote{Even though it is not known whether non-trivial CFTs exist in arbitrary $d$, holography suggests that their leading planar limit exists (at least for generalized free fields).} and although they have been known for many years, their specific form appears to have received little attention. 
In the following chapters we will investigate some of their properties from a point of view that directly connects them to (i) higher-derivative theories of gravity, and (ii) holographic c-theorems.

\begin{figure}[t]
        \centering
        \includegraphics[width=.3\textwidth]{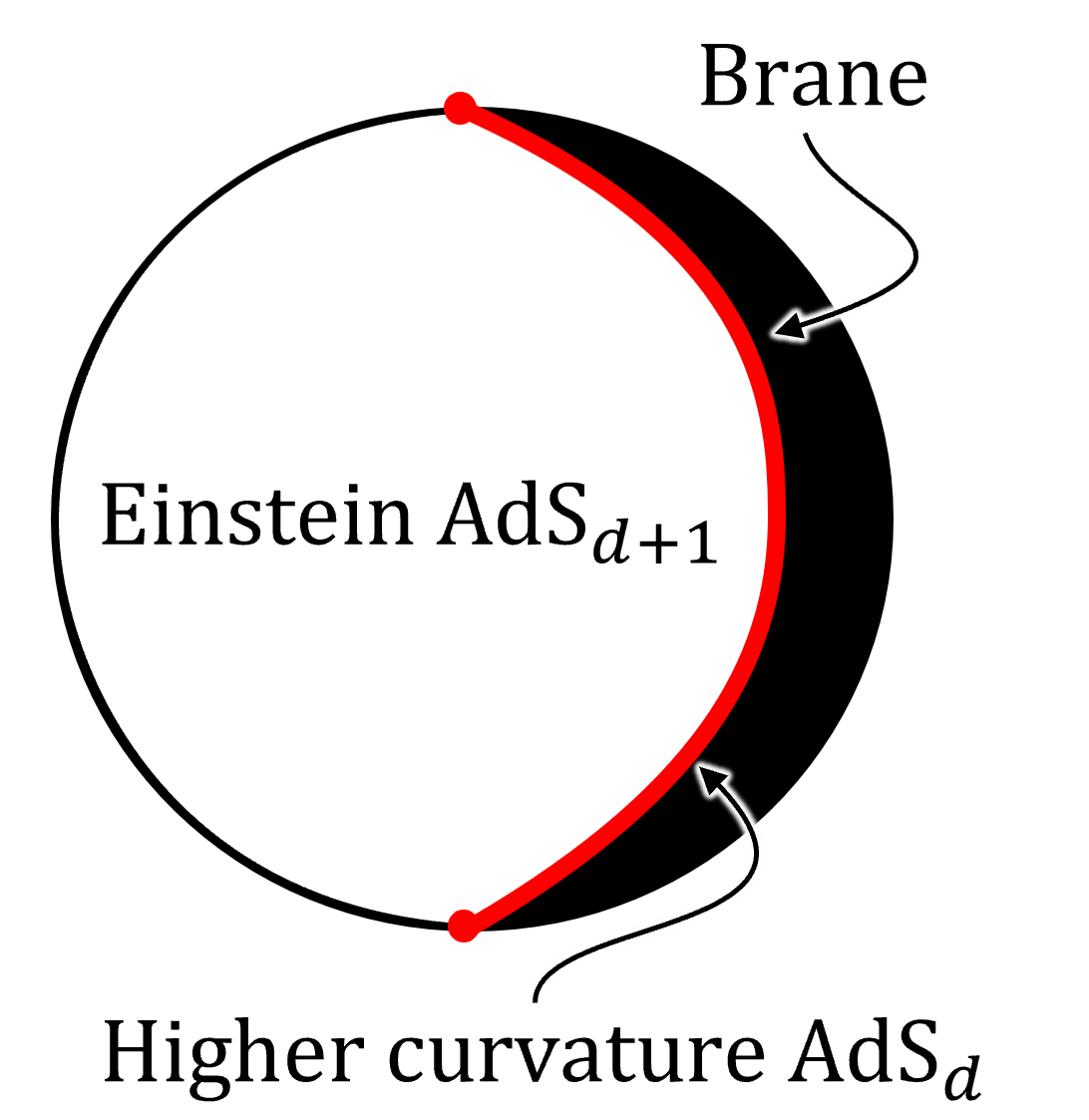}
        \caption{\small Brane-world gravity and holography. The bulk is described by Einstein-AdS$_{d+1}$ gravity. The black region is excluded by the introduction of a brane, where a gravitational theory with higher-derivative terms is induced. When the brane geometry is asymptotically AdS$_d$ (as in the figure), this higher-derivative gravitational theory can be dualized to a CFT$_{d-1}$ at its boundary (red dots). This leads to a doubly-holographic construction of boundary CFT, but this view will not be prominent in our article, where we regard the higher-derivative theory (and its dual CFT$_{d-1}$) on its own, regardless of its coupling to a CFT$_d$ dual to the AdS$_{d+1}$ bulk.}
        \label{fig:brane-world}
\end{figure}

For this purpose, we will introduce a brane near the boundary of the AdS bulk \cite{Randall:1999vf, Karch:2000ct}, as explained in Chapter \ref{chp:ReviewBWs}. 
The brane effectively acts as a cut-off that renders the action  \eqref{divact} finite, and furthermore, it makes the metric $g_{ab}$ dynamical. Then, \eqref{divact} is interpreted as the effective action of the gravitational theory that is induced on the $d$-dimensional brane, with a Newton's constant $G_{N,\rm{eff}}=(d-2)G_N/L$, and with the brane tension adding to the cosmological constant term \cite{deHaro:2000wj}.\footnote{See Section \ref{sec:BranePOV} or \eg \cite{Chen:2020uac,Neuenfeld:2021wbl} for more details. If we consider the brane to be two-sided, then \eqref{divact} will contribute twice to the effective action. Since we are only interested in the structure of the curvature terms, these considerations will be immaterial for us.} 
In effect, the Einstein-Hilbert term and all the higher-derivative operators in the effective action are generated when the bulk Einstein equations are solved in the region near the boundary excluded by the introduction of the brane. In dual terms, gravitational dynamics is induced from the integration of the ultraviolet degrees of freedom of the CFT above the cut-off. As a result, we obtain a holographic realization of `induced gravity' (figure~\ref{fig:brane-world}).

In this manner, we can view the brane-world construction as a means of generating a specific theory of higher-derivative gravity, which we will denote as $I_\text{bgrav}$. This $d$-dimensional action must be regarded as an effective theory with an infinite series of terms, each naturally smaller than the previous one. Since the $(d+1)$-dimensional Einstein bulk theory is well-defined, we expect that this good behaviour is inherited by the $d$-dimensional effective theory---at least for the entire series. However, one may also attempt to truncate the expansion at a finite order, and hope that the higher-derivative gravitational theory that results is, if not completely well-defined by itself, at least special in some respects. That is, we are proposing the holographic brane-world perspective as an appealing rationale motivating a class of higher-derivative theories with distinctive properties, which we shall investigate in the following chapters.

In this chapter, we will focus on showing how this induced higher-derivative theory of gravity on the brane originates.
First, we will show how to compute the divergent action \eqref{divact} à la Skenderis et al. \cite{Skenderis:2002wp}.
Then, we will argue that we can push this computation beyond the divergent terms, to obtain the full gravitational effective action on branes at a finite distance from the boundary $I_\text{bgrav}$, whose properties we will study in the following chapters.
Finally, we will present the KLS algorithm \cite{Kraus:1999di} to compute the terms in $I_\text{bgrav}$ and give the explicit results up to fifth order in curvature for general $d$, and up to sixth order for $d=3$.


\section{Holographic Renormalization and Induced Gravity}\label{sec:Counterterms}

\subsection{Review of Holographic Renormalization}

We begin with a sketch of how the action \eqref{divact} arises, following \cite{Skenderis:2002wp}. The starting point is the gravitational bulk action for a $(d+1)$-dimensional asymptotically AdS spacetime,
\begin{equation}\label{bulkaction}
    I_\text{bulk} = \frac{1}{16 \pi G_N} \left[ \int_{\mathcal{M}} \ \df^{d+1}x \sqrt{-G} \left( R[G] + \frac{d(d-1)}{L^2} \right) + 2 \int_{\partial\mathcal{M}} \df^dx \sqrt{-g} K \right]\,.
\end{equation}
Near the asymptotic boundary at $\rho = 0$, we can write the bulk metric in a Fefferman-Graham expansion as \cite{Fefferman:2007rka}
\begin{equation}
    G_{\mu\nu}(\rho,x) dy^\mu dy^\nu = \frac{L^2}{4\rho^2} d\rho^2 + \frac{L^2}{\rho} \tilde{g}_{ij}(\rho,x) dx^i dx^j\,,
\end{equation}
where
\begin{equation}\label{FGg2}
    \tilde{g}_{ij}(\rho,x) = \tilde{g}^{(0)}_{ij}(x) + \tilde{g}^{(2)}_{ij}(x) \rho + \cdots + \tilde{g}^{(d)}_{ij}(x) \rho^{d/2} + \tilde{h}^{(d)}_{ij}(x)  \rho^{d/2} \log (\rho) + \mathcal{O}(\rho^{d/2+1}) \,.
\end{equation}
The term containing $\tilde{h}^{(d)}_{ij}$ is only present when $d$ is even.

In this metric, the Einstein Equations read
\begin{align}
    & (ij): & \rho \left[ 2\tilde{g}''_{ij} - 2\tilde{g}'_{ik}\tilde{g}^{kl}\tilde{g}'_{lj} + \tilde{g}^{kl}\tilde{g}'_{kl}\tilde{g}'_{ij} \right] + \tilde{R}_{ij} - (d-2)\tilde{g}'_{ij} - \tilde{g}^{kl}\tilde{g}'_{kl}\tilde{g}_{ij} & = 0\,, \\
    & (i\rho): & \tilde{g}^{kl} \tilde{\nabla}_i \tilde{g}'_{kl} - \nabla^k\tilde{g}'_{ki} & = 0\,, \\
    & (\rho\rho): & \tilde{g}^{kl}\tilde{g}''_{kl} - \frac{1}{2} \tilde{g}_{kl}\tilde{g}'_{lm}\tilde{g}^{mn}\tilde{g}'_{nk} & = 0\,,
\end{align}
where the prime now denotes the partial derivative with respect to $\rho$, $\tilde{\nabla}$ is the Levi-Civita covariant derivative of $\tilde{g}_{ij}(x,\rho)$, and $\tilde{R}_{ij}$ is its Ricci tensor.

We then solve the Einstein equations, order by order in $\rho$, in terms of the `renormalized metric' $\tilde{g}^{(0)}_{ij}$ and its derivatives \cite{deHaro:2000vlm}. 
The coefficients $\tilde{g}^{(n)}_{ij}$ with $n < d$, as well as $\tilde{h}^{(d)}_{ij}$ for even $d$, can all be rewritten in terms of $\tilde{g}^{(0)}_{ij}$ and its derivatives.
In particular, one can show that all terms with odd $n < d$ identically vanish. That is why we did not even include them in eq. \eqref{FGg2}.
The equations also force $\tilde{g}^{(d)}_{ij}$ to be symmetric and covariantly conserved, and also fix the value of its trace $\tilde{g}^{(d)}$, but they do not determine it entirely.

The following terms $\tilde{g}^{(n)}_{ij}$ with $n > d$ can then be entirely determined by $\tilde{g}^{(0)}_{ij}$ and $\tilde{g}^{(d)}_{ij}$, and they are non-vanishing both for even and odd $n$.
Two tensors are needed because the Einstein Equations are second order. The tensor $\tilde{g}^{(0)}_{ij}$ corresponds to the metric at the boundary, while $\tilde{g}^{(d)}_{ij}$ is proportional to the dual CFT stress-energy tensor.

This series solution is then plugged into the bulk on-shell action, and, after introducing a cut-off at $\rho=\varepsilon$, the bulk coordinate $\rho$ is integrated between $\varepsilon$ and a finite value of $\rho>\varepsilon$. 
The result is a series expansion where the first terms diverge as $\varepsilon\to 0$ in the form
\begin{align}\label{evend}
    I_\text{div}
    = \frac{L}{16\pi G_N} \int \df^d x \sqrt{-\tilde{g}^{(0)}} \left( \varepsilon^{-d/2} \tilde{\mathcal{L}}_{(0)} +  \cdots + \varepsilon^{-1} \tilde{\mathcal{L}}_{(\lceil d/2 \rceil -1)} - \log (\varepsilon) \tilde{\mathcal{L}}_{(d/2)} \right) + \mathcal{O}(\varepsilon^{0})\,.
\end{align}
Here the $\tilde{\mathcal{L}}_{(i)}$ are invariants of $\tilde{g}^{(0)}_{ij}$ and its intrinsic curvature, but do not depend on $\tilde{g}^{(d)}_{ij}$. The  logarithmic term is present only in even $d$, due to the holographic Weyl anomaly \cite{Henningson:1998gx}. This means any given $d$ only the terms that diverge as $\varepsilon\to 0$ are uniquely determined by the boundary metric. In dual terms, they are fixed by the definition of the theory in the ultraviolet, and are independent of the state of the CFT$_d$. 

We can rewrite these divergent invariants in terms of the (physical) metric induced at $\rho=\varepsilon$,
\begin{equation}\label{indmetric}
    	g_{ij} (x,\varepsilon) = \frac{L^2}{\varepsilon} \tilde{g}_{ij} (x,\varepsilon) 
    	\,,
\end{equation}
which gives a divergent action of the form\footnote{Notice that we have absorbed a factor of $L/2$ in $\mathcal{L}$, in order to match the conventions in \cite{Kraus:1999di}, which we will follow in the next subsection.} 
\begin{equation}\label{divact2}
    I_{\rm{div}} = \frac{1}{8\pi G_N} \int \df^d x \sqrt{-g} \mathcal{L}\, , \quad \text{where}\quad  \mathcal{L}\equiv \mathcal{L}_{(0)} + \cdots + \mathcal{L}_{(\lceil d/2 \rceil - 1)} - \log (\varepsilon) \mathcal{L}_{(d/2)} \,,
\end{equation}
and where again the logarithmic term is present only in even $d$ (more about it below).
The first three terms of $I_{\rm{div}}$ were presented in \eqref{divact}, and a few following ones will be computed in Section \ref{sec:algo}.

Then, holographic renormalization is performed by adding a counterterm action $I_\text{ct}=-I_{\text{div}}$ to \eqref{bulkaction} in order to render the action finite when $\varepsilon\to 0$. The action that results is the quantum effective action of the CFT$_d$, and its variation with respect to the renormalized metric  $\tilde{g}^{(0)}_{ij}$ generates the expectation value of the renormalized stress tensor of the CFT$_d$.
Adding to the action higher curvature terms that are finite when $\varepsilon\to 0$ corresponds to changing the renormalization scheme. 


\subsection{``Finite'' Renormalization on the Brane}

 Our framework will, however, be different than that of holographic renormalization. Instead of regarding $\rho=\varepsilon$ as a regularization device to be eventually removed, we will keep it finite and non-zero, taking it to correspond to the location of a physical brane, and, as in Chapter \ref{chp:ReviewBWs}, adding to the action \eqref{bulkaction} a purely tensional term for the brane,
\begin{equation}\label{braneaction}
    I_\text{brane} \propto - \tau \int_{\rho = \varepsilon} \df^dx \sqrt{-g}\,.   
\end{equation}

Since the action is finite when $\varepsilon\neq 0$, no counterterms need to be added, and our theory will be completely well-defined by \eqref{bulkaction} and \eqref{braneaction}, without any other boundary terms.\footnote{This is the case for de Sitter or Minkowski branes, but in Karch-Randall models infinite renormalization must still be performed at the asymptotic boundary not removed by the brane. It will become clear that, for our purposes, we need not concern ourselves with this.} 

Neglecting $\tilde{g}^{(d)_{ij}}$, the expansion \eqref{divact2} can then be continued to arbitrarily high orders, producing additional densities $\mathcal{L}_{(n)}$ which only depend on the metric on the brane $g_{ij}$ and its curvature. 
This expansion now includes terms that would not diverge when $\varepsilon\to 0$. 
Such terms are necessary in order to correctly reproduce the dynamics of the brane in the bulk, which is determined by the Israel junction conditions \cite{Israel:1966rt} derived from the brane action \eqref{braneaction}  \cite{deHaro:2000wj}. 
The infinite series of these terms constitute an effective gravitational action $I_{\rm{bgrav}}$ in $d$ dimensions, and the fact that the action \eqref{divact2} is large for small $\varepsilon$ reflects the strong localization of gravity on the brane.

In practice, one obtains all the gravitational terms $\mathcal{L}_{(n)}$ in $ I_{\rm{bgrav}}$ in a unique manner by deriving them as terms of $I_{\rm{div}}$ for arbitrary $d$, without regard to whether they are finite or divergent in any specific dimension $d$, as we will do in the next section. 

Now the entire action, when evaluated on a generic bulk solution, will be
\begin{equation}\label{actdec}
    I_\text{bulk}+I_\text{brane} = I_{\rm{bgrav}}+I^{UV}_{\rm{CFT}}\,.
\end{equation}
We can think of $I^{UV}_{\rm{CFT}}$ as the finite-$\varepsilon$ counterpart of the bulk contribution that is not entirely determined by the boundary metric, but depends on $\tilde{g}^{(d)}_{ij}$.
Thus, $I^{UV}_{\rm{CFT}}$ accounts for the state of the CFT$_d$ on the brane, but some care must be exercised. 
The left-hand side of \eqref{actdec} is the action of a finite gravitational system with Einstein-Hilbert dynamics, plus a brane, in $d+1$ dimensions. The right-hand side recasts it in the form of a higher-derivative gravitational theory in $d$ dimensions, coupled to a cut-off CFT$_d$. This CFT$_d$ backreacts on the metric $g_{ij}$, so once the cut-off is introduced and the gravitational theory $I_{\rm{eff}}$ is defined, there is no more `renormalization scheme dependence' of the CFT$_d$. 

Note that the effective action $I_{\rm{bgrav}}$ is unambiguously determined (up to total derivatives) by the exact theory that it is derived from. This is not typically the case with effective theories, which can be subject to field redefinitions that change their form. For instance, the metric in an effective gravitational theory may be redefined as $g_{ij}\to g_{ij} +\varepsilon \alpha R_{ij}+ \mathcal{O}(\varepsilon^2)$, with some arbitrary coefficient $\alpha$.\footnote{Field redefinitions that involve the conformal fields reduce to the previous ones by using the lower-order effective equations of motion.} However, in our case the metric $g_{ij} (x,\varepsilon)$ is exactly determined for finite $\varepsilon$ by its bulk definition \eqref{indmetric}, and moreover its dynamics is also exactly specified by the Israel junction conditions in the bulk. So the effective gravitational theory for $g_{ij} (x,\varepsilon)$ is free from such ambiguities. A minor subtlety remains in even $d$ for the anomaly term $\mathcal{L}_{(d/2)}$, which we will discuss in the next subsection.

Then, in \eqref{actdec}, the terms $I_\text{bulk}$, $I_\text{brane}$ and $I_{\rm{bgrav}}$ are well-defined, but the action of the CFT$_d$ is only specified through \eqref{actdec}.\footnote{That is, unless we work in some specific version of AdS/CFT where the CFT is independently defined. We will not be assuming this.}
That is, the value of the CFT$_d$ action $I^{UV}_{\rm{CFT}}$, and of any other magnitude derived from it (stress tensor, entropy, etc), is obtained as the difference between the bulk action $I_\text{bulk}+I_\text{brane}$ and the $d$-dimensional action $I_{\rm{bgrav}}$, when these are evaluated on any solution of the theory.

All of these considerations simply set the stage for our statement that, in the following two chapters, we will not be concerned with $I^{UV}_{\rm{CFT}}$, but only with the gravitational action $I_{\rm{bgrav}}$.\footnote{
This is in contrast with our viewpoint in Chapters \ref{chp:ReviewBWs} and \ref{chp:BWsWithDGP} of this thesis, where we need $I^{UV}_{\rm{CFT}}$ to properly describe the physics from the dual brane picture.}
It is interpreted as the effective action of the gravity theory that is induced on the brane through the integration of the bulk degrees of freedom in the region $0\leq \rho \leq \varepsilon$. 
In dual terms, we integrate the ultraviolet degrees of freedom of the CFT$_{d}$ at energy scales above the cut-off.
Once we have obtained it this way, we will later study the effective gravitational theory on its own.


\section{Algorithm for Counterterms}\label{sec:algo}

The method of computing the effective action described in the previous section is cumbersome, but there exist iterative algorithms that greatly simplify the calculations \cite{Kraus:1999di, Papadimitriou:2004ap, Elvang:2016tzz}.
Here we will follow \cite{Kraus:1999di}.

Let us define $\Pi_{ab}$ as the stress-energy tensor associated to the full effective action $I_{\rm bgrav}$, with Lagrangian $\mathcal{L} \equiv \mathcal{L}_{(0)} + \mathcal{L}_{(1)} + \cdots\,$,
\begin{equation}\label{DefPi}
    \Pi^{ab} \equiv \frac{2}{\sqrt{-g}}\frac{\delta}{\delta g_{ab}} \int \df^d x \sqrt{-g} \, \mathcal{L}\,,
\end{equation}
and $\Pi$ as its trace, $\Pi \equiv g^{ab}\Pi_{ab}$.

The Gauss-Codazzi equations starting at the boundary are equivalent to the bulk Einstein equations in a Fefferman-Graham expansion.
The Gauss scalar constraint is
\begin{equation}\label{KLSeq}
    \frac{1}{d-1} \Pi^2 - \Pi_{ab} \Pi^{ab} = \frac{d(d-1)}{L^2} + R\,,
\end{equation}
where $R$ is the scalar curvature of the boundary metric $g_{ab}$. We will solve this equation order by order in the curvature, and then integrate \eqref{DefPi} to find the corresponding order-$n$ effective Lagrangian, $\mathcal{L}_{(n)}$. 

Two key observations were made in \cite{Kraus:1999di}.  First, that we can start by taking 
\begin{equation}
    \Pi^{ab}_{(0)} = \frac{d-1}{L} g^{ab}\,,
\end{equation}
since at the leading order the terms that are proportional to the curvature can be neglected, implying that $\Pi^{ab}_{(0)}$ must be proportional to the metric. Second, by studying the behaviour of the counterterms under Weyl rescalings, \cite{Kraus:1999di} found that the integration of \eqref{DefPi} must simply be
\begin{equation}\label{LPi}
    \mathcal{L}_{(n)} = \frac{1}{d-2n} \Pi_{(n)}\,,
\end{equation} up to total derivatives. This procedure then generates the corresponding order-$n$ term in the effective Lagrangian, and it can be iterated to compute the counterterms. 

The final algorithm is the following. 
We start from
\begin{equation}
    \Pi^{ab}_{(0)} = \frac{d-1}{L} g^{ab}, \qquad \Pi_{(0)} = \frac{d(d-1)}{L}, \qquad \mathcal{L}_{(0)} = \frac{d-1}{L}\,,
\end{equation}
and then we follow these steps iteratively:

\begin{enumerate}
    \item Knowing all $\Pi_{(i)}$ and $\Pi_{(i)}^{ab}$ of order less than $n$, solve for $\Pi_{(n)}$ using \eqref{KLSeq}.
    \item Compute $\mathcal{L}_{(n)}$ using \eqref{LPi}.
    \item Vary $\mathcal{L}_{(n)}$ to find $\Pi^{ab}_{(n)}$.
\end{enumerate}
In step 1, it is important to notice that at each order $n$, equation \eqref{KLSeq} involves terms of the form $\Pi_{(n)} \Pi_{(n-i)}$ and $\Pi_{ab}^{(i)} \Pi^{ab}_{(n-i)}$, with $i \leq n$. Since $\Pi_{(0)}^{ab}$ is proportional to $g^{ab}$, the term $\Pi_{ab}^{(0)} \Pi^{ab}_{(n)}$ is proportional to $\Pi_{(n)}$, and so indeed we find an equation for $\Pi_{(n)}$.
Moreover, for all orders $n \geq 2$, there are no other terms on the right-hand side of \eqref{KLSeq}, so we can directly solve for $\Pi_{(n)}$ to find
\begin{equation}\label{PiKLS}
    \Pi_{(n \geq 2)} = - \frac{L}{2} \sum_{i = 1}^{n-1} \left[ \frac{1}{d-1} \Pi_{(i)} \Pi_{(n-i)} - \Pi_{ab}^{(i)} \Pi^{ab}_{(n-i)} \right].
\end{equation}

Notice that when $d$ is even, the algorithm seems to break down for $n=d/2$ due to the divergence in \eqref{LPi}. The reason for this is the following. Even if, in our context, for $\varepsilon\neq 0$ the action $I_\text{bulk}+I_\text{brane}$ is finite, when we expand it in powers of $\varepsilon$, there appears a logarithmic term. It reflects the fact that the integration of conformal degrees of freedom produces non-local terms, and in the effective theory it shows up as the trace anomaly \cite{Henningson:1998gx}. In the algorithmic approaches to the computation of counterterms,  it was shown in \cite{Papadimitriou:2004ap} that one must effectively replace $1/(d-2n) \to \log\varepsilon$. Therefore, in a brane-world construction where $\varepsilon$ is finite, the apparent divergence in $\mathcal{L}_{(d/2)}$ for even $d$ is an artifact.
A similar argument would also work for the divergences appearing in $\mathcal{L}_{(n)}$ for $n \geq d/2$.

For our purposes in this final Part III of the thesis, we will not concern ourselves with these effects. On the one hand, the overall coefficients of each of the $\mathcal{L}_{(n)}$ terms will not play a role in our discussion in Chapter \ref{chp:cTheorem}, except in Sec.~\ref{BIse} of the following chapter, where we consider them in $d=3$ where there is no anomaly. 
On the other hand, in Chapter \ref{chp:Spectrum}, we will be able to resum the whole tower of $\mathcal{L_{(n)}}$ to quadratic order in perturbation around flat space, and the resulting expression will show no such problems.

The iterative procedure explained above gives for the first terms, already presented in \cite{Kraus:1999di}, the result
\begin{align}
    \mathcal{L}_{(0)} & = \frac{d-1}{L} \,, \label{L0}\\
    \mathcal{L}_{(1)} & = \frac{L}{2(d-2)} R \,, \label{L1} \\ 
    \mathcal{L}_{(2)} & = \frac{L^3}{2(d-2)^2(d-4)} \left[ R_{ab}R^{ab} - \frac{d}{4(d-1)} R^2 \right] \,,  \label{L2} \\
    \mathcal{L}_{(3)} & = - \frac{L^5}{(d-2)^3(d-4)(d-6)} \Bigg[ \frac{3d+2}{4(d-1)}RR_{ab}R^{ab} - \frac{d(d+2)}{16(d-1)^2}R^3 \nonumber \\
    \ & - 2R^{ab}R_{acbd}R^{cd} + \frac{d-2}{2(d-1)}R^{ab}\nabla_{a}\nabla_{b} R - R^{ab} \Box R_{ab} + \frac{1}{2(d-1)} R \Box R \Bigg] \,. \label{L3}
\end{align}
Since we are computing the brane effective action and not its counterterms, our results differ from those in \cite{Kraus:1999di} by an overall minus sign.

Using the \texttt{Mathematica} packages \texttt{xAct} \cite{xPerm:2008,Nutma:2013zea}, we have been able to extend these results to quartic and quintic order for general dimension $d$, and to sextic order for $d=3$.
For general dimension, the quartic term reads
\begin{align}
\mathcal{L}_{(4)} = - & \frac{L^7}{(d-2)^4(d-4)(d-6)(d-8)} \nonumber \\
\Bigg[ &  \frac{13 d^2 - 38 d - 80}{8 (d-1)(d-4)} R_{ab} R^{ab} R_{cd} R^{cd} + \frac{- 15 d^3 + 18 d^2 + 192 d + 64}{16 (d-4) (d-1)^2} R_{ab} R^{ab} R^2 \nonumber \\ 
& + \frac{d (5 d^3 + 10 d^2 - 112 d - 128)}{128 (d-4) (d-1)^3} R^4 + \frac{5 d^2 - 16 d - 24 }{(d-1)(d-4)} R^{ab} R^{cd} R R_{acbd} \nonumber \\
&- 12 R_{a}{}^{c} R^{ab} R^{de} R_{bdce} + 8 R^{ab} R^{cd} R_{ac}{}^{ef} R_{bdef} - 8 R^{ab} R^{cd} R_{a}{}^{e}{}_{c}{}^{f} R_{bedf}  \nonumber \\
&-\frac{2 (d-6)}{d-4}R^{ab} R^{cd} R_{a}{}^{e}{}_{b}{}^{f} R_{cedf} + \frac{d^2 + 4 d -36}{2 (d-4) (d-1)}  R_{bc} R^{bc} \nabla_{a}\nabla^{a}R  \nonumber \\ 
&+ \frac{- 7 d^2 + 22 d +32}{4 (d-4) (d-1)^2}R^2 \nabla_{a}\nabla^{a}R + \frac{4}{d-1} R^{bc} \nabla_{a}R_{bc} \nabla^{a}R - \frac{d+8 }{4 (d-1)^2} R \nabla_{a}R \nabla^{a}R \nonumber \\
& + \frac{3d-8}{d-1}R^{ab} \nabla_{a}R^{cd} \nabla_{b}R_{cd} + \frac{d(d-6) }{8 (d-4) (d-1)^2}\nabla_{a}\nabla^{a}R \nabla_{b}\nabla^{b}R \nonumber \\ 
& + \frac{1}{d-1} R \nabla_{b}\nabla^{b}\nabla_{a}\nabla^{a}R - \frac{(d-4)(d+2)}{4 (d-1)^2} R_{ab} \nabla^{a}R \nabla^{b}R + \frac{d-4}{d-1} R_{a}{}^{c} R_{bc} \nabla^{b}\nabla^{a}R \nonumber \\ 
&- \frac{5 d^3 - 38 d^2 + 64 d + 16}{4 (d-4) (d-1)^2}R_{ab} R \nabla^{b}\nabla^{a}R + \frac{3 d^2 - 20d + 28}{(d-1)(d-4)}R^{cd} R_{acbd} \nabla^{b}\nabla^{a}R \nonumber \\ 
&- \frac{(d-6)(d-2)^2}{8 (d-4) (d-1)^2} \nabla_{b}\nabla_{a}R \nabla^{b}\nabla^{a}R + \frac{d-4}{d-1} R^{bc}\nabla^{a}R \nabla_{c}R_{ab} - 8 R^{ab} \nabla_{e}R_{acbd} \nabla^{e}R^{cd}  \nonumber \\ 
&+ \frac{5 d^2 - 6 d - 64}{2 (d-1)(d-4)}R^{ab} R \nabla_{c}\nabla^{c}R_{ab} + \frac{(d-2)(d-6)}{2(d-1)(d-4)} \nabla^{b}\nabla^{a}R \nabla_{c}\nabla^{c}R_{ab} \nonumber \\ 
&+ \frac{(d-2)}{d-1} R_{ab} \nabla_{c}\nabla^{c}\nabla^{b}\nabla^{a}R + \frac{5 }{d-1}R \nabla_{c}R_{ab} \nabla^{c}R^{ab} + 12 R^{ab} R^{cd} \nabla_{d}\nabla_{b}R_{ac} \nonumber \\ 
&+ \frac{11d-6}{d-1} R^{ab} R^{cd} \nabla_{d}\nabla_{c}R_{ab} - \frac{d-6}{2 (d-4)}\nabla_{c}\nabla^{c}R^{ab} \nabla_{d}\nabla^{d}R_{ab} - 2 R^{ab} \nabla_{d}\nabla^{d}\nabla_{c}\nabla^{c}R_{ab} \nonumber \\ 
&- 4 R^{ab} \nabla_{b}R_{cd} \nabla^{d}R_{a}{}^{c} + 4 R^{ab} \nabla_{c}R_{bd} \nabla^{d}R_{a}{}^{c} + \frac{2(5d-22) }{d-4}R^{ab} R_{acbd} \nabla_{e}\nabla^{e}R^{cd} \Bigg].   \label{L4}
\end{align}
The quintic and sextic terms are too large to present here, and so we include them in Appendix \ref{chp:App-Counterterms}.

To finish, let us mention that the algorithm of \cite{Kraus:1999di} was improved in \cite{Papadimitriou:2004ap} into the dilatation operator method using a Hamiltonian formulation. This allowed to include matter fields, prove the equivalence of these algorithmic techniques to the holographic renormalization method of \cite{deHaro:2000vlm}, and rigorously recover the trace anomaly.
The method has been further explored \cite{Papadimitriou:2010as, Papadimitriou:2016yit}, and a practical implementation that circumvents the Hamiltonian framework has been presented in \cite{Elvang:2016tzz}.


\section{Conclusions}

In this chapter, we have reviewed holographic renormalization and its reinterpretation for brane-world holography, and we have also implemented the algorithm of \cite{Kraus:1999di} in \texttt{Mathematica} to obtain the quartic and quintic counterterms for pure AdS$_{d+1}$ gravity. 

It would be interesting to see if the methods of \cite{Papadimitriou:2004ap, Elvang:2016tzz} allow for an easier computation of the higher-order counterterms, or if the calculations can be simplified by writing the algorithm in a different basis of curvature invariants.

The theory of gravity $I_{\rm{bgrav}}$ that is induced on the brane may admit solutions that are asymptotically AdS, and indeed, this can always be achieved with a brane tension $\tau$ below a critical value. In this case, the theory may be thought of as putatively dual to a CFT$_{d-1}$ (at least at planar level). 
A necessary condition for this theory to be well-defined  is that it satisfies a c-theorem. 
The goal of our next chapter will be to show that, not only the CFT$_{d-1}$ dual to the theory $I_{\rm{bgrav}}$ satisfies this condition, but also that all the higher-derivative terms in this effective action separately do so. 

\cleardoublepage
\lhead{Chapter 6}
\rhead{Holographic c-Theorem}

\chapter{Brane-World Gravities and the Holographic c-Theorem}
\label{chp:cTheorem}


\section{Introduction}

Brane-worlds can have, as we have seen in Chapter \ref{chp:ReviewBWs}, an induced cosmological constant on the brane theory that can be positive, negative or zero \cite{Karch:2000ct}. 
The three cases give valid higher-derivative effective theories $I_\text{bgrav}$, as shown in eq. \eqref{Ibgrav} and explained in the previous chapter.
However, when the cosmological constant is negative, and the geometry on the brane is asymptotically AdS$_d$, we can perform one more holographic dualization. 
Namely, we can envisage that the gravitational theory on the brane is itself dual to a CFT$_{d-1}$ at its boundary. 

The usual interpretation of this doubly-holographic setup is in terms of a duality to a boundary CFT (BCFT), that is, a CFT$_{d}$ in a space with a boundary where a CFT$_{d-1}$ lives \cite{Karch:2000ct,Takayanagi:2011zk}. 
However, we will not adopt this view in this chapter. 
Now that we have obtained the gravitational theory $I_\text{bgrav}$, we will be considering it on its own, without regard to its possible coupling to the holographic cut-off CFT$_d$ on the brane.
If the theory is on AdS$_d$, then it may be thought of as putatively dual to a CFT$_{d-1}$ (at least at planar level). 
This CFT$_{d-1}$ to which our gravitational theory is dual will be different from the one that resides at the boundary of the CFT$_{d}$ in doubly-holographic setups. 
In other words, for us, the holographic construction is simply a means of generating a specific class of higher-derivative gravitational theories which are plausibly dual to conformal field theories, but these are not necessarily coupled to any other system.

In this chapter, we will prove that these holographic CFT$_{d-1}$ possess a basic property of well-defined conformal theories, namely, they satisfy c-theorems. Holographic theories incorporate renormalization group flows as bulk solutions that interpolate between two asymptotically AdS regions \cite{Freedman:1999gp,Girardello:1998pd}. These act as the UV and IR fixed points, while the bulk radial coordinate parametrizes the flow. 
Holographically, one expects that the c-function should be a rough measure of the curvature radius of the geometry, such that it monotonously decreases along the flow from the boundary into the bulk.

We will actually find a stronger result: the higher-derivative theories that are defined by the Lagrangian densities at each order in the expansion \eqref{divact} separately satisfy holographic c-theorems. 
Although this might not be unexpected given the good behaviour of the ``parent theory'' that gives rise to them, it is not a direct consequence of the c-theorem of the holographic CFT$_{d}$. 
Neither is it the same as the $g$-theorem for holographic boundary CFTs in \cite{Fujita:2011fp} since, as we mentioned above and will discuss later in more detail, our CFT$_{d-1}$ are differently defined, and our method of proof and bulk interpretation of the result are also very different.

The proof of these holographic c-theorems relies on particularities of the $d$-dimensional order-$n$ densities, but not in a very detailed way. 
Further examination of their structure, up to the highest order we have computed them, reveals finer features. In particular, we can decompose each order-$n$ density $\mathcal{L}_{(n)}$ appearing in the brane effective action $I_\text{bgrav}$ into a linear combination of a term $\mathcal{S}_n$, which gives a non-trivial c-function, and a term $T_n$ that does not contribute to it, since it identically vanishes on the renormalization group flow geometry.
We find evidence that this decomposition can always be made in such a way that all the $\mathcal{S}_n$ are algebraic in the Riemann tensor, with no derivatives of it. That is,
\begin{equation}\label{Lng}
     \mathcal{L}_{(n)} \propto \ \mathcal{S}_n[R_{abcd}]+T_n[R_{abcd},\nabla_a]\, .
\end{equation}
We have proven that this is possible for all $n$ in $d=3$, and strong evidence suggests that it should hold for all $n$ and $d$.

Using the decomposition \eqref{Lng}, we have then looked for other special properties of these densities. In most cases, we do not have proofs that apply to all orders and dimensions, but instead we have identified particular features by direct inspection of the terms that we have explicitly generated.

A first observation follows directly from the form of the first three orders in the effective action, shown in \eqref{divact}. In any dimension $d$, we have
\begin{equation}
    T_0=T_1=T_2=0\,.
\end{equation}
In particular, in $d=3$ the only quadratic order term is, up to an overall factor,
\begin{equation}\label{3dS2}
    \mathcal{S}_2 = R_{ab}R^{ab} - \frac{3}{8} R^2\, ,
\end{equation}
which, as noted in \cite{Emparan:2020znc}, is the same density as in the New Massive Gravity (NMG) of \cite{Bergshoeff:2009hq}.
At the next, cubic order, the $T_n$ make appearance in every $d$ (see \req{L3} in the previous chapter). In $d=3$, up to an overall factor, we find
\begin{equation}\label{3dS3}
    \mathcal{S}_3 = R_{a}^b R_{b}^c R_c^a + \frac{17}{64} R^3 - \frac{9}{8} R R_{ab}R^{ab}\, ,
\end{equation}
and
\begin{equation}
    T_3=\frac{1}{2}C_{abc}C^{abc}\,,
\end{equation}
where $C_{abc}$ is the Cotton tensor. Both these densities have featured in earlier literature:  $\mathcal{S}_3$ was proposed in \cite{Sinha:2010ai} as a cubic generalization of NMG that satisfies a holographic c-theorem, and $T_3$ defines the only cubic theory whose equations of motion have a third-order trace \cite{Oliva:2010zd}.

The appearance of \eqref{3dS2} and \eqref{3dS3} might point to a stronger link between the three-dimensional massive gravity theories of Karch-Randall brane-worlds and the generalized higher-curvature theories that satisfy holographic c-theorems \cite{Paulos:2010ke}. 
Remember, however, that the origin of the graviton mass in Karch-Randall brane-worlds is tightly linked to its coupling to the dual CFT \cite{Neuenfeld:2021wbl}, as we saw in Section \ref{sec:BranePOV}, which is in general absent in NMG and its generalizations.\footnote{Note also that the coefficient of the Einstein-Hilbert term in NMG is negative \cite{Bergshoeff:2009hq}, opposite to the `normal' sign it has in the brane-world, as seen in \eqref{Ibgrav}.}

For general higher dimensions, the cubic densities $\mathcal{S}_3$ and $T_3$ are also special in similar ways. We find that $\mathcal{S}_3$ can be identified with a linear combination of the cubic Quasi-topological gravity density \cite{Oliva:2010zd,Oliva:2010eb,Myers:2010ru}, which has second-order traced equations, plus a density which contributes trivially to the c-theorem. On the other hand, $T_3$ turns out to be given by another previously identified combination \cite{Oliva:2010zd}, distinguished, just like in three dimensions, by possessing third-order traced equations.

The reduced-order property of the traced equations is a rather stringent feature, but in holographically induced gravities it does not seem to generally hold beyond cubic terms. Indeed, the quartic term $T_4$ already does not satisfy it in $d=3$. 

Finally, also in three dimensions, we have found an intriguing connection between the full tower of counterterms and the Born-Infeld-like extension of NMG presented in \cite{Gullu:2010pc}. At present, we do not know whether this finding is fortuitous, or instead it has a deeper meaning. 

The remainder of this chapter proceeds as follows. 
First, we review the holographic c-theorem construction for general higher-derivative gravities, and also present a few new observations on the topic. 
Then, in section \ref{sec:GenProof}, we prove that all the terms in the effective action separately fulfil a holographic c-theorem. In sections \ref{sec:3DResults} and \ref{sec:GenD} we study the structure of each order-$n$ density in \eqref{divact}, in $d=3$ and in general dimensions, respectively. 
We end with comments on possible future directions.


\section{RG Flow Geometry and c-Function}\label{sec:cThmGen}

In this chapter, we review the holographic proof of the c-theorem, and the characterization of higher-derivative gravities which satisfy it.
Most of the content here is a compilation of previous results, but we also make a few observations which do not seem to have appeared explicitly in the literature before. 

The holographic c-theorem  involves a domain-wall type ansatz 
\begin{equation}\label{dwa}
    ds^2= e^{2A(r)} \left[-dt^2+ d\textbf{x} {}^2 \right]+dr^2\, ,
\end{equation}
which, in the presence of a matter
stress-energy tensor $T_{ab}$
satisfying the null energy condition (NEC), produces a profile for $A(r)$ which makes the solution interpolate between two asymptotically AdS$_{d}$ regions \cite{Freedman:1999gp,Girardello:1998pd}. From the dual CFT perspective, these correspond to UV and IR fixed points, where the metric function is asked to behave as 
\begin{equation}
    A(r \rightarrow +\infty)=\frac{r}{L_{{\rm AdS}_{\rm UV}}}\, , \qquad  A(r \rightarrow -\infty)=\frac{r}{L_{{\rm AdS}_{\rm IR}}}\, ,
\end{equation}
where $L_{{\rm AdS}_{\rm UV}}$, $L_{{\rm AdS}_{\rm IR}}$ characterize the AdS curvature radii at each end of the geometry. Since the central charge of a holographic CFT is in general proportional to a power of the AdS curvature radius measured in Planck units, these geometries appear to adequately represent holographic RG flows when going from $r\rightarrow +\infty$ to $r\rightarrow -\infty$.

The idea of the holographic c-theorem\footnote{Here we will use the term `c-theorem' to refer to monotonicity theorems in general dimensions, often called the `c-theorem', `F-theorem' and `a-theorem' in two-, three- and four-dimensional CFTs \cite{Zamolodchikov:1986gt,Casini:2012ei,Komargodski:2011vj,Casini:2017vbe}.} is then to construct a function $c(r)$---the RG monotone or `c-function'---which monotonously decreases along the flow. A weak version of the theorem would require that $c_{\rm UV} > c_{\rm IR}$, whereas a strong one (which we will aim for) demands monotonicity along the entire flow,
\begin{equation}
    c'(r) \geq 0 \quad \forall\, r\, .
\end{equation}

A natural way of constructing a candidate $c(r)$ is to find an expression for $c'(r)$ that is proportional to the combination $T_t^t-T_r^r$. Then, if the matter stress-tensor satisfies the NEC, this combination is negative semidefinite,
\begin{equation}
    T_t^t-T_r^r \overset{\rm \ssc NEC}{\leq} 0\, ,
\end{equation}
and hence any $c'(r)\propto -(T_t^t-T_r^r)$ with a non-negative proportionality constant does the job. In this chapter, we will always assume that matter is minimally coupled to gravity, and so the NEC does not involve any curvature terms.

If we denote the equations of motion of a given higher-derivative theory with Lagrangian $\mathcal{L}\left(g_{ab},R_{abcd},\nabla_a\right)$ by
\begin{equation}
    \mathcal{E}_{ab} \equiv \frac{1}{\sqrt{-g}}\frac{\delta}{\delta g_{ab}} \int \df^d x \sqrt{-g} \ \mathcal{L}\,,
\end{equation}
then the combination $\mathcal{E}_t^t-\mathcal{E}_r^r$ evaluated on \eqref{dwa} will in general be a complicated combination of terms involving $A(r)$ and its higher-order derivatives, making the identification of $c(r)$ cumbersome (or directly impossible).

An important simplification occurs for theories with equations of motion that become second-order in derivatives of $A(r)$ and are at most linear in $A''(r)$ when evaluated on \req{dwa}.  This condition can be most easily implemented, for general families of higher-derivative theories, at the level of the action \cite{Paulos:2010ke}. Indeed, let 
\begin{equation}
    I[A] = \int \df^d x \sqrt{-g} \mathcal{L} \left[A\right]
\end{equation} be the on-shell action from the evaluation of the corresponding higher-derivative action on the metric \req{dwa}. It is easy to show that the Euler-Lagrange equation of $A(r)$ is proportional to the $tt$ component of the field equations evaluated on \eqref{dwa}, namely, 
\begin{equation}\label{ttEq}
    \frac{\delta I[A]}{\delta A}=-2(d-1) e^{(d-1)A(r)}\left.\mathcal{E}_t^t\right|_A\,.
\end{equation}
Thus, whenever $I[A]$ is second-order in derivatives of $A(r)$ and linear in $A''(r)$, so is $\mathcal{E}_t^t$.

The additional independent equation, corresponding to $\mathcal{E}_r^r$, is related to $\mathcal{E}_t^t$ by the Bianchi identity
\begin{equation}\label{rrEq}
    \partial_r \left.\mathcal{E}_r^r\right|_A+(d-1)A'(r)\left.\mathcal{E}_r^r\right|_A=(d-1)A'(r)\left.\mathcal{E}_t^t\right|_A\, .
\end{equation}
This immediately implies that $\mathcal{E}_r^r$ does not contain terms involving derivatives of $A(r)$ higher than one (since it is the scalar constraint\footnote{The explicit form of the equation $\mathcal{E}_r^r$ can be obtained from the on-shell action of $d s^2= e^{2A(r)} \left[-d t^2+ d \bf{x}^2 \right]+N(r)^2d r^2$ by varying with respect to the lapse function $N(r)$ \cite{Arciniega:2018tnn}.}) and that the combination $\mathcal{E}_t^t-\mathcal{E}_r^r$ is second-order in derivatives and linear in $A''(r)$.    
Throughout the paper, when speaking about theories satisfying the holographic c-theorem, we will be referring to theories that satisfy these reduced-order properties.\footnote{These requirements are identical to the ones satisfied by higher-curvature gravities which produce generalized Friedman equations of second order for the scale factor when evaluated on a Friedmann-Lema\^itre-Robertson-Walker ansatz with flat spatial slices---see \eg \cite{Sinha:2010pm,Tekin:2016rdx,Arciniega:2018fxj,Arciniega:2018tnn,Cisterna:2018tgx,Moreno:2023arp}. } 

For this kind of theories, it is straightforward to construct a function $c(r)$ such that \cite{Freedman:1999gp,Myers:2010xs,Myers:2010tj} 
\begin{equation}
    c'(r)=-\frac{\pi^{\frac{d-3}{2}}}{8\Gamma\left[\tfrac{d-1}{2}\right] G_N }\, \frac{T_t^t-T_r^r}{A'(r)^{d-1}}\, ,
\end{equation}
where, as required, the right-hand side is positive semidefinite, including for even $d$ \cite{Myers:2010tj}. As observed in \cite{Sinha:2010ai,Myers:2010tj}, $c(r)$ can be obtained for these theories from the Wald-like \cite{Wald:1993nt} formula
\begin{equation}
    c(r)\equiv \frac{\pi^{\frac{d-1}{2}}}{2\Gamma[\tfrac{d-1}{2}] A'(r)^{d-2}}\frac{\partial \mathcal{L}}{\partial R^{tr}\,_{tr}}\, ,
\end{equation}
where the Lagrangian derivative components are evaluated on \req{dwa}. By construction, $c(r)$ coincides at the fixed points with the holographic central charges $c$.


\section{Constraints on Theories}

When trying to construct theories that satisfy simple holographic c-theorems, Lovelock gravities \cite{Lovelock1,Lovelock2,Padmanabhan:2013xyr} are natural candidates, as they have second-order equations on general backgrounds. The $n$-th order Lovelock density is
\begin{equation}
    \mathcal{L}^{(n)}_{\rm Lovelock}\equiv \mathcal{X}_{2n}\equiv \frac{(2n)!}{2^n} \delta_{a_1}^{[b_1}\delta_{a_2}^{b_2}\cdots \delta_{a_{2n}}^{b_{2n}]} R^{a_1 a_2}_{b_1b_2} \cdots R^{a_{2n-1} a_{2n}}_{b_{2n-1} b_{2n}}\, .
\end{equation}
When $d$ is even, the density with $n=d/2$ is a topological invariant. All the higher order densities (with $n> (d-1)/2$ when $d$ is odd, and with $n> d/2$  when $d$ is even) vanish identically.  
Hence, Lovelock theories are too restricted to provide a non-trivial family of order-$n$ densities in arbitrary dimensions.

A different set can be obtained using the Schouten tensor
\begin{equation}
    S_{ab}=\frac{1}{d-2} \left[R_{ab}- \frac{1}{2(d-1)}g_{ab} R \right] 
\end{equation}
as a building block. The general relation
\begin{equation}
    R_{abcd}=C_{abcd} - 2(g_{a[c} S_{d]b}+g_{b[d}S_{c]a})\,,
\end{equation}
and the fact that the Weyl tensor vanishes on the RG flow ansatz \req{dwa}, since it is a conformally flat metric, suggests considering the family \cite{Paulos:2012xe}
\begin{equation}\label{Pn}
    \mathcal{P}^{(n)}= \delta^{[b_1}_{a_1} \delta^{b_2}_{a_2}\cdots \delta^{b_n]}_{a_n} S^{a_1}_{b_1} \cdots S^{a_n}_{b_n} \, .
\end{equation}
This vanishes for $n>d$ because the totally antisymmetric product of Kronecker deltas is identically zero in that case,  but it has been shown that a simple limiting procedure\footnote{The idea involves computing $ \mathcal{P}^{(n)}$ for some $\bar d$ greater than the dimension of interest $d$, dividing by $(\bar d -d)$ and then taking the limit $\bar{d} \rightarrow d$ of the resulting expression.} can be applied to $ \mathcal{P}^{(n)}$, which gives non-trivial densities for additional orders and dimensions \cite{Alkac:2020zhg} (see also \cite{Gabadadze:2020tvt,Bergshoeff:2021tbz}).

One may also systematically consider all the densities of a given curvature order for fixed $d$, with arbitrary relative coefficients, and identify the combinations that satisfy the aforementioned conditions. At quadratic order, this selects the Gauss-Bonnet density 
\begin{equation}
    \mathcal{X}_{4}= R-4R_{ab}R^{ab}+R_{abcd}R^{abcd} \,,
\end{equation}
and the Weyl-squared term $C_{abcd}C^{abcd}$, which identically vanishes on \req{dwa}. The cubic case was studied in \cite{Myers:2010tj} for general $d$.  At that order, there exist eight independent densities (there are fewer for low enough $d$), and the holographic c-theorem imposes two constraints on them, leaving six independent densities that satisfy all the requirements. 

Hence, in general, for fixed $d$ and $n$ there will be several independent densities  satisfying the holographic c-theorem.
However, it is natural to expect that the functional on-shell dependence on $A(r)$ for fixed $d$ and $n$ is unique---in particular, given $j$ order-$n$ densities satisfying the c-theorem, $\sum_j \alpha_j \mathcal{L}_j$, we would have
\begin{equation}
   \mathcal{E}_t^t\left|_A - \mathcal{E}_r^r\right|_A = \biggl(\sum_j c_j \alpha_j \biggr)\cdot F_n(A,A',A'')\, ,
\end{equation}
where the dependence on the gravitational couplings fully factorizes. This always allows us to change the basis of densities so that a single one of them contributes non-trivially to 
$\mathcal{E}_t^t - \mathcal{E}_r^r $, while all the others produce a vanishing contribution---\eg the Weyl-squared density at quadratic order.

As for the explicit form of $\mathcal{E}_t^t$, $\mathcal{E}_r^r$ and   $F_n(A,A',A'')$ when evaluated on \req{dwa} for individual non-trivial densities, a quick inspection of various cases strongly suggests that these are always given by 
\begin{equation}
  \left.\mathcal{E}_t^t \right|_A \propto A'(r)^{2(n-1)}\left[(d-1)A'(r)^2+2nA''(r) \right]\, ,\quad \left.\mathcal{E}_r^r\right|_A \propto (d-1)A'(r)^{2n}\, ,
\end{equation}
up to an overall factor, and
\begin{equation}\label{FcThm}
    F_n(A,A',A'')= 2n A'(r)^{2(n-1)} A''(r)\, ,
\end{equation}
for general $n$ and $d$. The functional dependence of the c-function is then $c(r)\propto A'(r)^{2n-d}$.
Indeed, we proved in Section \ref{sec:ctheorem3D} of Chapter \ref{chp:3DHCGs} that this sort of `uniqueness' holds for general curvature orders in $d=3$, as initially hinted by \cite{Paulos:2010ke}.

Several other properties have been observed to hold for gravities in three dimensions that satisfy a c-theorem, as discussed in depth in Section \ref{sec:ctheorem3D}.  At quadratic order, the resulting theory is the New Massive Gravity of \cite{Bergshoeff:2009hq}---more on this below.
At higher curvature orders, theories of this kind arise from an order-by-order expansion \cite{Gullu:2010pc,Gullu:2010st,Bueno:2022lhf} of a Born-Infeld-type gravity \cite{Gullu:2010pc}, which in turn satisfies the holographic c-theorem by itself \cite{Gullu:2010st,Alkac:2018whk}. 
In addition, it has been found that certain theories that satisfy the holographic c-theorem---some of which involve explicit covariant derivatives---are equivalent to Chern-Simons gravities \cite{Afshar:2014ffa}. More recently, theories of this kind have been related to truncations of certain infinite-dimensional Lie algebras \cite{Bergshoeff:2021tbz}. 
Finally, recall that we saw that theories of this kind never propagate the scalar mode that is present in the linearized spectrum of generic higher-derivative theories \cite{Bueno:2022lhf}. This feature is likely valid for general $d$.


\section{Holographic c-Theorem for Induced Gravity}\label{sec:GenProof}

We will now prove one of our main results of this chapter: all the densities in the action of holographically induced gravity, at arbitrary order $n$ and in general dimension $d$, belong to the class of theories whose dual CFTs satisfy a holographic c-theorem.

Before we proceed, let us emphasize that this is not the same as the monotonicity theorem---the $g$-theorem---for the theory that is dual to the brane in the doubly-holographic construction. The latter is dual to the entire system of the induced gravity on the brane plus the cut-off CFT$_d$ coupled to it. The holographic $g$-theorem proven for this system in \cite{Fujita:2011fp} amounts to showing that, as the brane moves deeper into the bulk, its curvature decreases---in CFT terms, flowing to the IR reduces the number of degrees of freedom that are dual to the brane.
This is not what we are doing. After deriving the induced gravitational action $I_{\rm{bgrav}}$, we take this theory on its own and disregard its coupling to the CFT$_d$. Then, our proof of a c-theorem for the putative dual CFT$_{d-1}$ is no longer related to the properties of the brane moving in the bulk.  

To prove the c-theorem we shall assume that our gravitational theory is coupled to a matter sector that satisfies the NEC and that this condition can be readily translated, via  the field equations, into a condition on the curvature terms as shown in the previous section. For this purpose, we assume that matter is minimally coupled to gravity, so that no curvature terms enter the NEC. This assumption is consistent but technically unnatural, and it could be interesting to investigate if it can be relaxed.

That the entire theory $I_{\rm{bgrav}}$ might satisfy a holographic c-theorem might not be unexpected, given its origin in a `good' theory (Einstein-AdS in $d+1$ dimensions, plus a brane) but it is less obvious that the separate order-$n$ densities should also do it. 

We will give two proofs of this result, the first one applying an induction method to the algorithm described in Sec.~\ref{sec:Counterterms}, and the second one using the counterterms adapted for conformally flat boundaries obtained in \cite{Anastasiou:2020zwc}.

\subsection{Inductive algorithm proof}

An examination of the terms $\mathcal{L}_{(n)}$ obtained in Section \ref{sec:Counterterms}, evaluated on the RG-flow metric \eqref{dwa}, suggests that the following expression may be valid for general orders and dimensions,
\begin{equation}\label{LcThD}
    \mathcal{L}_{(n)}\big|_A = - C_n \frac{d-1}{d-2n} (A')^{2(n-1)} \left[ d (A')^2 + 2n A'' \right]\, , \quad \text{where} \quad C_n \equiv L^{2n-1} \frac{(2n - 3)!!}{(2n)!!}\, .
\end{equation}
Remarkably, this expression, if correct, directly implies that each and all of the $\mathcal{L}_{(n)}$ satisfy a holographic c-theorem. 
Recall that if $\mathcal{L}_{(n)}\big|_A$ is second-order in derivatives of $A(r)$ and linear in $A''(r)$, so is $\mathcal{E}_t^t\left|_A - \mathcal{E}_r^r\right|_A$, and then we can easily build a monotonous c-function, as shown in the previous section. 
We will now prove that \eqref{LcThD} is indeed correct.

We proceed by induction. We assume that \eqref{LcThD} is true for all orders $k < n$, and then we perform the KLS algorithm described in Section \ref{sec:Counterterms} to see that it is also valid for order $n$.

From \eqref{LPi}, the induction hypothesis implies that, for all $k < n$, we have
\begin{equation}\label{PicThD}
    \Pi_{(k)}\big|_A = - C_k (d-1) (A')^{2(k-1)} \left[ d (A')^2 + 2k A'' \right]\,.
\end{equation}
Then, following equations \eqref{ttEq} and \eqref{rrEq}, with $\mathcal{E}_{ab} = \Pi_{ab}/2$, we obtain  
\begin{align}
    \Pi_{(k)}^{tt} \ \big|_A & = - C_k e^{-2A} (A')^{2(n-1)} \left[ (d-1) (A')^2 + 2 n A'' \right] = - \Pi_{(k)}^{x_ix_i} \big|_A \,, \\
    \Pi_{(k)}^{rr} \ \big|_A & = C_k (d-1) (A')^{2n}.
\end{align}
Now, using equations \eqref{LPi} and \eqref{PiKLS} we can compute  $\mathcal{L}_{(n)}\big|_A$. The result reads
\begin{align}
    \mathcal{L}_{(n)} \big|_A & = \frac{1}{d-2n} \Pi_{(n)} \big|_A \\
    & = - \frac{L}{2(d-2n)} \sum_{k = 1}^{n-1} \left[ \frac{1}{d-1} \Pi_{(k)} \Pi_{(n-k)} - \Pi_{ab}^{(k)} \Pi^{ab}_{(n-k)} \right]_A \\
    & = - \frac{d-1}{d-2n} (A')^{2(n-1)} \left[ d(A')^2 + 2n A'' \right] \frac{L}{2} \sum_{k = 1}^{n-1} C_k C_{n-k}\, .
\end{align}
Finally, using the identity
\begin{equation}
     \frac{L}{2} \sum_{k = 1}^{n-1} C_k C_{n-k} = \frac{L^{2n-1}}{2} \sum_{k = 1}^{n-1} \frac{(2k-3)!!(2(n-k)-3)!!}{(2k)!!(2(n-k))!!} = L^{2n-1}\frac{(2n-3)!!}{(2n)!!} = C_n \, ,
\end{equation}
it follows that $\mathcal{L}_{(n)} \big|_A$ indeed reduces to the form \req{LcThD}, which means that all the order-$n$ Lagrangians appearing in the effective action $I_{\rm{eff}}$ satisfy holographic c-theorems.

It would appear that the proof breaks down at $n = d/2$ for even $d$, but as discussed in Sec.~\ref{sec:algo}, these divergences are easily avoided artifacts.

\subsection{Proof with conformally flat counterterms}

As we have seen in Section \ref{sec:cThmGen}, the proof of the holographic c-theorem relies on the evaluation of each density $\mathcal{L}_{(n)}$ on the conformally flat metric \eqref{dwa} on the brane. 
Instead of computing the general brane effective action, and then evaluating it on the conformally flat metric \eqref{dwa}, we may choose to compute the effective action directly for a conformally flat brane.
For this, we can use (minus) the counterterms for an AdS$_{d+1}$ bulk with a conformally flat boundary, recently obtained in \cite{Anastasiou:2020zwc}.\footnote{We are grateful to I.~Papadimitriou for bringing these results to our attention.} For $n \leq d/2$ these are
\begin{equation}\label{cflat}
    \mathcal{L}_{(n)} |_{\rm{c.flat}} = (-1)^{n} L^{2n-1} \frac{(2n-3)!!(d-n)!}{(d-2)!(d-2n)} \mathcal{P}^{(n)} ,
\end{equation}
where $\mathcal{P}^{(n)}$ is the product of Schouten tensors defined in \eqref{Pn}, along with the necessary dimensional regularization prescription for the $n = d/2$ term. 

Since we have seen in the previous section that the $\mathcal{P}^{(n)}$ satisfy the holographic c-theorem, \eqref{cflat} directly proves our result. Indeed, when evaluated on the metric \eqref{dwa}, the expression above coincides with \eqref{LcThD}, since
\begin{equation}
    \mathcal{P}^{(n)} |_{A} = \frac{(-1)^{n+1}} {(2n)!!} \frac{(d-1)!}{(d-n)!} (A')^{2(n-1)} \left[ d(A')^2 + 2n A'' \right] .
\end{equation}

For $n > d/2$, the limiting procedure of \cite{Alkac:2020zhg}, described in the previous section, gives non-trivial densities when applied to $\mathcal{P}^{(n)}$. When we evaluate these densities on \eqref{dwa}, they also match our results above.


\section{Counterterm Densities in Three Dimensions}\label{sec:3DResults}
Now we have a closer look at the explicit structure of the densities $\mathcal{L}_{(n)}$ for $n\geq 2$ that appear on the brane effective action $I_{bgrav}$. 
Recall that the explicit expressions for the densities $\mathcal{L}_{(n)}$ with $n \leq 4$ can be found in eqs. \eqref{L2}---\eqref{L4} of the previous chapter.
We shall first study them for $d=3$, making use of the results obtained in Chapter \ref{chp:3DHCGs}, and then, in the next section, we will move to $d\geq 4$.

\subsection{Structure of Counterterm Densities, Order by Order}

As argued in \eg \cite{Paulos:2010ke,Gurses:2011fv,Bueno:2022lhf} and in our first chapter, in $d=3$ the most general higher-curvature density constructed from contractions of the metric and the Riemann tensor is a function of the three densities\footnote{In \cite{Paulos:2010ke}, the notation $\mathcal{R}_2$, $\mathcal{R}_3$ is used for the same contractions as in \req{densi} but with the Ricci tensor replaced by its traceless part.} 
\begin{equation}\label{densi}
  R\equiv g^{ab}R_{ab}\, , \quad  \mathcal{R}_2 \equiv R_{ab}R^{ab}\, , \quad \mathcal{R}_3 \equiv R_{a}^b R_{b}^c R_c^a\, .
\end{equation}
This follows from the fact that all Riemann curvatures are Ricci curvatures due to the vanishing of the Weyl tensor, along with the existence of Schouten identities which relate terms involving higher-order contractions of the Ricci tensor to the ones above.
In three dimensions, conformal flatness is equivalent to the vanishing of the Cotton tensor,
\begin{equation}
C_{abc}\equiv 2 \nabla_{[c} R_{a|b]}+ \frac{1}{2} \nabla_{[b|} R g_{a|c]}\,.
\end{equation}
Then, the metric \req{dwa} used for holographic RG flows has $C_{abc}=0$.

 
\subsubsection*{Quadratic Order}
As mentioned in the introduction of this chapter, in $d=3$ the density $\mathcal{L}_{(n)}$ coincides, up to an overall factor, with the quadratic term in the New Massive Gravity \cite{Bergshoeff:2009hq}. This is given by
\begin{equation}
    \bar{\mathcal{L}}_2= R_{ab}R^{ab}-\frac{3}{8} R^2\,,
\end{equation}
where the overbar in $\bar{\mathcal{L}}$ simply indicates that we remove the overall factors containing $L$ from the expressions in \eqref{L2}--\eqref{L4}.
NMG is known to satisfy a holographic c-theorem \cite{Sinha:2010ai}. An additional property of $\bar{\mathcal{L}}_2$ is that, when linearized around maximally symmetric backgrounds, it propagates no scalar mode. Moreover, the equations of motion of $\bar{\mathcal{L}}_2$ have second-order trace \cite{Oliva:2010zd}.


\subsubsection*{Cubic Order}
To cubic order, and up to an overall factor, \eqref{L3} gives
\begin{equation}
    \bar{\mathcal{L}}_3=\frac{11}{8} R \mathcal{R}_2 - \frac{15}{64 }R^3 - 2 R^{ac} R^{bd} R_{abcd}+\frac{1}{4} R^{ab} \nabla_a \nabla_b R- R_{ab} \Box R^{ab}+ \frac{1}{4} R \Box R \, .
\end{equation}
Integrating by parts and substituting the three-dimensional Riemann tensor in terms of Ricci tensors, this can be rewritten as
 \begin{equation}\label{cubicupto}
  \bar{\mathcal{L}}_3\overset{\nabla}{=}-\frac{29}{8} R \mathcal{R}_2 + 4 \mathcal{R}_3+\frac{49}{64} R^3+ \frac{3}{8} R \Box R - R_{ab} \Box R^{ab} \, ,
  \end{equation}
 where we have introduced the notation 
 \begin{equation}
     \overset{\nabla}{=}\quad:  \textrm{equal up to total derivatives}\,.
 \end{equation}
If we use that
\begin{align}
  C_{abc} C^{abc}
   & \overset{\nabla}{=} - 2R_{ab} \Box R^{ab} +\frac{3}{4} R \Box R + 6 \mathcal{R}_3 -5 R \mathcal{R}_2+R^3 \, ,
 \end{align} 
then \eqref{cubicupto} can be further rewritten as
\begin{equation}\label{cubicc}
     \bar{\mathcal{L}}_3 \overset{\nabla}{=} \mathcal{S}_3[R_{ab}] +T_3[C_{abc}\cdots,\nabla_a]\, , 
\end{equation}     
where
\begin{equation}
    \mathcal{S}_3[R_{ab}] \equiv \mathcal{R}_3 + \frac{17}{64} R^3 - \frac{9}{8} R \mathcal{R}_2\, ,
\end{equation}
and
\begin{equation}
T_3[C_{abc}\cdots,\nabla_a]\equiv \frac{1}{2} C_{abc} C^{abc}\, .
\end{equation}
On the one hand, $\mathcal{S}_3$ is the cubic generalization of NMG identified in \cite{Sinha:2010ai} as the most general density of that order ---not involving covariant derivatives of the Ricci tensor--- which satisfies a holographic c-theorem. 
On the other hand, $T_3$ involves explicit covariant derivatives of the Ricci tensor. However, since it is proportional to the Cotton tensor, which identically vanishes on \eqref{dwa}, it has no effect on the holographic RG flow. Then, $\bar{\mathcal{L}}_3$ satisfies a holographic c-theorem. 

As it turns out, this density has interesting additional properties. On the one hand, as observed in \cite{Afshar:2014ffa}, the criterion that cubic extensions of NMG do not propagate a scalar mode usually present in the spectrum of higher-curvature gravities, and that they admit a Chern-Simons formulation, restricts them to a general linear combination  of $\mathcal{S}_3$ and $T_3$. Hence, $\bar{\mathcal{L}}_3$ satisfies these two requirements---the first one is in fact implied when the holographic c-theorem is required, as shown in Section \ref{sec:ctheorem3D}.

On the other hand, $T_3$ had been previously singled out in \cite{Oliva:2010zd} using yet a different criterion: it is the cubic density with the lowest-order traced field equations in three dimensions. Indeed, it is the only cubic theory whose equations of motion have a trace which only contains terms involving up to three derivatives of the metric.\footnote{Note that $\mathcal{S}_3$ does not have equations of motion with a reduced-order trace, which means that the c-theorem property and the reduced-order trace one are not directly connected, even though there are cases in which they do coincide, such as NMG itself and the $T_3$ density.  }


\subsubsection*{Quartic Order}
At quartic order, evaluating \eqref{L4} for $d=3$ gives
\begin{align}
\bar{\mathcal{L}}_4 =&
-\frac{83}{16}\mathcal{R}_2^2-17 R \mathcal{R}_3 +\frac{1155}{64} R^2 \mathcal{R}_2-\frac{3635}{1024} R^4 \
\nonumber \\ 
& -  \frac{31}{4} \mathcal{R}_2 \Box R + \frac{57}{16} R^2 \Box R - 10 R^{bc} \nabla_{a}R_{bc} \nabla^{a}R + \frac{53}{16} R \nabla_{a}R \nabla^{a}R \nonumber \\ 
& + \frac{1}{2} R^{ab} \nabla_{a}R^{cd} \nabla_{b}R_{cd} + \frac{9}{32} \Box R \Box R + \frac{1}{2} R \Box^2 R + \frac{5}{16} R_{ab} \nabla^{a}R \nabla^{b}R \nonumber \\ 
&-  \frac{11}{2} R_{a}{}^{c} R_{bc} \nabla^{b}\nabla^{a}R + \frac{61}{16} R_{ab} R \nabla^{b}\nabla^{a}R -  \frac{3}{32} \nabla_{b}\nabla_{a}R \nabla^{b}\nabla^{a}R \nonumber \\ 
&-  \frac{1}{2} R^{bc} \nabla^{a}R \nabla_{c}R_{ab}  -  \frac{47}{4} R^{ab} R \nabla_{c}\nabla^{c}R_{ab} + \frac{3}{4} \nabla^{b}\nabla^{a}R \Box R_{ab} + \frac{1}{2} R_{ab} \Box \nabla^{b}\nabla^{a}R \nonumber \
\\ 
& -  \frac{11}{2} R \nabla_{c}R_{ab} \nabla^{c}R^{ab} + 12 R^{ab} \
R^{cd} \nabla_{d}\nabla_{b}R_{ac} -  \frac{27}{2} R^{ab} R^{cd} \nabla_{d}\nabla_{c}R_{ab}  \nonumber \\ 
&-  \frac{3}{2} \Box R^{ab} \Box R_{ab} + 28 R_{a}{}^{c} R^{ab} \Box R_{bc} - 2 R^{ab} \Box^2 R_{ab} - 4 R^{ab} \nabla_{b}R_{cd} \nabla^{d}R_{a}{}^{c} \nonumber \\ 
&  + 4 R^{ab} \nabla_{c}R_{bd} \nabla^{d}R_{a}{}^{c} + 16 R^{ab} \nabla_{d}R_{bc} \nabla^{d}R_{a}{}^{c}\, .
\end{align}
Again, when we decompose it as
\begin{equation}
    \bar{\mathcal{L}}_4 \overset{\nabla}{=} \mathcal{S}_4[R_{ab}] +T_4[C_{abc}\cdots,\nabla_a]\,,
\end{equation}
where 
\begin{equation}
    \mathcal{S}_4[R_{ab}] \equiv \frac{5}{4} \mathcal{R}_3 R -\frac{15}{16} \mathcal{R}_2^2 - \frac{45}{64} R^2 \mathcal{R}_2 + \frac{205}{1024} R^4 \,,
    \end{equation}
and
\begin{align}
   & T_4[C_{abc}\cdots,\nabla_a]\equiv +R C_{abc}C^{abc}-\frac{11}{2}R_{b}^a C_{aef}C^{bef}+\frac{23}{4} R^ {ac}R^ {bd}\nabla_{a}C_{bcd} \notag \\ &-\frac{17}{2}R_{eb} R^{e}_c\nabla_a C^{bac}+\frac{5}{2}R R_{bc}\nabla_a C^{bac}-\frac{5}{4}C_{bcd}R^{ac}\nabla_a R^{bd}-\frac{11}{2} R_{c}^e C^{bac}\nabla_a R_{be}\,.
\end{align}
Similarly to the cubic case, we find that $\mathcal{S}_4$ is the quartic generalization of NMG---algebraic in the curvature---which non-trivially satisfies a holographic c-theorem \cite{Sinha:2010ai,Paulos:2010ke}. On the other hand, we see that $T_4$ is a linear combination of terms which always involve at least one Cotton tensor and therefore identically vanish when evaluated on the RG-flow metric \eqref{dwa}. Again, this makes evident that $\bar{\mathcal{L}}_4 $ satisfies a holographic c-theorem. 

Motivated by the cubic case, we have tried to express $T_4$ as one of the theories identified in \cite{Afshar:2014ffa} by the criterion that they admit a Chern-Simons description, but we have not succeeded in doing so. It seems that such identification only works for the quadratic and cubic terms. Similarly, while  $T_3$ had the property of possessing a reduced order for the trace of its equations of motion, this is no longer the case for $T_4$, whose traced equations are of order six.


\subsubsection*{Higher Orders}

It seems, then, that of all the special properties that we identified for $\mathcal{S}_3$ and $T_3$ in $d=3$, only those that refer to the holographic c-theorem extend to higher orders. Of course we have already given a general proof that all the ${\mathcal{L}}_{(n)}$ satisfy this theorem, but we can aim at distinguishing a finer structure of how this happens.

We decompose the $\bar{\mathcal{L}}_n$ into terms $\mathcal{S}_n$ and $T_n$ such that the $\mathcal{S}_n$ contain all of the non-vanishing contribution to the c-function, and the $T_n$ vanish identically on the RG-flow metric \eqref{dwa}. For the lowest orders we have seen that this separation  can be performed in such a way that $\mathcal{S}_n$ is algebraic in the curvature, that is,
\begin{equation}\label{SnTn}
    \bar{\mathcal{L}}_n \overset{\nabla}{=} \mathcal{S}_n[R_{ab}] + T_n[\nabla_a, R_{ab}]\, .
    \end{equation}
In fact, in $d=3$ this decomposition can be performed in all $n$. This follows from the results in Section \ref{sec:ctheorem3D}, which show that, at every $n$, there always exists a density $\mathcal{C}_n[R_{ab}]$ which non-trivially satisfies the c-theorem.

For the cubic and quartic terms, we have found that the $T_n$ are proportional to the Cotton tensor. It is unclear whether this is the case also for the quintic term, since the expressions are exceedingly complicated.
On the other hand, the structure of $\mathcal{S}_n[R_{ab}]$ is uniquely constrained not only in $n=3,4$, as we have seen, but also in $n=5$. Up to that order, there is a single order-$n$ algebraic density $\mathcal{C}_n$ which non-trivially satisfies the holographic c-theorem \cite{Paulos:2010ke}, and so $\mathcal{S}_n[R_{ab}]$ must be proportional to it. The proportionality constant can be found by evaluating both $\bar{\mathcal{L}}_n$ and $\mathcal{C}_{n}$ on the RG-flow metric \eqref{dwa}. For the quintic case, we obtain $\mathcal{S}_5 = \frac{5}{64} \mathcal{C}_5$, where
\begin{equation}
    \mathcal{C}_{5} = \frac{61R^5}{960} - \frac{7R^3\mathcal{R}_2}{12} + \frac{2R^2\mathcal{R}_3}{15} + \frac{7R\mathcal{R}_2^2}{5} - \frac{16\mathcal{R}_2\mathcal{R}_3}{15}\, .
\end{equation}

However, degeneracies start to appear at order $6$. From that order on, there exist densities that are algebraic in the Ricci tensor and which trivially satisfy the holographic c-theorem \cite{Paulos:2010ke}. These have been characterized in a precise manner. As shown in Sections \ref{sec:ctheorem3D} and \ref{sec:Ommm6} of our first chapter, there is a unique sextic density  of this type,\footnote{This is more easily written in terms of contractions of the traceless Ricci tensor $\tilde{\mathcal{R}}_2\equiv \tilde R_{a}^b \tilde R_b^a$, $\tilde{\mathcal{R}}_3\equiv \tilde R_{a}^b \tilde R_b^c \tilde R_c^a$, where $\tilde R_{ab}\equiv R_{ab}-\frac{1}{3}g_{ab}R$, namely, $\Omega_{(6)} = 6 \tilde{\mathcal{R}}_3 ^2-\tilde{\mathcal{R}}_2 ^3$. }
\begin{align}\label{Omeg6cT}
\Omega_{(6)} 
=\frac{1}{3}  \left[R^6- 9R^4\mathcal{R}_2+8R^3\mathcal{R}_3+21R^2\mathcal{R}_2^2-36R\mathcal{R}_2\mathcal{R}_3-3\mathcal{R}_2^3+18\mathcal{R}_3^2\right]\,,
\end{align}
with the important property that, at any order $n\geq 6$, all the densities algebraic in curvature that vanish on the RG flow geometry are proportional to $\Omega_{(6)}$. Then, by taking $L^{\rm general}_{n-6}$ to be the most general density that is algebraic in the curvature, we have that $L_{n-6}^{\rm general}\cdot \Omega_{(6)}$ is the most general density of that type at order $n$ that vanishes on  RG flows.

This implies that the characterization of the terms in \req{SnTn} is ambiguous for $n\geq 6$, since we can redefine
\begin{equation}
    \mathcal{S}_n' =\mathcal{S}_n + L_{n-6} \Omega_6\,, \qquad T_n'= T_n-L_{n-6} \Omega_6 \,,
\end{equation}
where $L_{n-6}$ is an arbitrary order-$(n-6)$ density algebraic in the curvature. Still, it is possible that a particular separation exists such that $T_{n\geq6}$ does not involve any $\Omega_6$ and vanishes exclusively due to the presence of Cotton tensors in all its terms. If that is the case,  one can use this criterion to  give a unique definition for $\mathcal{S}_{n\geq 6}$.

As far as we know, there are two different proposals for special order-$n$ densities that non-trivially satisfy the holographic c-theorem. The first results from the expansion of the Born-Infeld-like extension of NMG presented in \cite{Gullu:2010pc}, and in the following subsection we find hints that this may indeed coincide with $\mathcal{S}_{n\geq 6}$ as defined by the above criterion. The second corresponds to a basis of densities selected by the fact that they satisfy a simple recursive formula which relates the order-$n$ representative to the order-$(n-1)$ and order-$(n-2)$ ones described in Section \ref{sec:ctheorem3D}. 


\subsection{Born-Infeld Gravities and Counterterms}\label{BIse}

An interesting generalization of NMG with a Born-Infeld-type Lagrangian was proposed in \cite{Gullu:2010pc}. The Lagrangian is
\begin{equation}\label{BI-NMG}
    \mathcal{L}_{\text{BI-NMG}} =\alpha  \sqrt{\det \left( \delta_a^b + \beta  G_a^b \right) }\, ,
\end{equation}
where $G_{ab}$ is the Einstein tensor and $\alpha,\beta$ are constants.
This theory satisfies the holographic c-theorem \cite{Gullu:2010st}, and when expanded at low curvatures it also generates higher-derivative densities which non-trivially satisfy it at any truncated order \cite{Alkac:2018whk}. As we have seen, this property is shared by the effective gravitational action induced on the brane-world.

Following our results in Chapter \ref{chp:3DHCGs}, we can expand $\mathcal{L}_{\text{BI-NMG}} $ order by order, to find higher-curvature densities $\mathcal{B}_{(n)}$ which, on the RG flow metric \eqref{dwa}, give
\begin{equation}
    \mathcal{B}_{(n)}[\alpha,\beta] \big|_A =\alpha \left( -\beta  \right)^n \frac{(2n-5)!!}{(2n)!!} (A')^{2(n-1)} \left[ 3(A')^2 + 2n A'' \right].
\end{equation}
Remarkably, if we take $\alpha=2/L$ and $\beta=-L^2$, then this result coincides, for all $n$,
with the RG flow of the order-$n$ brane-world density \eqref{LcThD} in $d=3$, namely
\begin{equation}
 \left. \mathcal{L}_{(n)}\right|_A  =  \mathcal{B}_{(n)}[2/L,-L^2] \big|_A \, .
\end{equation}
This result is highly non-trivial, since the coincidence occurs also for the relative factors between the different order-$n$ Lagrangians, and not only for the functional dependence in $A$ and its derivatives, which might have been expected. It is then natural to conjecture that the $d=3$ counterterm Lagrangian may be resummed as
\begin{equation}
    \mathcal{L} = \frac{2}{L} \sqrt{\det \left( \delta_a^b - L^2 G_a^b \right) } \ + \ \mathcal{T} [C_{abc}\cdots,\nabla_a],
\end{equation}
where again
\begin{equation}
    \mathcal{T} [C_{abc}\cdots,\nabla_a] \big|_A = 0.
\end{equation}
An even stronger conjecture would be that the whole tower of counterterms (including $ \mathcal{T} $) could be written as a Born-Infeld-like action. The idea
that Born-Infeld type actions may act as suitable AdS counterterms has been considered before in \cite{Mann:1999bt,Jatkar:2011ue,Sen:2012fc}.


\section{Structure of Counterterm Densities in Higher Dimensions}
\label{sec:GenD}
Let us now move to $d\geq 4$. The expressions become considerably more involved than in three dimensions, but we can still infer a similar general structure based on the lowest orders.
For the following discussion, it will be useful to keep in mind that the Weyl tensor $C_{abcd}$ identically vanishes on the RG-flow geometry \req{dwa}.


\subsection*{Quadratic Order}
Up to an overall factor, the quadratic term reads 
\begin{equation}
    \bar{\mathcal{L}}_2= R_{ab}R^{ab}-\frac{d}{4(d-1)} R^2\,,
\end{equation}
which is the $d$-dimensional generalization of NMG. Since it can be rewritten as a linear combination of the Weyl tensor squared and the quadratic Lovelock density, namely,
\begin{equation}
    \bar{\mathcal{L}}_2 = \frac{d-2}{4(d-3)}\left[ C_{abcd}C^{abcd}-\mathcal{X}_4\right]\, ,
\end{equation}
it is easy to see why it also fulfils a holographic c-theorem. Similar to the $d=3$ case,  $\bar{\mathcal{L}}_2$ propagates no scalar mode when linearized around maximally symmetric backgrounds \cite{Hassan:2013pca,Bueno:2016ypa}. Moreover, $\bar{\mathcal{L}}_2$ also belongs to the set of quadratic theories which have the property of possessing equations of motion whose trace is second-order, since for $d\geq 4$, that set is given by an arbitrary linear combination of
$C_{abcd}C^{abcd}$ and 
the quadratic Lovelock density
$\mathcal{X}_4$
\cite{Nakasone:2009bn,Oliva:2010zd,Oliva:2010eb}.


\subsection*{Cubic Order}
The cubic density was written in \req{L3} above.
Observe first that integrating by parts this can be rewritten as
\begin{align}\label{L32}
    \bar{\mathcal{L}}_3\overset{\nabla}{=} &\notag +\frac{3d+2}{4(d-1)}R R_{ab}R^{ab}-\frac{d(d+2)}{16(d-1)^2}R^3-2 R^{ab}R_{acbd}R^{cd}\\ &-\frac{d}{4(d-1)} \nabla_a R \nabla^a R + \nabla^c R^{ab}\nabla_c R_{ab}\, .
\end{align}
Now, following inspiration from the three-dimensional case, we can try to rewrite $ \bar{\mathcal{L}}_3$ as a linear combination of densities with special properties. We find that, indeed, $ \bar{\mathcal{L}}_3$ can be written for general $d\geq 4$ as
\begin{equation}
\bar{\mathcal{L}}_3= \frac{d-2}{16(d-3)}\mathcal{N}_6+\Xi+\Delta \, ,
\end{equation}
where $\mathcal{N}_6$, $\Xi$ and $\Delta$ are distinguished for different reasons. On the one hand, $\mathcal{N}_6$, which is defined as
\begin{align}
    \mathcal{N}_6\equiv &-24 R^{abcd}R_{cdbe}R^{e}_a-\frac{3(d+2)}{d-1}R R^{abcd}R_{abcd}-\frac{24d}{d-2}R^{abcd}R_{ac}R_{bd}\\ \notag &-\frac{16d(d-1)}{(d-2)^2}R^{ab}R_{bc}R^c_a+\frac{12(d^3-2d^2+6d-8)}{(d-2)^2(d-1)}R R^{ab}R_{ab} \\ \notag &- \frac{d^4-3d^3+10d^2+4d-24}{(d-2)^2(d-1)^2}R^3\, ,
\end{align}
is the cubic Quasi-topological density \cite{Oliva:2010zd,Oliva:2010eb,Myers:2010ru}. This satisfies a number of interesting properties. Firstly, it can be written as 
\begin{equation}
   \mathcal{N}_6= \frac{d-2}{d-5} \left[4 W_1+8W_2-\mathcal{X}_6 \right]\, ,
\end{equation}
where $W_1\equiv C^{ab}{}_{cd}C^{cd}{}_{ef}C^{ef}{}_{ab}$, $W_2\equiv C_{abcd}C^{ebcf}C^a\,_{ef}\,^d$ and $\mathcal{X}_6$ is the cubic Lovelock density. This expression makes manifest that $ \mathcal{N}_6$ satisfies the holographic c-theorem \cite{Myers:2010tj}.  $ \mathcal{N}_6$ identically vanishes in $d= 4$ but it is non-trivial for $d\geq 5$. It is in fact the term involving $ \mathcal{N}_6$ (actually $\mathcal{X}_6$) the one which makes $\bar{\mathcal{L}}_3$ be non-trivial when evaluated on \req{dwa} for $d\geq 5$ ($d \geq 6$).  In addition, $ \mathcal{N}_6$ is one of the few cubic densities which possess second-order traced equations for general $d\geq 5$ \cite{Oliva:2010eb}.\footnote{For $d=5,6$ there are two independent densities which possess second-order traced equations whereas for $d\geq 7$ there exist three.} Finally, $ \mathcal{N}_6$ only propagates the usual massless graviton when linearized around maximally symmetric backgrounds and it admits particularly simple black hole solutions \cite{Oliva:2010eb,Myers:2010ru}.

On the other hand, $\Xi$ is the piece which contains the terms involving explicit covariant derivatives. It is explicitly given  by
\begin{equation}\label{sigom}
    \Xi\equiv \frac{(d-2)^2}{4(d-3)(d-6)}\left[ \Sigma + \frac{2(d-3)}{3(d-2)^2}\Theta\right]\, ,
\end{equation}
where $\Sigma$ and $\Theta$ were previously identified again in \cite{Oliva:2010zd} as the two only densities which possess field equations whose trace is third-order in derivatives for $d\geq 4$. They are given, respectively, by\footnote{Similarly to the case of $\mathcal{N}_6$ in $d=5$, the combination inside the brackets in \req{sigom} vanishes identically in $d=6$, and then one finds
\begin{align}\label{sigo9m}
    \Xi|_{d=6}\equiv &+\frac{2}{9} R^{abcd}R_{cdef}R^{ef}_{ab} - \frac{8}{9}R^{abcd}R_{ac}R_{bd}-\frac{4}{3}R^{abcd}R_{ac}R_{bd}+\frac{10}{9}R^{ab}R_{bc}R^c_a\\ \notag &+\frac{1}{450}R^3 -\frac{3}{10} \nabla_a R \nabla^a R+\nabla_a R_{bc}\nabla^a R^{bc}\, .
\end{align}}
\begin{align}
    \Sigma= & -\frac{3d-2}{2} R^{abcd}R_{cdef}R^{ef}_{ab} +\frac{8d}{3} R_{cd}^{ab}R_{bf}^{ce}R_{ae}^{df} +\frac{4d}{d-2}R^{abcd}R_{ac}R_{bd}\\ \notag &+ \frac{4(d-4)}{d-2}R^{ab}R_{bc}R^c_a-\frac{2d}{3(d-1)^2}R^3-\frac{d(d-3)}{(d-2)(d-1)}\nabla_a R \nabla^a R \\ \notag &+\frac{4(d-3)}{d-2}\nabla_a R_{bc}\nabla^a R^{bc} \, , 
    \end{align}
    and
    \begin{align}
    \Theta=& +2(d^2-4)R^{abcd}R_{cdef}R^{ef}_{ab}-4(d^2-4)R_{cd}^{ab}R_{bf}^{ce}R_{ae}^{df}-12(d-2)R^{abcd}R_{ac}R_{bd}\\ \notag &-16R^{ab}R_{bc}R^c_a+\frac{d^2-d+2}{(d-1)^2}R^3+\frac{6d}{d-1}\nabla_a R \nabla^a R-24\nabla_a R_{bc}\nabla^a R^{bc}   \, .
\end{align}
Both $\Sigma$ and $\Theta$ non-trivially fulfil the holographic c-theorem when evaluated on \req{dwa}. However, the combination appearing in the density $\Xi$ trivially satisfies the holographic c-theorem for general $d$, as it becomes a total derivative when evaluated on \req{dwa}.

Finally, $\Delta$ is a density which does not involve explicit covariant derivatives, which is trivial when evaluated on the holographic c-theorem ansatz for general $d$ and which does not satisfy any additional special property involving a reduced order for its traced equations. It is given by
\begin{align}
   \Delta &\equiv \frac{1}{d-3} \left[  \frac{(d-10)(d-2)}{24}R^{abcd}R_{cdef}R^{ef}_{ab}+\frac{36-d(10+7d)}{4(d-2)(d-1)}R R^{ab}R_{ab} \right. \\ \notag & +\frac{3(d-2)}{2} R^{abcd}R_{cdbe}R_a^e+\frac{3(d-2)(d+2)}{16(d-1)}R R^{abcd}R_{abcd}+\frac{d+8}{2}R^{abcd}R_{ac}R_{bd} \\ \notag & \left. +\frac{11d-16}{2}R^{ab}R_{bc}R^c_a+\frac{2(d-2)}{3}R_{cd}^{ab}R_{bf}^{ce}R_{ae}^{df} +\frac{-28+d(21d-16)}{24(d-2)(d-1)^2}R^3\right]\, .
\end{align}
As mentioned earlier,  the general set of cubic theories constructed from arbitrary contractions of the metric and the Riemann tensor satisfying the holographic c-theorem property was obtained in \cite{Myers:2010tj}. $ \Delta$ is one of the 5 independent densities which contribute trivially to the c-function.

In view of the three-dimensional case, it is natural to wonder whether all terms appearing in $\Xi$ and $\Delta$ may be rewritten in a simplified way in terms of the Weyl tensor---so that the fact that they vanish when evaluated on \req{dwa} becomes manifest.

An alternative decomposition of $\bar{\mathcal{L}}_{3}$, found in \cite{Anastasiou:2020zwc}, is
\begin{equation}
    \bar{\mathcal{L}}_{3} = S_{ab}\left(S_{cd} +  \frac{1}{d-3} \nabla_c \nabla_d \right)C^{acbd} + 3(d-4) \mathcal{P}^{(3)}.
\end{equation}
Since the Weyl tensor and $\mathcal{P}^{(3)}$ are explicit in this form, it makes manifest that $\bar{\mathcal{L}}_{3}$ satisfies the holographic c-theorem.


\subsection*{Higher Orders}
Going to higher orders complicates the expressions considerably. We presented the result for the general-$d$ quartic density in \req{L4}. 
We have verified that, analogously to the $d=3$ case, it is also possible to write $\bar{\mathcal{L}}_4$ as a sum of a term which does not involve explicit covariant derivatives and which non-trivially satisfies the c-theorem, plus another one which does contain covariant derivatives and is trivial when evaluated on \req{dwa}. It is then natural to expect that the $n$-th order density in $d$ dimensions can always be written as
\begin{equation}
   \bar{ \mathcal{L}}_n= \mathcal{S}_n[R_{abcd}]+T_n[R_{abcd},\nabla_a]\,,
\end{equation}
where $\mathcal{S}_n[R_{abcd}]$ is linear in $A''(r)$ when evaluated on \req{dwa} and does not involve higher-derivative terms, and where $T_n[R_{abcd},\nabla_a]$ vanishes (or it is a total derivative) for the same ansatz. 


\section{Conclusions and Outlook}

In this chapter, we have proven that the higher-derivative densities $\mathcal{L}_{(n)}$ that are induced from a brane-world construction fulfil a simple holographic c-theorem, and we have explored some of their other properties at low-enough order-$n$.

Let us close with a few observations and possible future directions.

\paragraph{Structure of Counterterms.}
We have just seen that some properties of the invariants $\mathcal{L}_{(n)}$ are more easily seen if they are written in terms of Weyl and Schouten tensors since, in particular, the Weyl tensor vanishes for the c-theorem ansatz \eqref{dwa}.
Formulating the algorithm described in the previous Chapter \ref{chp:Alg} on a basis of Weyl and Schouten tensors will perhaps allow us to push it to higher orders, or it may also reveal finer structures in the counterterms.

\paragraph{Higher-curvature gravities in the bulk.} We have seen that starting from Einstein gravity in the $(d+1)$-dimensional bulk, the effective $d$-dimensional  higher-derivative theories induced on the brane satisfy holographic c-theorems. What would happen if the bulk gravitational theory were itself a higher-curvature theory? It seems likely that the c-theorem we have proven is an imprint of the healthy dynamics of bulk Einstein gravity: {\it good parents raise good children}. In that case, we would expect it to fail for a general higher-curvature bulk theory. Natural exceptions to be expected are Lovelock gravities \cite{Lovelock1,Lovelock2}, which also have second order equations. In fact, it has been suggested in \cite{Brihaye:2008xu} that in that case the counterterm at a given order is a linear combination of the same Einstein gravity-induced counterterm plus a new piece proportional to the $d$-dimensional Lovelock density of the corresponding order. Hence, for instance, $\mathcal{L}_3$ would be a linear combination of \req{L3} plus the cubic Lovelock density $\mathcal{X}_6$, and so on. It would then follow that these modified brane actions also satisfy holographic c-theorems, since the Lovelock terms satisfy the required conditions---namely, second-order on-shell action and linearity in $A''(r)$ when evaluated on the \req{dwa} ansatz.

\paragraph{Counterterms as Born-Infeld gravities in higher-dimensions?} In Section \ref{BIse}, we showed that the order-$n$ counterterm Lagrangian $\mathcal{L}_{n}$ coincides, when evaluated on the holographic c-theorem metric ansatz \req{dwa}, with the general term resulting from the expansion of the Born-Infeld-type generalization of NMG \cite{Gullu:2010pc}. This suggests that the full three-dimensional counterterms Lagrangian might be rewritten in such a Born-Infeld form plus a possible term which would vanish when evaluated on the RG-ansatz metric \req{dwa}. A possible $d$-dimensional generalization of these observations is far from obvious at the moment, but a quick inspection of some low-dimensional cases suggests that the modified Born-Infeld-like Lagrangian 
\begin{equation}\label{BI-d}
    \mathcal{L}^{(d)}_{\text{BI}} =\alpha \left[\det \left( \delta_a^b + \beta  G_a^b \right) \right]^{\frac{1}{d-1}}
\end{equation}
also fulfils a simple holographic c-theorem. Moreover, when \req{BI-d} is evaluated on-shell (on \req{dwa}) and expanded order by order, we find densities $\mathcal{B}_{(n)}|_A$ with the same functional dependence on $A$ as in the on-shell counterterm Lagrangians \eqref{LcThD}. We have found, however, no straightforward way to define $\alpha$ and $\beta$ such that the relative (overall) coefficients match our findings in equation \eqref{LcThD}. It would be interesting to analyse this possibility in more detail and, more generally, to study the properties of the Lagrangian defined by \req{BI-d}.

\paragraph{Holographic c-theorem gravities and scalar modes.} We have seen that the counterterm Lagrangians of the lowest orders often satisfy additional properties besides the holographic c-theorem. One of them is the absence of the scalar mode that generically appears in the linearized spectrum around maximally symmetric backgrounds of higher-derivative theories---see \eg \cite{Bueno:2016ypa}. 
Many higher-curvature theories which satisfy the holographic c-theorem also seem to share this property. In fact, we have proven in Section \ref{sec:ctheorem3D} that, in $d=3$, all the higher-curvature theories that satisfy a holographic c-theorem propagate no scalar mode. 
It would be interesting to prove or disprove this for $d\geq 4$. Observe that the class of theories which do not propagate the scalar mode is larger than the class of theories that admit a holographic c-theorem, so the question is whether the latter class is fully contained within the former.  

In the case considered in this work, it seems natural that the higher-derivative gravities holographically induced on the brane should propagate no scalar mode when linearized around maximally symmetric backgrounds. This fact is true in $d=3$ to all orders, as we have just said, and in general $d$ at least for $n=2$. 
After all, these theories are induced from Einstein gravity in AdS$_{d+1}$. And from the bulk perspective and to linear order, it was shown already in \cite{Karch:2000ct} that one can choose an axial TT gauge for the (massless spin-2) $d+1$-dimensional graviton to induce an almost massless spin-2 $d$-dimensional graviton on the brane, plus an infinite tower of massive spin-2 modes.
We will prove that this is indeed the case in the following chapter, at least when $I_{\text{bgrav}}$ is linearized around flat space.

\cleardoublepage
\lhead{Chapter 7}
\rhead{The Spectrum of Brane-World Gravities}

\chapter{The Spectrum of Brane-World Gravities}
\label{chp:Spectrum}


In this chapter, we will study the linearization of the induced gravity on the brane $I_\text{bgrav}$ around flat space, making use of our results from Chapter \ref{chp:HDGs}.
We will first obtain its effective quadratic action to all orders in derivatives, which we will be able to resum into a compact expression. We will then obtain the linearized equations of motion. Finally, we will characterize the pole structure of the metric propagator in various dimensions.

\section{Introduction}

Recall that the effective gravitational action induced on the brane world volume is given by 
\cite{Kraus:1999di,Emparan:1999pm,Balasubramanian:1999re,Papadimitriou:2004ap,Papadimitriou:2010as,Elvang:2016tzz,Bueno:2022log,Anastasiou:2020zwc,Bueno:2022lhf}
\begin{equation}
I_{\rm bgrav}= \frac{1}{16\pi G_{N,\text{eff}}} \int \df^{d}x\sqrt{-g} \left[  R +\frac{L^2}{(d-2)(d-4)} R^{ab}\left( R_{ab}-\frac{d}{4(d-1)}g_{ab}R \right) +\dots\right]\, ,\label{totalactiondd}
\end{equation}
where $L$ is the AdS$_{d+1}$ radius of the ambient spacetime.
Notice that we have chosen $\Lambda_\text{eff} = 0$, since we will be interested in studying the linearization of $I_\text{bgrav}$ in flat space.
Starting at sixth order in derivatives, all the higher-derivative densities involve terms with covariant derivatives of the Riemann tensor, as seen in \req{L3} and \req{L4} of Chapter \ref{chp:Alg}, and so they fall into the kind of higher-derivative theories that we have studied in Chapter \ref{chp:HDGs}.

In this chapter, we will show that the effective curvature-squared action of the full brane-world gravity ---including the infinite tower of terms with covariant derivatives---, which fully specifies the linearized theory around flat space, can be written as
\begin{align}
    I_{\rm bgrav}^{(2)} = \frac{1}{16\pi G_{N,\text{eff}}} \int \df^{d}x\sqrt{-g}  \left[R+L^2 R^{ab}F_d \left(L^2\Box\right) \left( R_{ab}-\frac{d}{4(d-1)} g_{ab }R\right)\right]\,,
\end{align}
where
\begin{equation}
    F_{d}(L^2\Box)\equiv \frac{d(d-2)}{L^4\Box^2}-\frac{1}{L^2\Box}-\frac{(d-2)Y_{\frac{d+2}{2}}\left(L \sqrt{\Box}\right)}{L^3 \Box^{3/2}Y_{\frac{d}{2}}\left(L \sqrt{\Box}\right)}\, ,
\end{equation}
and $Y_{k}$ are Bessel functions of the second kind. Using this expression, we will study the linearized spectrum of the theory on Minkowski spacetime in various dimensions. Generically, the metric perturbations propagator includes poles of the form
\begin{equation}
    P_d(L^2 k^2) \sim \frac{1}{L^2 k^2}\, ,\qquad P_d(L^2 k^2) \sim -\frac{2}{(d-2)L^2 [k^2+m_j^2]} \, ,
\end{equation}
where the first is the usual Einstein gravity massless spin-2 mode, and the second corresponds to infinite towers of massive spin-2 modes (labelled by $j$) which always have negative kinetic energy. Depending on the dimension, some of those modes have positive squared masses, some of them have negative squared masses and some of them have imaginary squared masses.  
In our case, we will prove that brane-world gravities do not propagate a scalar mode, since their effective quadratic action fulfils condition \eqref{keyko}.
Moreover, we will see that they always present, besides the usual massless graviton, an infinite tower of spin-2 ghosts.

Recall that in Chapter \ref{chp:Alg} we saw how to compute $I_{\rm bgrav}$ algorithmically, by integrating the Gauss radial constraint. 
Since we are interested in studying the spectrum of $I_\text{bgrav}$ around flat space, we can fix the brane tension to be critical so that the theory has a vanishing cosmological constant.
The radial constraint reads, in this case,
\begin{equation}\label{GaussMink}
\Pi=\frac{L}{2}\left[R+\Pi_{ab}\Pi^{ab}-\frac{1}{d-1}\Pi^2\right]\, .
\end{equation}
Again, assuming that the Lagrangian allows for a derivative expansion of the form
\begin{equation}
\mathcal{L}=\sum_{n=1}^{\infty} L^{2n-1}\mathcal{L}_{(n)}\, ,
\end{equation}
the new algorithm that solves eq. \eqref{GaussMink} is
\begin{align}
\Pi_{(1)}&=\frac{R}{2}\, ,\\
\Pi_{(n)}&=\frac{1}{2}\sum_{i=1}^{n-1}\left[\Pi_{(i)\, ab}\Pi^{ab}_{(n-i)}-\frac{1}{d-1}\Pi_{(i)}\Pi_{(n-i)}\right]\, , \quad n\ge 2\, ,
\label{Pinrecursive1}
\end{align}
along with \cite{Kraus:1999di}
\begin{equation}\label{PiL}
\Pi_{(n)}=(d-2n)\mathcal{L}_{(n)}+\text{total derivative}\, .
\end{equation}
Since the total derivatives are irrelevant for the Lagrangian, this allows us to get $\mathcal{L}_{(n)}$ from the trace of the equation of motion $\Pi_{(n)}$.
Thus, we get
\begin{equation}
\mathcal{L}_{(1)}=\frac{R}{2(d-2)}\, ,\quad \Pi_{(1)\, ab}=-\frac{1}{d-2}G_{ab}\, .
\end{equation}
In a similar fashion, this process allows us to generate all the Lagrangian densities $\mathcal{L}_{(n)}$. 

Observe that all of these Lagrangians can be written in the form,
\begin{equation}
\mathcal{L}=\mathcal{L}(R_{ab},\nabla_{c}R_{ab}, \nabla_{c}\nabla_{d}R_{ab}, \ldots)\, ,
\end{equation}
since Riemann curvature appears nowhere in the process.\footnote{
Obviously, the same thing is true for our the general algorithm in Chapter \ref{chp:Alg}. There, however, we chose to commute some covariant derivatives to simplify expressions and match with known results in the literature.
} 


\section{Quadratic Action}
We are interested in studying the linearized equations of these theories around the Minkowski vacuum. As we saw in Chapter \ref{chp:HDGs}, the only higher-derivative terms that contribute to the linearized equations are those quadratic in the curvature (but with an arbitrary number of covariant derivatives) and, therefore, the only possible quadratic Lagrangians are $R\Box^{n} R$ and $R^{ab}\Box^n R_{ab}$.  Thus, at order $2n$ in derivatives, we will necessarily have
\begin{equation}\label{Ln}
\mathcal{L}_{(n)}=\alpha_{n} R\Box^{n-2} R+\beta_{n}R^{ab}\Box^{n-2}R_{ab}+\mathcal{O}(R^3)\, .
\end{equation} 
Our goal is to determine the coefficients $\alpha_{n}$ and $\beta_{n}$, for which we will use \req{Pinrecursive1}. First of all, in order to evaluate the left-hand-side of \req{Pinrecursive1}, we use \req{PiL}, so that we get
\begin{equation}\label{PiL2}
\Pi_{(n)}=(d-2n)\left(\alpha_{n} R\Box^{n-2} R+\beta_{n}R^{ab}\Box^{n-2}R_{ab}\right)+\ldots\, .
\end{equation}
Now we must evaluate the right-hand-side. The case $n=2$ must be considered independently, and it yields
\begin{equation}
\Pi_{(2)}=\frac{1}{2}\left[\Pi_{(1)\, ab}\Pi^{ab}_{(1)}-\frac{1}{d-1}\Pi_{(1)}^2\right]=-\frac{d}{8(d-1)(d-2)^2}R^2+\frac{1}{2(d-2)^2}R^{ab}R_{ab}\, ,
\end{equation}
so that we identify
\begin{equation}
\alpha_{2}=-\frac{d}{8(d-1)(d-2)^2(d-4)}\, ,\quad \beta_{2}=\frac{1}{2(d-2)^2(d-4)}\, .
\end{equation}
Now, for $n\ge 3$ we have
\begin{align}\notag
\Pi_{(n)}&=\Pi_{(1)\, ab}\Pi^{ab}_{(n-1)}-\frac{1}{d-1}\Pi_{(1)}\Pi_{(n-1)}+\frac{1}{2}\sum_{i=2}^{n-2}\left[\Pi_{(i)\, ab}\Pi^{ab}_{(n-i)}-\frac{1}{d-1}\Pi_{(i)}\Pi_{(n-i)}\right]\\
&=-\frac{R_{ab}\Pi^{ab}_{(n-1)}}{d-2}+\frac{R\Pi_{(n-1)}}{2(d-1)(d-2)}+\frac{1}{2}\sum_{i=2}^{n-2}\left[\Pi_{(i)\, ab}\Pi^{ab}_{(n-i)}-\frac{1}{d-1}\Pi_{(i)}\Pi_{(n-i)}\right] \, .\label{Pin2}
\end{align}
In order to evaluate this expression we need the equations of motion $\Pi_{(n)\, ab}$. Notice that we will compare the resulting expression with \req{PiL2}, which is quadratic in the curvature. Now, \req{Pin2} is already quadratic in the equations of motion, and this means that, in order to obtain the terms that are quadratic in the curvature we only need to obtain the terms in the equations of motion that are linear in the curvature. Fortunately, all of these come from the term
\begin{equation}
-4\nabla^{c}\nabla^{e}P_{acbe}\subset \Pi_{(n)\, ab}\, ,\quad \text{where}\quad P_{acbe}=\frac{\delta \mathcal{L}}{\delta R^{abcd}}\, .
\end{equation}
For a theory that only depends on Ricci curvatures this can be expressed as 
\begin{equation}
\Pi_{(n)\, ab}=-2 g_{ab}\nabla^{c}\nabla^{e} P_{ce}-2\Box P_{ab}+4\nabla^{c}\nabla_{(a}P_{b)c}+\ldots\, ,\quad  \text{where}\quad P_{ab}=\frac{\delta \mathcal{L}}{\delta R^{ab}}\, .
\end{equation}
Thus, for the Lagrangians (\ref{Ln}) we get
\begin{align}
\Pi_{(n)\, ab}&=-(4\alpha_n+\beta_n)g_{ab}\Box^{n-1}R+2(2\alpha_n+\beta_n)\nabla_{a}\nabla_{b}\Box^{n-2}R-2\beta_{n}\Box^{n-1}R_{ab}+\ldots\, ,\\
\Pi_{(n)}&=-\left(4(d-1)\alpha_n+d \beta_n\right)\Box^{n-1}R+\ldots \, 
\end{align}
Then, we can use these expressions to evaluate \req{Pin2}, and after some simplifications we find
\begin{equation}
\Pi_{(n)}=2\left(-\frac{d}{4(d-1)}R\Box^{n-2} R+R^{ab}\Box^{n-2}R_{ab}\right)\left[\frac{\beta_{n-1}}{(d-2)}+\sum_{i=2}^{n-2}\beta_{i}\beta_{n-i}\right]+\ldots\, ,
\end{equation}
where the ellipsis also contain total derivatives that arise when rearranging the derivatives. Therefore, comparing with \req{PiL2}, we conclude that 
\begin{equation}
\alpha_{n}=-\frac{d}{4(d-1)}\beta_{n}\, ,
\end{equation}
while $\beta_{n}$ satisfies the recursive relation 
\begin{equation}
\beta_{n}=\frac{2}{(d-2n)}\left[\frac{\beta_{n-1}}{(d-2)}+\sum_{i=2}^{n-2}\beta_{i}\beta_{n-i}\right]\, .
\end{equation}
We can transform this recursive relation into a differential equation by introducing the generating function 
\begin{equation}\label{smallf}
f(x)=\sum_{n=2}^{\infty}\beta_{n}x^{2n-d}\, .
\end{equation}
By taking the derivative and using the recursive relation for $\beta_{n\ge 3}$, we have 
\begin{equation}
\begin{aligned}
f'(x)&=\sum_{n=2}^{\infty}(2n-d)\beta_{n}x^{2n-d-1}=(4-d)\beta_{2} x^{3-d}+2\sum_{n=3}^{\infty}\left[\frac{\beta_{n-1}}{d-2}+\sum_{i=2}^{n-2}\beta_{i}\beta_{n-i}\right]x^{2n-d-1}\\
&=(4-d)\beta_{2} x^{3-d}-\frac{2}{d-2}x f(x)-2 x^{d-1} f(x)^2\, .
\end{aligned}
\end{equation}
Now, the action can in fact we written in terms of this function. The full action (at quadratic order) reads
\begin{align}\notag
I_{\rm eff}^{(2)}&=\frac{1}{16\pi G_{d+1}}\int  \df^{d}x\sqrt{-g}\left[\frac{L}{2(d-2)}R+\sum_{n=2}^{\infty}\beta_{n}L^{2n-1}\left(R^{ab}\Box^{n-2}R_{ab}-\frac{d}{4(d-1)}R\Box^{n-2} R\right)\right]\\
&=\frac{1}{16\pi G_{d}}\int  \df^{d}x\sqrt{-g}\left[R+L^2 R^{ab}F\left(L^2\Box\right)R_{ab}-\frac{d}{4(d-1)}L^2 RF\left(L^2\Box\right)R\right]\, ,\label{totalaction}
\end{align}
where 
\begin{equation}\label{bigF}
F(L^2\Box)=2(d-2)\sum_{n=2}^{\infty}\beta_{n}\left(L^2\Box\right)^{n-2}\, ,
\end{equation}
and $G_{d}=2(d-2)G_{d+1}/L$. We see that this $F$ is related to $f$ in \req{smallf} by 
\begin{equation}
f(x)=\frac{1}{2(d-2)}x^{4-d}F(x^2)\, .
\end{equation}
Thus, $F(x)$ satisfies the equation
\begin{equation}
F'(x)=(d-4)\frac{F(x)-F(0)}{2x}-\frac{1}{2(d-2)}\left(2F(x)+x F(x)^2\right)\, ,
\end{equation}
where 
\begin{equation}
F(0)=2(d-2)\beta_2=\frac{1}{(d-2)(d-4)}\, .
\end{equation}
Remarkably, this differential equation allows for a general solution in terms of Bessel functions. We find that the appropriate solution, that corresponds to the summation of the series \req{bigF}, is given by 
\begin{equation}
F_{d}(x)=\frac{d(d-2)}{x^2}-\frac{1}{x}-\frac{(d-2)Y_{\frac{d+2}{2}}\left(\sqrt{x}\right)}{x^{3/2}Y_{\frac{d}{2}}\left(\sqrt{x}\right)}\, ,
\end{equation}
where $Y_{k}$ are the Bessel functions of the second kind. Inserting $F_{d}(L^2 \Box)$ in \req{totalaction} we obtain our final expression for the quadratic action of the brane-world theory in general dimensions.

Despite the singular appearance of this function at $x=0$, it is actually analytic around that point for odd $d$. In fact, for odd $d$, $F_{d}$ can actually be written in terms of trigonometric functions. We have
\begin{align}
F_{3}(x)&=-\frac{\sin \left(\sqrt{x}\right)}{x \sin \left(\sqrt{x}\right)+\sqrt{x} \cos \left(\sqrt{x}\right)}\approx -1 +\frac{2x}{3}-\frac{7x^2}{15}+\frac{34x^3}{105}+\dots \, ,\\
F_{5}(x)&=\frac{\cos \left(\sqrt{x}\right)}{(3-x) \cos \left(\sqrt{x}\right)+3 \sqrt{x} \sin \left(\sqrt{x}\right)}\approx \frac{1}{3}-\frac{2x}{9}+\frac{x^2}{27}+\frac{2x^3}{405}+\dots\, ,\\
F_{7}(x)&=-\frac{\sqrt{x} \sin \left(\sqrt{x}\right)+\cos \left(\sqrt{x}\right)}{(x-15) \sqrt{x} \sin \left(\sqrt{x}\right)+3 (2 x-5) \cos \left(\sqrt{x}\right)} \approx \frac{1}{15}+\frac{2x}{75}-\frac{13x^2}{1125}+\frac{22x^3}{16875}+\dots\, ,
\end{align}
where we included the first terms in the expansions around $x=0$.
On the other hand, in even $d\geq 4$, the expansion around $x=0$ contains logarithmic divergences, which are the counterpart of the $1/(d-2n)$ divergences in the definition of these theories. For instance, for $d=4$ one finds
\begin{equation}
F_4(x)=\frac{8}{x^2}-\frac{1}{x}-\frac{2 Y_{3}(\sqrt{x})}{x^{3/2} Y_2(\sqrt{x})}\approx \frac{1}{4}\left[-2\gamma_{\rm E}-\log(x/4) \right]+\frac{1}{8}\left[ -1+\gamma_{\rm E}+\log(x/4)\right]x + \dots \,,
\end{equation}
where $\gamma_{\rm E}$ is the Euler-Mascheroni constant. Finally, the $d=2$ case is a bit different, as it simply yields 
\begin{equation}
F_2(x)=-\frac{1}{x}\, ,
\end{equation}
which means that the corresponding quadratic action is proportional to the Polyakov induced-gravity action \cite{Polyakov:1987zb} ---see also \cite{Chen:2020uac}.


\section{Linearized Equations and Modes}
It is obvious from \req{totalaction} that the brane-world theory belongs to the class of theories which satisfy the no-scalar condition (\ref{keyko}), as in this case we have $F_1=F$, $F_2=-d/(4(d-1))F$, $F_3=0$. As a consequence, the linearized equations of the theory impose the condition (\ref{oop}), namely,
\begin{equation}
-\frac{(d-2)}{64\pi G} R^{(1)}=0\, ,
\end{equation}
so the trace of the equation has no dynamics and we are left with
\begin{equation}
\frac{1}{32\pi G} \left[1+ F(L^2 \bar \Box) L^2 \bar \Box \right] G_{ab}^{(1)}=0\, .
\end{equation}
By going to the Lorentz gauge as in Section \ref{minko}, one finds
\begin{equation}
-\frac{1}{64\pi G}\left[1+ F\left(L^2\bar \Box\right) L^2 \bar \Box\right]\bar \Box h_{\langle ab \rangle}\,,
\end{equation}
and the corresponding propagator is given by
\begin{equation}\displaystyle
P_{d}(k)=\frac{64\pi G_{d}}{(d-2) } \left[\frac{i L k Y_{\frac{d+2}{2}}(i L k)}{Y_{\frac{d}{2}}(i L k)}-d\right]^{-1}\, .
\end{equation}
Using this we can analyze the pole structure in various dimensions.

\subsection*{Three dimensions}
In $d=3$ the propagator becomes
\begin{align}
\frac{P_{3}(k)}{64\pi G_3}=\frac{1}{L^2 k^2}-\frac{L k \tanh (L k)}{L^2 k^2}\, .
\end{align}
Studying its pole structure we find a massless mode as well as an infinite tower of massive gravitons. The massless mode is the same as the one appearing in the pure Einstein gravity spectrum and it is pure gauge in three dimensions. On the other hand, the massive gravitons have masses
\begin{equation}
 m_{n}=\frac{\pi}{2L }(2n-1)\, ,\quad n= 1,2, \ldots\, ,
\end{equation}
and all of them have negative kinetic energy. This can be seen by expanding the propagator around each of the poles and comparing the overall sign with the one of the positive-energy would-be massless mode. For this, one has
\begin{align}
\frac{P_{3}(k^2\rightarrow 0)}{64\pi G_3}=\frac{1}{L^2 k^2}+\mathcal{O}(1) \, .
\end{align}
For the new modes one finds, instead,
\begin{align}
\frac{P_{3}(k^2\rightarrow -m_n^2  )}{64\pi G_3}=-\frac{2}{L^2 [k^2+m_n^2]}+\mathcal{O}(1) \, .
\end{align}
Hence, all the new modes are ghosts. 

\subsection*{Four dimensions}
In $d=4$, the analysis of the propagator becomes more cumbersome. To begin with, there is no simplified way to write down the propagator in terms of trigonometric functions. Instead, are left with
\begin{equation}
\frac{P_{4}(k)}{64 \pi G_4}= \frac{i Y_{2}(i L k)}{2 L k Y_{1}(i L k)}\,.
\end{equation}
 Again, we find the Einstein-like massless graviton and an infinite tower of massive ghost gravitons, with masses
 \begin{equation}
 m_n\approx \frac{\pi}{L} (0.69937, 1.72832, 2.73619, 3.73987, 4.742, 5.74339, 6.74437, 7.7451,\dots)\,.
 \end{equation}
In this case, the masses are not equispaced, but the difference between pairs of modes tends to $\pi/L$ as $n \rightarrow \infty$. Indeed, the $m_n$ tend to $\frac{\pi}{L} \left( n - 1/4 \right)$ as $n \rightarrow \infty$. 
Moreover, we now find a tower of modes with complex squared masses which are conjugate of each other,
 \begin{equation}
 m_{n, \rm \pm} \approx \frac{\pi}{L} (\pm 0.1790 + 1.220 i, \pm 0.1762 + 2.233 i, \pm 0.1755 + 3.238 i, \cdots) \,.
 \end{equation}
These tend to $\frac{\pi}{L} (\pm 0.17485 + (n+1/4) i)$ as $n \rightarrow \infty$.
Again we find that all massive modes, including the complex ones, have negative kinetic energy, namely,
 \begin{align}
\frac{P_{4}(k^2\rightarrow -m_j^2  )}{64\pi G_4}=-\frac{1}{L^2 [k^2+m_j^2]}+\mathcal{O}(1) \, ,
\end{align}
$\forall j \in \{n, \pm \}$ so again they are all ghosts. 

\subsection*{Five dimensions}
In $d=5$ one finds
\begin{equation}
\frac{P_{5}(k)}{64\pi G_5}=\frac{1}{L^2 k^2}+\frac{1}{3-3 L k \tanh (L k)}\, ,
\end{equation}
In addition to the Einstein-like massless graviton, we again find an infinite tower of massive gravitons with masses
 \begin{equation}
 m_n\approx \frac{\pi}{L} (0.89075, 1.9485, 2.9660, 3.9746, 4.9797, 5.9831, 6.9855,\dots)\,.
 \end{equation}
Now, however, there is only one tachyonic mode with imaginary mass
 \begin{equation}
 m^2_{\rm t}\approx -\frac{1.43923}{L^2}\, .
 \end{equation}
Once again, we find that all the massive modes have negative kinetic energy, namely,
 \begin{align}
\frac{P_{5}(k^2\rightarrow -m_j^2  )}{64\pi G_5}=-\frac{2}{3L^2 [k^2+m_j^2]}+\mathcal{O}(1) \, ,
\end{align}
 $\forall j\in \{n, {\rm t}\}$, so they are all ghosts.

\subsection*{Six dimensions}
The case of $d=6$ is similar to the four-dimensional case. The propagator reads
\begin{equation}
\frac{P_{6}(k)}{64 \pi G_6}= \frac{i Y_{3}(i L k)}{4 L k Y_{2}(i L k)}\,,
\end{equation}
  and again, we find the Einstein-like massless graviton, an infinite tower of massive ghost gravitons, with masses
 \begin{equation}
 m_n\approx \frac{\pi}{L} (1.077, 2.163, 3.191, 4.205, 5.214, 6.220, 7.224,\dots)\,,
 \end{equation}
which tend to $\frac{\pi}{L} \left( n + 1/4 \right)$ as $n \rightarrow \infty$; and a tower of modes with complex squared masses which are conjugate of each other,
 \begin{equation}
 m_{n, \rm \pm} \approx \frac{\pi}{L} (\pm 0.3382 + 0.4711 i, \pm 0.1877 + 1.636 i, \pm 0.1795 + 2.680 i, \pm 0.1773 + 
 3.699 i, \cdots) \,.
 \end{equation}
These tend to $\frac{\pi}{L} (\pm 0.17485 + (n-1/4) i)$ as $n \rightarrow \infty$. Moreover, we find an extra conjugate pair,
 \begin{equation}
 m_{0, \rm \pm} \approx \frac{\pi}{L} \pm 0.4716 - 0.1503 i \,.
 \end{equation}
As before, all massive modes are ghosts, including the complex ones, since 
 \begin{align}
\frac{P_{6}(k^2\rightarrow -m_j^2  )}{64\pi G_6}=-\frac{1}{2 L^2 [k^2+m_j^2]}+\mathcal{O}(1) \,,
\end{align}
$\forall j \in \{n, \pm \}$, so they all have negative kinetic energy.

\subsection*{Seven dimensions}
Finally, in $d=7$, one finds
\begin{equation}
\frac{P_{7}(k)}{64 \pi G_7}=\frac{1}{15} + \frac{1}{L^2 k^2} - \frac{L^2 k^2}{15(3 + L^2 k^2 - 3 L k \tanh (L k))} \, ,
\end{equation}
 Again, we find the Einstein-like massless graviton, and an infinite tower of massive ghost gravitons with masses
 \begin{equation}
 m_n\approx \frac{\pi}{L} (1.2604, 2.3719, 3.4109,4.4314, 5.4442,6.4529,7.4593,\dots) \,,
 \end{equation}
 with the difference between pairs of modes tending to $\pi/L$ as $n \rightarrow \infty$. Now, there are only two extra modes with complex squared masses which are conjugate of each other, namely,
 \begin{equation}
 m^2_{\rm \pm}\approx -\frac{2.01933 \pm 3.19512 i}{L^2}\, .
 \end{equation} 
Once more, we find that all the massive modes, including the ones with complex squared-masses, have negative kinetic energy, namely,
 \begin{align}
\frac{P_{7}(k^2\rightarrow -m_j^2  )}{64\pi G_7}=-\frac{2}{5L^2 [k^2+m_j^2]}+\mathcal{O}(1) \, ,
\end{align}
$\forall j \in \{n, \pm \}$ so they are all ghosts.  


\section{Conclusions}

We have found that, regardless of the number of dimensions, there are always pathological modes appearing in the linearized spectrum of these brane-world gravities, with squared masses of order $\sim 1/L^2$.
Since the bulk theory is Einstein gravity, which is perfectly well-defined, the appearance of these pathological modes on the gravitational effective theory induced on the brane might seem worrisome at first. The bulk, however, is dual to this induced theory on the brane plus a cut-off CFT, which we have neglected in this analysis. The CFT cut-off is precisely $\sim 1/L^2$, and so it is not surprising that pathologies might appear at this order.

Indeed, when one takes the coupling between this cut-off CFT and the induced gravity on the brane into account, the observed pathologies disappear, as we saw in Chapter \ref{chp:ReviewBWs} from the perspective of the bulk. In a sense, coupling the induced action to the cut-off CFT allows one to ``UV-complete'' the theory by making it dual to the perfectly defined Einstein gravity in the bulk. 
It would be interesting, however, to perform this computation directly, without having to explain it through the dual bulk picture.


\cleardoublepage
\part{Final Remarks}\label{part:FinalRemarks}
\renewcommand{\headrulewidth}{0pt}

\lhead{}
\rhead{}

\chapter*{Conclusions}
\label{chp:Conclusions}
\addcontentsline{toc}{chapter}{Conclusions}

A detailed summary of the results in this thesis can be found at the end of the Introduction and at the end of each chapter, so let us just highlight the most important ones, and the connections between them.


In Chapter \ref{chp:3DHCGs}, we presented the equations of motion for general higher-curvature gravities in three-dimensional spacetimes, and we fully characterized their linear spectrum on maximally symmetric spacetimes. We also identified all three-dimensional higher-curvature gravities which satisfy a holographic c-theorem. Then, in Chapter \ref{chp:HDGs}, we studied the structure of the linearized equations of general higher-derivative gravities on maximally symmetric spacetimes of arbitrary dimension, and described the spectrum of gravitational perturbations around flat space. 


In Chapter \ref{chp:ReviewBWs}, we gave a review on brane-worlds, and generalized and expanded the previously known results, both from the bulk perspective and the dual brane perspective. In Chapter \ref{chp:BWsWithDGP}, we then added a DGP term on the brane, and put bounds on the allowed values for its coupling.


In Chapter \ref{chp:Alg}, we explicitly computed the first few terms of the higher-derivative theory of gravity that is induced on the brane, which we then studied in detail in the following two chapters, neglecting its coupling to the cut-off CFT on the brane. In Chapter \ref{chp:cTheorem}, we proved that, at each curvature order, the terms of this theory fulfil a holographic c-theorem. Lastly, in Chapter \ref{chp:Spectrum}, we studied the linearization of the induced gravity theory around flat space, while accounting for the full tower of higher-derivative terms. Besides the usual massless graviton, we found that there is always a tower of massive spin-2 ghosts.


Throughout this thesis, combining the study of higher-derivative theories of gravity with that of brane-world models, we have investigated the properties of the theory of gravity that is induced on the brane, which in turn reflect the UV structure of holographic CFTs.
Although the theory inherits some good properties from the well-defined Einstein bulk, such as fulfilling a holographic c-theorem, we have seen that it is not fully free of pathologies, presenting ghosts in the spectrum of linearized metric perturbations around flat space ---if one ignores its coupling to the holographic CFT on the brane.
Therefore, we have shown that brane-world holography is an interesting way of generating an appealing class of higher-derivative theories, but that one should be careful when considering them on their own, disregarding their coupling to the brane CFT and brane-world holographic origin.
Moreover, the results in this thesis allow us to better understand brane-world holography, and we hope that it will help expand and clarify their use as models for semiclassical gravity within AdS/CFT.


\thispagestyle{empty}
\cleardoublepage
\thispagestyle{empty}


\lhead{}
\rhead{}

\chapter*{Future Directions}
\label{chp:Future}

\addcontentsline{toc}{chapter}{Future Directions}

As it is often the case with scientific research, this thesis brings more questions than answers.
A first, natural question to ask is whether the induced brane-world gravity theory fulfils any other remarkable qualities, which could tell us more about the properties of brane-world holography or the UV structure of holographic CFTs.
Perhaps a rewriting of the algorithm generating the higher-derivative terms in the induced action, in terms of the Schouten tensor, could reveal new, undiscovered, structures.
For example, we saw that all theories that fulfil a holographic c-theorem in three-dimensions propagate no scalar mode when perturbed around maximally symmetric spacetimes, and we showed that the theory of induced gravity on the brane fulfils a holographic c-theorem and does not propagate a scalar mode when linearized around flat space.
An obvious next step would be proving that indeed the induced gravity on the brane propagates no scalar mode on all maximally symmetric spacetimes.
It would be even more interesting to prove that all  higher-derivative theories which fulfil a holographic c-theorem have no scalar mode in their spectrum on maximally symmetric spacetimes.

On another front, one can think of extending the brane-world constructions in different directions. One possibility would consist on studying brane-world holography with higher order operators on the brane beyond the DGP term, as we started doing at the end of Chapter \ref{chp:BWsWithDGP}. 
But one could also consider brane-world holographic models in which the bulk theory is not Einstein gravity but a higher-curvature theory of gravity.
A different route would be considering alternative boundary conditions on the brane, such as conformal boundary conditions.
Lastly, we still intend to further study brane-world models with a DGP term in the limit in which the brane is sent to the boundary, in order to understand models of AdS/CFT with dynamical boundary.

Finally, it would be interesting to extend the original holographic renormalization computation including the terms that do not diverge as the cut-off is removed.
Then, keeping the cut-off finite, we would clearly understand the separation between matter and geometric degrees of freedom in brane-world holographic models, and we would have an expression for the cut-off CFT on the brane in terms of the bulk metric in Fefferman-Graham coordinates.
These results could then be used to double-check the known properties of C-metric brane black holes interacting with strongly coupled CFTs.


\thispagestyle{empty}

\cleardoublepage
\part*{Appendices}\label{part:Appendices}
\addcontentsline{toc}{part}{Appendices}
\begin{appendices}

\renewcommand{\headrulewidth}{1pt}


\lhead{Appendix A}
\rhead{Basis of HDG Invariants}

\chapter{Basis of higher-derivative invariants}
\label{chp:App-basis}


We present here a complete list of the curvature invariants at each order in derivatives. The same list can be found in~\cite{Fulling:1992vm}. Our ordering also follows~\cite{Fulling:1992vm}: The invariants are ordered by the number of covariant derivatives acting on individual curvature tensors. We begin with those invariants that involve the largest number of derivatives acting on curvature, and end with the polynomial curvature invariants (those built exclusively from contractions of the Riemann tensor). 

\section*{Four derivatives}
There are four possible terms involving four derivatives of the metric:
\small{
\begin{align}
    \mathcal{R}_4^{(1)} &= \Box R \, , 
    \quad 
    \mathcal{R}_4^{(2)} = R^2 \, ,
    \quad 
    \mathcal{R}_4^{(3)} = R^{pq} R_{pq} \, ,
    \quad
    \mathcal{R}_4^{(4)} = R^{pqrs} R_{pqrs} \, .
\end{align}
}

\section*{Six derivatives}
There are 17 terms involving six derivatives of the metric:
\small{
\begin{align}
    \mathcal{R}_{6}^{(1)} &= \Box R^2 \, ,
    \quad
    \mathcal{R}_{6}^{(2)} = R \Box R \, ,
    \quad
    \mathcal{R}_{6}^{(3)} = R^{;pq} R_{;pq} \, ,
    \quad 
    \mathcal{R}_{6}^{(4)} = R^{pq} \Box R_{pq} \, ,
    \quad 
    \mathcal{R}_{6}^{(5)} = R^{pq;rs} R_{pqrs} \, ,
    \nonumber\\
    \mathcal{R}_{6}^{(6)} &= R^{;p} R_{;p} \, ,
    \quad 
    \mathcal{R}_{6}^{(7)} = R^{pq;r} R_{pq;r} \, ,
    \quad
    \mathcal{R}_{6}^{(8)} = R^{pq;r} R_{pr;q} \, ,
    \quad 
    \mathcal{R}_{6}^{(9)} = R^{pqrs;t} R_{pqrs;t} \, ,
    \quad 
    \mathcal{R}_{6}^{(10)} = R^3 \, ,
    \nonumber\\
    \mathcal{R}_{6}^{(11)} &= R R^{pq} R_{pq} \,,
    \quad
    \mathcal{R}_{6}^{(12)} = R^{pq} \tensor{R}{_p^r} R_{qr} \, ,
    \quad 
    \mathcal{R}_{6}^{(13)} = R^{pq} R^{rs} R_{prqs} \, ,
    \quad
    \mathcal{R}_{6}^{(14)} = R R^{pqrs} R_{pqrs} \, ,
    \nonumber\\
    \mathcal{R}_{6}^{(15)} &= R^{pq} \tensor{R}{^{rst}_p} R_{rstq} \, ,
    \quad
    \mathcal{R}_{6}^{(16)} = R^{pqrs} \tensor{R}{_{pq}^{tu}} R_{rstu} \, ,
    \quad
    \mathcal{R}_{6}^{(17)} = R^{pqrs} \tensor{R}{_p^t_r^u} R_{qtsu} \, .
\end{align}
}

\section*{Eight derivatives}
There are 92 terms involving eight derivatives of the metric:
\small{
\begin{align}
    \mathcal{R}_{8}^{(1)} &= \Box^3 R \, ,
    \quad
    \mathcal{R}_{8}^{(2)} = R \Box^2 R \, ,
    \quad
    \mathcal{R}_{8}^{(3)} = R_{pq} \Box\tensor{R}{^{;pq}} \,,
    \quad
    \mathcal{R}_{8}^{(4)} = R^{pq} \Box^2 \tensor{R}{_{pq}} \, ,
    \quad
    \mathcal{R}_{8}^{(5)} = \tensor{R}{^{pq;rs}} \tensor{R}{_{prqs}}
    \, ,
    \nonumber\\
    \mathcal{R}_{8}^{(6)} &= \tensor{R}{^{;p}} \Box \tensor{R}{_{;p}} \, ,
    \quad
    \mathcal{R}_{8}^{(7)} = \tensor{R}{^{pq;r}} \tensor{R}{_{pq;r}} \, ,
    \quad
    \mathcal{R}_{8}^{(8)} = \tensor{R}{^{pq;r}} \Box \tensor{R}{_{pq;r}} \, ,
    \quad
    \mathcal{R}_{8}^{(9)} = \tensor{R}{^{pq;r}} \Box \tensor{R}{_{pr;q}} \, ,
    \nonumber\\
    \mathcal{R}_{8}^{(10)} &= \tensor{R}{^{pq;rst}} \tensor{R}{_{prqs;t}} \, ,
    \quad
    \mathcal{R}_{8}^{(11)} = \left( \Box R \right)^2 \, ,
    \quad
    \mathcal{R}_{8}^{(12)} = \tensor{R}{^{;pq}} \tensor{R}{_{;pq}} \, ,
    \quad
    \mathcal{R}_{8}^{(13)} = \tensor{R}{^{;pq}} \Box \tensor{R}{_{pq}} \, ,
    \nonumber\\
    \mathcal{R}_{8}^{(14)} &= \Box R^{pq} \Box R_{pq} \, , 
    \quad
    \mathcal{R}_{8}^{(15)} = \tensor{R}{^{pq;rs}} \tensor{R}{_{pq;rs}} \, ,
    \quad
    \mathcal{R}_{8}^{(16)} = \tensor{R}{^{pq;rs}} \tensor{R}{_{pr;qs}} \, ,
    \quad
    \mathcal{R}_{8}^{(17)} = \tensor{R}{^{pq;rs}} \tensor{R}{_{rs;pq}} \, ,
    \nonumber\\
    \mathcal{R}_{8}^{(18)} &= \tensor{R}{^{pqrs;tu}} \tensor{R}{_{pqrs;tu}} \, ,
    \quad
    \mathcal{R}_{8}^{(19)} = R^2 \Box R \, , 
    \quad 
    \mathcal{R}_{8}^{(20)} = R \tensor{R}{^{;pq}} R_{pq} \, ,
    \quad 
    \mathcal{R}_{8}^{(21)} = \Box R R^{pq} R_{pq} \, ,
    \nonumber\\
    \mathcal{R}_{8}^{(22)} &= R R^{pq} \Box R_{pq} \, ,
    \quad
    \mathcal{R}_{8}^{(23)} = R^{;pq} \tensor{R}{_p^r} R_{qr} \, ,
    \quad
    \mathcal{R}_{8}^{(24)} = R^{pq}  \tensor{R}{_p^r} \Box R_{qr} \, ,
    \quad 
    \mathcal{R}_{8}^{(25)} = R^{pq} R^{rs} R_{pq;rs} \, ,
    \nonumber\\
    \mathcal{R}_{8}^{(26)} &= R^{pq} R^{rs} R_{pr;qs} \, ,
    \quad
    \mathcal{R}_{8}^{(27)} = R^{;pq} R^{rs} R_{prqs} \, ,
    \quad 
    \mathcal{R}_{8}^{(28)} = R R^{pq;rs} R_{prqs} \, ,
    \quad 
    \mathcal{R}_{8}^{(29)} = R^{pq} \Box R^{rs} R_{prqs} \, ,
    \nonumber\\
    \mathcal{R}_{8}^{(30)} &= R^{pq} \tensor{R}{_p^{r;st}} R_{qsrt} \, ,
    \quad 
    \mathcal{R}_{8}^{(31)} = R^{pq} \tensor{R}{^{rs}_{;q}^t} R_{prst} \, ,
    \quad 
    \mathcal{R}_{8}^{(32)} = \Box R R^{pqrs} R_{pqrs} \, ,
    \quad 
    \mathcal{R}_{8}^{(33)} = R^{;pq} \tensor{R}{^{rst}_q} R_{rstp} \, ,
    \nonumber\\
    \mathcal{R}_{8}^{(34)} &= \Box R^{pq} \tensor{R}{^{rst}_p} R_{rstq} \, ,
    \quad 
    \mathcal{R}_{8}^{(35)} = R^{pq;rs} \tensor{R}{^{tu}_{pr}} R_{tuqs} \, ,
    \quad
    \mathcal{R}_{8}^{(36)} = R^{;pqrs} \tensor{R}{^t_p^u_q}R_{trus} \, , 
    \nonumber\\
    \mathcal{R}_{8}^{(37)} &= R^{pq;rs} \tensor{R}{^t_p^u_r} R_{tqus} \, ,
    \quad
    \mathcal{R}_{8}^{(38)} = R^{pq} R^{rstu} R_{rstu;pq} \, ,
    \quad
    \mathcal{R}_{8}^{(39)} = R^{pqrs} \tensor{R}{_p^{tuv}} R_{qtru;sv} \, ,
    \nonumber\\
    \mathcal{R}_{8}^{(40)} &= R R^{;p} R_{;p} \, , 
    \quad 
    \mathcal{R}_{8}^{(41)} = R^{;p} R^{;q} R_{pq} \, ,
    \quad
    \mathcal{R}_{8}^{(42)} = R R^{pq;r} R_{pq;r} \, ,
    \quad
    \mathcal{R}_{8}^{(43)} = R R^{pq;r} R_{pr;q} \, ,
    \nonumber\\
    \mathcal{R}_{8}^{(44)} &= R^{;p} R^{qr} R_{qr;p} \, ,
    \quad
    \mathcal{R}_{8}^{(45)} = R^{;p} R^{qr} R_{pq;r} \, ,
    \quad 
    \mathcal{R}_{8}^{(46)} = R^{pq} \tensor{R}{_p^{r;s}} R_{qr;s} \,,
    \quad 
    \mathcal{R}_{8}^{(47)} = R^{pq} \tensor{R}{_p^{r;s}} R_{qs;r} \, ,
    \nonumber\\
    \mathcal{R}_{8}^{(48)} &=  R{^pq} \tensor{R}{^{rs}_{;p}} R_{rs;q} \, , 
    \quad 
    \mathcal{R}_{8}^{(49)} = R^{pq} \tensor{R}{^{rs}_{;p}} R_{rq;s} \, , 
    \quad
    \mathcal{R}_{8}^{(50)} = R^{;p} R^{qr;s} R_{pqrs} \, ,
    \nonumber\\
    \mathcal{R}_{8}^{(51)} &=   R^{pq;r} \tensor{R}{_p^{s;t}} R_{qrst} \, ,
    \quad
    \mathcal{R}_{8}^{(52)} = R^{pq;r} \tensor{R}{_p^{s;t}} R_{qsrt} \, ,
    \quad
    \mathcal{R}_{8}^{(53)} = R^{qr;p} \tensor{R}{^{st}_{;p}} R_{qsrt} \,,
    \nonumber\\
    \mathcal{R}_{8}^{(54)} &= R^{pq;r} \tensor{R}{^{st}_{;p}} \tensor{R}{_{qsrt}}  \, , 
    \quad
    \mathcal{R}_{8}^{(55)} =  R^{pq} R^{rs;t} R_{prqs;t} \, ,
    \quad
    \mathcal{R}_{8}^{(56)} =  R^{pq}R^{rs;t} R_{rtsp;q} \, ,
    \nonumber\\
    \mathcal{R}_{8}^{(57)} &=  R^{;p} R^{qrst} R_{qrst;p} \, ,
    \quad
    \mathcal{R}_{8}^{(58)} = R R^{pqrs;t} R_{pqrs;t} \, ,
    \quad 
    \mathcal{R}_{8}^{(59)} = R^{pq} \tensor{R}{^{rstu}_{;p}} R_{rstu;q} \, ,
    \nonumber\\
    \mathcal{R}_{8}^{(60)} &= R^{pq} \tensor{R}{^{rstu}_{;p}} R_{rstq;u} \, ,
    \quad
    \mathcal{R}_{8}^{(61)} = R^{pq;r} \tensor{R}{^{st}_r^u} R_{stpq;u} \, ,
    \quad
    \mathcal{R}_{8}^{(62)} = R^{pq;r} \tensor{R}{^{st}_p^u} R_{stqr;u} \, ,
    \nonumber\\
    \mathcal{R}_{8}^{(63)} &= R^{pq;r} \tensor{R}{^s_p^{tu}} R_{sqtr;u} \, ,
    \quad
    \mathcal{R}_{8}^{(64)} = R^{pqrs} \tensor{R}{^{tuv}_{p;q}} R_{tuvr;s} \, ,
    \quad
    \mathcal{R}_{8}^{(65)} = R^{pqrs} \tensor{R}{^{tuv}_{p;r}} R_{tuvq;s} \, ,
    \nonumber\\
    \mathcal{R}_{8}^{(66)} &= R^{pqrs} \tensor{R}{^t_p^u_r^{;v}} R_{tqus;v} \, ,
    \quad 
    \mathcal{R}_{8}^{(67)} = R^4 \, ,
    \quad 
    \mathcal{R}_{8}^{(68)} = R^2 R^{pq} R_{pq} \, ,
    \quad
    \mathcal{R}_{8}^{(69)} = R R^{pq} \tensor{R}{_p^r} R_{qr} \, ,
    \nonumber\\
    \mathcal{R}_{8}^{(70)} &= \left( R^{pq} R_{pq} \right)^2 \, ,
    \quad
    \mathcal{R}_{8}^{(71)} = R^{pq} \tensor{R}{_p^r} \tensor{R}{_q^s} R_{rs} \, ,
    \quad
    \mathcal{R}_{8}^{(72)} = R R^{pq} R^{rs} R_{prqs} \, ,
    \quad
    \mathcal{R}_{8}^{(73)} = R^{pq} R^{rs} \tensor{R}{_r^t} R_{psqt} \, ,
    \nonumber\\
    \mathcal{R}_{8}^{(74)} &= R^2 R^{pqrs} R_{pqrs} \, ,
    \quad
    \mathcal{R}_{8}^{(75)} = R R^{pq} \tensor{R}{^{rst}_p} R_{rstq} \, ,
    \quad
    \mathcal{R}_{8}^{(76)} = R^{pq} R_{pq} R^{rstu} R_{rstu} \, ,
    \nonumber\\
    \mathcal{R}_{8}^{(77)} &= R^{pq} \tensor{R}{_p^r} \tensor{R}{^{stu}_q} R_{stur} \, ,
    \quad 
    \mathcal{R}_{8}^{(78)} = R^{pq} R^{rs} \tensor{R}{^{tu}_{pr}} R_{tuqs} \, , 
    \quad
    \mathcal{R}_{8}^{(79)} = R^{pq} R^{rs} \tensor{R}{^t_p^u_q} R_{trus} \, ,
    \nonumber\\
    \mathcal{R}_{8}^{(80)} &= R^{pq} R^{rs} \tensor{R}{^t_p^u_r} R_{tqus} \, ,
    \quad
    \mathcal{R}_{8}^{(81)} = R R^{psrs} \tensor{R}{_{pq}^{tu}} R_{rstu} \, ,
    \quad
    \mathcal{R}_{8}^{(82)} = R R^{pqrs} \tensor{R}{_p^t_r^u} R_{qtsu} \, ,
    \nonumber\\
    \mathcal{R}_{8}^{(83)} &= R^{pq} \tensor{R}{_p^r_q^s} \tensor{R}{^{tuv}_r} \tensor{R}{_{tuvs}} \, ,
    \quad
    \mathcal{R}_{8}^{(84)} = R^{pq} R^{rstu} \tensor{R}{_{rs}^v_p}R_{tuvq} \, ,
    \quad 
    \mathcal{R}_{8}^{(85)} = R^{pq} R^{rstu} \tensor{R}{_r^v_{tp}} R_{svuq} \, ,
    \nonumber\\
    \mathcal{R}_{8}^{(86)} &= \left( R^{pqrs} R_{pqrs} \right)^2 \, ,
    \quad 
    \mathcal{R}_{8}^{(87)} = R^{pqrs} \tensor{R}{_{pq}^{tu}} \tensor{R}{_{tu}^{vw}} R_{rsvw} \, ,
    \quad
    \mathcal{R}_{8}^{(88)} = R^{pqrs} \tensor{R}{_{pq}^{tu}} \tensor{R}{_{tu}^{vw}} R_{rsvw} \, ,
    \nonumber\\
    \mathcal{R}_{8}^{(89)} &= R^{pqrs} \tensor{R}{_{pq}^{tu}} \tensor{R}{_{rt}^{vw}} R_{suvw} \, ,
    \quad
    \mathcal{R}_{8}^{(90)} = R^{pqrs} \tensor{R}{_{pq}^{tu}} \tensor{R}{_r^v_t^w} R_{svuw} \, ,
    \quad 
    \mathcal{R}_{8}^{(91)} = R^{pqrs} \tensor{R}{_p^t_r^u} \tensor{R}{_t^v_u^w} R_{qvsw} \, ,
    \nonumber\\
    \mathcal{R}_{8}^{(92)} &= R^{pqrs} \tensor{R}{_p^t_r^u} \tensor{R}{_t^v_q^w} R_{uvsw}  \, .
\end{align}
}

\section*{Ten derivatives}
The number of independent invariants grows rapidly with an increasing number of derivatives.
To the best of our knowledge, a complete classification of terms involving more than eight-derivatives of the metric has not been completed. However, for example, at ten-derivative order it is known that there are 668 invariants. The set of ten-derivative invariants we have used consists of $180 = 20 + 92 + 4\times 17$ elements, and so it is necessarily \textit{very} incomplete. Out of the $180$ densities that we use, only 20 are not built from products of lower-order densities. These are
\small{
\begin{align}
    \nonumber
    \mathcal{R}_{10}^{(1)} &= \tensor{C}{_{abcd}} \tensor{C}{^{abcd}}  \tensor{C}{^{efgh;i}} \tensor{C}{_{efgh;i}}\, 
    \nonumber\\
    \mathcal{R}_{10}^{(2)} &= R_b^a R_d^b R_f^c \tensor{R}{_{ag}^{de}} \tensor{R}{_{ce}^{fg}} \, ,
    \quad
    \mathcal{R}_{10}^{(3)}  = R_b^a R_d^b R_f^c \tensor{R}{_{cg}^{de}} \tensor{R}{_{ae}^{fg}} \, ,
    \quad 
    \mathcal{R}_{10}^{(4)}  = R_b^a R_c ^b \tensor{R}{_{ae}^{cd}} \tensor{R}{_{gh}^{ef}} \tensor{R}{_{df}^{gh}} \, ,
    \nonumber
    \\
    \mathcal{R}_{10}^{(5)} &= R_b^a R_c^b \tensor{R}{_{ef}^{cd}} \tensor{R}{_{gh}^{ef}} \tensor{R}{_{ad}^{gh}} \, ,
    \quad
    \mathcal{R}_{10}^{(6)} = R_b^a R_c^b \tensor{R}{_{eg}^{cd}} \tensor{R}{_{ah}^{ef}} \tensor{R}{_{df}^{gh}} \, ,
    \quad
    \mathcal{R}_{10}^{(7)} = R_c^a R_d^b \tensor{R}{_{ab}^{cd}}\tensor{R}{_{gh}^{ef}}\tensor{R}{_{ef}^{gh}} \, ,
    \nonumber\\
    \mathcal{R}_{10}^{(8)} &= R_c^a R_d^b \tensor{R}{_{ae}^{cd}}\tensor{R}{_{gh}^{ef}}\tensor{R}{_{bf}^{gh}} \, ,
    \quad 
    \mathcal{R}_{10}^{(9)} = R_c^a R_d ^b \tensor{R}{_{ef}^{cd}}\tensor{R}{_{gh}^{ef}} \tensor{R}{_{ab}^{gh}} \, ,
    \quad 
    \mathcal{R}_{10}^{(10)} = R_c^a R_d^b \tensor{R}{_{eg}^{cd}} \tensor{R}{_{ah}^{ef}} \tensor{R}{_{bf}^{gh}} \, ,
    \nonumber\\
    \mathcal{R}_{10}^{(11)} &= R_c^a R_e^b \tensor{R}{_{af}^{cd}} \tensor{R}{_{gh}^{ef}} \tensor{R}{_{bd}^{gh}} \, ,
    \quad
    \mathcal{R}_{10}^{(12)} = R_b^a \tensor{R}{_{ad}^{bc}} \tensor{R}{_{fh}^{de}}\tensor{R}{_{c i}^{fg}} \tensor{R}{_{eg}^{hi}} \, ,
    \quad
    \mathcal{R}_{10}^{(13)} = R_b^a \tensor{R}{_{de}^{bc}}\tensor{R}{_{cf}^{de}} \tensor{R}{_{hi}^{fg}} \tensor{R}{_{ag}^{hi}} \, ,
    \nonumber\\
    \mathcal{R}_{10}^{(14)} &= R_b^a \tensor{R}{_{df}^{bc}} \tensor{R}{_{ac}^{de}} \tensor{R}{_{hi}^{fg}} \tensor{R}{_{eg}^{hi}} \, ,
    \quad 
    \mathcal{R}_{10}^{(15)} = R_b^a \tensor{R}{_{df}^{bc}}\tensor{R}{_{ah}^{de}}\tensor{R}{_{ei}^{fg}}\tensor{R}{_{cg}^{hi}} \, ,
    \quad 
    \mathcal{R}_{10}^{(16)} = R_b^a \tensor{R}{_{df}^{bc}} \tensor{R}{_{gh}^{de}} \tensor{R}{_{ei}^{fg}} \tensor{R}{_{ac}^{hi}} \, ,
    \nonumber\\
    \mathcal{R}_{10}^{(17)} &= \tensor{R}{_{cd}^{ab}} \tensor{R}{_{eg}^{cd}} \tensor{R}{_{ai}^{ef}} \tensor{R}{_{fj}^{gh}} \tensor{R}{_{bh}^{ij}} \, ,
    \quad
    \mathcal{R}_{10}^{(18)} = \tensor{R}{_{ce}^{ab}} \tensor{R}{_{af}^{cd}} \tensor{R}{_{gi}^{ef}} \tensor{R}{_{bj}^{gh}} \tensor{R}{_{dh}^{ij}} \, ,
    \nonumber\\
    \mathcal{R}_{10}^{(19)} &= \tensor{R}{_{ce}^{ab}} \tensor{R}{_{ag}^{cd}} \tensor{R}{_{bi}^{ef}} \tensor{R}{_{fj}^{gh}} \tensor{R}{_{dh}^{ij}} \, ,
    \quad
    \mathcal{R}_{10}^{(20)} = \tensor{R}{_{ce}^{ab}} \tensor{R}{_{fg}^{cd}} \tensor{R}{_{hi}^{ef}} \tensor{R}{_{aj}^{gh}} \tensor{R}{_{bd}^{ij}} \,  .
\end{align}
}
\normalsize{}
\cleardoublepage

\lhead{Appendix B}
\rhead{AdS Slicing Metric}

\chapter{AdS Slicing Metric}
\label{chp:App-SlicingMetric}


This appendix includes some explicit results necessary to follow Chapter \ref{chp:ReviewBWs}. We start by writing our AdS$_{d+1}$ spacetime in slicing Poincaré-like coordinates,
\begin{equation}
    ds^2_{d+1} = G_{\mu \nu} (x,z) dx^\mu dx^\nu = e^{2 A(z)} \left[ dz^2 + \hat{g}_{i j} (x) dx^i dx^j \right]\,,
\end{equation}
where the $d$-dimensional metric $\hat{g}_{ij}$ is either flat, or an (A)dS$_d$ metric with unit curvature radius.
To match with the expressions in Chapter \ref{chp:ReviewBWs}, we will simply take
\begin{equation}\label{e2Af}
    e^{A(z)} = \frac{L}{f(z)}\,,
\end{equation}
where $L$ is the bulk AdS radius. In the following, primes will denote $z$ derivatives, tensors with no hats will be built from the bulk metric $G_{\mu\nu}$, and tensors with hats from the metric $\hat{g}_{ij}(x)$.

\section*{Background Tensors}

Metric components:
\begin{align}
        G_{i j} &= e^{2 A}\hat{g}_{i j}\,, & G_{zz} &= e^{2 A}\,, \nonumber \\
    G^{i j} &= e^{-2 A}\hat{g}^{i j}\,, & G^{zz} &= e^{- 2 A}\,.
\end{align}

Christoffel Symbols:
\begin{align}
    \Gamma^k_{i j} & = \hat{\Gamma}^k_{i j}\,, & \Gamma^z_{zz} = A'\,, \nonumber \\
    \Gamma^z_{i j} & = -A' \hat{g}_{i j}\,, & \Gamma^k_{zz} = 0\,, \nonumber \\
    \Gamma^k_{i z} & = A' \hat{\delta}^k_i \,, & \Gamma^z_{i z} = 0\,. 
\end{align}

Ricci Tensor:
\begin{align}
    R_{i j} & = \hat{R}_{i j} - \left( (d-1)(A')^2 + A'' \right) \hat{g}_{i j}\,, \nonumber \\
    R_{zz} & = -d A''\,.
\end{align}



\section*{Background Einstein Equations}
The bulk AdS$_{d+1}$ Einstein Equations are
\begin{equation}
    R_{\mu \nu} = \frac{2}{d-1} \Lambda G_{\mu \nu}\,,
\end{equation}
where the bulk cosmological constant is
\begin{equation}
    \Lambda = -\frac{d(d-1)}{2L^2}\,.
\end{equation}
Substituting in the results from the previous section, we find that the background equations are
\begin{align}
    \hat{R}_{ij} - \left( (d-1)(A')^2 + A'' \right) \hat{g}_{ij} & = \frac{2}{d-1} e^{2A} \Lambda \hat{g}_{ij}, \label{e2AEEij} \\
    -d A'' & = \frac{2}{d-1} e^{2 A} \Lambda. \label{e2AEEzz}
\end{align}
Recall that the brane is a maximally symmetric spacetime with unit curvature radius, so
\begin{equation}
    \hat{R}_{ij} = \frac{2}{d-2} \hat{\Lambda} \hat{g}_{ij}\,,
\end{equation}
with
\begin{equation}
    \hat{\Lambda} = \sigma\frac{(d-1)(d-2)}{2}\,,
\end{equation}
where
\begin{equation}
    \sigma \ = \  
    \begin{cases}
        -1 \quad \text{for AdS branes}\,,\\
         \ 0 \quad \ \text{for flat branes}\,,\\
        +1 \quad \text{for dS branes}\,.
    \end{cases}
\end{equation}
It is now easy to see that the bulk Einstein Equations are fulfilled, using \eqref{e2Af} and
\begin{equation}
    f(z) \ = \  
    \begin{cases}
        \sin(z) \quad \text{for AdS branes}\,,\\
         \ \ z \quad \ \text{for flat branes}\,,\\
        \sinh(z) \quad \text{for dS branes}\,.
    \end{cases}
\end{equation}
Notice also that we can factor out the metric $\hat{g}_{ij}$ from equation \eqref{e2AEEij} to get the following relation, which will be useful later,
\begin{equation}\label{rel2A}
    \frac{2}{d-2} \hat{\Lambda} - \left( (d-1)(A')^2 + A'' \right) = \frac{2}{d-1} e^{2A} \Lambda\,.
\end{equation}

\section*{Brane Hypersurface}

If we now put a brane at $z = z_b$ and excise the part of the spacetime with $0 < z < z_b$, the induced metric on the brane is
\begin{equation}
    g_{ij} = e^{2A} \hat{g}_{ij}\,,
\end{equation}
evaluated at $z = z_b$.
Since the metric is block diagonal, the outward-directed unit normal metric to the brane is
\begin{equation}
    \partial_n = - e^{-A} \partial_z\,.
\end{equation}
Then, the extrinsic curvature on the brane is
\begin{equation}
    K_{ij} = 
    \frac{1}{2} \partial_n g_{ij} = -\frac{1}{2} e^{-A} \partial_z \left(e^{2A} \hat{g}_{ij}  \right) = -A' e^A \hat{g}_{ij} = 
    - A' e^{-A} g_{ij}\,,
\end{equation}
and its trace reads
\begin{equation}
    K = -d A' e^{-A}\,.
\end{equation}

\section*{Linear Perturbations}

We now perturb the metric in the following way,
\begin{equation}
    ds^2_{d+1} = e^{2 A(z)} \left[ dz^2 + \left( \hat{g}_{ij}(x) + \delta \hat{g}_{ij}(x,z) \right) dx^i dx^j \right]\,,
\end{equation}
that is, we have chosen an axial gauge with $\delta G_{\mu z} = 0$.
We raise and lower indices using the unperturbed brane metric $\hat{g}_{ij}$, and denote its Levi-Civita covariant derivative as $\hat{\nabla}$.
\\

Metric components:
\begin{align}
    \delta G_{ij} &= e^{2 A(z)} \delta \hat{g}_{ij}(x,z)\,, & \delta G_{zz} &= 0\,, \nonumber \\
    \delta G^{ij} &= - e^{-2 A(z)} \delta \hat{g}^{ij}(x,z) \,, & \delta G^{zz} &= 0\,.
\end{align}

Christoffel Symbols:
\begin{align}
    \delta \Gamma^k_{ij} & = \frac{1}{2}\hat{g}^{kl} \left[ \hat{\nabla}_i \delta \hat{g}_{lj} + \hat{\nabla}_j \delta \hat{g}_{il} - \hat{\nabla}_l \delta \hat{g}_{ij} \right]\,, & \delta \Gamma^z_{zz} = 0\,, \nonumber \\
    \delta \Gamma^z_{ij} & = - \left[ A'\delta \hat{g}_{ij} + \frac{1}{2}\delta \hat{g}'_{ij} \right]\,,  & \delta \Gamma^k_{zz} = 0\,, \nonumber \\
    \delta \Gamma^k_{i z} & = \frac{1}{2}\delta \hat{g}'^k_i & \delta \Gamma^z_{i z} = 0\,.
\end{align}

Ricci Tensor:
\begin{align}
    \delta R_{ij} = & \ \frac{1}{2}\left[ \hat{\nabla}^k \hat{\nabla}_i \delta \hat{g}_{kj} + \hat{\nabla}^k \hat{\nabla}_j \delta \hat{g}_{ik} - \hat{\nabla}^k \hat{\nabla}_k \delta \hat{g}_{ij} - \hat{\nabla}_i \hat{\nabla}_j \delta \hat{g} \right] \nonumber \\
    & \ - \left[ \frac{1}{2}\delta \hat{g}''_{ij} + \frac{d-1}{2}A'\delta \hat{g}'_{ij} + (A'' + (d-1)(A')^2)\delta \hat{g}_{ij} + \frac{1}{2}A'\delta \hat{g}'\hat{g}_{ij} \right]\,, \nonumber \\
    \delta R_{iz} = & \ \frac{1}{2} \left[ \hat{\nabla}^k\delta \hat{g}'_{ik} - \hat{\nabla}_i \delta \hat{g}' \right]\,, \nonumber \\
    \delta R_{zz} = & -\frac{1}{2} \left[ \delta \hat{g}'' + A'\delta \hat{g}' \right]\,.
\end{align}


\section*{Linearized Einstein Equations}
\noindent The linearized Einstein Equations are
\begin{equation}
    \delta R_{\mu \nu} = \frac{2}{d-1} \Lambda \delta G_{\mu \nu}\,,
\end{equation}
therefore,
\begin{align}
    \delta R_{ij} = & \ \frac{1}{2}\left[ \hat{\nabla}^k \hat{\nabla}_i \delta \hat{g}_{kj} + \hat{\nabla}^k \hat{\nabla}_j \delta \hat{g}_{ik} - \hat{\nabla}^k \hat{\nabla}_k \delta \hat{g}_{ij} - \hat{\nabla}_i \hat{\nabla}_j \delta \hat{g} \right] \nonumber \\ \quad & - \left[ \frac{1}{2}\delta \hat{g}''_{ij} + \frac{d-1}{2}A'\delta \hat{g}'_{ij} + (A'' + (d-1)(A')^2)\delta \hat{g}_{ij} + \frac{1}{2}A'\delta \hat{g}'\hat{g}_{ij} \right] \nonumber \\ = & \ \frac{2}{d-1}e^{2A} \Lambda \delta \hat{g}_{ij}\,, \label{lin2Aij} \\
    \delta R_{iz} = & \ \frac{1}{2} \left[ \hat{\nabla}^k\delta \hat{g}'_{ik} - \hat{\nabla}_i \delta \hat{g}' \right] = 0\,, \label{lin2Aiz}\\
    \delta R_{zz} = & -\frac{1}{2} \left[\delta \hat{g}'' + A'\delta \hat{g}'\right] = 0 \iff (e^{A}\delta \hat{g}')' = 0\,. \label{lin2Azz}
\end{align}
Using the relation \eqref{rel2A},
we can rewrite the ($ij$) equation \eqref{lin2Aij} as
\begin{align}\label{lin2Aijv2}
    \frac{1}{2} & \left[ \hat{\nabla}^k \hat{\nabla}_i \delta \hat{g}_{kj} + \hat{\nabla}^k \hat{\nabla}_j \delta \hat{g}_{ik} - \hat{\nabla}^k \hat{\nabla}_k \delta \hat{g}_{ij} - \hat{\nabla}_i \hat{\nabla}_j \delta \hat{g} \right] \nonumber \\ & - \left[ \frac{1}{2}\delta \hat{g}''_{ij} + \frac{d-1}{2}A'\delta \hat{g}'_{ij} + \frac{1}{2}A'\delta \hat{g}'\hat{g}_{ij} \right] - \frac{2}{d-2} \hat{\Lambda} \delta \hat{g}_{ij} = 0\,.
\end{align}
Furthermore, using the Ricci identity and the fact that the background metric $\hat{g}_{ij}$ on the slice is a maximally symmetric metric with unit curvature radius, we can write
\begin{equation}
    \hat{\nabla}^k \hat{\nabla}_i \delta \hat{g}_{kj} = \hat{\nabla}_i \hat{\nabla}^k \delta \hat{g}_{kj} + \sigma d \delta \hat{g}_{ij} - \sigma \hat{g}_{ij} \delta \hat{g},
\end{equation}
and similarly the term with $i \leftrightarrow j$. Substituting these into \eqref{lin2Aijv2}, trading $\hat{\Lambda}$ for $\sigma$, and regrouping, we obtain
\begin{align}\label{lin2Aijv3}
    \frac{1}{2} & \left[ \hat{\nabla}_i \hat{\nabla}^k \delta \hat{g}_{kj} + \hat{\nabla}_j \hat{\nabla}^k \delta \hat{g}_{ik} - \hat{\nabla}^k \hat{\nabla}_k \delta \hat{g}_{ij} - \hat{\nabla}_i \hat{\nabla}_j \delta \hat{g} \right] \nonumber \\ - & \left[ \frac{1}{2}\delta \hat{g}''_{ij} + \frac{d-1}{2}A'\delta \hat{g}'_{ij} + \frac{1}{2}A' \hat{g}_{ij} \delta \hat{g}' \right] + \sigma \delta \hat{g}_{ij} - \sigma \hat{g}_{ij} \delta \hat{g} = 0.
\end{align}

From equations \eqref{lin2Aiz} and \eqref{lin2Azz} one can argue that the only dynamical degrees of freedom are the transverse and traceless perturbations \cite{Karch:2000ct,Karch:2001jb}, \ie perturbations fulfilling
\begin{align}
    \delta \hat{g}^{TT} & = \hat{g}^{ij} \delta \hat{g}^{TT}_{ij} = 0\,, & \hat{\nabla}^i \delta \hat{g}^{TT}_{ij} & = 0\,.
\end{align}
Then, the ($iz$) and ($zz$) Einstein equations \eqref{lin2Aiz} and \eqref{lin2Azz} vanish identically,
and equation \eqref{lin2Aijv3} becomes, after multiplying by a (-2) factor,
\begin{equation}
    \left[ \partial_z^2 + (d-1)A'\partial_z + \left( \hat{\square} - 2\sigma \right) \right]\delta \hat{g}^{TT}_{ij} = 0,
\end{equation}
where $\hat{\square} = \hat{\nabla}_k \hat{\nabla}^k$.

\section*{Linear Perturbations on the Brane Hypersurface}

On the brane at $z = z_b$, the perturbation on the induced metric is
\begin{equation}
    \delta g_{ij} = e^{2A} \delta \hat{g}_{ij}\,.
\end{equation}
Then, the perturbation on the extrinsic curvature reads
\begin{equation}
    \delta K_{ij} = 
    - A' e^{A} \delta \hat{g}_{ij} - \frac{1}{2} e^A\delta \hat{g}'_{ij}\,.
\end{equation}
Notice that its trace is zero for transverse and traceless perturbations,
\begin{equation}
    \delta K^{TT} = 0\,.
\end{equation}
\cleardoublepage

\lhead{Appendix C}
\rhead{Brane-World Graviton Modes}

\chapter{Brane-World Graviton Modes}
\label{chp:App-BWresults}


Here is an extended list of our results from Chapters \ref{chp:ReviewBWs} and \ref{chp:BWsWithDGP}.

\section*{AdS Brane}

We write our background bulk metric as
\begin{equation}
    d\hat{s}^2_{d+1} = G_{\mu \nu}(z,x) dy^\mu dy^\nu = \frac{L^2}{\sin^2(z)} \left[ dz^2 + \hat{g}_{ij}(x) dx^i dx^j \right]\,,
\end{equation}
and perturb it with a separable linear perturbation 
\begin{equation}
    G_{\mu \nu} (x,z) = L^2 \chi(z) h_{\mu \nu}(x)\,, 
\end{equation}
that is transverse and traceless, and where we have chosen an axial gauge,
\begin{equation}
    h_{\mu z} = 0\,, \quad \quad \nabla^i h_{ij} = 0\,, \quad \quad \hat{g}^{ij} h_{ij} = 0\,,
\end{equation}
where $\hat{g}_{ij}(x)$ is the background brane metric, which is an AdS$_d$ spacetime with unit curvature radius, and $\hat{\nabla}$ is its covariant derivative.

After imposing the bulk Einstein Equations, we find the following.

\subsection*{Brane equation}
\begin{equation}
    (\Box + 2) h_{ij}(x) = E^2 h_{ij}(x)\,.
\end{equation}

\subsection*{Radial equation}
There are many ways to write down the radial function,
\begin{equation}
    L^2 \chi(z) = \frac{L^2}{\sin^2 z} H(z) = \left( \frac{\sin z}{L} \right)^{\frac{d-5}{2}} \tilde{H}(z) = \frac{L^2}{\rho} H(\rho)\,.
\end{equation}
The resulting radial equations are
\begin{align}
    -E^2 \chi(z) & = \left[ \partial_z^2 - (d-5) \cot(z) \partial_z + \left( 2(d-3)-2(d-2)\csc^2(z) \right) \right ] \chi(z)\,, \\
    -E^2 H(z) & = \left[ \partial_z^2 - (d-1)\cot(z)\partial_z \right ] H(z)\,, \\
    -E^2 \tilde{H}(z) & = \left[ \partial_z^2 - \left( \frac{d^2-1}{4} \csc^2(z) - \frac{(d-1)^2}{4} \right) \right ] \tilde{H}(z)\,, \\
    -E^2 H(\rho) & = \left[ \rho(1+\rho)^2\partial_\rho^2 + \frac{1}{2} (1+\rho) \left( 2-d+(2+d)\rho \right) \partial_\rho \right ] H(\rho)\,.
\end{align}
\subsection*{Junction conditions}
Imposing the Israel junction condition on the brane relates the position of the brane and the brane tension as
\begin{equation}
    \tau = \frac{d-1}{8 \pi G L} \cos (z_b)\,,
\end{equation}
and fixes the brane boundary condition for the radial equations to be
\begin{align}
    0 & = \chi'(z_b) + 2 \cot(z_b) \chi(z_b)\,, \\
    0 & = H'(z_b)\,, \\
    0 & = \tilde{H}'(z_b) + \frac{d-1}{2} \cot(z_b) \tilde{H}(z_b)\,, \\
    0 & = H'(\rho_B)\,.
\end{align}
\subsection*{Junction conditions with DGP}
When we add a DGP term on the brane, the brane tension is given by
\begin{equation}
    \tau = \frac{d-1}{8\pi G L} \left[ \cos(z_b) - \alpha \frac{(d-2)}{L} \sin^2 (z_b) \right]\,,
\end{equation}
while the boundary conditions for the radial equation become
\begin{align}
    0 & = \chi'(z_b) + 2 \cot(z_b) \chi(z_b)\ + \frac{2 \alpha}{L} E^2 \sin(z_b) \chi(z_b)\,, \\
    0 & = H'(z_b)+ \frac{2 \alpha}{L} E^2 \sin(z_b) H(z_b)\,, \\
    0 & = \tilde{H}'(z_b) + \frac{d-1}{2} \cot(z_b) \tilde{H}(z_b)+ \frac{2 \alpha}{L} E^2 \sin(z_b) \tilde{H}(z_b)\,, \\
    0 & = (1+\rho_B^2)H'(\rho_B) + \frac{4 \alpha}{L} E^2 H(\rho_B)\,.
\end{align}

\subsection*{Eigenvalues}

For small values of $\rho_b$ and $A$, we find
\begin{equation}
    E^2_{(n,d)} \simeq n(n+d-1) + \frac{1}{2}(d-2)(2n+d-1) \frac{\Gamma(n+d-1)}{(\Gamma(d/2))^2\Gamma(n+1)} \frac{\rho_B^{d/2-1}}{(1+2A)}\,,
\end{equation}
where
\begin{equation}
    A = \frac{\alpha(d-2)}{L}\,.
\end{equation}
In odd dimensions, 
we were able to find improved expressions. For $d=3$ we found a general expression, even with the presence of the DGP term, while for $d=5$ and $d=7$ we were only able to find them case-by-case in the case with no DGP term.
\medskip

$d = 3$

\begin{equation}
    E^2_n = \frac{n(n+2)(1+2A)\pi + (n^2+2n+4)\sqrt{\rho_B}}{(1+2A)\pi - 3\sqrt{\rho_B}}\,.
\end{equation}

\smallskip

\begin{equation}
    E^2_0 \simeq 0 + \frac{4 \sqrt{\rho_B}}{(1+2A)\pi - 3\sqrt{\rho_B}} \simeq 0 + \frac{4}{(1+2A)\pi}\sqrt{\rho_B}\,,
\end{equation}
\begin{equation}
    E^2_1 \simeq \frac{3(1+2A)\pi + 7\sqrt{\rho_B}}{(1+2A)\pi - 3\sqrt{\rho_B}} \simeq 3 + \frac{16}{(1+2A)\pi}\sqrt{\rho_B}\,,
\end{equation}
\begin{equation}
    E^2_2 \simeq \frac{8(1+2A)\pi + 12\sqrt{\rho_B}}{(1+2A)\pi - 3\sqrt{\rho_B}} \simeq 8 + \frac{36}{(1+2A)\pi}\sqrt{\rho_B}\,,
\end{equation}
\begin{equation}
    E^2_3 \simeq \frac{15(1+2A)\pi + 19\sqrt{\rho_B}}{(1+2A)\pi - 3\sqrt{\rho_B}} \simeq 15 + \frac{64}{(1+2A)\pi}\sqrt{\rho_B}\,.
\end{equation}

\textbf{$d = 5 \ (A = 0)$}

\begin{equation}
    E^2_0 \simeq 0 + \frac{192 \rho_B^{3/2}}{3\pi + 12\pi\rho_B - 100\rho_B^{3/2}} \simeq 0 + \frac{64}{\pi}\rho_B^{3/2}\,,
\end{equation}
\begin{equation}
    E^2_1 \simeq \frac{5\pi + 70\pi\rho_B -16\rho_B^{3/2}}{\pi + 14\pi\rho_B  - 80\rho_B^{3/2}} \simeq 5 + \frac{384}{\pi}\rho_B^{3/2}\,,
\end{equation}
\begin{equation}
    E^2_2 \simeq \frac{36\pi + 1008\pi\rho_B  - 1392\rho_B^{3/2}}{3\pi + 84\pi\rho_B  - 436\rho_B^{3/2}} \simeq 12 + \frac{1280}{\pi}\rho_B^{3/2}\,,
\end{equation}
\begin{equation}
    E^2_3 \simeq \frac{63\pi +2898\pi\rho_B  -4848\rho_B^{3/2}}{3\pi + 138\pi\rho_B  - 688\rho_B^{3/2}} \simeq 21 + \frac{3200}{\pi}\rho_B^{3/2}\,.
\end{equation}

\textbf{$d = 7 \ (A = 0)$}

\begin{equation}
    E^2_0 \simeq 0 + \frac{7680 \rho_B^{5/2}}{15\pi + 20\pi\rho_B + 295\pi\rho_B^2 - 3136\rho_B^{3/2}} \simeq 0 + \frac{512}{\pi}\rho_B^{5/2}\,.
\end{equation}

\bigskip

\section*{dS Brane}

We write our background bulk metric as
\begin{equation}
    d\hat{s}^2_{d+1} = G_{\mu \nu}(z,x) dy^\mu dy^\nu = \frac{L^2}{\sinh^2(z)} \left[ dz^2 + \hat{g}_{ij}(x) dx^i dx^j \right]\,,
\end{equation}
and perturb it with a separable linear perturbation 
\begin{equation}
    G_{\mu \nu} (x,z) = L^2 \chi(z) h_{\mu \nu}(x)\,, 
\end{equation}
that is transverse and traceless, and where we have chosen an axial gauge,
\begin{equation}
    h_{\mu z} = 0\,, \quad \quad \nabla^i h_{ij} = 0\,, \quad \quad \hat{g}^{ij} h_{ij} = 0\,,
\end{equation}
where $\hat{g}_{ij}(x)$ is the background brane metric, which is an dS$_d$ spacetime with unit curvature radius, and $\hat{\nabla}$ is its covariant derivative.

After imposing the bulk Einstein Equations, we find the following.

\subsection*{Brane equation}
\begin{equation}
    (\Box - 2) h_{ij}(x) = E^2 h_{ij}(x)\,.
\end{equation}

\subsection*{Radial equation}
There are many ways to write down the radial function,
\begin{equation}
    L^2 \chi(z) = \frac{L^2}{\sinh^2 z} H(z) = \left( \frac{\sinh z}{L} \right)^{\frac{d-5}{2}} \tilde{H}(z) = \frac{L^2}{\rho} H(\rho)\,.
\end{equation}
The resulting radial equations are
\begin{align}
    -E^2 \chi(z) & = \left[ \partial_z^2 - (d-5) \coth(z) \partial_z - \left( 2(d-3)+2(d-2)\csch^2(z) \right) \right ] \chi(z)\,, \\
    -E^2 H(z) & = \left[ \partial_z^2 - (d-1)\coth(z)\partial_z \right ] H(z)\,, \\
    -E^2 \tilde{H}(z) & = \left[ \partial_z^2 - \left( \frac{d^2-1}{4} \csch^2(z) + \frac{(d-1)^2}{4} \right) \right ] \tilde{H}(z)\,, \\
    -E^2 H(\rho) & = \left[ \rho(\rho-1)^2\partial_\rho^2 + \frac{1}{2} (\rho-1) \left( d-2+(d+2)\rho \right) \partial_\rho \right ] H(\rho)\,.
\end{align}
\subsection*{Junction conditions}
Imposing the Israel junction conditions relates the position of the brane and the brane tension as
\begin{equation}
    \tau = \frac{d-1}{8 \pi G L} \cosh(z_b)\,,
\end{equation}
and gives the following boundary condition on the brane for the radial equations
\begin{align}
    0 & = \chi'(z_b) + 2 \coth(z_b) \chi(z_b)\,, \\
    0 & = H'(z_b)\,, \\
    0 & = \tilde{H}'(z_b) + \frac{d-1}{2} \coth(z_b) \tilde{H}(z_b)\,, \\
    0 & = H'(\rho_B)\,.
\end{align}
\subsection*{Junction conditions with DGP}
When we add a DGP term on the brane, the brane tension is given by
\begin{equation}
    \tau = \frac{d-1}{8\pi G L} \left[ \cosh(z_b) + \alpha \frac{(d-2)}{L} \sinh^2 (z_b) \right]\,,
\end{equation}
while the boundary conditions for the radial equation become
\begin{align}
    0 & = \chi'(z_b) + 2 \coth(z_b) \chi(z_b)\ + \frac{2 \alpha}{L} E^2 \sinh(z_b) \chi(z_b)\,, \\
    0 & = H'(z_b)+ \frac{2 \alpha}{L} E^2 \sinh(z_b) H(z_b)\,, \\
    0 & = \tilde{H}'(z_b) + \frac{d-1}{2} \coth(z_b) \tilde{H}(z_b)+ \frac{2 \alpha}{L} E^2 \sinh(z_b) \tilde{H}(z_b)\,, \\
    0 & = (1-\rho_B^2)H'(\rho_B) + \frac{4 \alpha}{L} E^2 H(\rho_B)\,.
\end{align}

\bigskip

\section*{Flat Brane}

We write our background bulk metric as
\begin{equation}
    d\hat{s}^2_{d+1} = G_{\mu \nu}(z,x) dy^\mu dy^\nu = \frac{L^2}{z^2} \left[ dz^2 + \eta_{ij}(x) dx^i dx^j \right]\,,
\end{equation}
and perturb it with a separable linear perturbation 
\begin{equation}
    G_{\mu \nu} (x,z) = L^2 \chi(z) h_{\mu \nu}(x)\,, 
\end{equation}
that is transverse and traceless, and where we have chosen an axial gauge,
\begin{equation}
    h_{\mu z} = 0\,, \quad \quad \partial^i h_{ij} = 0\,, \quad \quad \eta^{ij} h_{ij} = 0\,,
\end{equation}
where $\eta_{ij}(x)$ is the background flat brane metric, $\partial$ is its (covariant) derivative.

After imposing the bulk Einstein Equations, we find the following.

\subsection*{Brane equation}
\begin{equation}
    \partial^k \partial_k h_{ij}(x) = E^2 h_{ij}(x)\,.
\end{equation}

\subsection*{Radial equation}
There are many ways to write down the radial function,
\begin{equation}
    L^2\chi(z) = \frac{L^2}{z^2} H(z) = \left(\frac{z}{L}\right)^{\frac{d-5}{2}} \tilde{H}(z) = \frac{L^2}{\rho} H(\rho)\,.
\end{equation}
The resulting radial equations are
\begin{align}
    -E^2 \chi(z) & = \left[ \partial_z^2 - (d-5) \frac{1}{z} \partial_z + 2(d-2)\frac{1}{z^2} \right ] \chi(z)\,, \\
    -E^2 H(z) & = \left[ \partial_z^2 - (d-1)\frac{1}{z}\partial_z \right ] H(z)\,, \\
    -E^2 \tilde{H}(z) & = \left[ \partial_z^2 - \frac{d^2-1}{4z^2} \right ] \tilde{H}(z)\,, \\
    -E^2 H(\rho) & = \left[ 4\rho\partial_\rho^2 -  2(d-2) \partial_\rho \right ] H(\rho)\,.
\end{align}
\subsection*{Junction condition}
Imposing the Israel junction condition on the brane we see that the brane tension must be fixed at
\begin{equation}
    \tau = \frac{d-1}{8 \pi G L}\,,
\end{equation}
and that the boundary condition on the brane for the radial equations is
\begin{align}
    0 & = \chi'(z_b) + \frac{2}{z_b} \chi(z_b)\,, \\
    0 & = H'(z_b)\,, \\
    0 & = \tilde{H}'(z_b) + \frac{d-1}{2 z_b} \tilde{H}(z_b)\,, \\
    0 & = H'(\rho_B)\,.
\end{align}
\subsection*{Junction condition with DGP}
When we add a DGP term on the brane, this does not affect the brane tension, which remains at
\begin{equation}
    \tau = \frac{d-1}{8 \pi G L}\,,
\end{equation}
but the boundary conditions on the brane change, becoming
\begin{align}
    0 & = \chi'(z_b) + \frac{2}{z_b} \chi(z_b) + \frac{2 \alpha}{L} E^2 z \chi(z_b)\,, \\
    0 & = H'(z_b) + \frac{2 \alpha}{L} E^2 z H(z_b)\,, \\
    0 & = \tilde{H}'(z_b) + \frac{d-1}{2} \cot(z_b) \tilde{H}(z_b) + \frac{2 \alpha}{L} E^2 z \tilde{H}(z_b)\,, \\
    0 & = (1-\rho^2) H'(\rho_B) +  4\frac{\alpha}{L} E^2 H(\rho_B)\,.
\end{align}

\cleardoublepage

\lhead{Appendix D}
\rhead{Counterterms}

\chapter{Counterterm Results}
\label{chp:App-Counterterms}

The first terms in the gravitational effective action on the brane, which were already known in the literature before this work \cite{deHaro:2000vlm,Kraus:1999di,Emparan:1999pm}, read
\begin{align}
    \mathcal{L}_{(0)} & = \frac{d-1}{L} , \\
    \mathcal{L}_{(1)} & = \frac{L}{2(d-2)} R ,  \\ 
    \mathcal{L}_{(2)} & = \frac{L^3}{2(d-2)^2(d-4)} \left[ R_{ab}R^{ab} - \frac{d}{4(d-1)} R^2 \right] ,  \\
    \mathcal{L}_{(3)} & = - \frac{L^5}{(d-2)^3(d-4)(d-6)} \Bigg[ \frac{3d+2}{4(d-1)}RR_{ab}R^{ab} - \frac{d(d+2)}{16(d-1)^2}R^3 \nonumber \\
    \ & - 2R^{ab}R_{acbd}R^{cd} + \frac{d-2}{2(d-1)}R^{ab}\nabla_{a}\nabla_{b} R - R^{ab} \Box R_{ab} + \frac{1}{2(d-1)} R \Box R \Bigg] . 
\end{align}

\noindent The quartic term, first computed by the author in \cite{Bueno:2022log}, is
\begin{align}
\mathcal{L}_{(4)} = - & \frac{L^7}{(d-2)^4(d-4)(d-6)(d-8)} \nonumber \\
\Bigg[ &  \frac{13 d^2 - 38 d - 80}{8 (d-1)(d-4)} R_{ab} R^{ab} R_{cd} R^{cd} + \frac{- 15 d^3 + 18 d^2 + 192 d + 64}{16 (d-4) (d-1)^2} R_{ab} R^{ab} R^2 \nonumber \\ 
& + \frac{d (5 d^3 + 10 d^2 - 112 d - 128)}{128 (d-4) (d-1)^3} R^4 + \frac{5 d^2 - 16 d - 24 }{(d-1)(d-4)} R^{ab} R^{cd} R R_{acbd} \nonumber \\
&- 12 R_{a}{}^{c} R^{ab} R^{de} R_{bdce} + 8 R^{ab} R^{cd} R_{ac}{}^{ef} R_{bdef} - 8 R^{ab} R^{cd} R_{a}{}^{e}{}_{c}{}^{f} R_{bedf}  \nonumber \\
&-\frac{2 (d-6)}{d-4}R^{ab} R^{cd} R_{a}{}^{e}{}_{b}{}^{f} R_{cedf} + \frac{d^2 + 4 d -36}{2 (d-4) (d-1)}  R_{bc} R^{bc} \nabla_{a}\nabla^{a}R  \nonumber \\ 
&+ \frac{- 7 d^2 + 22 d +32}{4 (d-4) (d-1)^2}R^2 \nabla_{a}\nabla^{a}R + \frac{4}{d-1} R^{bc} \nabla_{a}R_{bc} \nabla^{a}R - \frac{d+8 }{4 (d-1)^2} R \nabla_{a}R \nabla^{a}R \nonumber \\
& + \frac{3d-8}{d-1}R^{ab} \nabla_{a}R^{cd} \nabla_{b}R_{cd} + \frac{d(d-6) }{8 (d-4) (d-1)^2}\nabla_{a}\nabla^{a}R \nabla_{b}\nabla^{b}R \nonumber \\ 
& + \frac{1}{d-1} R \nabla_{b}\nabla^{b}\nabla_{a}\nabla^{a}R - \frac{(d-4)(d+2)}{4 (d-1)^2} R_{ab} \nabla^{a}R \nabla^{b}R + \frac{d-4}{d-1} R_{a}{}^{c} R_{bc} \nabla^{b}\nabla^{a}R \nonumber \\ 
&- \frac{5 d^3 - 38 d^2 + 64 d + 16}{4 (d-4) (d-1)^2}R_{ab} R \nabla^{b}\nabla^{a}R + \frac{3 d^2 - 20d + 28}{(d-1)(d-4)}R^{cd} R_{acbd} \nabla^{b}\nabla^{a}R \nonumber \\ 
&- \frac{(d-6)(d-2)^2}{8 (d-4) (d-1)^2} \nabla_{b}\nabla_{a}R \nabla^{b}\nabla^{a}R + \frac{d-4}{d-1} R^{bc}\nabla^{a}R \nabla_{c}R_{ab} - 8 R^{ab} \nabla_{e}R_{acbd} \nabla^{e}R^{cd}  \nonumber \\ 
&+ \frac{5 d^2 - 6 d - 64}{2 (d-1)(d-4)}R^{ab} R \nabla_{c}\nabla^{c}R_{ab} + \frac{(d-2)(d-6)}{2(d-1)(d-4)} \nabla^{b}\nabla^{a}R \nabla_{c}\nabla^{c}R_{ab} \nonumber \\ 
&+ \frac{(d-2)}{d-1} R_{ab} \nabla_{c}\nabla^{c}\nabla^{b}\nabla^{a}R + \frac{5 }{d-1}R \nabla_{c}R_{ab} \nabla^{c}R^{ab} + 12 R^{ab} R^{cd} \nabla_{d}\nabla_{b}R_{ac} \nonumber \\ 
&+ \frac{11d-6}{d-1} R^{ab} R^{cd} \nabla_{d}\nabla_{c}R_{ab} - \frac{d-6}{2 (d-4)}\nabla_{c}\nabla^{c}R^{ab} \nabla_{d}\nabla^{d}R_{ab} - 2 R^{ab} \nabla_{d}\nabla^{d}\nabla_{c}\nabla^{c}R_{ab} \nonumber \\ 
&- 4 R^{ab} \nabla_{b}R_{cd} \nabla^{d}R_{a}{}^{c} + 4 R^{ab} \nabla_{c}R_{bd} \nabla^{d}R_{a}{}^{c} + \frac{2(5d-22) }{d-4}R^{ab} R_{acbd} \nabla_{e}\nabla^{e}R^{cd} \Bigg].  
\end{align}



\bigskip

\noindent The quintic term is too large, so we have decided to reduce the font size. It reads,
\tiny{
 
}













\bigskip

\normalsize
\noindent Finally, we have also been able to compute the sextic term in $d=3$,
\scriptsize{
%
}
\cleardoublepage
\end{appendices}


\lhead{}
\rhead{}
\renewcommand{\headrulewidth}{0pt}

\bibliography{ThesisBib.bib}


\end{document}


\typeout{get arXiv to do 4 passes: Label(s) may have changed. Rerun}